\documentclass[11pt]{article}
\usepackage{amsmath}
\usepackage{graphicx,psfrag,epsf}
\usepackage{enumerate}
\usepackage[numbers, sort&compress]{natbib}
\usepackage[hyphens]{url} 
\usepackage{soul}
\newcommand{\blind}{0}

\usepackage{geometry}
\usepackage{makecell}
\usepackage[table,dvipsnames]{xcolor}
\usepackage{amsthm}
\usepackage{bm}
\usepackage{bbm}
\usepackage{multirow}
\usepackage{mathtools}
\usepackage{collcell}
\usepackage{dsfont}
\usepackage{rotating}
\usepackage[skip=0pt]{subcaption}
\usepackage{wrapfig}
\usepackage[utf8]{inputenc}
\usepackage{tikz}
\usepackage{lettrine}
\usepackage{dblfloatfix}
\usepackage[english]{babel} 
\usepackage{amsfonts}
\graphicspath{ {fig/} }
\usepackage[hyperfootnotes=false]{hyperref}
\usepackage[normalem]{ulem}
\usepackage[skip=3pt]{caption}
\usepackage{lipsum}
\usetikzlibrary{arrows,decorations.markings}
\usepackage{standalone}

\newtheorem{theorem}{Theorem}

\definecolor{gray}{rgb}{0.5,0.5,0.5}
\definecolor{red}{rgb}{0.8,0,0}
\definecolor{dred}{rgb}{0.5,0,0}
\definecolor{blue}{rgb}{0,0.1,1}
\definecolor{dblue}{rgb}{0,0.1,0.6}
\definecolor{cyan}{rgb}{0,0.5,.5}
\definecolor{dcyan}{rgb}{0,0.3,.3}

\definecolor{b}{rgb}{0,0,.8}	
\definecolor{g}{rgb}{0,.6,0}	
\definecolor{n}{rgb}{0,0,0}	
\definecolor{h}{rgb}{0.4,0.2,0.2}	
\definecolor{v}{rgb}{0.2,0.6,0}

\newcommand{\C}{{\mathbb C}}

\newcommand{\E}{{\mathbb E}}

\newcommand{\Q}{{\mathbb Q}}
\newcommand{\R}{{\mathbb R}}

\newcommand{\V}{{\mathbb V}}

\newcommand{\MM}{{\mathcal{M}}}

\newcommand{\PP}{{\mathcal{P}}}
\newcommand{\QQ}{{\mathcal{Q}}}
\newcommand{\RR}{{\mathcal{R}}}

\newcommand{\TT}{{\mathcal{T}}}

\newcommand{\bsa}{\boldsymbol a}
\newcommand{\bsb}{\boldsymbol b}
\newcommand{\bsc}{\boldsymbol c}

\newcommand{\bsp}{\boldsymbol p}

\newcommand{\bss}{\boldsymbol s}

\newcommand{\bsv}{\boldsymbol v}

\newcommand{\bsz}{\boldsymbol z}
\newcommand{\bsA}{\boldsymbol A}
\newcommand{\bsB}{\boldsymbol B}
\newcommand{\bsC}{\boldsymbol C}

\newcommand{\bsI}{\boldsymbol I}

\newcommand{\bsP}{\boldsymbol P}

\newcommand{\bsR}{\boldsymbol R}
\newcommand{\bsS}{\boldsymbol S}

\newcommand{\bsV}{\boldsymbol V}

\newcommand{\bsone}{\boldsymbol 1}

\newcommand{\bszero}{\boldsymbol 0}

\newcommand{\bsnull}{\boldsymbol 0}

\newcommand{\bPP}{\boldsymbol{\mathcal{P}}}

\newcommand{\bsbeta}{\boldsymbol \beta}

\newcommand{\bsnu}{\boldsymbol \nu}
\newcommand{\bseps}{\boldsymbol \varepsilon}

\newcommand{\bstau}{\boldsymbol \tau}
\newcommand{\bsxi}{\boldsymbol \xi}

\DeclareMathOperator*{\argmin}{arg\,min}
\DeclareMathOperator*{\argmax}{arg\,max}

\newcommand{\abs}{|\cdot|}

\DeclareMathOperator{\var}{\V ar}
\DeclareMathOperator{\cov}{\C ov}
\DeclareMathOperator{\cor}{\C or}

\newcommand{\ov}\overline
\newcommand{\what}{\widehat}
\newcommand{\wtilde}{\widetilde}

\newcommand{\rig}\right
\newcommand{\lef}\left
\newcommand{\nf}\normalfont

\newcommand{\bsDelta}{\boldsymbol \Delta}

\newcommand{\Sup}{\text{S}} 
\newcommand{\Dem}{\text{D}}

\begin{document}

	\def\spacingset#1{\renewcommand{\baselinestretch}%
		{#1}\small\normalsize} \spacingset{1}

	
	\if0\blind
	{
		\title{\bf Optimal bidding in hourly and quarter-hourly electricity price auctions: trading large volumes of power with market impact and transaction costs}
		\author{Michał Narajewski\hspace{.2cm}\\
			University of Duisburg-Essen\\
			and \\
			Florian Ziel \\
			University of Duisburg-Essen}
		\maketitle
	} \fi
	
	\if1\blind
	{
		
		\bigskip
		
		\begin{center}
			{\LARGE\bf Title}
		\end{center}
	
	} \fi
	\begin{abstract}
		This paper addresses the question of how much to bid to maximize the profit when trading in two electricity markets: the hourly Day-Ahead Auction and the quarter-hourly Intraday Auction. For optimal coordinated bidding many price scenarios are examined,
		the own non-linear market impact is estimated by considering empirical supply and demand curves,
		and a number of trading strategies is used.
		Additionally, we provide theoretical results for risk neutral agents.
		The application study is conducted using the German market data, but the presented methods can be easily utilized with other two
		consecutive auctions.
	    This paper contributes to the existing literature by evaluating the costs of electricity trading, i.e. the price impact and the transaction costs. The empirical results for the  German EPEX market show that it is far more profitable to minimize the price impact rather than maximize the arbitrage.
	\end{abstract}
	
	\noindent%
	\vfill
	
	\newpage
	\spacingset{1.45} 
	

	\newpage

	\section{Introduction and motivation}
	
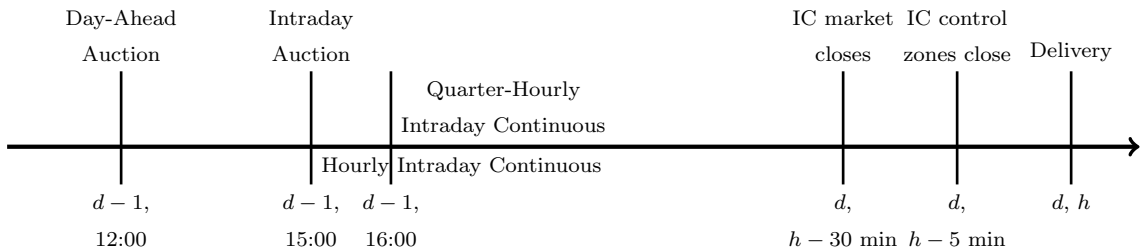
\begin{figure*}[!b]
		\begin{tikzpicture}[scale=1]
			\draw [->] [ultra thick] (0,0) -- (14.9,0);
			\draw [line width = 1] (1.5,1) -- (1.5, -0.5);
			\node [align = center, below, font = \scriptsize] at (1.5, -0.5) {$d-1$,\\ 12:00};
			\node [align = center, above, font = \scriptsize] at (1.5,1) {Day-Ahead\\Auction};
			\draw [line width = 1] (4,1) -- (4, -0.5);
			\node [align = center, below, font = \scriptsize] at (4, -0.5) {$d-1$,\\ 15:00};
			\node [align = center, above, font = \scriptsize] at (4,1) {Intraday\\Auction};
			\node [align = center, below right, font = \scriptsize] at (4,0) {Hourly Intraday Continuous};
			\draw [line width = 1] (5.05,1) -- (5.05, -0.5);
			\node [align = center, below, font = \scriptsize] at (5.05, -0.5) {$d-1$,\\ 16:00};
			\node [align = center, above right, font = \scriptsize] at (5.05,0) {Quarter-Hourly\\ Intraday Continuous};
			
			\draw [line width = 1] (11,1) -- (11, -0.5);
			\node [align = center, above, font = \scriptsize] at (11,1) {IC market\\ closes};
			\node [align = center, below, font = \scriptsize] at (11, -0.5) {$d$,\\ $h - 30$ min};
			
			\draw [line width = 1] (12.5,1) -- (12.5, -0.5);
			\node [align = center, above, font = \scriptsize] at (12.5,1) {IC control \\ zones close};
			\node [align = center, below, font = \scriptsize] at (12.5, -0.5) {$d$,\\ $h - 5$ min};
			\draw [line width = 1] (14,1) -- (14, -0.5);
			\node [align = center, above, font = \scriptsize] at (14,1) {Delivery};
			\node [align = center, below, font = \scriptsize] at (14, -0.5) {$d$, $h$};
		\end{tikzpicture}
		\caption{The daily routine of the German spot electricity market. $d, h$ correspond to the day and hour of the delivery, respectively.}
		\label{fig:market}
	\end{figure*}

	Since the deregulation of the electricity markets the energy exchanges like the European Energy Exchange (EEX) have created many trading possibilities to account for various market challenges. The electricity trading in Europe consists of futures, spot and balancing markets. Here, we deal with the biggest and the most important one -- the spot market. A brief description of the German electricity spot market can be seen in Figure~\ref{fig:market}, for more details on the German market see e.g. \citet{Viehmann2017}. We present the German spot electricity market as this is the biggest one in Europe, and we will also perform our empirical study based on the data from this market.
	
	The diversity of the trading possibilities in the market has raised new very important questions and challenges, e.g. when and how to trade the electricity in order to maximize the expected gain. In a perfectly efficient market with risk neutral agents this problem would become irrelevant as the expected gain should be the same, disregarding the market part and product type that one would use to trade the electricity. 
	However, this is not the case due to the fact that the market participants do not possess the full information, and they are highly dependent on the quality of their forecasts. Additionally, a very important role is being played by the own price impact of the market participants, especially for the large ones. Furthermore, some market agents may be not perfectly risk neutral but may be risk averse.
	
    The following paper raises the issue considering two European auction-based spot markets: the hourly EPEX Day-Ahead Auction (DA) and the quarter-hourly Intraday Auction (IA), as they are currently used in Germany, Netherlands, Belgium and Austria. The DA market is the main spot market and often serves as a reference price \cite{Viehmann2017}. On the other hand, the IA was introduced in the purpose of balancing the ramping effects of demand and power generation \cite{kiesel2017econometric, kremer2020intraday, kremer2021econometric}. Let us note that there are countries with other settings as e.g. France and Great Britain who use half-hourly Intraday Auctions. After a slight adjustment, the presented analysis can be also applied to these markets.
	
	In the analysis, we assume that a market participant wants to trade volume of electricity in a given hour, and they split it between the two markets, ignoring all other trading possibilities, as well as not speculating against the balancing market. 
	Limiting ourselves only to the two auctions is a simplification to some extent, but we discuss in the paper that it could be also generalized for usage with other markets and with a higher number of them. Additionally, we put emphasis on large trades and the price impact they make to the auctions. The existing studies have mostly disregarded this problem \cite{fleten2007stochastic, lohndorf2013optimizing, lohndorf2020value, finnah2021optimal, finnah2021integrated} or used simplified settings as e.g. linear impact assumption \cite{boomsma2014bidding, kongelf2019portfolio}. Due to this novelty, we decided to start with a smaller setting for a better understanding of the problem. Moreover, estimation of the impact in the continuous and balancing markets is very complex \cite{kath2020optimal} and deserves a separate study. We also assume that the market player places bids that are unlimited in prices. That is to say, they bid the minimum price on the supply side and the maximum price on the demand side. Both unlimited bidding \cite{kim2011optimal} and price-volume bidding have previously been used in the literature \cite{fleten2007stochastic, lohndorf2013optimizing, boomsma2014bidding, kongelf2019portfolio, lohndorf2020value, finnah2021optimal}.

	Even though we assume some restrictions, we take into account other major features in the markets.
	As mentioned, we do consider the non-linear market impact and the transaction costs that the trader must account for. 
	Moreover, we assume multiple trading strategies like minimization of the transaction costs, risk neutral and risk averse agents. For the latter one we utilize arbitrary, but well-known in the literature and practice risk functions such as the mean-variance utility, the value-at-risk (VaR) and the expected shortfall, also known as the conditional value-at-risk (CVaR).

	The portfolio optimization approach to the trading of the produced electricity has already been taken into consideration in the literature. A significant amount of the existing papers consider the setting with futures market, spot market, and bilateral contracts \cite{liu2007portfolio,garcia2017applying,odeh2018portfolio, canelas2020electricity}. The authors utilize the modern portfolio theory and do not estimate the own price impact, assuming that the market participant is a price-taker. Another stream of the literature name the problem an offering strategy, and they concern the spot day-ahead and intraday markets as well as the balancing one \cite{dai2015optimal, baringo2015offering, mazzi2017price, kath2018value, rintamaki2020strategic, wozabal2020optimal}. These studies are similar to our one, but their main downside is the fact that they assume the market player to be a price-taker. This assumption automatically makes these studies inapplicable for  market players that trade medium-sized or large volumes and impact the price with their bids significantly.
	An important part of the literature compares the coordinated and sequential bidding, especially for storages \cite{fleten2007stochastic, kim2011optimal, lohndorf2013optimizing, boomsma2014bidding, kongelf2019portfolio, lohndorf2020value}. The authors consider a multi-market setting, however they also simplify the market impact issue. A detailed review of coordinated bidding literature was prepared by \citet{aasgaard2019hydropower}.
	
	To the best of our knowledge, the problem of portfolio optimization in auction-based spot markets with market impact and trading costs has not been addressed in the literature so far. \citet{kath2020optimal} investigate the optimal order execution in the intraday continuous market accounting for the market impact. The work that is the closest to our setting is the paper of \citet{ayon2017aggregators} who investigate the optimal bidding curves in day-ahead and intraday electricity markets. The authors, however, consider a flexible demand setting and again assume not to make any price impact in the market.
	A big part of the literature concerning the optimal trading problem or the bidding behaviour in the spot markets, focuses only on one part of the market \cite{narajewski2019estimation, kiesel2017econometric, graf2020modeling, glas2020intraday}. These papers investigate multiple aspects of the trading in the intraday continuous market. Also the renewable energy forecasting plays a crucial role in the decision process for optimal power trading \cite{kozlova2020optimal, li2021modelling}.

		An important factor in the strategy optimization for large volumes is the price impact estimation. We approach the problem using the aggregated curves data. Bidding in the auction-based markets causes shifts in the demand or supply curves. We use the fact to calculate the non-linear price impact of the market participant's own bids. Here we also need a forecast for the curves. Multiple papers look into the issue \cite{ziel2016electricity, ziel2018probabilistic, mestre2020forecasting, kulakov2020x, soloviova2021efficient}. The models are, however, very time-consuming as they estimate the full supply and demand curves. The resulting
		intersection of the curves, can be regarded as an electricity price forecast. However, usually this is not as accurate as  electricity price forecasting (EPF) models that are designed only for the purpose providing accurate price predictions.
		To avoid the 
		aforementioned problem we actually consider a modelling approach that does not need to have curve forecast that provide accurate price predictions, but only gives reasonable curve forecast for the neighbourhood of the expected price.
It is then compared to using the perfect forecast of both curves and prices to see the possible gain of having better, more sophisticated models.
As mentioned, we need suitable EPF models for our trading approach, see \cite{weron2014electricity, nowotarski2018recent, ziel2018probabilistic} for reviews.  
However, as the objective of this study is not to develop a new EPF model we chose to use the two well-known models often called the naive and the expert.

Now, let us summarize the major contributions of the manuscript:
	\begin{enumerate}
	\setlength\itemsep{-1mm}
		\item It is the first work concerning the electricity trading of large volumes with market impact in the auction-based markets.
		\item The first manuscript which considers transaction costs that may vary across the considered markets.
		\item We provide an extensive analysis and discussion of the price formation and trading problem in European price auctions.
		\item The paper presents theoretical results on optimal bidding for risk neutral agents under linear market impact and transactions.
		\item The trading setting is thoroughly examined with all the issues considered including risk averse agents, and the possible extensions and generalization are discussed.
		\item The predictive performance of the utilized methods are compared in a forecasting study for different market players (e.g. wind and solar power traders or retailers that just buy electricity), multiple trading strategies, and forecasts' qualities. 
		\item We provide insights on importance of the overall price impact reduction and evidence of irrelevance of the arbitrage between the DA and IA markets which indicates market efficiency.
		\item The importance of this research is emphasized by the fact of launching intraday auctions in further European countries \cite{epexIAintroduction}.
	\end{enumerate}

	The remainder of this manuscript has the following structure. Section~\ref{sec:price_formation} describes the price formation in European price auctions. The trading setting and objective in the day-ahead and intraday auctions are discussed in Section~\ref{sec:trading}. Trading strategies are described in Section~\ref{sec:trading_strategies}. Section~\ref{sec:forecasting_model} presents the models used for the forecasting of the electricity prices and market impact using auction curve predictions.  The application including the data description, evaluation measures and results is presented in Section~\ref{sec:application}.
	Section~\ref{sec:discussion} discusses the limitations and generalizations of the study where we focus on potential relaxation of the assumptions.
	Finally, Section~\ref{sec:conclusion} concludes the paper. In the Appendix we present the important abbreviations, the notation used in Sections~\ref{sec:price_formation}-\ref{sec:forecasting_model}, the evaluation of the EPF models and additional figures.

	\section{Price formation in European price auctions}\label{sec:price_formation}
We consider two consecutive auctions for the day-ahead (DA) power market and the intraday opening auction (IA) in Germany. The former one offers trading of hourly products, the latter one trading of quarter-hourly products. 
For a selected delivery time point for day $d$ and hour $h$
we have in total five products traded in five corresponding auctions. All auctions find the market clearing price by matching supply and demand such that welfare (consumer and producer rent) is maximized. For the day-ahead this is based on the EUPHEMIA algorithm which incorporates the market coupling of the region to allow cross-border trading and increase of the overall welfare.

\begin{figure}[b!]
	\centering
	\includegraphics[width = 1\linewidth]{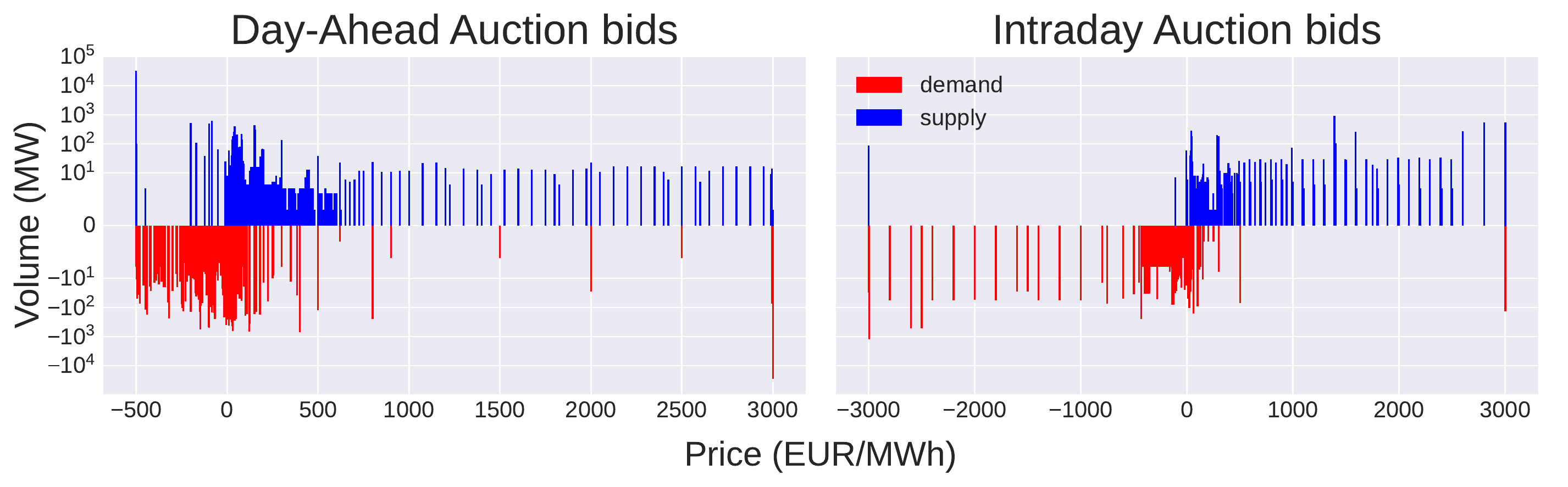}	
	\caption{An example of the bids 
		$B^{\Sup}$ and $B^{\Dem}$ in the German market with delivery on 01.06.2017. The delivery periods are 12:00 to 13:00 for the DA and 12:00 to 12:15 for the IA. The demand bids are plotted as negative for better comparability and the volumes are given in a symmetric logarithmic scale}
	\label{fig:bids_example}
\end{figure}

On all markets we have non-negative volume bids 
$B^{\Sup}(p)$ and $B^{\Dem}(p)$ for $p\in \PP$ with $\PP$ as potential price grid on the considered auction on the supply and demand side. Here $\Dem$ represent ask/buy/demand/purchase and $\Sup$ represent bid/sell/supply/sale. Further, $B^{\Sup}(p)$ and $B^{\Dem}(p)$ are aggregates of all bids at price $p$, so if multiple market participants bid volumes at $p$ they are aggregated in $B^{\Sup}(p)$.
On both markets we have a minimal bid price increment of $0.1$ EUR/MWh. For the considered markets we have
$p_{\min,\text{DA}}=-500$, $p_{\min,\text{IA}}=-3000$
and 
$p_{\max,\text{DA}}=p_{\max,\text{IA}}=3000$. Figure~\ref{fig:bids_example} shows an example of the bids.

Using the bids $B^{\Sup}$ and $B^{\Dem}$ on the supply and demand side we can compute the supply and demand curves $A^{\Sup}$ and $A^{\Dem}$ by aggregation.
More precisely, the aggregated curves $A^{\Sup}$ and $A^{\Dem}$ are then defined by a linear interpolation
of all aggregated bids at all bidden prices $\PP^{\Sup}$ and $\PP^{\Dem}$.
This is for the supply curve
$A^{\Sup}(p) = \sum_{x\in \PP^{\Sup}\cap (-\infty, p]} B^{\Sup}(x)$
for $p\in\PP^{\Sup}$ and for the demand curve
$A^{\Dem}(p) = \sum_{x\in \PP^{\Dem}\cap [p,\infty) } B^{\Dem}(x)$
for $p\in \PP^{\Dem}$.
By construction, it is clear that the curves are strictly monotonic. Moreover, 
their inverse $(A^{\Sup})^{-1}$ and $(A^{\Dem})^{-1}$
can be regarded as continuous supply and demand curves.
\begin{figure}[t!]
	\centering
	\includegraphics[width = 1\linewidth]{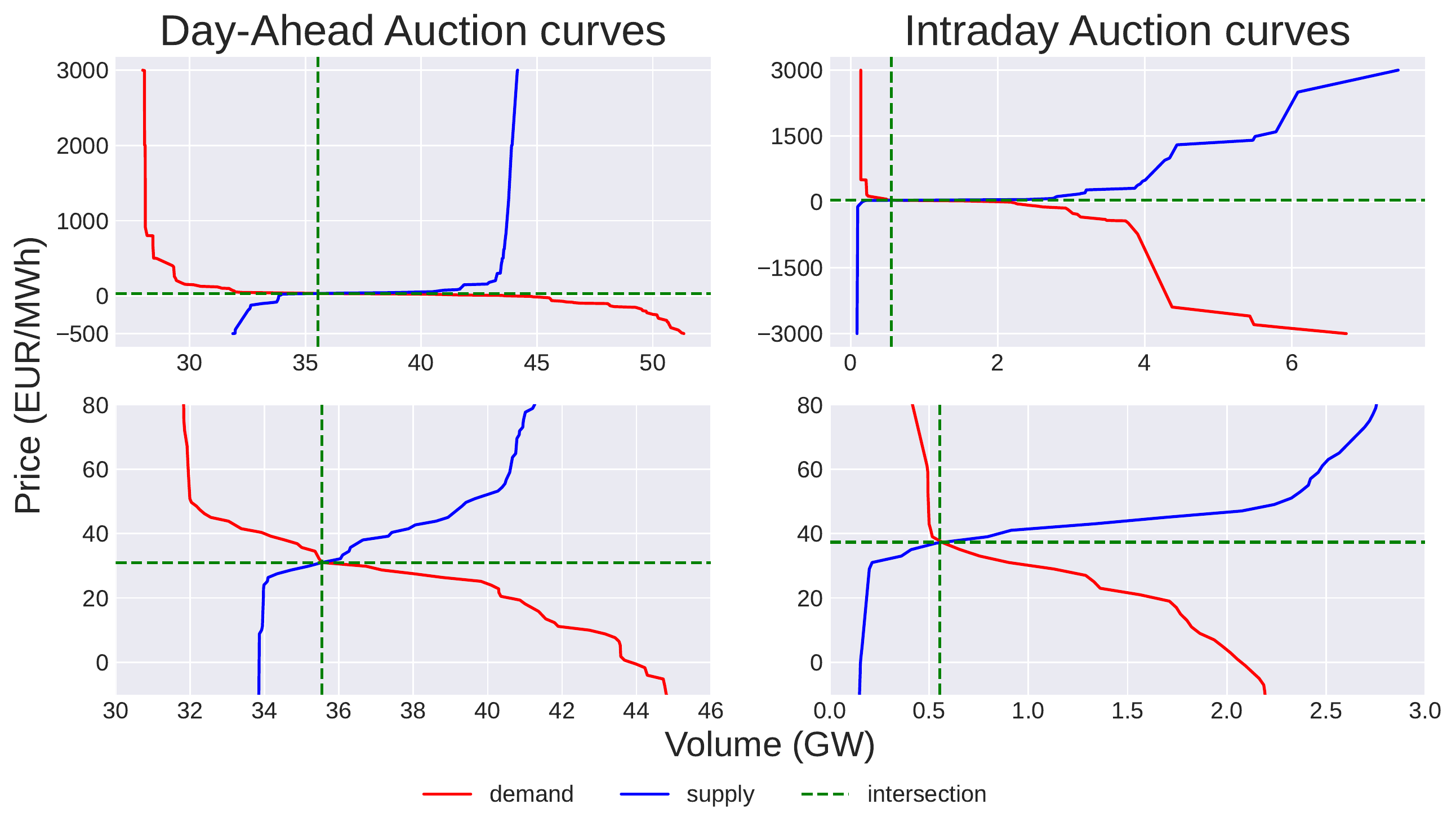}	
	\caption{An example of the supply and demand curves $(A^{\Sup})^{-1}$ and $(A^{\Dem})^{-1}$
	in the German market with delivery on 01.06.2017. The delivery periods are 12:00 to 13:00 for the DA and 12:00 to 12:15 for the IA. The bottom plots are the zoomed-in versions of the top ones}
	\label{fig:auction_curves_example}
\end{figure}
The unique intersection of $A^{\Sup}$ and $A^{\Dem}$ (resp. their graphs) or their inverse 
yields the market clearing volume and price $(V^{*},P^{*})$.\footnote{ 
Theoretically, it may happen that the graphs have no intersection. This happens if either $A^{\Sup}(p_{\max})< A^{\Dem}(p_{\max}) $ or $A^{\Dem}(p_{\min})<A^{\Sup}(p_{\min})$. In those extreme event scenarios, we define the intersections $(V^*,P^*)$ by $(A^{\Sup}(p_{\max}),p_{\max})$ and  $(A^{\Dem}(p_{\min}),p_{\min})$. Note that in the considered German market, any of those events never happened since 2010.
}
Finally, denote $A^{\Sup}_i$ and $A^{\Dem}_i$ for $i\in \{0,\ldots,4\}$ 
	the curves in the corresponding markets ($0=\text{DA}$, $1,\ldots,4=\text{IA}$), analogue $B^{\Sup}_i$ and $B^{\Dem}_i$, $\PP_i^{\Sup}$ and $\PP_i^{\Dem}$. Further, 
	let $\bsA^{\Sup}$, $\bsA^{\Dem}$, $\bsB^{\Sup}$, $\bsB^{\Dem}$, $\bPP^{\Sup}$, $\bPP^{\Dem}$, $\bsp_{\min}$ and $\bsp_{\max}$
	be the corresponding vectors, e.g. $\bsA^{\Sup}=(A^{\Sup}_0,\ldots,A^{\Sup}_4)'$.

 An example of the curves is presented in Figure~\ref{fig:auction_curves_example}. 
For reporting purpose the exchange rounds to a $0.01$ EUR/MWh price increment, and volumes to $0.1$ MW. 
Further, note that especially for the day-ahead market clearing results of the intersection $(V^{*},P^{*})$ does not always equal exactly the reported market clearing volume and price of the exchange. There are sometimes small deviations, which we ignore within this paper.\footnote{The deviation results mainly due to the handling of multiple and complex orders at the money solving the market clearing optimization problem, for more details see e.g. \cite{ziel2016electricity}.}

\section{Trading/Bidding in DA and IA markets}\label{sec:trading}

\subsection{Setting} 

To simplify the trading problem, we consider a market player which only submits unlimited bids, often referred as volume bids.
Thus, we want to trade 
volumes $\bsv=(v_{1}, v_{2}, v_3, v_4)'$ in MWh in the corresponding market, this could be the planned wind power to be generated in the four quarter hours of the considered hour.
The trading problem is to find bids
$\bsb = (b_0,b_1,\ldots, b_4 )'$ in MW for the five markets. We use the convention that the signs of $\bsb$ indicate the market side. Thus, $b_i>0$ are sell bids which shift the supply curve and  $b_i<0$ are bids on the demand side.
Obviously, the market participant may influence only their own bids $\bsb$. Therefore, we introduce the notation
$\bsA^{\Sup}_{\bsb}$, $\bsA^{\Dem}_{\bsb}$, $\bsB^{\Sup}_{\bsb}$, $\bsB^{\Dem}_{\bsb}$ 
 which reflect the agent's bidding behaviour.
 Note that the sets of bidded prices $\bPP^{\Sup}$ and $\bPP^{\Dem}$ are not impacted by $\bsb$ as we assume that even without the agent's market impact there is at least one further unlimited bid on the relevant market side. 
 
 The intersections of
$\bsA_{\bsb}^{\Sup}(\bsp)$ and demand
$\bsA_{\bsb}^{\Dem}(\bsp)$ define the market clearing volumes and prices
$(\bsV_{\bsb}^*,\bsP_{\bsb}^*)$
which also depend on the own bid~$\bsb$.
Moreover, we want to remind that the markets have a sequential order, i.e. first the DA auction is realized, and then the IA auctions, see Figure \ref{fig:market}. Thus, only $b_0$ impacts the DA price, whereas next to $b_1,\ldots, b_4$ also 
 $b_0$ may have an influence on the IA auctions.
This is because other market participants may react on the IA auctions due to $b_0$-influenced DA auction results. We model this impact in Section~\ref{sec:market_impact}.

Of fundamental importance are situations without own market impact, i.e. $\bsb=\bsnull$. As it is relevant for us in further analysis, we 
summarize some characteristics.
Obviously it holds for arbitrary unlimited bids $\bsb$ that
$\bsB^{\Sup}_{\bsnull}(\bsp_{\min}) 
= \bsB^{\Sup}_{\bsb}(\bsp_{\min}) - \bsb^+$
and
$\bsB^{\Dem}_{\bsnull}(\bsp_{\max}) 
= \bsB^{\Dem}_{\bsb}(\bsp_{\max})  - \bsb^-$
where $\bsb^+$ and $\bsb^-$ are the element-wise positive and negative part of $\bsb$.
Further, it holds
that 
$\bsB^{\Sup}_{\bsnull}(\bsp) 
= \bsB^{\Sup}_{\bsb}(\bsp)$ for $\bsp>\bsp_{\min}$
and $\bsB^{\Dem}_{\bsnull}(\bsp) 
= \bsB^{\Dem}_{\bsb}(\bsp)$ for $\bsp<\bsp_{\max}$.
In conclusion, we receive 
\begin{equation}
(\bsA_{\bsnull}^{\Sup})^{-1}(\bsz) =  
(\bsA_{\bsb}^{\Sup})^{-1} ( \bsz- \bsb^{+})
\text{ and }(\bsA_{\bsnull}^{\Dem})^{-1}(\bsz) =  
(\bsA_{\bsb}^{\Dem})^{-1} ( \bsz- \bsb^{-}).
\label{eq_shift_supply_demand} 
\end{equation}

\subsection{Trading objective}

Now, the trader has the gain of
\begin{align}
G(\bsb;\bsv)&= \underbrace{(\bsP_{\bsb}^*)'(\bss \odot \bsb)}_{\text{trading revenue}}- 
\underbrace{\bstau'(\bss \odot \bsb^{\abs})}_{\text{transaction costs}}
- \underbrace{ ((\underbrace{\bsv-\bsS'\bsb}_{\text{ imbalance}})^{\abs})'\bsR }_{\text{imbalance penalty}}\label{eq_gain_v_rebap} \\
&= \sum_{i=0}^4 P^*_{\bsb,i}s_i b_i - \sum_{i=0}^4 \tau_{i}s_i |b_i| + \sum_{j=1}^4
\left|v_j- \sum_{i=0}^4 s_{i,j} b_i \right|R_j
\end{align}
where $\odot$ is the element-wise multiplication (also known as Hadamard product), $\bstau$ is a transaction cost vector, $\bsS =(S_{i,j})= (\bsone_4,\bsI_4)'$ is a 5x4 dimensional summation matrix
and $\bss=\bsS'\bsone_4/4 = (s_0,\ldots,s_4)'= (1,.25,.25,.25,.25)'$ a summation vector which transfers MW to MWh for the corresponding markets (it contains the length of the delivery product period for each auction). 
$\bsR=(R_1,\ldots, R_4)$ is the imbalance penalty price
and $\bsz^{\abs}$ a element-wise absolute value, i.e.
$\bsz^{\abs} = \bsz^+ + \bsz^-$. The imbalance price is a cross-control area uniform balancing energy price (in German named REBAP: Regelzonenübergreifender Einheitlicher BilanzAusgleichsnergiePreis).
In practice $\bstau = (\tau_0,\ldots,\tau_4)' = (\tau_{\text{DA}}, \tau_{\text{IA}},\ldots,\tau_{\text{IA}})'$ satisfies 
$\tau_{\text{DA}}\leq\tau_{\text{IA}}$, thus trading in the hourly day-ahead market is not more expensive than trading the same volume in the intraday opening auction.
Nowadays, the EPEX negotiate with all market participants its own trading fees.

Note that strict market regulations require that
market participants have to avoid system imbalance. Thus, they
have to satisfy the linear \emph{imbalance constraint}
\begin{equation}
\bsv-\bsS'\bsb = \bsnull
\label{eq_rabap_constr}
\end{equation}
as we are not allowed to speculate against the imbalance price.
Thus, under constraint \eqref{eq_rabap_constr}
the gain equation \eqref{eq_gain_v_rebap} simplifies to
\begin{align}
G(\bsb)&= (\bsP_{\bsb}^*)'(\bss \odot \bsb) 
 - \bstau'(\bss \odot \bsb^{\abs}) 
= \sum_{i=0}^4 P^*_{\bsb,i}s_i b_i - \sum_{i=0}^4 \tau_{i}s_i |b_i|.
\label{eq_gain_wo_rebap}
\end{align}
$G$ does not depend on $\bsv$ any more, but $\bsv$ is contained in
the constraint \eqref{eq_rabap_constr}.

In practice, we want to maximize $G$ with respect to $\bsb$.
However, one of the key challenges is to describe adequately the price $\bsP_{\bsb}^*$. This is a multivariate random variable which depends on the own bidding impact due to the bid $\bsb$. To simplify this task
we consider the 'no bidding' situation with $\bsb=\bsnull$ as baseline. 
Hence, we define
$\bsDelta_{\bsb} = \bsP_{\bsb}^* - \bsP_{\bsnull}^*$
as the price impact due to the trading of volume $\bsb$.
The hope is that we can easier access $\bsDelta_{\bsb}$ 
and $\bsP_{\bsnull}^*$ than $\bsP_{\bsb}^*$.

We rewrite $G$ to
\begin{align}
G(\bsb)&= 
(\bsP_{\bsnull}^* + \bsDelta_{\bsb})'(\bss \odot \bsb) 
 - \bstau'(\bss \odot \bsb^{\abs}) 
= \sum_{i=0}^4 (P^*_{\bsnull,i} + \Delta_{\bsb,i})s_i b_i - \sum_{i=0}^4 \tau_{i}s_i |b_i|.
\label{eq_gain_delta} 
\end{align}
Note that $\bsDelta_{\bsb}$ is a highly non-linear function in 
$\bsb$ as the supply and demand curves are non-linear what can be observed in Figure~\ref{fig:auction_curves_example}. 
Note that even under $\bsDelta_{\bsb}=\bsnull$ the remaining equation is non-linear in $\bsb$, as the absolute value is a non-linear function.

Now, let us assume that the market participant wants to maximize a risk functional $\RR(G)$.
Typically, this could be $\RR(G)=\E[G]$ or $\mu$-$\sigma$-utility $\RR(G)=\E[G] - \gamma\V ar[G]$, but expected shortfall (CVaR) or value-at-risk (VaR) measures are plausible options as well. 
Thus, the maximization problem is
\begin{align}
\bsb_{\text{opt}} = \argmax_{\bsb \in \R^5 \text{ with }
{\bsv=\bsS'\bsb} } \RR(G(\bsb)).
\label{eq_risk_optimization}
\end{align}
Note that choosing non-linear risk measures does not seriously 
increase the complexity of the trading problem.
Thus, even in this relatively simple setting we are facing a non-linear optimization problem due to the non-linearity of $G$. This holds even if we choose $\RR$ as a linear functional, e.g. $\RR=\E$.

The linear imbalance constraint \eqref{eq_rabap_constr}
allows us to simplify the optimization problem 
\eqref{eq_risk_optimization}
significantly.
As $\bsv-\bsS'\bsb=\bsnull$ yields immediately that $b_i = v_i-b_0$ for $i>0$, that is to say
\begin{align}
\widetilde{\bsb} = (2b_0,\bsv )'-b_0\bsone = (b_0, v_1-b_0,\ldots,v_4-b_0)',
\label{eq_v0}
\end{align}
and we highlight that $\widetilde{\bsb}$ is a linear function in $b_0$
under the imbalance constraint \eqref{eq_rabap_constr}.
We receive
the one-dimensional optimization problem
\begin{align}
\bsb^{\text{opt}} =  (2b_0^{\text{opt}},\bsv )'-b_0^{\text{opt}}\bsone \, \,
\text{ with } \, \, b^{\text{opt}}_{0} =
\argmax_{b_0 \in \R } \RR(G(\wtilde{\bsv}+b_0\bsc \widetilde{\bsb}))
\label{eq_risk_optimization2}
\end{align}
For $\widetilde{\bsb}$ the latter term in \eqref{eq_gain_delta} are the transaction costs $\TT(b_0)$ which can be simplified to
\begin{align}
\TT(b_0) = \bstau'\left(\bss \odot \widetilde{\bsb}^{\abs} \right) = \tau_0 |b_0| + \frac{1}{4}\sum_{i=1}^{4} \tau_i|v_i - b_0|.
\label{eq_transaction_costs_init}
\end{align} 

In addition, we want to present a decomposition of 
$G(\widetilde{\bsb})$ into four interpretable components.
Disentangling the DA and IA part by remembering the definitions of $\widetilde{\bsb}$ and $\bss$ gives
\begin{equation}
\begin{aligned}
G(\widetilde{\bsb})
=&
\left(P^*_{\bsnull,0}+\Delta_{\widetilde{\bsb},0}\right) b_0
+
\frac{1}{4} \sum_{i=1}^4 \left(P^*_{\bsnull,i}+\Delta_{\widetilde{\bsb},i}\right) (v_i-b_0) -   \TT(b_0)  \\
=& 
\underbrace{
\frac{1}{4} \sum_{i=1}^4  P^*_{\bsnull,i}v_i}_{\text{IA revenue}}  + \underbrace{\left( P^*_{\bsnull,0}  - \frac{1}{4}\sum_{i=1}^4 P^*_{\bsnull,i}\right) b_0 }_{ \text{DA-IA arbitrage}}  
 + \underbrace{\Delta_{\widetilde{\bsb},0} b_0
+ \frac{1}{4} \sum_{i=1}^4 \Delta_{\widetilde{\bsb},i} (v_i-b_0)
 }_{ \text{DA\&IA market impact} }
  -   \underbrace{\TT(b_0).}_{\text{Transaction costs}}
\label{eq_gain_gen_transaction_costs_decomp_interpret}
\end{aligned}
\end{equation}
The four interpretable components are: 
a revenue term, an arbitrage term, a market impact term and the transactions costs $\TT$. We will interpret them in more detail in the next section.
Here, we only want to point out that the IA revenue term does not depend on the bid $b_0$. However, in practice it usually contributes the most to the gain $G$, but it cannot be influenced by a trader.

\section{Trading strategies}\label{sec:trading_strategies}

\subsection{Intraday Auction only}

A straightforward strategy is to bid the volume $\bsv$ only in the IA market.
\begin{equation}
	\bsb_{\text{IA-only}} = (0,\bsv) = (0, v_1,\ldots,v_4)
\end{equation}
Obviously, \textbf{IA-only} seems odd if $\tau_{\text{DA}}<\tau_{\text{IA}}$ holds. Moreover, we observe that the DA auctions have much larger volumes than the IA auctions.
However, a simple counterpart strategy \textbf{DA-only} that bids only at the DA auction and zero volume at the IA auctions is only possible if $\bsv$ is constant, i.e. $v_{1}=\ldots =v_{4}$. Once we start having ramps in at least one asset, we may face ramps in the accepted bids as well. Under the imbalance constraint~\eqref{eq_rabap_constr} this imbalance is removed. In our setting this forces us to bid at the IA auction in such situations.

\subsection{Minimal transaction costs}
\label{subsec_min_trans}

From our point of view the intuitive DA-only counterpart to IA-only is the bidding strategy that minimizes transaction costs under the $\tau_{\text{DA}}\leq \tau_{\text{IA}}$ assumption. Intuitively this approach trades as much volume in the cheaper (from the transaction cost point of view) DA market, and balances the remaining power in the IA auction.
Now, we derive the minimal transaction cost strategy. 
The transaction costs $\TT(b_0)$ in \eqref{eq_transaction_costs_init}
is a convex, piecewise linear function. Thus, there exists a minimum.  
The minimizer of $\TT$ is the $\bstau \odot \bss$-weighted median of $(0,\bsv)'$ which we define as the \textbf{TC-min} strategy
\begin{equation}
\bsb_{\text{TC-min}} = \wtilde{\bsb}_{\text{TC-min}} =  (b_{\text{TC-min},0}, v_1 - b_{\text{TC-min},0},\ldots, v_4 - b_{\text{TC-min},0}).
\label{eq_b_tcmin}
\end{equation}

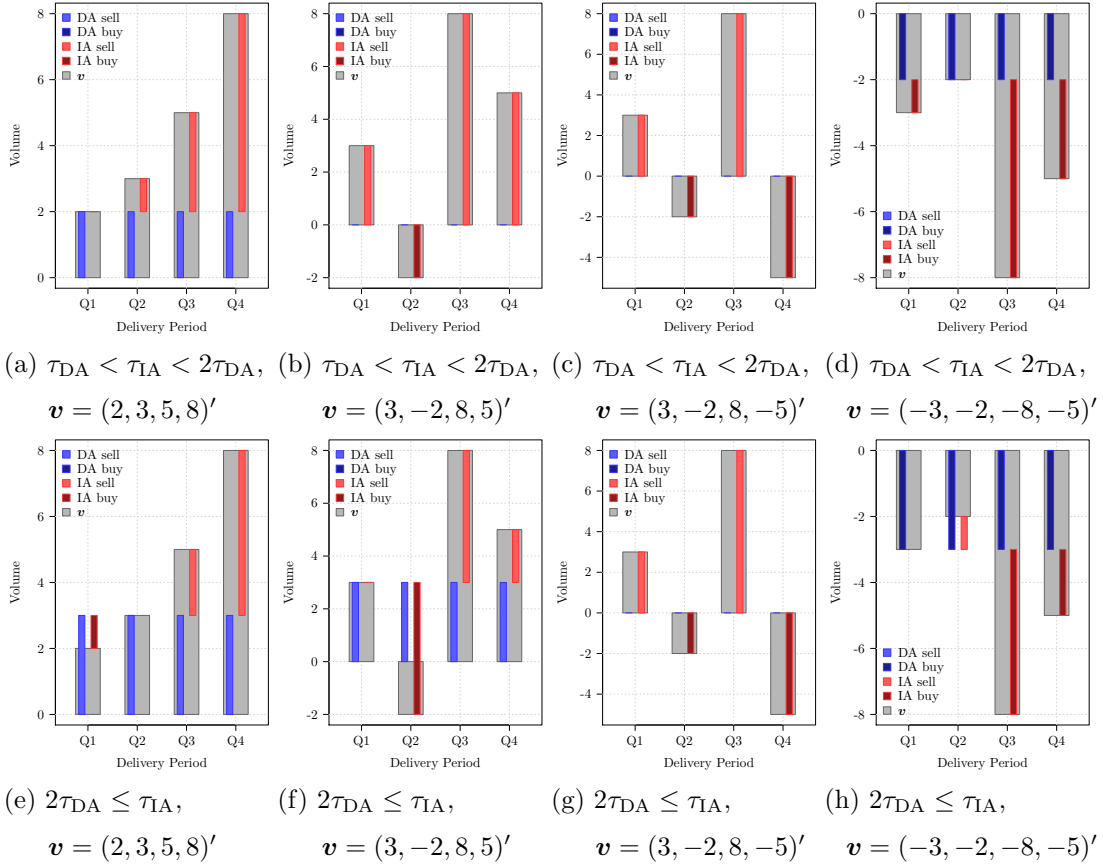
\begin{figure}[hbt!]
\centering
 \subfloat[$\tau_{\text{DA}}<\tau_{\text{IA}}<2\tau_{\text{DA}}$,\newline \indent $\quad \ \ \bsv=(2,3,5,8)'$]{\resizebox{0.24\textwidth}{!}{
\begin{tikzpicture}[x=1pt,y=1pt]
\definecolor{fillColor}{RGB}{255,255,255}
\path[use as bounding box,fill=fillColor,fill opacity=0.00] (0,0) rectangle (289.08,368.58);
\begin{scope}
\path[clip] ( 55.20, 55.20) rectangle (283.08,362.58);
\definecolor{drawColor}{RGB}{255,255,255}

\path[draw=drawColor,line width= 0.4pt,line join=round,line cap=round] ( 90.02,102.16) circle (  2.25);

\path[draw=drawColor,line width= 0.4pt,line join=round,line cap=round] (142.76,137.74) circle (  2.25);

\path[draw=drawColor,line width= 0.4pt,line join=round,line cap=round] (195.51,173.31) circle (  2.25);

\path[draw=drawColor,line width= 0.4pt,line join=round,line cap=round] (248.26,208.89) circle (  2.25);
\end{scope}
\begin{scope}
\path[clip] (  0.00,  0.00) rectangle (289.08,368.58);
\definecolor{drawColor}{RGB}{0,0,0}

\path[draw=drawColor,line width= 0.4pt,line join=round,line cap=round] ( 55.20, 66.58) -- ( 55.20,351.19);

\path[draw=drawColor,line width= 0.4pt,line join=round,line cap=round] ( 55.20, 66.58) -- ( 49.20, 66.58);

\path[draw=drawColor,line width= 0.4pt,line join=round,line cap=round] ( 55.20,137.74) -- ( 49.20,137.74);

\path[draw=drawColor,line width= 0.4pt,line join=round,line cap=round] ( 55.20,208.89) -- ( 49.20,208.89);

\path[draw=drawColor,line width= 0.4pt,line join=round,line cap=round] ( 55.20,280.04) -- ( 49.20,280.04);

\path[draw=drawColor,line width= 0.4pt,line join=round,line cap=round] ( 55.20,351.19) -- ( 49.20,351.19);

\node[text=drawColor,anchor=base east,inner sep=0pt, outer sep=0pt, scale=  1.30] at ( 43.20, 62.11) {0};

\node[text=drawColor,anchor=base east,inner sep=0pt, outer sep=0pt, scale=  1.30] at ( 43.20,133.26) {2};

\node[text=drawColor,anchor=base east,inner sep=0pt, outer sep=0pt, scale=  1.30] at ( 43.20,204.41) {4};

\node[text=drawColor,anchor=base east,inner sep=0pt, outer sep=0pt, scale=  1.30] at ( 43.20,275.56) {6};

\node[text=drawColor,anchor=base east,inner sep=0pt, outer sep=0pt, scale=  1.30] at ( 43.20,346.72) {8};

\path[draw=drawColor,line width= 0.4pt,line join=round,line cap=round] ( 55.20, 55.20) --
	(283.08, 55.20) --
	(283.08,362.58) --
	( 55.20,362.58) --
	( 55.20, 55.20);
\end{scope}
\begin{scope}
\path[clip] (  0.00,  0.00) rectangle (289.08,368.58);
\definecolor{drawColor}{RGB}{0,0,0}

\node[text=drawColor,anchor=base,inner sep=0pt, outer sep=0pt, scale=  1.30] at (169.14,  9.60) {Delivery Period};

\node[text=drawColor,rotate= 90.00,anchor=base,inner sep=0pt, outer sep=0pt, scale=  1.30] at ( 16.80,208.89) {Volume};
\end{scope}
\begin{scope}
\path[clip] (  0.00,  0.00) rectangle (289.08,368.58);
\definecolor{drawColor}{RGB}{0,0,0}

\path[draw=drawColor,line width= 0.4pt,line join=round,line cap=round] ( 90.02, 55.20) -- (248.26, 55.20);

\path[draw=drawColor,line width= 0.4pt,line join=round,line cap=round] ( 90.02, 55.20) -- ( 90.02, 49.20);

\path[draw=drawColor,line width= 0.4pt,line join=round,line cap=round] (142.76, 55.20) -- (142.76, 49.20);

\path[draw=drawColor,line width= 0.4pt,line join=round,line cap=round] (195.51, 55.20) -- (195.51, 49.20);

\path[draw=drawColor,line width= 0.4pt,line join=round,line cap=round] (248.26, 55.20) -- (248.26, 49.20);

\node[text=drawColor,anchor=base,inner sep=0pt, outer sep=0pt, scale=  1.30] at ( 90.02, 33.60) {Q1};

\node[text=drawColor,anchor=base,inner sep=0pt, outer sep=0pt, scale=  1.30] at (142.76, 33.60) {Q2};

\node[text=drawColor,anchor=base,inner sep=0pt, outer sep=0pt, scale=  1.30] at (195.51, 33.60) {Q3};

\node[text=drawColor,anchor=base,inner sep=0pt, outer sep=0pt, scale=  1.30] at (248.26, 33.60) {Q4};
\end{scope}
\begin{scope}
\path[clip] ( 55.20, 55.20) rectangle (283.08,362.58);
\definecolor{drawColor}{RGB}{211,211,211}

\path[draw=drawColor,line width= 0.4pt,dash pattern=on 1pt off 3pt ,line join=round,line cap=round] ( 90.02, 55.20) -- ( 90.02,362.58);

\path[draw=drawColor,line width= 0.4pt,dash pattern=on 1pt off 3pt ,line join=round,line cap=round] (142.76, 55.20) -- (142.76,362.58);

\path[draw=drawColor,line width= 0.4pt,dash pattern=on 1pt off 3pt ,line join=round,line cap=round] (195.51, 55.20) -- (195.51,362.58);

\path[draw=drawColor,line width= 0.4pt,dash pattern=on 1pt off 3pt ,line join=round,line cap=round] (248.26, 55.20) -- (248.26,362.58);

\path[draw=drawColor,line width= 0.4pt,dash pattern=on 1pt off 3pt ,line join=round,line cap=round] ( 55.20, 66.58) -- (283.08, 66.58);

\path[draw=drawColor,line width= 0.4pt,dash pattern=on 1pt off 3pt ,line join=round,line cap=round] ( 55.20,137.74) -- (283.08,137.74);

\path[draw=drawColor,line width= 0.4pt,dash pattern=on 1pt off 3pt ,line join=round,line cap=round] ( 55.20,208.89) -- (283.08,208.89);

\path[draw=drawColor,line width= 0.4pt,dash pattern=on 1pt off 3pt ,line join=round,line cap=round] ( 55.20,280.04) -- (283.08,280.04);

\path[draw=drawColor,line width= 0.4pt,dash pattern=on 1pt off 3pt ,line join=round,line cap=round] ( 55.20,351.19) -- (283.08,351.19);
\definecolor{drawColor}{gray}{0.40}
\definecolor{fillColor}{RGB}{183,183,183}

\path[draw=drawColor,line width= 0.4pt,line join=round,line cap=round,fill=fillColor] (103.20, 66.58) --
	(103.20,137.74) --
	( 76.83,137.74) --
	( 76.83, 66.58) --
	cycle;

\path[draw=drawColor,line width= 0.4pt,line join=round,line cap=round,fill=fillColor] (155.95, 66.58) --
	(155.95,173.31) --
	(129.58,173.31) --
	(129.58, 66.58) --
	cycle;

\path[draw=drawColor,line width= 0.4pt,line join=round,line cap=round,fill=fillColor] (208.70, 66.58) --
	(208.70,244.46) --
	(182.33,244.46) --
	(182.33, 66.58) --
	cycle;

\path[draw=drawColor,line width= 0.4pt,line join=round,line cap=round,fill=fillColor] (261.45, 66.58) --
	(261.45,351.19) --
	(235.08,351.19) --
	(235.08, 66.58) --
	cycle;
\definecolor{drawColor}{RGB}{51,51,255}
\definecolor{fillColor}{RGB}{91,91,255}

\path[draw=drawColor,line width= 0.8pt,line join=round,line cap=round,fill=fillColor] ( 86.72, 66.58) --
	( 86.72,137.74) --
	( 80.12,137.74) --
	( 80.12, 66.58) --
	cycle;

\path[draw=drawColor,line width= 0.8pt,line join=round,line cap=round,fill=fillColor] (139.47, 66.58) --
	(139.47,137.74) --
	(132.87,137.74) --
	(132.87, 66.58) --
	cycle;

\path[draw=drawColor,line width= 0.8pt,line join=round,line cap=round,fill=fillColor] (192.22, 66.58) --
	(192.22,137.74) --
	(185.62,137.74) --
	(185.62, 66.58) --
	cycle;

\path[draw=drawColor,line width= 0.8pt,line join=round,line cap=round,fill=fillColor] (244.97, 66.58) --
	(244.97,137.74) --
	(238.37,137.74) --
	(238.37, 66.58) --
	cycle;
\definecolor{drawColor}{RGB}{255,51,51}
\definecolor{fillColor}{RGB}{255,91,91}

\path[draw=drawColor,line width= 0.8pt,line join=round,line cap=round,fill=fillColor] ( 99.91,137.74) --
	( 99.91,137.74) --
	( 93.31,137.74) --
	( 93.31,137.74) --
	cycle;

\path[draw=drawColor,line width= 0.8pt,line join=round,line cap=round,fill=fillColor] (152.66,137.74) --
	(152.66,173.31) --
	(146.06,173.31) --
	(146.06,137.74) --
	cycle;

\path[draw=drawColor,line width= 0.8pt,line join=round,line cap=round,fill=fillColor] (205.41,137.74) --
	(205.41,244.46) --
	(198.81,244.46) --
	(198.81,137.74) --
	cycle;

\path[draw=drawColor,line width= 0.8pt,line join=round,line cap=round,fill=fillColor] (258.16,137.74) --
	(258.16,351.19) --
	(251.56,351.19) --
	(251.56,137.74) --
	cycle;

\path[] ( 55.20,362.58) rectangle (129.40,268.98);
\definecolor{drawColor}{RGB}{51,51,255}
\definecolor{fillColor}{RGB}{91,91,255}

\path[draw=drawColor,line width= 0.4pt,line join=round,line cap=round,fill=fillColor] ( 62.91,342.99) rectangle ( 70.89,350.97);
\definecolor{fillColor}{RGB}{28,28,141}

\path[draw=drawColor,line width= 0.4pt,line join=round,line cap=round,fill=fillColor] ( 62.91,327.39) rectangle ( 70.89,335.37);
\definecolor{drawColor}{RGB}{255,51,51}
\definecolor{fillColor}{RGB}{255,91,91}

\path[draw=drawColor,line width= 0.4pt,line join=round,line cap=round,fill=fillColor] ( 62.91,311.79) rectangle ( 70.89,319.77);
\definecolor{fillColor}{RGB}{141,28,28}

\path[draw=drawColor,line width= 0.4pt,line join=round,line cap=round,fill=fillColor] ( 62.91,296.19) rectangle ( 70.89,304.17);
\definecolor{drawColor}{gray}{0.40}
\definecolor{fillColor}{RGB}{183,183,183}

\path[draw=drawColor,line width= 0.4pt,line join=round,line cap=round,fill=fillColor] ( 62.91,280.59) rectangle ( 70.89,288.57);
\definecolor{drawColor}{RGB}{0,0,0}

\node[text=drawColor,anchor=base west,inner sep=0pt, outer sep=0pt, scale=  1.30] at ( 78.60,342.50) {DA sell };

\node[text=drawColor,anchor=base west,inner sep=0pt, outer sep=0pt, scale=  1.30] at ( 78.60,326.90) {DA buy};

\node[text=drawColor,anchor=base west,inner sep=0pt, outer sep=0pt, scale=  1.30] at ( 78.60,311.30) {IA sell};

\node[text=drawColor,anchor=base west,inner sep=0pt, outer sep=0pt, scale=  1.30] at ( 78.60,295.70) {IA buy};

\node[text=drawColor,anchor=base west,inner sep=0pt, outer sep=0pt, scale=  1.30] at ( 78.60,280.10) {$\bm{v}$};
\end{scope}
\end{tikzpicture}}}
 \subfloat[$\tau_{\text{DA}}<\tau_{\text{IA}}<2\tau_{\text{DA}}$,\newline \indent $\quad \ \ \bsv=(3,-2,8,5)'$]{\resizebox{0.24\textwidth}{!}{
\begin{tikzpicture}[x=1pt,y=1pt]
\definecolor{fillColor}{RGB}{255,255,255}
\path[use as bounding box,fill=fillColor,fill opacity=0.00] (0,0) rectangle (289.08,368.58);
\begin{scope}
\path[clip] ( 55.20, 55.20) rectangle (283.08,362.58);
\definecolor{drawColor}{RGB}{255,255,255}

\path[draw=drawColor,line width= 0.4pt,line join=round,line cap=round] ( 90.02,151.97) circle (  2.25);

\path[draw=drawColor,line width= 0.4pt,line join=round,line cap=round] (142.76,180.43) circle (  2.25);

\path[draw=drawColor,line width= 0.4pt,line join=round,line cap=round] (195.51,208.89) circle (  2.25);

\path[draw=drawColor,line width= 0.4pt,line join=round,line cap=round] (248.26,237.35) circle (  2.25);
\end{scope}
\begin{scope}
\path[clip] (  0.00,  0.00) rectangle (289.08,368.58);
\definecolor{drawColor}{RGB}{0,0,0}

\path[draw=drawColor,line width= 0.4pt,line join=round,line cap=round] ( 55.20, 66.58) -- ( 55.20,351.19);

\path[draw=drawColor,line width= 0.4pt,line join=round,line cap=round] ( 55.20, 66.58) -- ( 49.20, 66.58);

\path[draw=drawColor,line width= 0.4pt,line join=round,line cap=round] ( 55.20,123.51) -- ( 49.20,123.51);

\path[draw=drawColor,line width= 0.4pt,line join=round,line cap=round] ( 55.20,180.43) -- ( 49.20,180.43);

\path[draw=drawColor,line width= 0.4pt,line join=round,line cap=round] ( 55.20,237.35) -- ( 49.20,237.35);

\path[draw=drawColor,line width= 0.4pt,line join=round,line cap=round] ( 55.20,294.27) -- ( 49.20,294.27);

\path[draw=drawColor,line width= 0.4pt,line join=round,line cap=round] ( 55.20,351.19) -- ( 49.20,351.19);

\node[text=drawColor,anchor=base east,inner sep=0pt, outer sep=0pt, scale=  1.30] at ( 43.20, 62.11) {-2};

\node[text=drawColor,anchor=base east,inner sep=0pt, outer sep=0pt, scale=  1.30] at ( 43.20,119.03) {0};

\node[text=drawColor,anchor=base east,inner sep=0pt, outer sep=0pt, scale=  1.30] at ( 43.20,175.95) {2};

\node[text=drawColor,anchor=base east,inner sep=0pt, outer sep=0pt, scale=  1.30] at ( 43.20,232.87) {4};

\node[text=drawColor,anchor=base east,inner sep=0pt, outer sep=0pt, scale=  1.30] at ( 43.20,289.79) {6};

\node[text=drawColor,anchor=base east,inner sep=0pt, outer sep=0pt, scale=  1.30] at ( 43.20,346.72) {8};

\path[draw=drawColor,line width= 0.4pt,line join=round,line cap=round] ( 55.20, 55.20) --
	(283.08, 55.20) --
	(283.08,362.58) --
	( 55.20,362.58) --
	( 55.20, 55.20);
\end{scope}
\begin{scope}
\path[clip] (  0.00,  0.00) rectangle (289.08,368.58);
\definecolor{drawColor}{RGB}{0,0,0}

\node[text=drawColor,anchor=base,inner sep=0pt, outer sep=0pt, scale=  1.30] at (169.14,  9.60) {Delivery Period};

\node[text=drawColor,rotate= 90.00,anchor=base,inner sep=0pt, outer sep=0pt, scale=  1.30] at ( 16.80,208.89) {Volume};
\end{scope}
\begin{scope}
\path[clip] (  0.00,  0.00) rectangle (289.08,368.58);
\definecolor{drawColor}{RGB}{0,0,0}

\path[draw=drawColor,line width= 0.4pt,line join=round,line cap=round] ( 90.02, 55.20) -- (248.26, 55.20);

\path[draw=drawColor,line width= 0.4pt,line join=round,line cap=round] ( 90.02, 55.20) -- ( 90.02, 49.20);

\path[draw=drawColor,line width= 0.4pt,line join=round,line cap=round] (142.76, 55.20) -- (142.76, 49.20);

\path[draw=drawColor,line width= 0.4pt,line join=round,line cap=round] (195.51, 55.20) -- (195.51, 49.20);

\path[draw=drawColor,line width= 0.4pt,line join=round,line cap=round] (248.26, 55.20) -- (248.26, 49.20);

\node[text=drawColor,anchor=base,inner sep=0pt, outer sep=0pt, scale=  1.30] at ( 90.02, 33.60) {Q1};

\node[text=drawColor,anchor=base,inner sep=0pt, outer sep=0pt, scale=  1.30] at (142.76, 33.60) {Q2};

\node[text=drawColor,anchor=base,inner sep=0pt, outer sep=0pt, scale=  1.30] at (195.51, 33.60) {Q3};

\node[text=drawColor,anchor=base,inner sep=0pt, outer sep=0pt, scale=  1.30] at (248.26, 33.60) {Q4};
\end{scope}
\begin{scope}
\path[clip] ( 55.20, 55.20) rectangle (283.08,362.58);
\definecolor{drawColor}{RGB}{211,211,211}

\path[draw=drawColor,line width= 0.4pt,dash pattern=on 1pt off 3pt ,line join=round,line cap=round] ( 90.02, 55.20) -- ( 90.02,362.58);

\path[draw=drawColor,line width= 0.4pt,dash pattern=on 1pt off 3pt ,line join=round,line cap=round] (142.76, 55.20) -- (142.76,362.58);

\path[draw=drawColor,line width= 0.4pt,dash pattern=on 1pt off 3pt ,line join=round,line cap=round] (195.51, 55.20) -- (195.51,362.58);

\path[draw=drawColor,line width= 0.4pt,dash pattern=on 1pt off 3pt ,line join=round,line cap=round] (248.26, 55.20) -- (248.26,362.58);

\path[draw=drawColor,line width= 0.4pt,dash pattern=on 1pt off 3pt ,line join=round,line cap=round] ( 55.20, 66.58) -- (283.08, 66.58);

\path[draw=drawColor,line width= 0.4pt,dash pattern=on 1pt off 3pt ,line join=round,line cap=round] ( 55.20,123.51) -- (283.08,123.51);

\path[draw=drawColor,line width= 0.4pt,dash pattern=on 1pt off 3pt ,line join=round,line cap=round] ( 55.20,180.43) -- (283.08,180.43);

\path[draw=drawColor,line width= 0.4pt,dash pattern=on 1pt off 3pt ,line join=round,line cap=round] ( 55.20,237.35) -- (283.08,237.35);

\path[draw=drawColor,line width= 0.4pt,dash pattern=on 1pt off 3pt ,line join=round,line cap=round] ( 55.20,294.27) -- (283.08,294.27);

\path[draw=drawColor,line width= 0.4pt,dash pattern=on 1pt off 3pt ,line join=round,line cap=round] ( 55.20,351.19) -- (283.08,351.19);
\definecolor{drawColor}{gray}{0.40}
\definecolor{fillColor}{RGB}{183,183,183}

\path[draw=drawColor,line width= 0.4pt,line join=round,line cap=round,fill=fillColor] (103.20,123.51) --
	(103.20,208.89) --
	( 76.83,208.89) --
	( 76.83,123.51) --
	cycle;

\path[draw=drawColor,line width= 0.4pt,line join=round,line cap=round,fill=fillColor] (155.95,123.51) --
	(155.95, 66.58) --
	(129.58, 66.58) --
	(129.58,123.51) --
	cycle;

\path[draw=drawColor,line width= 0.4pt,line join=round,line cap=round,fill=fillColor] (208.70,123.51) --
	(208.70,351.19) --
	(182.33,351.19) --
	(182.33,123.51) --
	cycle;

\path[draw=drawColor,line width= 0.4pt,line join=round,line cap=round,fill=fillColor] (261.45,123.51) --
	(261.45,265.81) --
	(235.08,265.81) --
	(235.08,123.51) --
	cycle;
\definecolor{drawColor}{RGB}{51,51,255}
\definecolor{fillColor}{RGB}{91,91,255}

\path[draw=drawColor,line width= 0.8pt,line join=round,line cap=round,fill=fillColor] ( 86.72,123.51) --
	( 86.72,123.51) --
	( 80.12,123.51) --
	( 80.12,123.51) --
	cycle;

\path[draw=drawColor,line width= 0.8pt,line join=round,line cap=round,fill=fillColor] (139.47,123.51) --
	(139.47,123.51) --
	(132.87,123.51) --
	(132.87,123.51) --
	cycle;

\path[draw=drawColor,line width= 0.8pt,line join=round,line cap=round,fill=fillColor] (192.22,123.51) --
	(192.22,123.51) --
	(185.62,123.51) --
	(185.62,123.51) --
	cycle;

\path[draw=drawColor,line width= 0.8pt,line join=round,line cap=round,fill=fillColor] (244.97,123.51) --
	(244.97,123.51) --
	(238.37,123.51) --
	(238.37,123.51) --
	cycle;
\definecolor{drawColor}{RGB}{255,51,51}
\definecolor{fillColor}{RGB}{255,91,91}

\path[draw=drawColor,line width= 0.8pt,line join=round,line cap=round,fill=fillColor] ( 99.91,123.51) --
	( 99.91,208.89) --
	( 93.31,208.89) --
	( 93.31,123.51) --
	cycle;
\definecolor{fillColor}{RGB}{141,28,28}

\path[draw=drawColor,line width= 0.8pt,line join=round,line cap=round,fill=fillColor] (152.66,123.51) --
	(152.66, 66.58) --
	(146.06, 66.58) --
	(146.06,123.51) --
	cycle;
\definecolor{fillColor}{RGB}{255,91,91}

\path[draw=drawColor,line width= 0.8pt,line join=round,line cap=round,fill=fillColor] (205.41,123.51) --
	(205.41,351.19) --
	(198.81,351.19) --
	(198.81,123.51) --
	cycle;

\path[draw=drawColor,line width= 0.8pt,line join=round,line cap=round,fill=fillColor] (258.16,123.51) --
	(258.16,265.81) --
	(251.56,265.81) --
	(251.56,123.51) --
	cycle;

\path[] ( 55.20,362.58) rectangle (129.40,268.98);
\definecolor{drawColor}{RGB}{51,51,255}
\definecolor{fillColor}{RGB}{91,91,255}

\path[draw=drawColor,line width= 0.4pt,line join=round,line cap=round,fill=fillColor] ( 62.91,342.99) rectangle ( 70.89,350.97);
\definecolor{fillColor}{RGB}{28,28,141}

\path[draw=drawColor,line width= 0.4pt,line join=round,line cap=round,fill=fillColor] ( 62.91,327.39) rectangle ( 70.89,335.37);
\definecolor{drawColor}{RGB}{255,51,51}
\definecolor{fillColor}{RGB}{255,91,91}

\path[draw=drawColor,line width= 0.4pt,line join=round,line cap=round,fill=fillColor] ( 62.91,311.79) rectangle ( 70.89,319.77);
\definecolor{fillColor}{RGB}{141,28,28}

\path[draw=drawColor,line width= 0.4pt,line join=round,line cap=round,fill=fillColor] ( 62.91,296.19) rectangle ( 70.89,304.17);
\definecolor{drawColor}{gray}{0.40}
\definecolor{fillColor}{RGB}{183,183,183}

\path[draw=drawColor,line width= 0.4pt,line join=round,line cap=round,fill=fillColor] ( 62.91,280.59) rectangle ( 70.89,288.57);
\definecolor{drawColor}{RGB}{0,0,0}

\node[text=drawColor,anchor=base west,inner sep=0pt, outer sep=0pt, scale=  1.30] at ( 78.60,342.50) {DA sell };

\node[text=drawColor,anchor=base west,inner sep=0pt, outer sep=0pt, scale=  1.30] at ( 78.60,326.90) {DA buy};

\node[text=drawColor,anchor=base west,inner sep=0pt, outer sep=0pt, scale=  1.30] at ( 78.60,311.30) {IA sell};

\node[text=drawColor,anchor=base west,inner sep=0pt, outer sep=0pt, scale=  1.30] at ( 78.60,295.70) {IA buy};

\node[text=drawColor,anchor=base west,inner sep=0pt, outer sep=0pt, scale=  1.30] at ( 78.60,280.10) {$\bm{v}$};
\end{scope}
\end{tikzpicture}}}
 \subfloat[$\tau_{\text{DA}}<\tau_{\text{IA}}<2\tau_{\text{DA}}$,\newline \indent $\quad \ \ \bsv=(3,-2,8,-5)'$]{\resizebox{0.24\textwidth}{!}{
\begin{tikzpicture}[x=1pt,y=1pt]
\definecolor{fillColor}{RGB}{255,255,255}
\path[use as bounding box,fill=fillColor,fill opacity=0.00] (0,0) rectangle (289.08,368.58);
\begin{scope}
\path[clip] ( 55.20, 55.20) rectangle (283.08,362.58);
\definecolor{drawColor}{RGB}{255,255,255}

\path[draw=drawColor,line width= 0.4pt,line join=round,line cap=round] ( 90.02,197.94) circle (  2.25);

\path[draw=drawColor,line width= 0.4pt,line join=round,line cap=round] (142.76,219.83) circle (  2.25);

\path[draw=drawColor,line width= 0.4pt,line join=round,line cap=round] (195.51,241.73) circle (  2.25);

\path[draw=drawColor,line width= 0.4pt,line join=round,line cap=round] (248.26,263.62) circle (  2.25);
\end{scope}
\begin{scope}
\path[clip] (  0.00,  0.00) rectangle (289.08,368.58);
\definecolor{drawColor}{RGB}{0,0,0}

\path[draw=drawColor,line width= 0.4pt,line join=round,line cap=round] ( 55.20, 88.48) -- ( 55.20,351.19);

\path[draw=drawColor,line width= 0.4pt,line join=round,line cap=round] ( 55.20, 88.48) -- ( 49.20, 88.48);

\path[draw=drawColor,line width= 0.4pt,line join=round,line cap=round] ( 55.20,132.26) -- ( 49.20,132.26);

\path[draw=drawColor,line width= 0.4pt,line join=round,line cap=round] ( 55.20,176.05) -- ( 49.20,176.05);

\path[draw=drawColor,line width= 0.4pt,line join=round,line cap=round] ( 55.20,219.83) -- ( 49.20,219.83);

\path[draw=drawColor,line width= 0.4pt,line join=round,line cap=round] ( 55.20,263.62) -- ( 49.20,263.62);

\path[draw=drawColor,line width= 0.4pt,line join=round,line cap=round] ( 55.20,307.41) -- ( 49.20,307.41);

\path[draw=drawColor,line width= 0.4pt,line join=round,line cap=round] ( 55.20,351.19) -- ( 49.20,351.19);

\node[text=drawColor,anchor=base east,inner sep=0pt, outer sep=0pt, scale=  1.30] at ( 43.20, 84.00) {-4};

\node[text=drawColor,anchor=base east,inner sep=0pt, outer sep=0pt, scale=  1.30] at ( 43.20,127.79) {-2};

\node[text=drawColor,anchor=base east,inner sep=0pt, outer sep=0pt, scale=  1.30] at ( 43.20,171.57) {0};

\node[text=drawColor,anchor=base east,inner sep=0pt, outer sep=0pt, scale=  1.30] at ( 43.20,215.36) {2};

\node[text=drawColor,anchor=base east,inner sep=0pt, outer sep=0pt, scale=  1.30] at ( 43.20,259.14) {4};

\node[text=drawColor,anchor=base east,inner sep=0pt, outer sep=0pt, scale=  1.30] at ( 43.20,302.93) {6};

\node[text=drawColor,anchor=base east,inner sep=0pt, outer sep=0pt, scale=  1.30] at ( 43.20,346.72) {8};

\path[draw=drawColor,line width= 0.4pt,line join=round,line cap=round] ( 55.20, 55.20) --
	(283.08, 55.20) --
	(283.08,362.58) --
	( 55.20,362.58) --
	( 55.20, 55.20);
\end{scope}
\begin{scope}
\path[clip] (  0.00,  0.00) rectangle (289.08,368.58);
\definecolor{drawColor}{RGB}{0,0,0}

\node[text=drawColor,anchor=base,inner sep=0pt, outer sep=0pt, scale=  1.30] at (169.14,  9.60) {Delivery Period};

\node[text=drawColor,rotate= 90.00,anchor=base,inner sep=0pt, outer sep=0pt, scale=  1.30] at ( 16.80,208.89) {Volume};
\end{scope}
\begin{scope}
\path[clip] (  0.00,  0.00) rectangle (289.08,368.58);
\definecolor{drawColor}{RGB}{0,0,0}

\path[draw=drawColor,line width= 0.4pt,line join=round,line cap=round] ( 90.02, 55.20) -- (248.26, 55.20);

\path[draw=drawColor,line width= 0.4pt,line join=round,line cap=round] ( 90.02, 55.20) -- ( 90.02, 49.20);

\path[draw=drawColor,line width= 0.4pt,line join=round,line cap=round] (142.76, 55.20) -- (142.76, 49.20);

\path[draw=drawColor,line width= 0.4pt,line join=round,line cap=round] (195.51, 55.20) -- (195.51, 49.20);

\path[draw=drawColor,line width= 0.4pt,line join=round,line cap=round] (248.26, 55.20) -- (248.26, 49.20);

\node[text=drawColor,anchor=base,inner sep=0pt, outer sep=0pt, scale=  1.30] at ( 90.02, 33.60) {Q1};

\node[text=drawColor,anchor=base,inner sep=0pt, outer sep=0pt, scale=  1.30] at (142.76, 33.60) {Q2};

\node[text=drawColor,anchor=base,inner sep=0pt, outer sep=0pt, scale=  1.30] at (195.51, 33.60) {Q3};

\node[text=drawColor,anchor=base,inner sep=0pt, outer sep=0pt, scale=  1.30] at (248.26, 33.60) {Q4};
\end{scope}
\begin{scope}
\path[clip] ( 55.20, 55.20) rectangle (283.08,362.58);
\definecolor{drawColor}{RGB}{211,211,211}

\path[draw=drawColor,line width= 0.4pt,dash pattern=on 1pt off 3pt ,line join=round,line cap=round] ( 90.02, 55.20) -- ( 90.02,362.58);

\path[draw=drawColor,line width= 0.4pt,dash pattern=on 1pt off 3pt ,line join=round,line cap=round] (142.76, 55.20) -- (142.76,362.58);

\path[draw=drawColor,line width= 0.4pt,dash pattern=on 1pt off 3pt ,line join=round,line cap=round] (195.51, 55.20) -- (195.51,362.58);

\path[draw=drawColor,line width= 0.4pt,dash pattern=on 1pt off 3pt ,line join=round,line cap=round] (248.26, 55.20) -- (248.26,362.58);

\path[draw=drawColor,line width= 0.4pt,dash pattern=on 1pt off 3pt ,line join=round,line cap=round] ( 55.20, 88.48) -- (283.08, 88.48);

\path[draw=drawColor,line width= 0.4pt,dash pattern=on 1pt off 3pt ,line join=round,line cap=round] ( 55.20,132.26) -- (283.08,132.26);

\path[draw=drawColor,line width= 0.4pt,dash pattern=on 1pt off 3pt ,line join=round,line cap=round] ( 55.20,176.05) -- (283.08,176.05);

\path[draw=drawColor,line width= 0.4pt,dash pattern=on 1pt off 3pt ,line join=round,line cap=round] ( 55.20,219.83) -- (283.08,219.83);

\path[draw=drawColor,line width= 0.4pt,dash pattern=on 1pt off 3pt ,line join=round,line cap=round] ( 55.20,263.62) -- (283.08,263.62);

\path[draw=drawColor,line width= 0.4pt,dash pattern=on 1pt off 3pt ,line join=round,line cap=round] ( 55.20,307.41) -- (283.08,307.41);

\path[draw=drawColor,line width= 0.4pt,dash pattern=on 1pt off 3pt ,line join=round,line cap=round] ( 55.20,351.19) -- (283.08,351.19);
\definecolor{drawColor}{gray}{0.40}
\definecolor{fillColor}{RGB}{183,183,183}

\path[draw=drawColor,line width= 0.4pt,line join=round,line cap=round,fill=fillColor] (103.20,176.05) --
	(103.20,241.73) --
	( 76.83,241.73) --
	( 76.83,176.05) --
	cycle;

\path[draw=drawColor,line width= 0.4pt,line join=round,line cap=round,fill=fillColor] (155.95,176.05) --
	(155.95,132.26) --
	(129.58,132.26) --
	(129.58,176.05) --
	cycle;

\path[draw=drawColor,line width= 0.4pt,line join=round,line cap=round,fill=fillColor] (208.70,176.05) --
	(208.70,351.19) --
	(182.33,351.19) --
	(182.33,176.05) --
	cycle;

\path[draw=drawColor,line width= 0.4pt,line join=round,line cap=round,fill=fillColor] (261.45,176.05) --
	(261.45, 66.58) --
	(235.08, 66.58) --
	(235.08,176.05) --
	cycle;
\definecolor{drawColor}{RGB}{51,51,255}
\definecolor{fillColor}{RGB}{91,91,255}

\path[draw=drawColor,line width= 0.8pt,line join=round,line cap=round,fill=fillColor] ( 86.72,176.05) --
	( 86.72,176.05) --
	( 80.12,176.05) --
	( 80.12,176.05) --
	cycle;

\path[draw=drawColor,line width= 0.8pt,line join=round,line cap=round,fill=fillColor] (139.47,176.05) --
	(139.47,176.05) --
	(132.87,176.05) --
	(132.87,176.05) --
	cycle;

\path[draw=drawColor,line width= 0.8pt,line join=round,line cap=round,fill=fillColor] (192.22,176.05) --
	(192.22,176.05) --
	(185.62,176.05) --
	(185.62,176.05) --
	cycle;

\path[draw=drawColor,line width= 0.8pt,line join=round,line cap=round,fill=fillColor] (244.97,176.05) --
	(244.97,176.05) --
	(238.37,176.05) --
	(238.37,176.05) --
	cycle;
\definecolor{drawColor}{RGB}{255,51,51}
\definecolor{fillColor}{RGB}{255,91,91}

\path[draw=drawColor,line width= 0.8pt,line join=round,line cap=round,fill=fillColor] ( 99.91,176.05) --
	( 99.91,241.73) --
	( 93.31,241.73) --
	( 93.31,176.05) --
	cycle;
\definecolor{fillColor}{RGB}{141,28,28}

\path[draw=drawColor,line width= 0.8pt,line join=round,line cap=round,fill=fillColor] (152.66,176.05) --
	(152.66,132.26) --
	(146.06,132.26) --
	(146.06,176.05) --
	cycle;
\definecolor{fillColor}{RGB}{255,91,91}

\path[draw=drawColor,line width= 0.8pt,line join=round,line cap=round,fill=fillColor] (205.41,176.05) --
	(205.41,351.19) --
	(198.81,351.19) --
	(198.81,176.05) --
	cycle;
\definecolor{fillColor}{RGB}{141,28,28}

\path[draw=drawColor,line width= 0.8pt,line join=round,line cap=round,fill=fillColor] (258.16,176.05) --
	(258.16, 66.58) --
	(251.56, 66.58) --
	(251.56,176.05) --
	cycle;

\path[] ( 55.20,362.58) rectangle (129.40,268.98);
\definecolor{drawColor}{RGB}{51,51,255}
\definecolor{fillColor}{RGB}{91,91,255}

\path[draw=drawColor,line width= 0.4pt,line join=round,line cap=round,fill=fillColor] ( 62.91,342.99) rectangle ( 70.89,350.97);
\definecolor{fillColor}{RGB}{28,28,141}

\path[draw=drawColor,line width= 0.4pt,line join=round,line cap=round,fill=fillColor] ( 62.91,327.39) rectangle ( 70.89,335.37);
\definecolor{drawColor}{RGB}{255,51,51}
\definecolor{fillColor}{RGB}{255,91,91}

\path[draw=drawColor,line width= 0.4pt,line join=round,line cap=round,fill=fillColor] ( 62.91,311.79) rectangle ( 70.89,319.77);
\definecolor{fillColor}{RGB}{141,28,28}

\path[draw=drawColor,line width= 0.4pt,line join=round,line cap=round,fill=fillColor] ( 62.91,296.19) rectangle ( 70.89,304.17);
\definecolor{drawColor}{gray}{0.40}
\definecolor{fillColor}{RGB}{183,183,183}

\path[draw=drawColor,line width= 0.4pt,line join=round,line cap=round,fill=fillColor] ( 62.91,280.59) rectangle ( 70.89,288.57);
\definecolor{drawColor}{RGB}{0,0,0}

\node[text=drawColor,anchor=base west,inner sep=0pt, outer sep=0pt, scale=  1.30] at ( 78.60,342.50) {DA sell };

\node[text=drawColor,anchor=base west,inner sep=0pt, outer sep=0pt, scale=  1.30] at ( 78.60,326.90) {DA buy};

\node[text=drawColor,anchor=base west,inner sep=0pt, outer sep=0pt, scale=  1.30] at ( 78.60,311.30) {IA sell};

\node[text=drawColor,anchor=base west,inner sep=0pt, outer sep=0pt, scale=  1.30] at ( 78.60,295.70) {IA buy};

\node[text=drawColor,anchor=base west,inner sep=0pt, outer sep=0pt, scale=  1.30] at ( 78.60,280.10) {$\bm{v}$};
\end{scope}
\end{tikzpicture}}}
 \subfloat[$\tau_{\text{DA}}<\tau_{\text{IA}}<2\tau_{\text{DA}}$,\newline \indent $\ \ \ \bsv=(-3,-2,-8,-5)'$]{\resizebox{0.24\textwidth}{!}{
\begin{tikzpicture}[x=1pt,y=1pt]
\definecolor{fillColor}{RGB}{255,255,255}
\path[use as bounding box,fill=fillColor,fill opacity=0.00] (0,0) rectangle (289.08,368.58);
\begin{scope}
\path[clip] (  0.00,  0.00) rectangle (289.08,368.58);
\definecolor{drawColor}{RGB}{0,0,0}

\path[draw=drawColor,line width= 0.4pt,line join=round,line cap=round] ( 55.20, 66.58) -- ( 55.20,351.19);

\path[draw=drawColor,line width= 0.4pt,line join=round,line cap=round] ( 55.20, 66.58) -- ( 49.20, 66.58);

\path[draw=drawColor,line width= 0.4pt,line join=round,line cap=round] ( 55.20,137.74) -- ( 49.20,137.74);

\path[draw=drawColor,line width= 0.4pt,line join=round,line cap=round] ( 55.20,208.89) -- ( 49.20,208.89);

\path[draw=drawColor,line width= 0.4pt,line join=round,line cap=round] ( 55.20,280.04) -- ( 49.20,280.04);

\path[draw=drawColor,line width= 0.4pt,line join=round,line cap=round] ( 55.20,351.19) -- ( 49.20,351.19);

\node[text=drawColor,anchor=base east,inner sep=0pt, outer sep=0pt, scale=  1.30] at ( 43.20, 62.11) {-8};

\node[text=drawColor,anchor=base east,inner sep=0pt, outer sep=0pt, scale=  1.30] at ( 43.20,133.26) {-6};

\node[text=drawColor,anchor=base east,inner sep=0pt, outer sep=0pt, scale=  1.30] at ( 43.20,204.41) {-4};

\node[text=drawColor,anchor=base east,inner sep=0pt, outer sep=0pt, scale=  1.30] at ( 43.20,275.56) {-2};

\node[text=drawColor,anchor=base east,inner sep=0pt, outer sep=0pt, scale=  1.30] at ( 43.20,346.72) {0};

\path[draw=drawColor,line width= 0.4pt,line join=round,line cap=round] ( 55.20, 55.20) --
	(283.08, 55.20) --
	(283.08,362.58) --
	( 55.20,362.58) --
	( 55.20, 55.20);
\end{scope}
\begin{scope}
\path[clip] (  0.00,  0.00) rectangle (289.08,368.58);
\definecolor{drawColor}{RGB}{0,0,0}

\node[text=drawColor,anchor=base,inner sep=0pt, outer sep=0pt, scale=  1.30] at (169.14,  9.60) {Delivery Period};

\node[text=drawColor,rotate= 90.00,anchor=base,inner sep=0pt, outer sep=0pt, scale=  1.30] at ( 16.80,208.89) {Volume};
\end{scope}
\begin{scope}
\path[clip] (  0.00,  0.00) rectangle (289.08,368.58);
\definecolor{drawColor}{RGB}{0,0,0}

\path[draw=drawColor,line width= 0.4pt,line join=round,line cap=round] ( 90.02, 55.20) -- (248.26, 55.20);

\path[draw=drawColor,line width= 0.4pt,line join=round,line cap=round] ( 90.02, 55.20) -- ( 90.02, 49.20);

\path[draw=drawColor,line width= 0.4pt,line join=round,line cap=round] (142.76, 55.20) -- (142.76, 49.20);

\path[draw=drawColor,line width= 0.4pt,line join=round,line cap=round] (195.51, 55.20) -- (195.51, 49.20);

\path[draw=drawColor,line width= 0.4pt,line join=round,line cap=round] (248.26, 55.20) -- (248.26, 49.20);

\node[text=drawColor,anchor=base,inner sep=0pt, outer sep=0pt, scale=  1.30] at ( 90.02, 33.60) {Q1};

\node[text=drawColor,anchor=base,inner sep=0pt, outer sep=0pt, scale=  1.30] at (142.76, 33.60) {Q2};

\node[text=drawColor,anchor=base,inner sep=0pt, outer sep=0pt, scale=  1.30] at (195.51, 33.60) {Q3};

\node[text=drawColor,anchor=base,inner sep=0pt, outer sep=0pt, scale=  1.30] at (248.26, 33.60) {Q4};
\end{scope}
\begin{scope}
\path[clip] ( 55.20, 55.20) rectangle (283.08,362.58);
\definecolor{drawColor}{RGB}{211,211,211}

\path[draw=drawColor,line width= 0.4pt,dash pattern=on 1pt off 3pt ,line join=round,line cap=round] ( 90.02, 55.20) -- ( 90.02,362.58);

\path[draw=drawColor,line width= 0.4pt,dash pattern=on 1pt off 3pt ,line join=round,line cap=round] (142.76, 55.20) -- (142.76,362.58);

\path[draw=drawColor,line width= 0.4pt,dash pattern=on 1pt off 3pt ,line join=round,line cap=round] (195.51, 55.20) -- (195.51,362.58);

\path[draw=drawColor,line width= 0.4pt,dash pattern=on 1pt off 3pt ,line join=round,line cap=round] (248.26, 55.20) -- (248.26,362.58);

\path[draw=drawColor,line width= 0.4pt,dash pattern=on 1pt off 3pt ,line join=round,line cap=round] ( 55.20, 66.58) -- (283.08, 66.58);

\path[draw=drawColor,line width= 0.4pt,dash pattern=on 1pt off 3pt ,line join=round,line cap=round] ( 55.20,137.74) -- (283.08,137.74);

\path[draw=drawColor,line width= 0.4pt,dash pattern=on 1pt off 3pt ,line join=round,line cap=round] ( 55.20,208.89) -- (283.08,208.89);

\path[draw=drawColor,line width= 0.4pt,dash pattern=on 1pt off 3pt ,line join=round,line cap=round] ( 55.20,280.04) -- (283.08,280.04);

\path[draw=drawColor,line width= 0.4pt,dash pattern=on 1pt off 3pt ,line join=round,line cap=round] ( 55.20,351.19) -- (283.08,351.19);
\definecolor{drawColor}{gray}{0.40}
\definecolor{fillColor}{RGB}{183,183,183}

\path[draw=drawColor,line width= 0.4pt,line join=round,line cap=round,fill=fillColor] (103.20,351.19) --
	(103.20,244.46) --
	( 76.83,244.46) --
	( 76.83,351.19) --
	cycle;

\path[draw=drawColor,line width= 0.4pt,line join=round,line cap=round,fill=fillColor] (155.95,351.19) --
	(155.95,280.04) --
	(129.58,280.04) --
	(129.58,351.19) --
	cycle;

\path[draw=drawColor,line width= 0.4pt,line join=round,line cap=round,fill=fillColor] (208.70,351.19) --
	(208.70, 66.58) --
	(182.33, 66.58) --
	(182.33,351.19) --
	cycle;

\path[draw=drawColor,line width= 0.4pt,line join=round,line cap=round,fill=fillColor] (261.45,351.19) --
	(261.45,173.31) --
	(235.08,173.31) --
	(235.08,351.19) --
	cycle;
\definecolor{drawColor}{RGB}{51,51,255}
\definecolor{fillColor}{RGB}{28,28,141}

\path[draw=drawColor,line width= 0.8pt,line join=round,line cap=round,fill=fillColor] ( 86.72,351.19) --
	( 86.72,280.04) --
	( 80.12,280.04) --
	( 80.12,351.19) --
	cycle;

\path[draw=drawColor,line width= 0.8pt,line join=round,line cap=round,fill=fillColor] (139.47,351.19) --
	(139.47,280.04) --
	(132.87,280.04) --
	(132.87,351.19) --
	cycle;

\path[draw=drawColor,line width= 0.8pt,line join=round,line cap=round,fill=fillColor] (192.22,351.19) --
	(192.22,280.04) --
	(185.62,280.04) --
	(185.62,351.19) --
	cycle;

\path[draw=drawColor,line width= 0.8pt,line join=round,line cap=round,fill=fillColor] (244.97,351.19) --
	(244.97,280.04) --
	(238.37,280.04) --
	(238.37,351.19) --
	cycle;
\definecolor{drawColor}{RGB}{255,51,51}
\definecolor{fillColor}{RGB}{141,28,28}

\path[draw=drawColor,line width= 0.8pt,line join=round,line cap=round,fill=fillColor] ( 99.91,280.04) --
	( 99.91,244.46) --
	( 93.31,244.46) --
	( 93.31,280.04) --
	cycle;
\definecolor{fillColor}{RGB}{255,91,91}

\path[draw=drawColor,line width= 0.8pt,line join=round,line cap=round,fill=fillColor] (152.66,280.04) --
	(152.66,280.04) --
	(146.06,280.04) --
	(146.06,280.04) --
	cycle;
\definecolor{fillColor}{RGB}{141,28,28}

\path[draw=drawColor,line width= 0.8pt,line join=round,line cap=round,fill=fillColor] (205.41,280.04) --
	(205.41, 66.58) --
	(198.81, 66.58) --
	(198.81,280.04) --
	cycle;

\path[draw=drawColor,line width= 0.8pt,line join=round,line cap=round,fill=fillColor] (258.16,280.04) --
	(258.16,173.31) --
	(251.56,173.31) --
	(251.56,280.04) --
	cycle;

\path[] ( 55.20,148.80) rectangle (129.40, 55.20);
\definecolor{drawColor}{RGB}{51,51,255}
\definecolor{fillColor}{RGB}{91,91,255}

\path[draw=drawColor,line width= 0.4pt,line join=round,line cap=round,fill=fillColor] ( 62.91,129.21) rectangle ( 70.89,137.19);
\definecolor{fillColor}{RGB}{28,28,141}

\path[draw=drawColor,line width= 0.4pt,line join=round,line cap=round,fill=fillColor] ( 62.91,113.61) rectangle ( 70.89,121.59);
\definecolor{drawColor}{RGB}{255,51,51}
\definecolor{fillColor}{RGB}{255,91,91}

\path[draw=drawColor,line width= 0.4pt,line join=round,line cap=round,fill=fillColor] ( 62.91, 98.01) rectangle ( 70.89,105.99);
\definecolor{fillColor}{RGB}{141,28,28}

\path[draw=drawColor,line width= 0.4pt,line join=round,line cap=round,fill=fillColor] ( 62.91, 82.41) rectangle ( 70.89, 90.39);
\definecolor{drawColor}{gray}{0.40}
\definecolor{fillColor}{RGB}{183,183,183}

\path[draw=drawColor,line width= 0.4pt,line join=round,line cap=round,fill=fillColor] ( 62.91, 66.81) rectangle ( 70.89, 74.79);
\definecolor{drawColor}{RGB}{0,0,0}

\node[text=drawColor,anchor=base west,inner sep=0pt, outer sep=0pt, scale=  1.30] at ( 78.60,128.72) {DA sell };

\node[text=drawColor,anchor=base west,inner sep=0pt, outer sep=0pt, scale=  1.30] at ( 78.60,113.12) {DA buy};

\node[text=drawColor,anchor=base west,inner sep=0pt, outer sep=0pt, scale=  1.30] at ( 78.60, 97.52) {IA sell};

\node[text=drawColor,anchor=base west,inner sep=0pt, outer sep=0pt, scale=  1.30] at ( 78.60, 81.92) {IA buy};

\node[text=drawColor,anchor=base west,inner sep=0pt, outer sep=0pt, scale=  1.30] at ( 78.60, 66.32) {$\bm{v}$};
\end{scope}
\end{tikzpicture}}}\\
  \subfloat[$2\tau_{\text{DA}} \leq \tau_{\text{IA}}$,\newline \indent $\quad \ \ \bsv=(2,3,5,8)'$]{\resizebox{0.24\textwidth}{!}{
\begin{tikzpicture}[x=1pt,y=1pt]
\definecolor{fillColor}{RGB}{255,255,255}
\path[use as bounding box,fill=fillColor,fill opacity=0.00] (0,0) rectangle (289.08,368.58);
\begin{scope}
\path[clip] ( 55.20, 55.20) rectangle (283.08,362.58);
\definecolor{drawColor}{RGB}{255,255,255}

\path[draw=drawColor,line width= 0.4pt,line join=round,line cap=round] ( 90.02,102.16) circle (  2.25);

\path[draw=drawColor,line width= 0.4pt,line join=round,line cap=round] (142.76,137.74) circle (  2.25);

\path[draw=drawColor,line width= 0.4pt,line join=round,line cap=round] (195.51,173.31) circle (  2.25);

\path[draw=drawColor,line width= 0.4pt,line join=round,line cap=round] (248.26,208.89) circle (  2.25);
\end{scope}
\begin{scope}
\path[clip] (  0.00,  0.00) rectangle (289.08,368.58);
\definecolor{drawColor}{RGB}{0,0,0}

\path[draw=drawColor,line width= 0.4pt,line join=round,line cap=round] ( 55.20, 66.58) -- ( 55.20,351.19);

\path[draw=drawColor,line width= 0.4pt,line join=round,line cap=round] ( 55.20, 66.58) -- ( 49.20, 66.58);

\path[draw=drawColor,line width= 0.4pt,line join=round,line cap=round] ( 55.20,137.74) -- ( 49.20,137.74);

\path[draw=drawColor,line width= 0.4pt,line join=round,line cap=round] ( 55.20,208.89) -- ( 49.20,208.89);

\path[draw=drawColor,line width= 0.4pt,line join=round,line cap=round] ( 55.20,280.04) -- ( 49.20,280.04);

\path[draw=drawColor,line width= 0.4pt,line join=round,line cap=round] ( 55.20,351.19) -- ( 49.20,351.19);

\node[text=drawColor,anchor=base east,inner sep=0pt, outer sep=0pt, scale=  1.30] at ( 43.20, 62.11) {0};

\node[text=drawColor,anchor=base east,inner sep=0pt, outer sep=0pt, scale=  1.30] at ( 43.20,133.26) {2};

\node[text=drawColor,anchor=base east,inner sep=0pt, outer sep=0pt, scale=  1.30] at ( 43.20,204.41) {4};

\node[text=drawColor,anchor=base east,inner sep=0pt, outer sep=0pt, scale=  1.30] at ( 43.20,275.56) {6};

\node[text=drawColor,anchor=base east,inner sep=0pt, outer sep=0pt, scale=  1.30] at ( 43.20,346.72) {8};

\path[draw=drawColor,line width= 0.4pt,line join=round,line cap=round] ( 55.20, 55.20) --
	(283.08, 55.20) --
	(283.08,362.58) --
	( 55.20,362.58) --
	( 55.20, 55.20);
\end{scope}
\begin{scope}
\path[clip] (  0.00,  0.00) rectangle (289.08,368.58);
\definecolor{drawColor}{RGB}{0,0,0}

\node[text=drawColor,anchor=base,inner sep=0pt, outer sep=0pt, scale=  1.30] at (169.14,  9.60) {Delivery Period};

\node[text=drawColor,rotate= 90.00,anchor=base,inner sep=0pt, outer sep=0pt, scale=  1.30] at ( 16.80,208.89) {Volume};
\end{scope}
\begin{scope}
\path[clip] (  0.00,  0.00) rectangle (289.08,368.58);
\definecolor{drawColor}{RGB}{0,0,0}

\path[draw=drawColor,line width= 0.4pt,line join=round,line cap=round] ( 90.02, 55.20) -- (248.26, 55.20);

\path[draw=drawColor,line width= 0.4pt,line join=round,line cap=round] ( 90.02, 55.20) -- ( 90.02, 49.20);

\path[draw=drawColor,line width= 0.4pt,line join=round,line cap=round] (142.76, 55.20) -- (142.76, 49.20);

\path[draw=drawColor,line width= 0.4pt,line join=round,line cap=round] (195.51, 55.20) -- (195.51, 49.20);

\path[draw=drawColor,line width= 0.4pt,line join=round,line cap=round] (248.26, 55.20) -- (248.26, 49.20);

\node[text=drawColor,anchor=base,inner sep=0pt, outer sep=0pt, scale=  1.30] at ( 90.02, 33.60) {Q1};

\node[text=drawColor,anchor=base,inner sep=0pt, outer sep=0pt, scale=  1.30] at (142.76, 33.60) {Q2};

\node[text=drawColor,anchor=base,inner sep=0pt, outer sep=0pt, scale=  1.30] at (195.51, 33.60) {Q3};

\node[text=drawColor,anchor=base,inner sep=0pt, outer sep=0pt, scale=  1.30] at (248.26, 33.60) {Q4};
\end{scope}
\begin{scope}
\path[clip] ( 55.20, 55.20) rectangle (283.08,362.58);
\definecolor{drawColor}{RGB}{211,211,211}

\path[draw=drawColor,line width= 0.4pt,dash pattern=on 1pt off 3pt ,line join=round,line cap=round] ( 90.02, 55.20) -- ( 90.02,362.58);

\path[draw=drawColor,line width= 0.4pt,dash pattern=on 1pt off 3pt ,line join=round,line cap=round] (142.76, 55.20) -- (142.76,362.58);

\path[draw=drawColor,line width= 0.4pt,dash pattern=on 1pt off 3pt ,line join=round,line cap=round] (195.51, 55.20) -- (195.51,362.58);

\path[draw=drawColor,line width= 0.4pt,dash pattern=on 1pt off 3pt ,line join=round,line cap=round] (248.26, 55.20) -- (248.26,362.58);

\path[draw=drawColor,line width= 0.4pt,dash pattern=on 1pt off 3pt ,line join=round,line cap=round] ( 55.20, 66.58) -- (283.08, 66.58);

\path[draw=drawColor,line width= 0.4pt,dash pattern=on 1pt off 3pt ,line join=round,line cap=round] ( 55.20,137.74) -- (283.08,137.74);

\path[draw=drawColor,line width= 0.4pt,dash pattern=on 1pt off 3pt ,line join=round,line cap=round] ( 55.20,208.89) -- (283.08,208.89);

\path[draw=drawColor,line width= 0.4pt,dash pattern=on 1pt off 3pt ,line join=round,line cap=round] ( 55.20,280.04) -- (283.08,280.04);

\path[draw=drawColor,line width= 0.4pt,dash pattern=on 1pt off 3pt ,line join=round,line cap=round] ( 55.20,351.19) -- (283.08,351.19);
\definecolor{drawColor}{gray}{0.40}
\definecolor{fillColor}{RGB}{183,183,183}

\path[draw=drawColor,line width= 0.4pt,line join=round,line cap=round,fill=fillColor] (103.20, 66.58) --
	(103.20,137.74) --
	( 76.83,137.74) --
	( 76.83, 66.58) --
	cycle;

\path[draw=drawColor,line width= 0.4pt,line join=round,line cap=round,fill=fillColor] (155.95, 66.58) --
	(155.95,173.31) --
	(129.58,173.31) --
	(129.58, 66.58) --
	cycle;

\path[draw=drawColor,line width= 0.4pt,line join=round,line cap=round,fill=fillColor] (208.70, 66.58) --
	(208.70,244.46) --
	(182.33,244.46) --
	(182.33, 66.58) --
	cycle;

\path[draw=drawColor,line width= 0.4pt,line join=round,line cap=round,fill=fillColor] (261.45, 66.58) --
	(261.45,351.19) --
	(235.08,351.19) --
	(235.08, 66.58) --
	cycle;
\definecolor{drawColor}{RGB}{51,51,255}
\definecolor{fillColor}{RGB}{91,91,255}

\path[draw=drawColor,line width= 0.8pt,line join=round,line cap=round,fill=fillColor] ( 86.72, 66.58) --
	( 86.72,173.31) --
	( 80.12,173.31) --
	( 80.12, 66.58) --
	cycle;

\path[draw=drawColor,line width= 0.8pt,line join=round,line cap=round,fill=fillColor] (139.47, 66.58) --
	(139.47,173.31) --
	(132.87,173.31) --
	(132.87, 66.58) --
	cycle;

\path[draw=drawColor,line width= 0.8pt,line join=round,line cap=round,fill=fillColor] (192.22, 66.58) --
	(192.22,173.31) --
	(185.62,173.31) --
	(185.62, 66.58) --
	cycle;

\path[draw=drawColor,line width= 0.8pt,line join=round,line cap=round,fill=fillColor] (244.97, 66.58) --
	(244.97,173.31) --
	(238.37,173.31) --
	(238.37, 66.58) --
	cycle;
\definecolor{drawColor}{RGB}{255,51,51}
\definecolor{fillColor}{RGB}{141,28,28}

\path[draw=drawColor,line width= 0.8pt,line join=round,line cap=round,fill=fillColor] ( 99.91,173.31) --
	( 99.91,137.74) --
	( 93.31,137.74) --
	( 93.31,173.31) --
	cycle;
\definecolor{fillColor}{RGB}{255,91,91}

\path[draw=drawColor,line width= 0.8pt,line join=round,line cap=round,fill=fillColor] (152.66,173.31) --
	(152.66,173.31) --
	(146.06,173.31) --
	(146.06,173.31) --
	cycle;

\path[draw=drawColor,line width= 0.8pt,line join=round,line cap=round,fill=fillColor] (205.41,173.31) --
	(205.41,244.46) --
	(198.81,244.46) --
	(198.81,173.31) --
	cycle;

\path[draw=drawColor,line width= 0.8pt,line join=round,line cap=round,fill=fillColor] (258.16,173.31) --
	(258.16,351.19) --
	(251.56,351.19) --
	(251.56,173.31) --
	cycle;

\path[] ( 55.20,362.58) rectangle (129.40,268.98);
\definecolor{drawColor}{RGB}{51,51,255}
\definecolor{fillColor}{RGB}{91,91,255}

\path[draw=drawColor,line width= 0.4pt,line join=round,line cap=round,fill=fillColor] ( 62.91,342.99) rectangle ( 70.89,350.97);
\definecolor{fillColor}{RGB}{28,28,141}

\path[draw=drawColor,line width= 0.4pt,line join=round,line cap=round,fill=fillColor] ( 62.91,327.39) rectangle ( 70.89,335.37);
\definecolor{drawColor}{RGB}{255,51,51}
\definecolor{fillColor}{RGB}{255,91,91}

\path[draw=drawColor,line width= 0.4pt,line join=round,line cap=round,fill=fillColor] ( 62.91,311.79) rectangle ( 70.89,319.77);
\definecolor{fillColor}{RGB}{141,28,28}

\path[draw=drawColor,line width= 0.4pt,line join=round,line cap=round,fill=fillColor] ( 62.91,296.19) rectangle ( 70.89,304.17);
\definecolor{drawColor}{gray}{0.40}
\definecolor{fillColor}{RGB}{183,183,183}

\path[draw=drawColor,line width= 0.4pt,line join=round,line cap=round,fill=fillColor] ( 62.91,280.59) rectangle ( 70.89,288.57);
\definecolor{drawColor}{RGB}{0,0,0}

\node[text=drawColor,anchor=base west,inner sep=0pt, outer sep=0pt, scale=  1.30] at ( 78.60,342.50) {DA sell };

\node[text=drawColor,anchor=base west,inner sep=0pt, outer sep=0pt, scale=  1.30] at ( 78.60,326.90) {DA buy};

\node[text=drawColor,anchor=base west,inner sep=0pt, outer sep=0pt, scale=  1.30] at ( 78.60,311.30) {IA sell};

\node[text=drawColor,anchor=base west,inner sep=0pt, outer sep=0pt, scale=  1.30] at ( 78.60,295.70) {IA buy};

\node[text=drawColor,anchor=base west,inner sep=0pt, outer sep=0pt, scale=  1.30] at ( 78.60,280.10) {$\bm{v}$};
\end{scope}
\end{tikzpicture}}}
 \subfloat[$2\tau_{\text{DA}} \leq \tau_{\text{IA}}$,\newline \indent $\quad \ \ \bsv=(3,-2,8,5)'$]{\resizebox{0.24\textwidth}{!}{
\begin{tikzpicture}[x=1pt,y=1pt]
\definecolor{fillColor}{RGB}{255,255,255}
\path[use as bounding box,fill=fillColor,fill opacity=0.00] (0,0) rectangle (289.08,368.58);
\begin{scope}
\path[clip] ( 55.20, 55.20) rectangle (283.08,362.58);
\definecolor{drawColor}{RGB}{255,255,255}

\path[draw=drawColor,line width= 0.4pt,line join=round,line cap=round] ( 90.02,151.97) circle (  2.25);

\path[draw=drawColor,line width= 0.4pt,line join=round,line cap=round] (142.76,180.43) circle (  2.25);

\path[draw=drawColor,line width= 0.4pt,line join=round,line cap=round] (195.51,208.89) circle (  2.25);

\path[draw=drawColor,line width= 0.4pt,line join=round,line cap=round] (248.26,237.35) circle (  2.25);
\end{scope}
\begin{scope}
\path[clip] (  0.00,  0.00) rectangle (289.08,368.58);
\definecolor{drawColor}{RGB}{0,0,0}

\path[draw=drawColor,line width= 0.4pt,line join=round,line cap=round] ( 55.20, 66.58) -- ( 55.20,351.19);

\path[draw=drawColor,line width= 0.4pt,line join=round,line cap=round] ( 55.20, 66.58) -- ( 49.20, 66.58);

\path[draw=drawColor,line width= 0.4pt,line join=round,line cap=round] ( 55.20,123.51) -- ( 49.20,123.51);

\path[draw=drawColor,line width= 0.4pt,line join=round,line cap=round] ( 55.20,180.43) -- ( 49.20,180.43);

\path[draw=drawColor,line width= 0.4pt,line join=round,line cap=round] ( 55.20,237.35) -- ( 49.20,237.35);

\path[draw=drawColor,line width= 0.4pt,line join=round,line cap=round] ( 55.20,294.27) -- ( 49.20,294.27);

\path[draw=drawColor,line width= 0.4pt,line join=round,line cap=round] ( 55.20,351.19) -- ( 49.20,351.19);

\node[text=drawColor,anchor=base east,inner sep=0pt, outer sep=0pt, scale=  1.30] at ( 43.20, 62.11) {-2};

\node[text=drawColor,anchor=base east,inner sep=0pt, outer sep=0pt, scale=  1.30] at ( 43.20,119.03) {0};

\node[text=drawColor,anchor=base east,inner sep=0pt, outer sep=0pt, scale=  1.30] at ( 43.20,175.95) {2};

\node[text=drawColor,anchor=base east,inner sep=0pt, outer sep=0pt, scale=  1.30] at ( 43.20,232.87) {4};

\node[text=drawColor,anchor=base east,inner sep=0pt, outer sep=0pt, scale=  1.30] at ( 43.20,289.79) {6};

\node[text=drawColor,anchor=base east,inner sep=0pt, outer sep=0pt, scale=  1.30] at ( 43.20,346.72) {8};

\path[draw=drawColor,line width= 0.4pt,line join=round,line cap=round] ( 55.20, 55.20) --
	(283.08, 55.20) --
	(283.08,362.58) --
	( 55.20,362.58) --
	( 55.20, 55.20);
\end{scope}
\begin{scope}
\path[clip] (  0.00,  0.00) rectangle (289.08,368.58);
\definecolor{drawColor}{RGB}{0,0,0}

\node[text=drawColor,anchor=base,inner sep=0pt, outer sep=0pt, scale=  1.30] at (169.14,  9.60) {Delivery Period};

\node[text=drawColor,rotate= 90.00,anchor=base,inner sep=0pt, outer sep=0pt, scale=  1.30] at ( 16.80,208.89) {Volume};
\end{scope}
\begin{scope}
\path[clip] (  0.00,  0.00) rectangle (289.08,368.58);
\definecolor{drawColor}{RGB}{0,0,0}

\path[draw=drawColor,line width= 0.4pt,line join=round,line cap=round] ( 90.02, 55.20) -- (248.26, 55.20);

\path[draw=drawColor,line width= 0.4pt,line join=round,line cap=round] ( 90.02, 55.20) -- ( 90.02, 49.20);

\path[draw=drawColor,line width= 0.4pt,line join=round,line cap=round] (142.76, 55.20) -- (142.76, 49.20);

\path[draw=drawColor,line width= 0.4pt,line join=round,line cap=round] (195.51, 55.20) -- (195.51, 49.20);

\path[draw=drawColor,line width= 0.4pt,line join=round,line cap=round] (248.26, 55.20) -- (248.26, 49.20);

\node[text=drawColor,anchor=base,inner sep=0pt, outer sep=0pt, scale=  1.30] at ( 90.02, 33.60) {Q1};

\node[text=drawColor,anchor=base,inner sep=0pt, outer sep=0pt, scale=  1.30] at (142.76, 33.60) {Q2};

\node[text=drawColor,anchor=base,inner sep=0pt, outer sep=0pt, scale=  1.30] at (195.51, 33.60) {Q3};

\node[text=drawColor,anchor=base,inner sep=0pt, outer sep=0pt, scale=  1.30] at (248.26, 33.60) {Q4};
\end{scope}
\begin{scope}
\path[clip] ( 55.20, 55.20) rectangle (283.08,362.58);
\definecolor{drawColor}{RGB}{211,211,211}

\path[draw=drawColor,line width= 0.4pt,dash pattern=on 1pt off 3pt ,line join=round,line cap=round] ( 90.02, 55.20) -- ( 90.02,362.58);

\path[draw=drawColor,line width= 0.4pt,dash pattern=on 1pt off 3pt ,line join=round,line cap=round] (142.76, 55.20) -- (142.76,362.58);

\path[draw=drawColor,line width= 0.4pt,dash pattern=on 1pt off 3pt ,line join=round,line cap=round] (195.51, 55.20) -- (195.51,362.58);

\path[draw=drawColor,line width= 0.4pt,dash pattern=on 1pt off 3pt ,line join=round,line cap=round] (248.26, 55.20) -- (248.26,362.58);

\path[draw=drawColor,line width= 0.4pt,dash pattern=on 1pt off 3pt ,line join=round,line cap=round] ( 55.20, 66.58) -- (283.08, 66.58);

\path[draw=drawColor,line width= 0.4pt,dash pattern=on 1pt off 3pt ,line join=round,line cap=round] ( 55.20,123.51) -- (283.08,123.51);

\path[draw=drawColor,line width= 0.4pt,dash pattern=on 1pt off 3pt ,line join=round,line cap=round] ( 55.20,180.43) -- (283.08,180.43);

\path[draw=drawColor,line width= 0.4pt,dash pattern=on 1pt off 3pt ,line join=round,line cap=round] ( 55.20,237.35) -- (283.08,237.35);

\path[draw=drawColor,line width= 0.4pt,dash pattern=on 1pt off 3pt ,line join=round,line cap=round] ( 55.20,294.27) -- (283.08,294.27);

\path[draw=drawColor,line width= 0.4pt,dash pattern=on 1pt off 3pt ,line join=round,line cap=round] ( 55.20,351.19) -- (283.08,351.19);
\definecolor{drawColor}{gray}{0.40}
\definecolor{fillColor}{RGB}{183,183,183}

\path[draw=drawColor,line width= 0.4pt,line join=round,line cap=round,fill=fillColor] (103.20,123.51) --
	(103.20,208.89) --
	( 76.83,208.89) --
	( 76.83,123.51) --
	cycle;

\path[draw=drawColor,line width= 0.4pt,line join=round,line cap=round,fill=fillColor] (155.95,123.51) --
	(155.95, 66.58) --
	(129.58, 66.58) --
	(129.58,123.51) --
	cycle;

\path[draw=drawColor,line width= 0.4pt,line join=round,line cap=round,fill=fillColor] (208.70,123.51) --
	(208.70,351.19) --
	(182.33,351.19) --
	(182.33,123.51) --
	cycle;

\path[draw=drawColor,line width= 0.4pt,line join=round,line cap=round,fill=fillColor] (261.45,123.51) --
	(261.45,265.81) --
	(235.08,265.81) --
	(235.08,123.51) --
	cycle;
\definecolor{drawColor}{RGB}{51,51,255}
\definecolor{fillColor}{RGB}{91,91,255}

\path[draw=drawColor,line width= 0.8pt,line join=round,line cap=round,fill=fillColor] ( 86.72,123.51) --
	( 86.72,208.89) --
	( 80.12,208.89) --
	( 80.12,123.51) --
	cycle;

\path[draw=drawColor,line width= 0.8pt,line join=round,line cap=round,fill=fillColor] (139.47,123.51) --
	(139.47,208.89) --
	(132.87,208.89) --
	(132.87,123.51) --
	cycle;

\path[draw=drawColor,line width= 0.8pt,line join=round,line cap=round,fill=fillColor] (192.22,123.51) --
	(192.22,208.89) --
	(185.62,208.89) --
	(185.62,123.51) --
	cycle;

\path[draw=drawColor,line width= 0.8pt,line join=round,line cap=round,fill=fillColor] (244.97,123.51) --
	(244.97,208.89) --
	(238.37,208.89) --
	(238.37,123.51) --
	cycle;
\definecolor{drawColor}{RGB}{255,51,51}
\definecolor{fillColor}{RGB}{255,91,91}

\path[draw=drawColor,line width= 0.8pt,line join=round,line cap=round,fill=fillColor] ( 99.91,208.89) --
	( 99.91,208.89) --
	( 93.31,208.89) --
	( 93.31,208.89) --
	cycle;
\definecolor{fillColor}{RGB}{141,28,28}

\path[draw=drawColor,line width= 0.8pt,line join=round,line cap=round,fill=fillColor] (152.66,208.89) --
	(152.66, 66.58) --
	(146.06, 66.58) --
	(146.06,208.89) --
	cycle;
\definecolor{fillColor}{RGB}{255,91,91}

\path[draw=drawColor,line width= 0.8pt,line join=round,line cap=round,fill=fillColor] (205.41,208.89) --
	(205.41,351.19) --
	(198.81,351.19) --
	(198.81,208.89) --
	cycle;

\path[draw=drawColor,line width= 0.8pt,line join=round,line cap=round,fill=fillColor] (258.16,208.89) --
	(258.16,265.81) --
	(251.56,265.81) --
	(251.56,208.89) --
	cycle;

\path[] ( 55.20,362.58) rectangle (129.40,268.98);
\definecolor{drawColor}{RGB}{51,51,255}
\definecolor{fillColor}{RGB}{91,91,255}

\path[draw=drawColor,line width= 0.4pt,line join=round,line cap=round,fill=fillColor] ( 62.91,342.99) rectangle ( 70.89,350.97);
\definecolor{fillColor}{RGB}{28,28,141}

\path[draw=drawColor,line width= 0.4pt,line join=round,line cap=round,fill=fillColor] ( 62.91,327.39) rectangle ( 70.89,335.37);
\definecolor{drawColor}{RGB}{255,51,51}
\definecolor{fillColor}{RGB}{255,91,91}

\path[draw=drawColor,line width= 0.4pt,line join=round,line cap=round,fill=fillColor] ( 62.91,311.79) rectangle ( 70.89,319.77);
\definecolor{fillColor}{RGB}{141,28,28}

\path[draw=drawColor,line width= 0.4pt,line join=round,line cap=round,fill=fillColor] ( 62.91,296.19) rectangle ( 70.89,304.17);
\definecolor{drawColor}{gray}{0.40}
\definecolor{fillColor}{RGB}{183,183,183}

\path[draw=drawColor,line width= 0.4pt,line join=round,line cap=round,fill=fillColor] ( 62.91,280.59) rectangle ( 70.89,288.57);
\definecolor{drawColor}{RGB}{0,0,0}

\node[text=drawColor,anchor=base west,inner sep=0pt, outer sep=0pt, scale=  1.30] at ( 78.60,342.50) {DA sell };

\node[text=drawColor,anchor=base west,inner sep=0pt, outer sep=0pt, scale=  1.30] at ( 78.60,326.90) {DA buy};

\node[text=drawColor,anchor=base west,inner sep=0pt, outer sep=0pt, scale=  1.30] at ( 78.60,311.30) {IA sell};

\node[text=drawColor,anchor=base west,inner sep=0pt, outer sep=0pt, scale=  1.30] at ( 78.60,295.70) {IA buy};

\node[text=drawColor,anchor=base west,inner sep=0pt, outer sep=0pt, scale=  1.30] at ( 78.60,280.10) {$\bm{v}$};
\end{scope}
\end{tikzpicture}}}
 \subfloat[$2\tau_{\text{DA}} \leq \tau_{\text{IA}}$,\newline \indent $\quad \ \ \bsv=(3,-2,8,-5)'$]{\resizebox{0.24\textwidth}{!}{
\begin{tikzpicture}[x=1pt,y=1pt]
\definecolor{fillColor}{RGB}{255,255,255}
\path[use as bounding box,fill=fillColor,fill opacity=0.00] (0,0) rectangle (289.08,368.58);
\begin{scope}
\path[clip] ( 55.20, 55.20) rectangle (283.08,362.58);
\definecolor{drawColor}{RGB}{255,255,255}

\path[draw=drawColor,line width= 0.4pt,line join=round,line cap=round] ( 90.02,197.94) circle (  2.25);

\path[draw=drawColor,line width= 0.4pt,line join=round,line cap=round] (142.76,219.83) circle (  2.25);

\path[draw=drawColor,line width= 0.4pt,line join=round,line cap=round] (195.51,241.73) circle (  2.25);

\path[draw=drawColor,line width= 0.4pt,line join=round,line cap=round] (248.26,263.62) circle (  2.25);
\end{scope}
\begin{scope}
\path[clip] (  0.00,  0.00) rectangle (289.08,368.58);
\definecolor{drawColor}{RGB}{0,0,0}

\path[draw=drawColor,line width= 0.4pt,line join=round,line cap=round] ( 55.20, 88.48) -- ( 55.20,351.19);

\path[draw=drawColor,line width= 0.4pt,line join=round,line cap=round] ( 55.20, 88.48) -- ( 49.20, 88.48);

\path[draw=drawColor,line width= 0.4pt,line join=round,line cap=round] ( 55.20,132.26) -- ( 49.20,132.26);

\path[draw=drawColor,line width= 0.4pt,line join=round,line cap=round] ( 55.20,176.05) -- ( 49.20,176.05);

\path[draw=drawColor,line width= 0.4pt,line join=round,line cap=round] ( 55.20,219.83) -- ( 49.20,219.83);

\path[draw=drawColor,line width= 0.4pt,line join=round,line cap=round] ( 55.20,263.62) -- ( 49.20,263.62);

\path[draw=drawColor,line width= 0.4pt,line join=round,line cap=round] ( 55.20,307.41) -- ( 49.20,307.41);

\path[draw=drawColor,line width= 0.4pt,line join=round,line cap=round] ( 55.20,351.19) -- ( 49.20,351.19);

\node[text=drawColor,anchor=base east,inner sep=0pt, outer sep=0pt, scale=  1.30] at ( 43.20, 84.00) {-4};

\node[text=drawColor,anchor=base east,inner sep=0pt, outer sep=0pt, scale=  1.30] at ( 43.20,127.79) {-2};

\node[text=drawColor,anchor=base east,inner sep=0pt, outer sep=0pt, scale=  1.30] at ( 43.20,171.57) {0};

\node[text=drawColor,anchor=base east,inner sep=0pt, outer sep=0pt, scale=  1.30] at ( 43.20,215.36) {2};

\node[text=drawColor,anchor=base east,inner sep=0pt, outer sep=0pt, scale=  1.30] at ( 43.20,259.14) {4};

\node[text=drawColor,anchor=base east,inner sep=0pt, outer sep=0pt, scale=  1.30] at ( 43.20,302.93) {6};

\node[text=drawColor,anchor=base east,inner sep=0pt, outer sep=0pt, scale=  1.30] at ( 43.20,346.72) {8};

\path[draw=drawColor,line width= 0.4pt,line join=round,line cap=round] ( 55.20, 55.20) --
	(283.08, 55.20) --
	(283.08,362.58) --
	( 55.20,362.58) --
	( 55.20, 55.20);
\end{scope}
\begin{scope}
\path[clip] (  0.00,  0.00) rectangle (289.08,368.58);
\definecolor{drawColor}{RGB}{0,0,0}

\node[text=drawColor,anchor=base,inner sep=0pt, outer sep=0pt, scale=  1.30] at (169.14,  9.60) {Delivery Period};

\node[text=drawColor,rotate= 90.00,anchor=base,inner sep=0pt, outer sep=0pt, scale=  1.30] at ( 16.80,208.89) {Volume};
\end{scope}
\begin{scope}
\path[clip] (  0.00,  0.00) rectangle (289.08,368.58);
\definecolor{drawColor}{RGB}{0,0,0}

\path[draw=drawColor,line width= 0.4pt,line join=round,line cap=round] ( 90.02, 55.20) -- (248.26, 55.20);

\path[draw=drawColor,line width= 0.4pt,line join=round,line cap=round] ( 90.02, 55.20) -- ( 90.02, 49.20);

\path[draw=drawColor,line width= 0.4pt,line join=round,line cap=round] (142.76, 55.20) -- (142.76, 49.20);

\path[draw=drawColor,line width= 0.4pt,line join=round,line cap=round] (195.51, 55.20) -- (195.51, 49.20);

\path[draw=drawColor,line width= 0.4pt,line join=round,line cap=round] (248.26, 55.20) -- (248.26, 49.20);

\node[text=drawColor,anchor=base,inner sep=0pt, outer sep=0pt, scale=  1.30] at ( 90.02, 33.60) {Q1};

\node[text=drawColor,anchor=base,inner sep=0pt, outer sep=0pt, scale=  1.30] at (142.76, 33.60) {Q2};

\node[text=drawColor,anchor=base,inner sep=0pt, outer sep=0pt, scale=  1.30] at (195.51, 33.60) {Q3};

\node[text=drawColor,anchor=base,inner sep=0pt, outer sep=0pt, scale=  1.30] at (248.26, 33.60) {Q4};
\end{scope}
\begin{scope}
\path[clip] ( 55.20, 55.20) rectangle (283.08,362.58);
\definecolor{drawColor}{RGB}{211,211,211}

\path[draw=drawColor,line width= 0.4pt,dash pattern=on 1pt off 3pt ,line join=round,line cap=round] ( 90.02, 55.20) -- ( 90.02,362.58);

\path[draw=drawColor,line width= 0.4pt,dash pattern=on 1pt off 3pt ,line join=round,line cap=round] (142.76, 55.20) -- (142.76,362.58);

\path[draw=drawColor,line width= 0.4pt,dash pattern=on 1pt off 3pt ,line join=round,line cap=round] (195.51, 55.20) -- (195.51,362.58);

\path[draw=drawColor,line width= 0.4pt,dash pattern=on 1pt off 3pt ,line join=round,line cap=round] (248.26, 55.20) -- (248.26,362.58);

\path[draw=drawColor,line width= 0.4pt,dash pattern=on 1pt off 3pt ,line join=round,line cap=round] ( 55.20, 88.48) -- (283.08, 88.48);

\path[draw=drawColor,line width= 0.4pt,dash pattern=on 1pt off 3pt ,line join=round,line cap=round] ( 55.20,132.26) -- (283.08,132.26);

\path[draw=drawColor,line width= 0.4pt,dash pattern=on 1pt off 3pt ,line join=round,line cap=round] ( 55.20,176.05) -- (283.08,176.05);

\path[draw=drawColor,line width= 0.4pt,dash pattern=on 1pt off 3pt ,line join=round,line cap=round] ( 55.20,219.83) -- (283.08,219.83);

\path[draw=drawColor,line width= 0.4pt,dash pattern=on 1pt off 3pt ,line join=round,line cap=round] ( 55.20,263.62) -- (283.08,263.62);

\path[draw=drawColor,line width= 0.4pt,dash pattern=on 1pt off 3pt ,line join=round,line cap=round] ( 55.20,307.41) -- (283.08,307.41);

\path[draw=drawColor,line width= 0.4pt,dash pattern=on 1pt off 3pt ,line join=round,line cap=round] ( 55.20,351.19) -- (283.08,351.19);
\definecolor{drawColor}{gray}{0.40}
\definecolor{fillColor}{RGB}{183,183,183}

\path[draw=drawColor,line width= 0.4pt,line join=round,line cap=round,fill=fillColor] (103.20,176.05) --
	(103.20,241.73) --
	( 76.83,241.73) --
	( 76.83,176.05) --
	cycle;

\path[draw=drawColor,line width= 0.4pt,line join=round,line cap=round,fill=fillColor] (155.95,176.05) --
	(155.95,132.26) --
	(129.58,132.26) --
	(129.58,176.05) --
	cycle;

\path[draw=drawColor,line width= 0.4pt,line join=round,line cap=round,fill=fillColor] (208.70,176.05) --
	(208.70,351.19) --
	(182.33,351.19) --
	(182.33,176.05) --
	cycle;

\path[draw=drawColor,line width= 0.4pt,line join=round,line cap=round,fill=fillColor] (261.45,176.05) --
	(261.45, 66.58) --
	(235.08, 66.58) --
	(235.08,176.05) --
	cycle;
\definecolor{drawColor}{RGB}{51,51,255}
\definecolor{fillColor}{RGB}{91,91,255}

\path[draw=drawColor,line width= 0.8pt,line join=round,line cap=round,fill=fillColor] ( 86.72,176.05) --
	( 86.72,176.05) --
	( 80.12,176.05) --
	( 80.12,176.05) --
	cycle;

\path[draw=drawColor,line width= 0.8pt,line join=round,line cap=round,fill=fillColor] (139.47,176.05) --
	(139.47,176.05) --
	(132.87,176.05) --
	(132.87,176.05) --
	cycle;

\path[draw=drawColor,line width= 0.8pt,line join=round,line cap=round,fill=fillColor] (192.22,176.05) --
	(192.22,176.05) --
	(185.62,176.05) --
	(185.62,176.05) --
	cycle;

\path[draw=drawColor,line width= 0.8pt,line join=round,line cap=round,fill=fillColor] (244.97,176.05) --
	(244.97,176.05) --
	(238.37,176.05) --
	(238.37,176.05) --
	cycle;
\definecolor{drawColor}{RGB}{255,51,51}
\definecolor{fillColor}{RGB}{255,91,91}

\path[draw=drawColor,line width= 0.8pt,line join=round,line cap=round,fill=fillColor] ( 99.91,176.05) --
	( 99.91,241.73) --
	( 93.31,241.73) --
	( 93.31,176.05) --
	cycle;
\definecolor{fillColor}{RGB}{141,28,28}

\path[draw=drawColor,line width= 0.8pt,line join=round,line cap=round,fill=fillColor] (152.66,176.05) --
	(152.66,132.26) --
	(146.06,132.26) --
	(146.06,176.05) --
	cycle;
\definecolor{fillColor}{RGB}{255,91,91}

\path[draw=drawColor,line width= 0.8pt,line join=round,line cap=round,fill=fillColor] (205.41,176.05) --
	(205.41,351.19) --
	(198.81,351.19) --
	(198.81,176.05) --
	cycle;
\definecolor{fillColor}{RGB}{141,28,28}

\path[draw=drawColor,line width= 0.8pt,line join=round,line cap=round,fill=fillColor] (258.16,176.05) --
	(258.16, 66.58) --
	(251.56, 66.58) --
	(251.56,176.05) --
	cycle;

\path[] ( 55.20,362.58) rectangle (129.40,268.98);
\definecolor{drawColor}{RGB}{51,51,255}
\definecolor{fillColor}{RGB}{91,91,255}

\path[draw=drawColor,line width= 0.4pt,line join=round,line cap=round,fill=fillColor] ( 62.91,342.99) rectangle ( 70.89,350.97);
\definecolor{fillColor}{RGB}{28,28,141}

\path[draw=drawColor,line width= 0.4pt,line join=round,line cap=round,fill=fillColor] ( 62.91,327.39) rectangle ( 70.89,335.37);
\definecolor{drawColor}{RGB}{255,51,51}
\definecolor{fillColor}{RGB}{255,91,91}

\path[draw=drawColor,line width= 0.4pt,line join=round,line cap=round,fill=fillColor] ( 62.91,311.79) rectangle ( 70.89,319.77);
\definecolor{fillColor}{RGB}{141,28,28}

\path[draw=drawColor,line width= 0.4pt,line join=round,line cap=round,fill=fillColor] ( 62.91,296.19) rectangle ( 70.89,304.17);
\definecolor{drawColor}{gray}{0.40}
\definecolor{fillColor}{RGB}{183,183,183}

\path[draw=drawColor,line width= 0.4pt,line join=round,line cap=round,fill=fillColor] ( 62.91,280.59) rectangle ( 70.89,288.57);
\definecolor{drawColor}{RGB}{0,0,0}

\node[text=drawColor,anchor=base west,inner sep=0pt, outer sep=0pt, scale=  1.30] at ( 78.60,342.50) {DA sell };

\node[text=drawColor,anchor=base west,inner sep=0pt, outer sep=0pt, scale=  1.30] at ( 78.60,326.90) {DA buy};

\node[text=drawColor,anchor=base west,inner sep=0pt, outer sep=0pt, scale=  1.30] at ( 78.60,311.30) {IA sell};

\node[text=drawColor,anchor=base west,inner sep=0pt, outer sep=0pt, scale=  1.30] at ( 78.60,295.70) {IA buy};

\node[text=drawColor,anchor=base west,inner sep=0pt, outer sep=0pt, scale=  1.30] at ( 78.60,280.10) {$\bm{v}$};
\end{scope}
\end{tikzpicture}}}
 \subfloat[$2\tau_{\text{DA}} \leq \tau_{\text{IA}}$,\newline \indent $\ \ \ \bsv=(-3,-2,-8,-5)'$]{\resizebox{0.24\textwidth}{!}{
\begin{tikzpicture}[x=1pt,y=1pt]
\definecolor{fillColor}{RGB}{255,255,255}
\path[use as bounding box,fill=fillColor,fill opacity=0.00] (0,0) rectangle (289.08,368.58);
\begin{scope}
\path[clip] (  0.00,  0.00) rectangle (289.08,368.58);
\definecolor{drawColor}{RGB}{0,0,0}

\path[draw=drawColor,line width= 0.4pt,line join=round,line cap=round] ( 55.20, 66.58) -- ( 55.20,351.19);

\path[draw=drawColor,line width= 0.4pt,line join=round,line cap=round] ( 55.20, 66.58) -- ( 49.20, 66.58);

\path[draw=drawColor,line width= 0.4pt,line join=round,line cap=round] ( 55.20,137.74) -- ( 49.20,137.74);

\path[draw=drawColor,line width= 0.4pt,line join=round,line cap=round] ( 55.20,208.89) -- ( 49.20,208.89);

\path[draw=drawColor,line width= 0.4pt,line join=round,line cap=round] ( 55.20,280.04) -- ( 49.20,280.04);

\path[draw=drawColor,line width= 0.4pt,line join=round,line cap=round] ( 55.20,351.19) -- ( 49.20,351.19);

\node[text=drawColor,anchor=base east,inner sep=0pt, outer sep=0pt, scale=  1.30] at ( 43.20, 62.11) {-8};

\node[text=drawColor,anchor=base east,inner sep=0pt, outer sep=0pt, scale=  1.30] at ( 43.20,133.26) {-6};

\node[text=drawColor,anchor=base east,inner sep=0pt, outer sep=0pt, scale=  1.30] at ( 43.20,204.41) {-4};

\node[text=drawColor,anchor=base east,inner sep=0pt, outer sep=0pt, scale=  1.30] at ( 43.20,275.56) {-2};

\node[text=drawColor,anchor=base east,inner sep=0pt, outer sep=0pt, scale=  1.30] at ( 43.20,346.72) {0};

\path[draw=drawColor,line width= 0.4pt,line join=round,line cap=round] ( 55.20, 55.20) --
	(283.08, 55.20) --
	(283.08,362.58) --
	( 55.20,362.58) --
	( 55.20, 55.20);
\end{scope}
\begin{scope}
\path[clip] (  0.00,  0.00) rectangle (289.08,368.58);
\definecolor{drawColor}{RGB}{0,0,0}

\node[text=drawColor,anchor=base,inner sep=0pt, outer sep=0pt, scale=  1.30] at (169.14,  9.60) {Delivery Period};

\node[text=drawColor,rotate= 90.00,anchor=base,inner sep=0pt, outer sep=0pt, scale=  1.30] at ( 16.80,208.89) {Volume};
\end{scope}
\begin{scope}
\path[clip] (  0.00,  0.00) rectangle (289.08,368.58);
\definecolor{drawColor}{RGB}{0,0,0}

\path[draw=drawColor,line width= 0.4pt,line join=round,line cap=round] ( 90.02, 55.20) -- (248.26, 55.20);

\path[draw=drawColor,line width= 0.4pt,line join=round,line cap=round] ( 90.02, 55.20) -- ( 90.02, 49.20);

\path[draw=drawColor,line width= 0.4pt,line join=round,line cap=round] (142.76, 55.20) -- (142.76, 49.20);

\path[draw=drawColor,line width= 0.4pt,line join=round,line cap=round] (195.51, 55.20) -- (195.51, 49.20);

\path[draw=drawColor,line width= 0.4pt,line join=round,line cap=round] (248.26, 55.20) -- (248.26, 49.20);

\node[text=drawColor,anchor=base,inner sep=0pt, outer sep=0pt, scale=  1.30] at ( 90.02, 33.60) {Q1};

\node[text=drawColor,anchor=base,inner sep=0pt, outer sep=0pt, scale=  1.30] at (142.76, 33.60) {Q2};

\node[text=drawColor,anchor=base,inner sep=0pt, outer sep=0pt, scale=  1.30] at (195.51, 33.60) {Q3};

\node[text=drawColor,anchor=base,inner sep=0pt, outer sep=0pt, scale=  1.30] at (248.26, 33.60) {Q4};
\end{scope}
\begin{scope}
\path[clip] ( 55.20, 55.20) rectangle (283.08,362.58);
\definecolor{drawColor}{RGB}{211,211,211}

\path[draw=drawColor,line width= 0.4pt,dash pattern=on 1pt off 3pt ,line join=round,line cap=round] ( 90.02, 55.20) -- ( 90.02,362.58);

\path[draw=drawColor,line width= 0.4pt,dash pattern=on 1pt off 3pt ,line join=round,line cap=round] (142.76, 55.20) -- (142.76,362.58);

\path[draw=drawColor,line width= 0.4pt,dash pattern=on 1pt off 3pt ,line join=round,line cap=round] (195.51, 55.20) -- (195.51,362.58);

\path[draw=drawColor,line width= 0.4pt,dash pattern=on 1pt off 3pt ,line join=round,line cap=round] (248.26, 55.20) -- (248.26,362.58);

\path[draw=drawColor,line width= 0.4pt,dash pattern=on 1pt off 3pt ,line join=round,line cap=round] ( 55.20, 66.58) -- (283.08, 66.58);

\path[draw=drawColor,line width= 0.4pt,dash pattern=on 1pt off 3pt ,line join=round,line cap=round] ( 55.20,137.74) -- (283.08,137.74);

\path[draw=drawColor,line width= 0.4pt,dash pattern=on 1pt off 3pt ,line join=round,line cap=round] ( 55.20,208.89) -- (283.08,208.89);

\path[draw=drawColor,line width= 0.4pt,dash pattern=on 1pt off 3pt ,line join=round,line cap=round] ( 55.20,280.04) -- (283.08,280.04);

\path[draw=drawColor,line width= 0.4pt,dash pattern=on 1pt off 3pt ,line join=round,line cap=round] ( 55.20,351.19) -- (283.08,351.19);
\definecolor{drawColor}{gray}{0.40}
\definecolor{fillColor}{RGB}{183,183,183}

\path[draw=drawColor,line width= 0.4pt,line join=round,line cap=round,fill=fillColor] (103.20,351.19) --
	(103.20,244.46) --
	( 76.83,244.46) --
	( 76.83,351.19) --
	cycle;

\path[draw=drawColor,line width= 0.4pt,line join=round,line cap=round,fill=fillColor] (155.95,351.19) --
	(155.95,280.04) --
	(129.58,280.04) --
	(129.58,351.19) --
	cycle;

\path[draw=drawColor,line width= 0.4pt,line join=round,line cap=round,fill=fillColor] (208.70,351.19) --
	(208.70, 66.58) --
	(182.33, 66.58) --
	(182.33,351.19) --
	cycle;

\path[draw=drawColor,line width= 0.4pt,line join=round,line cap=round,fill=fillColor] (261.45,351.19) --
	(261.45,173.31) --
	(235.08,173.31) --
	(235.08,351.19) --
	cycle;
\definecolor{drawColor}{RGB}{51,51,255}
\definecolor{fillColor}{RGB}{28,28,141}

\path[draw=drawColor,line width= 0.8pt,line join=round,line cap=round,fill=fillColor] ( 86.72,351.19) --
	( 86.72,244.46) --
	( 80.12,244.46) --
	( 80.12,351.19) --
	cycle;

\path[draw=drawColor,line width= 0.8pt,line join=round,line cap=round,fill=fillColor] (139.47,351.19) --
	(139.47,244.46) --
	(132.87,244.46) --
	(132.87,351.19) --
	cycle;

\path[draw=drawColor,line width= 0.8pt,line join=round,line cap=round,fill=fillColor] (192.22,351.19) --
	(192.22,244.46) --
	(185.62,244.46) --
	(185.62,351.19) --
	cycle;

\path[draw=drawColor,line width= 0.8pt,line join=round,line cap=round,fill=fillColor] (244.97,351.19) --
	(244.97,244.46) --
	(238.37,244.46) --
	(238.37,351.19) --
	cycle;
\definecolor{drawColor}{RGB}{255,51,51}
\definecolor{fillColor}{RGB}{255,91,91}

\path[draw=drawColor,line width= 0.8pt,line join=round,line cap=round,fill=fillColor] ( 99.91,244.46) --
	( 99.91,244.46) --
	( 93.31,244.46) --
	( 93.31,244.46) --
	cycle;

\path[draw=drawColor,line width= 0.8pt,line join=round,line cap=round,fill=fillColor] (152.66,244.46) --
	(152.66,280.04) --
	(146.06,280.04) --
	(146.06,244.46) --
	cycle;
\definecolor{fillColor}{RGB}{141,28,28}

\path[draw=drawColor,line width= 0.8pt,line join=round,line cap=round,fill=fillColor] (205.41,244.46) --
	(205.41, 66.58) --
	(198.81, 66.58) --
	(198.81,244.46) --
	cycle;

\path[draw=drawColor,line width= 0.8pt,line join=round,line cap=round,fill=fillColor] (258.16,244.46) --
	(258.16,173.31) --
	(251.56,173.31) --
	(251.56,244.46) --
	cycle;

\path[] ( 55.20,148.80) rectangle (129.40, 55.20);
\definecolor{drawColor}{RGB}{51,51,255}
\definecolor{fillColor}{RGB}{91,91,255}

\path[draw=drawColor,line width= 0.4pt,line join=round,line cap=round,fill=fillColor] ( 62.91,129.21) rectangle ( 70.89,137.19);
\definecolor{fillColor}{RGB}{28,28,141}

\path[draw=drawColor,line width= 0.4pt,line join=round,line cap=round,fill=fillColor] ( 62.91,113.61) rectangle ( 70.89,121.59);
\definecolor{drawColor}{RGB}{255,51,51}
\definecolor{fillColor}{RGB}{255,91,91}

\path[draw=drawColor,line width= 0.4pt,line join=round,line cap=round,fill=fillColor] ( 62.91, 98.01) rectangle ( 70.89,105.99);
\definecolor{fillColor}{RGB}{141,28,28}

\path[draw=drawColor,line width= 0.4pt,line join=round,line cap=round,fill=fillColor] ( 62.91, 82.41) rectangle ( 70.89, 90.39);
\definecolor{drawColor}{gray}{0.40}
\definecolor{fillColor}{RGB}{183,183,183}

\path[draw=drawColor,line width= 0.4pt,line join=round,line cap=round,fill=fillColor] ( 62.91, 66.81) rectangle ( 70.89, 74.79);
\definecolor{drawColor}{RGB}{0,0,0}

\node[text=drawColor,anchor=base west,inner sep=0pt, outer sep=0pt, scale=  1.30] at ( 78.60,128.72) {DA sell };

\node[text=drawColor,anchor=base west,inner sep=0pt, outer sep=0pt, scale=  1.30] at ( 78.60,113.12) {DA buy};

\node[text=drawColor,anchor=base west,inner sep=0pt, outer sep=0pt, scale=  1.30] at ( 78.60, 97.52) {IA sell};

\node[text=drawColor,anchor=base west,inner sep=0pt, outer sep=0pt, scale=  1.30] at ( 78.60, 81.92) {IA buy};

\node[text=drawColor,anchor=base west,inner sep=0pt, outer sep=0pt, scale=  1.30] at ( 78.60, 66.32) {$\bm{v}$};
\end{scope}
\end{tikzpicture}}}
  \caption{Illustration of optimal transaction cost minimal strategies $\bsb$ for different transaction costs $\tau_{\text{DA}}$ and $\tau_{\text{IA}}$, and target trade volume $\bsv$.}
\label{fig_trading_mincost}
  \end{figure}

We see that the minimal transaction costs strategy depends on the transaction costs, so in fact on $\tau_{\text{DA}}$ and $\tau_{\text{IA}}$. For example, it is easy to derive that if we have $\bsv$ with
$v_{i}\leq v_{i+1}$ and $v_1 \geq 0$ then the optimal volume $b_{\text{TC-min},0}$ is 
\begin{align}
 b_{\text{TC-min},0} = \begin{cases}
                   0, & \text{ if } \tau_{\text{DA}} > \tau_{\text{IA}} \\
                   v_1, & \text{ if } \tau_{\text{DA}} \leq \tau_{\text{IA}} < 2 \tau_{\text{DA}} \\
                   v_2, & \text{ if } 2\tau_{\text{DA}} \leq \tau_{\text{IA}}
                  \end{cases}
                  \label{eq_example_tc}
\end{align}
where the limiting cases $\tau_{\text{IA}} = \tau_{\text{DA}}$
and $\tau_{\text{IA}} = 2\tau_{\text{DA}}$  are not unique. 
Then, any element in $[0,v_1]$ and $[v_1,v_2]$ is optimal, respectively.
The first case in \eqref{eq_example_tc} is not realistic in practice. 
For the remaining cases Figure~\ref{fig_trading_mincost} illustrates the optimal bidding strategy for selected volume settings $\bsv$.
Those remaining cases in \eqref{eq_example_tc} are realistic and may occur in practice. 
A trader with relatively cheap IA transaction trading costs that faces case 2 should trade according to the minimum transaction cost strategy only the lowest production among the four quarter hours. This would avoid any buy transactions in the IA market.

\subsection{Optimal expected profit trading}
\label{sec:optimal_expected_profit_no_imp}
In this section, we analyse the optimal strategy for a risk neutral trader which maximizes the objective 
\eqref{eq_risk_optimization}
with respect to the expectation of $\RR = \E$.
In this case, for equation~\eqref{eq_gain_gen_transaction_costs_decomp_interpret} we receive
\begin{equation}
\begin{aligned}
\E[G(\widetilde{\bsb})]
=& 
\underbrace{
\frac{1}{4} \sum_{i=1}^4  \E[P^*_{\bsnull,i}]v_i}_{\text{Expected IA revenue}}  + \underbrace{\left( \E[P^*_{\bsnull,0}]  - \frac{1}{4}\sum_{i=1}^4 \E[P^*_{\bsnull,i}]\right) b_0 }_{ \text{Expected DA-IA arbitrage}}  \\
& + \underbrace{\E[\Delta_{\widetilde{\bsb},0}] b_0
+ \frac{1}{4} \sum_{i=1}^4 \E[\Delta_{\widetilde{\bsb},i}] (v_i-b_0)
 }_{ \text{Expected DA\&IA market impact} }
  -   \underbrace{\TT(b_0).}_{\text{Transaction costs}}
  \label{eq_expected_gain_gen_transaction_costs_decomp_interpret}
\end{aligned}
\end{equation}
This decomposition into four interpretable components
corresponds to the expectations in \eqref{eq_gain_gen_transaction_costs_decomp_interpret}.
Only the transactions costs $\TT$ as studied in Section~\ref{subsec_min_trans} remain untouched.

The second term in~\eqref{eq_expected_gain_gen_transaction_costs_decomp_interpret} is the expected arbitrage opportunity between the day-ahead and the intraday auctions. If the price difference for trading 1 MWh
$\MM(\bsP^*_{\bsnull}) = \E[P^*_{\bsnull,0}] - \frac{1}{4}\sum_{i=1}^4 \E[P^*_{\bsnull,i}]$ is large, there are large arbitrage opportunities. In efficient markets
the \emph{market efficiency assumption}
\begin{align}
\MM(\bsP_{\bsnull}^*)=0\Leftrightarrow  \E[P_{\bsb,0}^*] = \frac{1}{4}\sum_{i=1}^4 \E[P_{\bsb,i}^*]
\label{eq_market_efficient}
\end{align}
holds. Thus, if \eqref{eq_market_efficient} holds the trader cannot expect any improvement of the expected gain $\E[G]$ from the second term.
The third term in   \eqref{eq_expected_gain_gen_transaction_costs_decomp_interpret}
represents the market impact due to the trading. If $\bsv$ is close to $\bsnull$ then we do not expect a big impact. However, for large volumes this should be relevant.

The expected market impact $\E[\Delta_{\widetilde{\bsb},i}]$
in   \eqref{eq_expected_gain_gen_transaction_costs_decomp_interpret}
depends non-linearly on $b_0$. The non-linear behaviour
results from the non-linearity of the auction curves (see e.g. Figure~\ref{fig:auction_curves_example}).
It cannot be analysed in more detail without making further assumptions. Therefore, we introduce the \emph{linear expected market impact assumption}
\begin{align}
\E[ \bsDelta_{\widetilde{\bsb}}] = \bsa \odot \widetilde{\bsb}
\label{eq_linear_market_impact_assumption} 
\end{align}
where $\bsa=(a_0,\ldots,a_4)$ is the expected linear impact
to analyze a mathematically tractable special case. Note that 
$\bsa$ is the expected linear impact and may depend on the prices $\bsP_{\bsnull}^*$, because $ \bsDelta_{\widetilde{\bsb}}$ may depend on $\bsP_{\bsnull}^*$.

In addition to \eqref{eq_linear_market_impact_assumption}, we may also consider the \emph{no expected market impact assumption} given by 
\begin{align}
\E[ \bsDelta_{\widetilde{\bsb}}] = 
\bsnull .
\label{eq_no_market_impact_assumption} 
\end{align}
Choosing $\bsa=\bsnull$ in 
the linear market impact case \eqref{eq_linear_market_impact_assumption} leads to the no market impact assumption. It is clear, that the no expected market impact assumption is not realistic when trading large volumes, but it helps us understanding the optimal trading behaviour of rational agents that trade small volumes.

\begin{figure}
\begin{tikzpicture}[->,>=stealth',shorten >=1pt,auto,node distance=3cm,
  thick,main node/.style={rectangle,rounded corners, fill=blue!10,draw,
  font=\Large,minimum size=8mm}]
  \node[main node] (root) at (0,0) {\normalsize $\bsb_{\E}$};
  \node[main node] (m) at (-4,-2) {\normalsize $\bsb_{\E\text{-Meff}}$};
  \node[main node] (l) at (4,-2) {\normalsize $\bsb_{\E\text{-LinImp}}$};
  \node[main node] (ml) at (0,-4) {\normalsize $\bsb_{\E\text{-LinImpMeff}}$};
  \node[main node] (n) at (8,-4) {\normalsize $\bsb_{\E\text{-NoImp}}$};
  \node[main node] (mn) at (4,-6) {\normalsize $\bsb_{\E\text{-NoImpMeff}}
  = \bsb_{\text{TC-min}}$ \eqref{eq_b_tcmin}};

  \path[every node/.style={font=\sffamily\small,
  		,inner sep=1pt}]
    (root) 
        edge [bend left=0, blue] node[right=1mm] {Market efficiency \eqref{eq_market_efficient} } (m)
    (root)
        edge [bend left=0, orange] node[right=1mm] {Linear market impact \eqref{eq_linear_market_impact_assumption}} (l)
    (l) 
        edge [bend left=0, blue] node[right=1mm] {Market efficiency \eqref{eq_market_efficient}} (ml)
    (l) 
        edge [bend left=0, red] node[right=1mm] {No market impact \eqref{eq_no_market_impact_assumption}} (n)
    (n) 
        edge [bend left=0, blue] node[right=1mm] {Market efficiency \eqref{eq_market_efficient}} (mn)
    ;
\end{tikzpicture}
\caption{Special cases of analyzed models for a risk neutral trader $\RR = \E$.}
\label{fig_tree}
\end{figure}
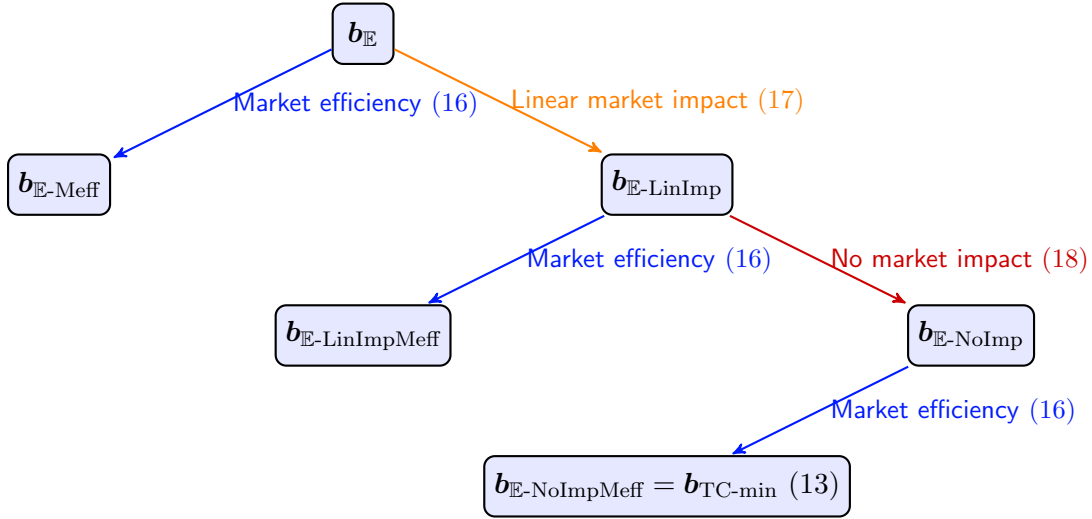

The three assumptions
\eqref{eq_market_efficient},
\eqref{eq_linear_market_impact_assumption} and
\eqref{eq_no_market_impact_assumption} open multiple combination options that lead all to different special cases that can be analyzed. They are visualized in Figure \ref{fig_tree}. As discussed, the $\bsb_{\E}$ and $\bsb_{\E\text{-Meff}}$ have the non-linear impact part which does not allow us to draw further analytical conclusions and has to be solved numerically. We will analyze the remaining solutions of Figure \ref{fig_tree} in the next paragraphs.

Under the no market impact assumption \eqref{eq_no_market_impact_assumption} the expected gain equation   \eqref{eq_expected_gain_gen_transaction_costs_decomp_interpret} simplifies to
\begin{align}
\E[G(\widetilde{\bsb})]
&= \frac{1}{4}\sum_{i=1}^4 \E[P^*_{\bsnull,i}] v_i + 
\underbrace{ \left( \E[P^*_{\bsnull,0}] - \frac{1}{4}\sum_{i=1}^4 \E[P^*_{\bsnull,i}] \right)}_{=\MM(\bsP_{\bsnull}^*)} b_0  
 -   \TT(b_0) .
 \label{eq_G_E}
\end{align}
The first term is the trading revenue of volume $\bsv$ in the IA markets.
The second term characterizes the arbitrage opportunity of the DA and IA markets.
It is linear in $b_0$.
The corresponding slope $\MM(\bsP_{\bsnull}^*)$ represents the 
expected price relationship between the DA and IA markets.
Thus, if $\TT(b_0)$ would be zero we would choose either $b_0\to \infty$ or $b_0 \to -\infty$ depending on sign of $\MM(\bsP_{\bsnull}^*)$. 
Anyway, if the second term in \eqref{eq_G_E} is zero
then the optimum depends only on the transaction costs 
$\TT(b_0)$.
This is the case if $b_0 = 0$ or the market efficiency assumption
\eqref{eq_market_efficient}
holds.
Thus, we have the following theorem.
\begin{theorem}
\label{theorem}
If the 
no expected market impact assumption \eqref{eq_no_market_impact_assumption} with $\bsa=\bsnull$ and the market efficiency assumption 
\eqref{eq_market_efficient} hold
then the optimal trading strategy of a risk neutral trader (i.e. $\RR=\E$) $\bsb_{\E\text{-NoImp}}$
is the minimal transaction cost strategy
$\bsb_{\text{TC-min}}$.
\end{theorem}

Now, let us discuss the more general minimum 
$\bsb_{\E\text{-NoImp}}$ which is characterized by
$b_{\E\text{-NoImp},0}$ of \eqref{eq_G_E}.
For the $\bsa=\bsnull$ case,   \eqref{eq_G_E}
is concave and piecewise linear
 as $-\TT$ is concave and piecewise linear.
 Thus, if a global minimum $b_{\E\text{-NoImp},0}$ exists, it is at one of the 5 corners of the graph of $\TT$ which are in $(0, \bsv)'$. 
 Hence, in practice we can simply evaluate the function in 
 all elements of $(0, \bsv)'$ such as $\max((0, \bsv)') + 1$ or
 $\min((0, \bsv)') - 1$. If the optimum is at the increased maximum
 or decreased minimum then there is no global optimum, otherwise the minimum correspond to the global minimum.
In this case, we can express $b_{\E\text{-NoImp},0}$ as an 
augmented weighted median.

Now, we assume the linear expected market impact assumption \eqref{eq_linear_market_impact_assumption} for \eqref{eq_expected_gain_gen_transaction_costs_decomp_interpret} to analyze the corresponding solution
$\bsb_{\E\text{-LinImp}}$.
It holds for the expected market impact 
that
\begin{align}
 \E[\Delta_{\widetilde{\bsb},0}] b_0
+ \frac{1}{4} \sum_{i=1}^4 \E[\Delta_{\widetilde{\bsb},i}] (v_i-b_0)
= \bsa'\bsone b_0^2 + \sum_{i=1}^4 a_i(v_i^2 + 2v_ib_0).
\end{align}
Then we receive for \eqref{eq_expected_gain_gen_transaction_costs_decomp_interpret}:
\begin{align}
\E[G(\widetilde{\bsb})]
&= 
\underbrace{
\frac{1}{4} \sum_{i=1}^4 \left( \E[P^*_{\bsnull,i}]+a_iv_i \right) v_i + \left( \E[P^*_{\bsnull,0}]  - \frac{1}{4}\sum_{i=1}^4 \E[P^*_{\bsnull,i}]+2a_iv_i \right) b_0
 + \bsa'\bsone b_0^2}_{=\QQ(b_0;\bsP_{\bsnull}^*,\bsv,\bsa)
 = \QQ_0(\bsP_{\bsnull}^*,\bsv,\bsa) + \QQ_1(\bsP_{\bsnull}^*,\bsv,\bsa)b_0 + \QQ_2(\bsa)b_0^2 }
  -   \TT(b_0) 
  \label{eq_gain_lin_transaction_costs}
\end{align}
where we introduce the quadratic polynomial $\QQ(b_0)=
\QQ_0+\QQ_1b_0 + \QQ_2b_0^2$
for the first three terms. 
 The coefficient $\bsa'\bsone$ in front of the quadratic term in  \eqref{eq_gain_lin_transaction_costs} is always non-positive. 
As the transaction costs $\TT(b_0)$ are a piecewise linear function the quadratic term $\bsa'\bsone b_0^2$ will always dominate for $b_0\to\pm \infty$
if $\bsa'\bsone < 0$. Thus, we have a unique maximum if there is linear, non-zero market impact.

\begin{figure}[t!]
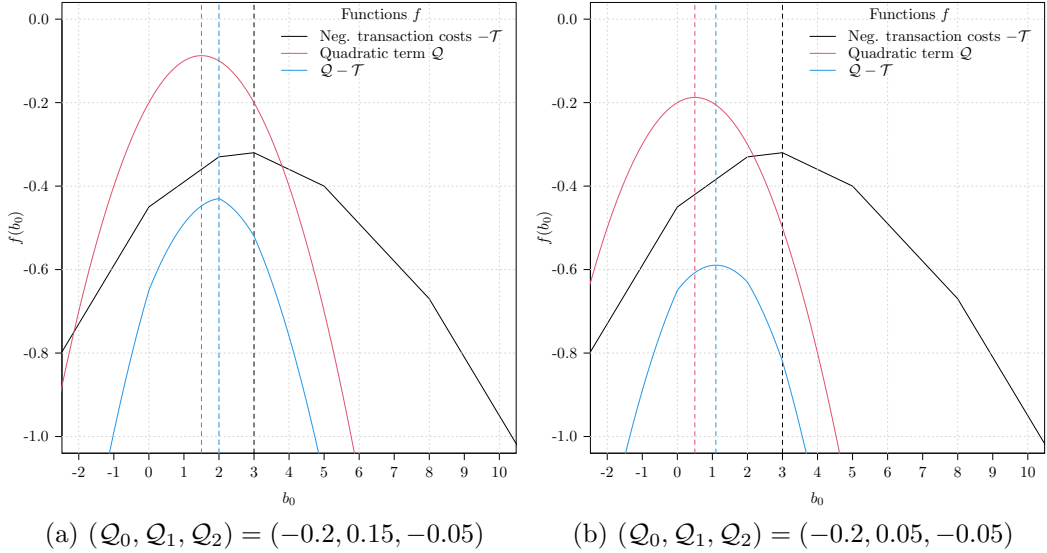

\centering
	\subfloat[$(\QQ_0,\QQ_1,\QQ_2)=(-0.2,0.15,-0.05)$]{\resizebox{0.455\textwidth}{!}{\input{fig/lin_impact1.tex}}} \label{fig_lin_imp1}
	\subfloat[$(\QQ_0,\QQ_1,\QQ_2)=(-0.2,0.05,-0.05)$]{\resizebox{0.455\textwidth}{!}{\input{fig/lin_impact2.tex}}} \label{fig_lin_imp2}
  \caption{Illustration of equation \eqref{eq_gain_lin_transaction_costs} for $\bsv=(2,3,5,8)'$, $\tau_{\text{DA}}=0.04$ and $\tau_{\text{IA}}=0.1$ for different values of $(\QQ_0,\QQ_1,\QQ_2)$ with dashed lines that highlight the maxima values.}
\label{fig_solution}
  \end{figure}

 This maximum has an explicit solution. 
 It may be computed by evaluating all maxima of the piecewise quadratic functions choosing the corresponding maximum. Therefore, remember that $\TT(b_0)$ is a piecewise linear function with at most 6 different slopes. Thus, we have to compute at most 6 solutions of quadratic functions to receive the optimum.
 We want to highlight that the maximum is not necessarily the maximum of 
 $\QQ$ or $-\TT$. Figure~\ref{fig_solution} illustrates two possible cases. In the first example, the optimum is exactly at an element of the $(0,\bsv)'$, namely at $2$. In the second example, the optimum is not an element of $\wtilde{\bsv}$.

Further, we want to remark that $\bsb_{\E\text{-LinImpMeff}}$
has to be computed in the same way as $\bsb_{\E\text{-LinImp}}$. There is no structural simplification possible.
 Finally, note that the linear market impact coefficients $\bsa$ are unknown in practice and have to be estimated. 
 Realistically, it should depend on the price $\bsP_{\bsnull}^*$ as well. So that the market impact in spiky price regions is larger than in common market situations.

\subsection{Risk-Averse strategies}
We also consider numerical solutions of several risk averse 
agents.
In detail, we consider:
mean-variance utility, value-at-risk (VaR) and expected shortfall (CVaR). These risk measures are well-known both in practice and in the literature \cite{liu2007portfolio, garcia2017applying, odeh2018portfolio, canelas2020electricity, dai2015optimal, baringo2015offering}. In the mean-variance utility, we maximize the following risk function
\begin{equation}
\RR(G)=\E[G] - \gamma\V ar[G]
\end{equation}
to estimate the optimal bidding vector $\bsb_{\E\text{-}\var\text{-}\textbf{U}}$. Here, a very important feature is the risk aversion parameter $\gamma$. In our analysis we set arbitrarily $\gamma =0.25$. The second risk averse function that we use to optimize the bidding vector is the value-at-risk (VaR)
\begin{equation}
	\RR(G) = \text{VaR}_{\alpha}(G) = \inf\{x \in \R: F_G(x) > \alpha\} = \Q_{\alpha}(G)
\end{equation}
which with the risk aversion parameter $\alpha$ can be interpreted as an $\alpha$-quantile of the predicted gain. The last utilized risk measure is the expected shortfall, also known as conditional value-at-risk (CVaR) with the following formula
\begin{equation}
	\RR(G) = \text{CVaR}_{\alpha}(G) = - \frac{1}{\alpha}\int_{0}^{\alpha}\text{VaR}_{\gamma}(G)d\gamma.
\end{equation}
In the case of CVaR also the $\alpha$ parameter takes the role of risk aversion parameter. Both for VaR and CVaR we assume arbitrarily that $\alpha = 0.05$.

\section{Forecasting models for the price and market impact}\label{sec:forecasting_model}

Due to the non-linearity of the purchase and sale curves, it is pretty hard to model directly the impacted prices $\bsP_{\bsb}^*$. Instead, we decided to model separately the not impacted price vector $\bsP_{\bsnull}^*$ and the price impact $\bsDelta_{\bsb}$ due to the trading of volume $\bsb$. In this section, we describe the utilized models.

\subsection{Price models}

Let us remind that $\bsP_{\bsnull}^* = (P_{\bsnull,0}^*, P_{\bsnull,1}^*, \dots, P_{\bsnull,4}^*)$ and thus $\bsP_{\bsnull, d,h}^* = \left(P_{\bsnull, d,h}^{*,\text{DA}}, P_{\bsnull, d,h}^{*,\text{IAq1}}, \dots, P_{\bsnull, d,h}^{*,\text{IAq4}}\right)$. For the scenario optimization we need many price trajectories, and we obtain them by forecasting the expected prices and by bootstrapping then the in-sample errors. The first expected price model that we consider is the well-known and widely utilized \cite{nogales2002forecasting, weron2014electricity, ziel2018day} the \textbf{naive} model. 
It predicts todays prices by prices of yesterday on Tuesday, Wednesday, Thursday and Friday, and the last week's prices on other weekdays.
Its formula is as follows
\begin{equation}
	\E\left(\bsP_{\bsnull, d,h}^*\right) = \begin{cases}
		\bsP_{\bsnull, d-7,h}^*,  & \textbf{DoW}_{d,h}^k = 1 \,\, \text{for} \,\, k = 1,6,7,\\
		\bsP_{\bsnull, d-1,h}^*, & \text{otherwise},
	\end{cases}
\end{equation}
where $d$ and $h$ indicate the day and the hour of the delivery, and $\text{DoW}_{d,h}^k$ is the day-of-the-week dummy. 

The second considered model is the autoregressive with exogenous variables estimated using the ordinary least-squares, in the literature often called the \textbf{expert} model \cite{maciejowska2020pca, weron2014electricity, ziel2018day, uniejewski2017variance, uniejewski2018efficient, narajewski2020econometric, uniejewski2019understanding}. It is given by
\begingroup
\small
\begin{equation}
	\begin{aligned}
		\E\left(\bsP_{\bsnull, d,h}^*\right) \, =&\,\, \underbrace{\bsbeta_{h,1}\odot\bsP_{\bsnull, d-1,h}^* + \bsbeta_{h,2}\odot\bsP_{\bsnull, d-2,h}^* + \bsbeta_{h,3}\odot\bsP_{\bsnull, d-7,h}^*}_{\text{autoregressive effects}} + \underbrace{\bsbeta_{h,4}\odot\bsP_{\bsnull, d-1,24}^*}_{\text{yesterday's last hour price}} \\ &+ \underbrace{\bsbeta_{h,5}\odot\bsP_{\bsnull, d-1,\min}^* + \bsbeta_{h,6}\odot\bsP_{\bsnull, d-1,\max}^*}_{\text{non-linear effects}}  + \underbrace{\sum_{i=1}^{7} \bsbeta_{h,6+i}\odot\textbf{DoW}_{d,h}^i}_{\text{weekday dummies}} \\
		&+ \underbrace{\bsbeta_{h,14}\odot\textbf{Load}_{d,h} + \bsbeta_{h,15}\odot\textbf{Solar}_{d,h} + \bsbeta_{h,16}\odot\textbf{WindOn}_{d,h} + \bsbeta_{h,17}\odot\textbf{WindOff}_{d,h}}_{\text{day-ahead forecasts of electricity generation/consumption}}\\
		&+  \underbrace{\bsbeta_{h,18}\odot\textbf{EUA}_{d-2}}_{\text{CO}_2 \text{e price}} + \underbrace{\bsbeta_{h,19}\odot\textbf{Coal}_{d-2} + \bsbeta_{h,20}\odot\textbf{Gas}_{d-2} + \bsbeta_{h,21}\odot\textbf{Oil}_{d-2}}_{\text{most recent fuel prices}},
	\end{aligned}
\end{equation}
\endgroup
where $\odot$ is the element-wise multiplication. The model is estimated separately for each of the 5 markets, however for convenience we use the vector notation.
The regressors considered in the model are not different to the ones utilized in the broad EPF literature. We use autoregressive effects of lag 1, 2 and 7, the last hour's price of the previous day, the element-wise minimum and maximum price of the previous day, and the weekday dummies. Additionally, we use the day-ahead forecasts of electricity load, solar, wind onshore, and wind offshore production. Each of the vectors contains the total forecasted power (MW) in the given time interval (hour or quarter-hour). We also feed the model with 
the EUA (European Union Allowance) price which represents emission costs and
the fuel prices: API2 coal, TTF natural gas and Brent oil. Here we use the settle price lagged by 2 as at the time of forecasting for day $d$, which is around 11:30 on day $d-1$, the settle price for day $d-1$ is not yet available.

The price trajectories are obtained using the bootstrap method \citet{efron1979bootstrap} which delivers very satisfying results \cite{nowotarski2018recent, uniejewski2019importance}. One could use more complicated probabilistic models, but as we already mentioned in this manuscript, it is out of scope of our research. Thus, we receive the trajectories by adding the in-sample bootstrapped errors to the forecasted expected price
\begin{equation}
	\widehat{\bsP}_{\bsnull, d,h}^{*,m} = \widehat{\E\left(\bsP_{\bsnull, d,h}^*\right)} + \widehat{\bseps}_{d,h}^m \, \, \text{for} \, \, m = 1, \dots, M
\end{equation}
where $\widehat{\bseps}_{d,h}^m$ are drawn with replacement in-sample residuals for day $d$ and hour $h$, i.e. we sample from the set of $\widehat{\bseps}_{j,h} = \bsP_{\bsnull, j,h}^{*} - \widehat{\bsP}_{\bsnull, j,h}^{*}$ for $j = 1, \dots, D$. $M$ is the number of predicted trajectories and $D$ is the number of in-sample days. Naturally $M>D$ is possible.

\subsection{Market impact models}\label{sec:market_impact}

The market impact is modelled using the aggregated supply $\bsA_{\bsnull}^{\Sup}(\bsp)$ and demand $\bsA_{\bsnull}^{\Dem}(\bsp)$ curves. Here we make use of the fact that bidding in the auction-based markets causes shifts in the respective curves. The curves are naturally unavailable at the time of forecasting, and thus we need to model them. The modelling and forecasting of the bidding curves has been already approached in the literature \cite{ziel2016electricity, ziel2018probabilistic, mestre2020forecasting, kulakov2020x, soloviova2021efficient}, but the models are rather complicated and time-consuming to estimate. Moreover, the obtained forecast of clearing price $\bsP_{\bsnull}^*$ is not that accurate as the one obtained using e.g. the \textbf{expert} model. Hence, we model the curves using a functional simple moving average of the aggregated curves. 
For the recent $K$ days, this is 
\begin{equation}
	\widehat{\bsA}_{\bsnull,d,h}^{\Sup}(\bsp) = \frac{1}{K}\sum_{k=1}^{K} \bsA_{\bsnull,d-k,h}^{\Sup}(\bsp)
\end{equation}
and
\begin{equation}
	\widehat{\bsA}_{\bsnull,d,h}^{\Dem}(\bsp) = \frac{1}{K}\sum_{k=1}^{K} \bsA_{\bsnull,d-k,h}^{\Sup}(\bsp).
\end{equation}

\begin{figure}[t!]
	\centering
	\includegraphics[width = 1\linewidth]{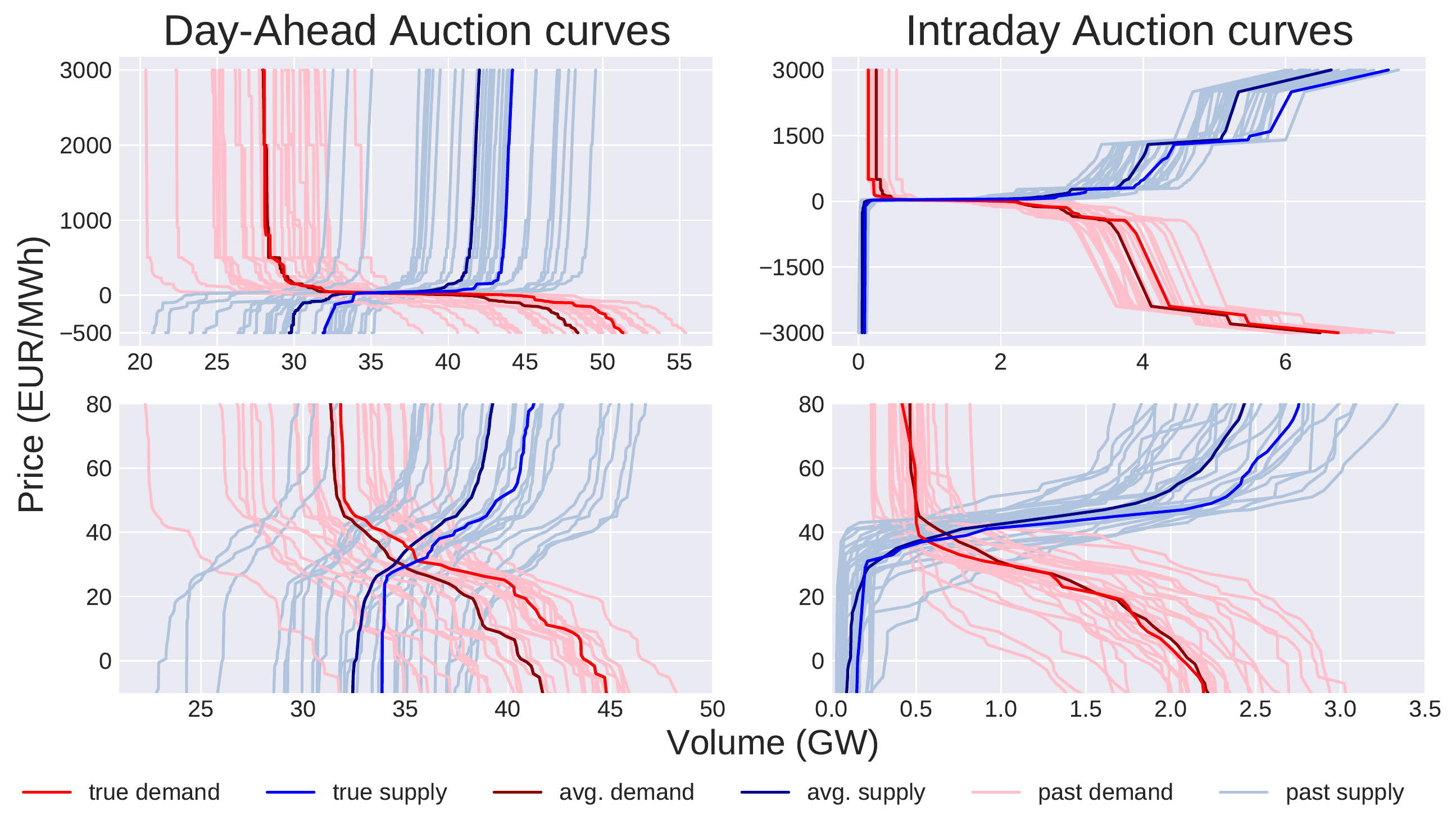}	
	\caption{An example of the average curves $\left(\widehat{\bsA}_{\bsnull,d,h}^{\Sup}\right)^{-1}$ and $\left(\widehat{\bsA}_{\bsnull,d,h}^{\Dem}\right)^{-1}$ in the German market calculated using $K=28$ days before 01.06.2017. The delivery periods are 12:00 to 13:00 for the DA and 12:00 to 12:15 for the IA. The bottom plots are the zoomed-in versions of the top ones}
	\label{fig:curve_model_example}
\end{figure}

An example of such modelling and forecasting of the aggregated curves is presented in Figure~\ref{fig:curve_model_example}. We see that the model does not forecast the true curves perfectly, and especially the location of the intersection of the forecasted curves may be very inaccurate. Still, remember we are only interested in the impacts which is essentially given by the shape of the curves around the intersection.

Moreover, in general the forecasted price induced as intersection of 
$\widehat{\bsA}_{\bsnull,d,h}^{\Sup}$ and $\widehat{\bsA}_{\bsnull,d,h}^{\Dem}$ does not coincide with the forecasted price 
$\widehat{\bsP}_{\bsnull, d,h}^{*,m}$ of the considered price forecasting model, e.g.  \textbf{naive} or \textbf{expert}.
Therefore, we shift our curve predictions so that the resulting intersection coincides with $\widehat{\bsP}_{\bsnull, d,h}^{*,m}$. More precisely, without loss of generality we will shift the supply side curve $\widehat{\bsA}_{\bsnull,d,h}^{\Sup}$.
As $\widehat{\bsP}_{\bsnull, d,h}^{*,m}$ differs for all $m=1,\ldots,M$
this curve shift will depend on $m$ as well.
Thus, we define the shift for the intersection adjustment by
\begin{equation}
\what{\bsxi}_{\bsnull,d,h}^{*,m} = \widehat{\bsA}_{\bsnull,d,h}^{\Dem}\left(\widehat{\bsP}_{\bsnull, d,h}^{*,m}\right) - \widehat{\bsA}_{\bsnull,d,h}^{\Sup}\left(\widehat{\bsP}_{\bsnull, d,h}^{*,m}\right).
\end{equation} 
Let us note that taking $-\what{\bsxi}_{\bsnull,d,h}^{*,m}$ we could shift the demand curve and get the same result. 

Now, remember that we are interested in the estimated impact
$\what{\bsDelta}_{\bsb,d,h}^m = \widehat{\bsP}_{\bsb,d,h}^{*,m} - \widehat{\bsP}_{\bsnull, d,h}^{*,m}$.
The $\bsb$-impacted price $\widehat{\bsP}_{\bsb,d,h}^{*,m}$
results from the intersection of the shifted $\widehat{\bsA}_{\bsb,d,h}^{\Sup}$ with $\widehat{\bsA}_{\bsb,d,h}^{\Dem}$.
To compute this intersection we may find the root of
\begin{align}
 \widehat{\bsA}_{\bsb,d,h}^{\Sup}\left(\bsp\right) + \what{\bsxi}_{\bsnull,d,h}^{*,m} -	\widehat{\bsA}_{\bsb,d,h}^{\Dem}\left(\bsp\right) 
= \widehat{\bsA}_{\bsnull,d,h}^{\Sup}\left(\bsp\right) + \what{\bsxi}_{\bsnull,d,h}^{*,m} -	\widehat{\bsA}_{\bsnull,d,h}^{\Dem}\left(\bsp\right) + \bsb
\label{eq_to_find_root}
\end{align}
where the equality holds by \eqref{eq_shift_supply_demand} and the fact that $\bsb = \bsb^+ - \bsb^-$.
The non-linearity of the function makes it impossible to find the root of \eqref{eq_to_find_root} in an analytical way. This leads to a need of finding it numerically. 
Using an optimizer is possible, but it is not very optimal solution, as we require the solution for all $m=1,\ldots,M$ and we later on have to optimize with respect to $\bsb$. This would result in using the inner optimization each iteration of the outer optimization with high computational costs.
Therefore, we 
decided to consider 
\begin{align}
 \what{\bsC}_{d,h}(\bsp) = \widehat{\bsA}_{\bsnull,d,h}^{\Sup}\left(\bsp\right) -	\widehat{\bsA}_{\bsnull,d,h}^{\Dem}\left(\bsp\right) 
\end{align}
and want to compute the intersection curve $\what{\bsC}_{d,h}^{-1}$.

\begin{figure}[b!]
	\centering
	\includegraphics[width = 1\linewidth]{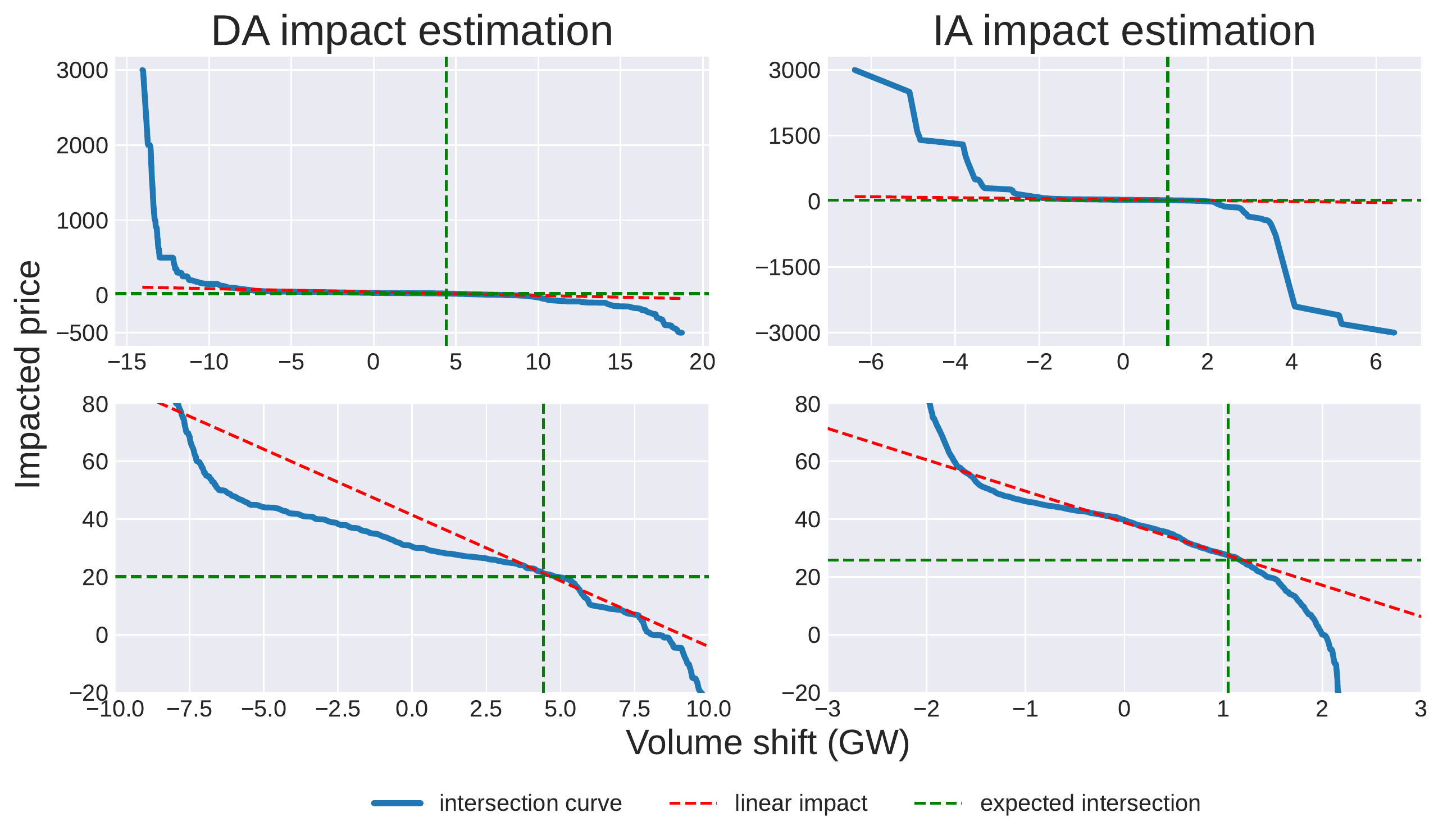}	
	\caption{An example of the predicted intersection curves $\bsC^{-1}_{d,h}$ in the German market with delivery on 01.06.2017. The delivery periods are 12:00 to 13:00 for the DA and 12:00 to 12:15 for the IA. The bottom plots are the zoomed-in versions of the top ones. In addition, the estimated linear impact $\widehat{\bsa}$ is visualized (red).}
	\label{fig:intersection_curve_example}
\end{figure}

For the calculation of $\what{\bsC}_{d,h}^{-1}$ we take the full price grid and compute the volumes.
For model
\eqref{eq_gain_lin_transaction_costs} with linear market impact assumption, the linear market impact coefficients $\bsa$ have to be estimated as well.
Obviously, this should be the slope of the intersection curves at the expected intersection as illustrated in Figure~\ref{fig:intersection_curve_example}. 

In the application study, we estimate 
$\bsa$ by central difference of the inverse impact curves $\what{\bsC}_{d,h}^{-1}$
with incremental slope of average 5\% of the market clearing volume of the past $K$ days. 
Formally, this is
$$	\what{\bsa}_{d,h} = \what{\bsC}_{d,h}^{-1}\left(\frac{1}{M}\sum_{m=1}^{M}\what{\bsxi}_{\bsnull,d,h}^{*,m}+\bsnu\right)-\what{\bsC}_{d,h}^{-1}\left(\frac{1}{M}\sum_{m=1}^{M}\what{\bsxi}_{\bsnull,d,h}^{*,m}-\bsnu\right) $$
with
$\bsnu = \frac{0.05}{K} \sum_{k=1}^K  \widehat{\bsA}_{\bsnull,d,h}^{\Dem} \left(\widehat{\E\left(\bsP_{\bsnull, d,h}^*\right)}\right) $.
Obviously, the $5\%$ is an ad hoc choice, and might be improved. However, our empirical study yield plausible results,
see Figure~\ref{fig:intersection_curve_example}.

Given the intersection curves, we can easily calculate the $\widehat{\bsP}_{\bsb,d,h}^{*,m}$ for $m = 1,\dots, M$ by evaluating 
$\what{\bsC}_{d,h}^{-1}(\what{\bsxi}_{\bsnull,d,h}^{*,m} + \bsb
)$, and thus also the $\what{\bsDelta}_{\bsb,d,h}^m$. The latter one is, however, not the final price impact. This is due to the fact that the DA and IA markets are in a sequential order. Therefore, the $b_0$ bid in the DA market can also influence the prices in the IA market. We refer to it with the following impact adjustment
\begin{equation}
	\what{\bsDelta}_{\bsb,d,h}^{*,m} = (\what{\bsDelta}_{b_0,d,h}^m, \what{\bsDelta}_{b_1,d,h}^m + \delta \what{\bsDelta}_{b_0,d,h}^m, \dots, \what{\bsDelta}_{b_4,d,h}^m + \delta \what{\bsDelta}_{b_0,d,h}^m)
	\label{eq:deltas}
\end{equation}
where $\delta \geq 0$ is a market efficiency factor. If the two markets were inefficient and fully independent, we would use $\delta$ close to 0. However, in our study we assume a more realistic scenario of $\delta = 1$. In other words, we assume that the markets are efficient and a 1 EUR price shift in the day-ahead auction results also in a 1 EUR price shift in the intraday auction.

In order to evaluate the quality of the relatively simple curve forecasts, we consider also a setting where we know the true curves in advance. This is naturally unrealistic, but can help us understand the possible gain of using better curve forecasts. For the same reason we consider also an instance of a perfect price forecast what allows us to inspect the highest possible and at the same time highly unreachable gain rate.

\section{Application: Forecasting and trading study}\label{sec:application}

	\subsection{Data and setting}
	
	For the purpose of application, we use the German market data from January 01, 2016 to December 31, 2020. We conduct a rolling window forecasting study, which is a standard procedure in the EPF literature. The initial in-sample data consists of $D=730$ days, i.e. 2 years and the out-of-sample of 3 years, i.e. $N = 1097$ days.. Every day of the out-of-sample dates we simulate a realistic situation: we estimate the price models based on the most recent $D=730$ days, bootstrap the in-sample residuals to obtain $M=1000$ price trajectories, and forecast the aggregated curves using $K=28$ last days. Based on them, we optimize the assumed risk functions using the sequential least squares programming (SLSQP) algorithm implemented in scipy package in Python to derive the trading strategy $\widehat{\bsb}_{d,h}$. For the transaction costs $\tau_{\text{DA}} = 0.05$~EUR/MWh and $\tau_{\text{IA}} = 0.10$~EUR/MWh are assumed.
	
	In the optimization, multiple settings and volumes $\bsv$ are considered. We start with $\bsv = v\bsone_4$ which assume constant electricity generation or consumption over all hours with
	$v \in \{1,10,100,1000\}$ both on the supply and demand sides. This allows us to observe the impact of growing volumes. Definitely more realistic are the $\bsv$ assumed to be 1\% or 5\% of the day-ahead predicted German wind or solar generation and 1\% or 5\% of the day-ahead predicted German load. These portfolios are far more possible and of high concern for practitioners. Basic summary statistics of the utilized $\bsv$ are presented in Table~\ref{tab:v_description}. We show only the 5\% values as the 1\% and constant are easy to derive. In total, we consider 14 different portfolios $\bsv$ and additionally we assume 2 settings concerning the past participation in the market. 
	In the first setting, we have a new market player that bids the portfolio $\bsv$ as new in the market. In the second one, the market player is already in the market bidding the minimal transaction cost strategy, but they would like to evaluate their current strategy. It means that in this setting the market player is rebidding the portfolio $\bsv$ in the market. 
	
	\begin{table}[b!]
		\centering
		\begingroup\small
		\begin{tabular}{rrrrrrrr}
			\hline
		$\bsv$	& \textbf{mean} & \textbf{std} & \textbf{min} & \textbf{25\%} &\textbf{ 50\%} & \textbf{75\%} & \textbf{max} \\ 
  \hline
\textbf{5\% of wind} & 551 & 427 & 15 & 221 & 429 & 764 & 2105 \\ 
\textbf{5\% of solar} & 245 & 372 & 0 & 0 & 9 & 394 & 1624 \\ 
\textbf{5\% of load} & -2753 & 472 & -3796 & -3151 & -2748 & -2368 & -1621 \\ 
\hline
		\end{tabular}
		\endgroup
		\caption{Basic summary statistics of selected hourly volumes $\bsv_{d,h}' \bsone_4/4$ (MWh). The values are derived using the data from January 01, 2016 to December 31, 2020.} 
		\label{tab:v_description}
	\end{table}

	It is worth to mention the forecasting of $\bsP_{\bsnull}^*$ in both settings. That is to say, in the first one the original historical price series are used as the market player is new in the market and did not impact the prices before with their own bids. It means that the original price series are the $\bsP_{\bsnull}^*$. In the second setting however, the market player was already bidding the $\bsv$ in the past and impacted the prices with their strategy. Here, the original price series are the $\bsP_{\bsb_{\text{TC-min}}}^*$. Thus, for every $\bsv$ in the second setting we subtract from the prices the impact of the trader caused by trading according to the $\bsb_{\text{TC-min}}$ strategy. Then, we conduct the forecasting using the newly acquired artificial price series $\bsP_{\bsnull}^*$ which depends on the past trading path.
	
	To summarize, let us remind that we use 2 models for the price forecasting: the \textbf{naive} and \textbf{expert} and one model for the aggregated curves. Additionally, we use perfect forecasts for prices and the curves. Then, we trade 14 various portfolios in 2 aforementioned settings. The portfolios are traded using 10 strategies. Three of them require no optimization: \textbf{IA-only}, \textbf{TC-min}, $\E$-\textbf{NoImp}. The other seven: $\E$-\textbf{LinImp}, $\E$-\textbf{LinImpMeff}, $\E$-\textbf{Meff}, $\E$, $\E$-$\var$-\textbf{U}, \textbf{VaR} and \textbf{CVaR} are optimized using the SLSQP algorithm.
	
	\subsection{Evaluation}
	
	As the objective of this paper is not the electricity price forecasting, we present a detailed evaluation of the forecasting accuracy price models in Appendix~B. Here, we only want to mention that for all accuracy measures 
	the expert model shows clearly better predictive accuracy than the naive model. Thus, we expect the trading strategies based on the expert model to perform better than those based on the naive forecasting model.

	For the evaluation of the bidding strategies $\widehat{\bsb}_{d,h}$ we calculate an actual gain
	\begin{align}
		\wtilde{G}(\widehat{\bsb}_{d,h}) &= 
		(\bsP_{\bsnull}^* + \bsDelta_{\widehat{\bsb}_{d,h}})'(\bss \odot \widehat{\bsb}_{d,h}) 
		- \bstau'(\bss \odot \widehat{\bsb}_{d,h}^{\abs}) \\
		&= \sum_{i=0}^4 (P^*_{\bsnull,i} + \Delta_{\widehat{\bsb}_{d,h},i})s_i \widehat{b}_{i,d,h} - \sum_{i=0}^4 \tau_{i}s_i |\widehat{b}_{i,d,h}|
	\end{align}
and for convenience and a better comparability among various $\bsv$ we report the average gain in EUR/MWh
	\begin{equation}
	\overline{\wtilde{G}} =  \frac{1}{24N }   \sum_{h = 1}^{24} \sum_{d = 1}^{N} \frac{\wtilde{G}(\widehat{\bsb}_{d,h})}{\bsv_{d,h}' \bsone_4/4}.
\end{equation}
To draw statistically significant conclusions, we perform additionally a two sample bootstrap test to compare the performance of $\widehat{\bsb}_{d,h}$ obtained using different models and strategies. Let $A$ and $B$ denote two strategies $\widehat{\bsb}_{d,h}^A$ and $\widehat{\bsb}_{d,h}^B$. For each model pair, we compute the p-value of two one-sided tests. In the first one we consider the null hypothesis $\mathcal{H}_0: \E\left(\wtilde{G}\left(\widehat{\bsb}_{d,h}^A\right)\right) > \E\left(\wtilde{G}\left(\widehat{\bsb}_{d,h}^B\right)\right)$, and in the second the reverse $\mathcal{H}_0: \E\left(\wtilde{G}\left(\widehat{\bsb}_{d,h}^A\right)\right)  \le  \E\left(\wtilde{G}\left(\widehat{\bsb}_{d,h}^B\right)\right) $.

Let us note that such constructed evaluation measure of the bidding strategies may favour the $\E$ strategy as it actually optimizes the expected gain. This, however, cannot be avoided as the $\E$-$\var$-\textbf{U} and \textbf{CVaR} are not elicitable \cite{gneiting2011making} which means that they cannot be evaluated in a one step decision approach. We could additionally evaluate the $\alpha = 5\%$-quantile of the actual gain what is the optimization goal of the \textbf{VaR} strategy, but we do not do so for the sake of brevity.
	
\subsection{Results}

\begin{table}[b!]
	\centering
	\begingroup\small
	\begin{adjustbox}{max width=1\textwidth}
	\setlength{\tabcolsep}{1pt}
		\begin{tabular}{|cr|rrrrrrrr||rrrrrr|}
			\hline
			&  & \multicolumn{8}{c||}{Supply/Sell (the higher the price the better)} & \multicolumn{6}{c|}{Demand/Buy (the lower the price the better)} \\
			\rotatebox[origin=c]{90}{Model} & Strategy & \thead{1\\MW} & \thead{10\\MW} & \thead{100\\MW} & \thead{1000\\MW} & \thead{1\% of\\ wind} & \thead{5\% of\\ wind} & \thead{1\% of\\ solar} & \thead{5\% of\\ solar} & \thead{1\\MW} & \thead{10\\MW} & \thead{100\\MW} & \thead{1000\\MW} & \thead{1\% of\\ load} & \thead{5\% of\\ load} \\ 
			\hline
			&IA-only & \cellcolor[rgb]{1,0.84,0.5} {37.20} & \cellcolor[rgb]{1,0.691,0.5} {36.87} & \cellcolor[rgb]{1,0.5,0.55} {33.68} & \cellcolor[rgb]{1,0.5,0.55} {-61.20} & \cellcolor[rgb]{1,0.5,0.55} {21.78} & \cellcolor[rgb]{1,0.5,0.55} {-214.07} & \cellcolor[rgb]{1,0.5,0.55} {26.69} & \cellcolor[rgb]{1,0.5,0.55} {-84.74} & \cellcolor[rgb]{1,0.969,0.5} {37.49} & \cellcolor[rgb]{1,0.832,0.5} {37.84} & \cellcolor[rgb]{1,0.5,0.55} {40.94} & \cellcolor[rgb]{1,0.5,0.55} {117.18} & \cellcolor[rgb]{1,0.5,0.55} {66.31} & \cellcolor[rgb]{1,0.5,0.55} {1055.64} \\ 
			&TC-min & \cellcolor[rgb]{0.843,1,0.5} {37.43} & \cellcolor[rgb]{0.5,0.9,0.5} {\textbf{37.34}} & \cellcolor[rgb]{0.5,0.9,0.5} {\textbf{36.78}} & \cellcolor[rgb]{1,0.981,0.5} { 32.55} & \cellcolor[rgb]{0.566,0.922,0.5} {29.73} & \cellcolor[rgb]{1,0.96,0.5} {  25.48} & \cellcolor[rgb]{0.5,0.9,0.5} {\textbf{32.69}} & \cellcolor[rgb]{0.821,1,0.5} { 29.38} & \cellcolor[rgb]{1,0.9,0.5} {37.57} & \cellcolor[rgb]{1,0.979,0.5} {37.67} & \cellcolor[rgb]{0.602,0.934,0.5} {38.23} & \cellcolor[rgb]{1,0.932,0.5} { 42.82} & \cellcolor[rgb]{0.94,1,0.5} {42.13} & \cellcolor[rgb]{1,0.5,0.55} {  63.10} \\ \hline
			 \parbox[t]{2mm}{\multirow{8}{*}{\rotatebox[origin=c]{90}{naive}}} 
			 & $\E$-NoImp & \cellcolor[rgb]{0.708,0.969,0.5} {37.46} & \cellcolor[rgb]{0.913,1,0.5} {37.25} & \cellcolor[rgb]{1,0.5,0.55} {35.41} & \cellcolor[rgb]{1,0.5,0.55} {-13.95} & \cellcolor[rgb]{1,0.5,0.55} {26.21} & \cellcolor[rgb]{1,0.5,0.55} { -85.85} & \cellcolor[rgb]{1,0.5,0.55} {30.30} & \cellcolor[rgb]{1,0.5,0.55} {-24.90} & \cellcolor[rgb]{0.761,0.987,0.5} {37.39} & \cellcolor[rgb]{0.912,1,0.5} {37.62} & \cellcolor[rgb]{1,0.5,0.55} {39.45} & \cellcolor[rgb]{1,0.5,0.55} { 79.09} & \cellcolor[rgb]{1,0.5,0.55} {53.30} & \cellcolor[rgb]{1,0.5,0.55} { 558.32} \\ 
			& $\E$-LinImp & \cellcolor[rgb]{0.708,0.969,0.5} {37.46} & \cellcolor[rgb]{0.809,1,0.5} {37.28} & \cellcolor[rgb]{1,0.932,0.5} {36.59} & \cellcolor[rgb]{1,0.991,0.5} { 32.56} & \cellcolor[rgb]{1,0.86,0.5} {29.52} & \cellcolor[rgb]{1,0.871,0.5} {  25.41} & \cellcolor[rgb]{1,0.961,0.5} {32.55} & \cellcolor[rgb]{1,0.99,0.5} { 29.33} & \cellcolor[rgb]{0.761,0.987,0.5} {37.39} & \cellcolor[rgb]{0.842,1,0.5} {37.60} & \cellcolor[rgb]{1,0.964,0.5} {38.37} & \cellcolor[rgb]{0.905,1,0.5} { 42.70} & \cellcolor[rgb]{0.971,1,0.5} {42.14} & \cellcolor[rgb]{1,0.766,0.5} {  61.99} \\ 
			& $\E$-LinImpMeff & \cellcolor[rgb]{0.843,1,0.5} {37.43} & \cellcolor[rgb]{0.552,0.917,0.5} {37.33} & \cellcolor[rgb]{0.712,0.971,0.5} {36.74} & \cellcolor[rgb]{1,0.871,0.5} { 32.44} & \cellcolor[rgb]{0.762,0.987,0.5} {29.70} & \cellcolor[rgb]{1,0.694,0.5} {  25.27} & \cellcolor[rgb]{0.918,1,0.5} {32.61} & \cellcolor[rgb]{1,0.879,0.5} { 29.23} & \cellcolor[rgb]{1,0.9,0.5} {37.57} & \cellcolor[rgb]{1,0.979,0.5} {37.67} & \cellcolor[rgb]{0.551,0.917,0.5} {38.22} & \cellcolor[rgb]{1,0.94,0.5} { 42.81} & \cellcolor[rgb]{0.94,1,0.5} {42.13} & \cellcolor[rgb]{1,0.5,0.55} {  64.44} \\ 
			& $\E$-Meff & \cellcolor[rgb]{0.843,1,0.5} {37.43} & \cellcolor[rgb]{0.552,0.917,0.5} {37.33} & \cellcolor[rgb]{0.606,0.935,0.5} {36.76} & \cellcolor[rgb]{0.56,0.92,0.5} { 32.66} & \cellcolor[rgb]{0.5,0.9,0.5} {\textbf{29.74}} & \cellcolor[rgb]{0.5,0.9,0.5} {\textbf{25.59}} & \cellcolor[rgb]{0.739,0.98,0.5} {32.65} & \cellcolor[rgb]{0.5,0.9,0.5} {\textbf{29.43}} & \cellcolor[rgb]{1,0.9,0.5} {37.57} & \cellcolor[rgb]{1,0.987,0.5} {37.66} & \cellcolor[rgb]{0.5,0.9,0.5} {\textbf{38.21}} & \cellcolor[rgb]{0.592,0.931,0.5} { 42.62} & \cellcolor[rgb]{0.546,0.915,0.5} {42.03} & \cellcolor[rgb]{1,0.983,0.5} {  61.58} \\ 
			& $\E$ & \cellcolor[rgb]{0.708,0.969,0.5} {37.46} & \cellcolor[rgb]{0.761,0.987,0.5} {37.29} & \cellcolor[rgb]{1,0.976,0.5} {36.64} & \cellcolor[rgb]{0.739,0.98,0.5} { 32.63} & \cellcolor[rgb]{1,0.947,0.5} {29.60} & \cellcolor[rgb]{0.854,1,0.5} {  25.54} & \cellcolor[rgb]{0.958,1,0.5} {32.60} & \cellcolor[rgb]{0.633,0.944,0.5} { 29.41} & \cellcolor[rgb]{0.761,0.987,0.5} {37.39} & \cellcolor[rgb]{0.808,1,0.5} {37.59} & \cellcolor[rgb]{1,0.998,0.5} {38.33} & \cellcolor[rgb]{0.814,1,0.5} { 42.67} & \cellcolor[rgb]{0.817,1,0.5} {42.09} & \cellcolor[rgb]{1,0.957,0.5} {  61.63} \\ 
			& $\E$-$\var$-U & \cellcolor[rgb]{1,0.987,0.5} {37.37} & \cellcolor[rgb]{1,0.969,0.5} {37.19} & \cellcolor[rgb]{1,0.764,0.5} {36.40} & \cellcolor[rgb]{1,0.941,0.5} { 32.51} & \cellcolor[rgb]{1,0.772,0.5} {29.44} & \cellcolor[rgb]{1,0.833,0.5} {  25.38} & \cellcolor[rgb]{1,0.802,0.5} {32.39} & \cellcolor[rgb]{1,0.945,0.5} { 29.29} & \cellcolor[rgb]{1,0.935,0.5} {37.53} & \cellcolor[rgb]{1,0.987,0.5} {37.66} & \cellcolor[rgb]{1,0.947,0.5} {38.39} & \cellcolor[rgb]{1,0.5,0.55} { 99.25} & \cellcolor[rgb]{1,0.5,0.55} {56.88} & \cellcolor[rgb]{1,0.5,0.55} { 414.28} \\ 
			& VaR & \cellcolor[rgb]{0.808,1,0.5} {37.44} & \cellcolor[rgb]{0.709,0.97,0.5} {37.30} & \cellcolor[rgb]{1,0.985,0.5} {36.65} & \cellcolor[rgb]{1,0.911,0.5} { 32.48} & \cellcolor[rgb]{1,0.925,0.5} {29.58} & \cellcolor[rgb]{1,0.833,0.5} {  25.38} & \cellcolor[rgb]{1,0.971,0.5} {32.56} & \cellcolor[rgb]{1,0.912,0.5} { 29.26} & \cellcolor[rgb]{0.948,1,0.5} {37.44} & \cellcolor[rgb]{0.912,1,0.5} {37.62} & \cellcolor[rgb]{0.974,1,0.5} {38.32} & \cellcolor[rgb]{1,0.902,0.5} { 42.86} & \cellcolor[rgb]{1,0.953,0.5} {42.21} & \cellcolor[rgb]{1,0.5,0.525} {  62.54} \\ 
			& CVaR & \cellcolor[rgb]{0.808,1,0.5} {37.44} & \cellcolor[rgb]{0.844,1,0.5} {37.27} & \cellcolor[rgb]{1,0.967,0.5} {36.63} & \cellcolor[rgb]{1,0.981,0.5} { 32.55} & \cellcolor[rgb]{1,0.947,0.5} {29.60} & \cellcolor[rgb]{1,0.909,0.5} {  25.44} & \cellcolor[rgb]{1,0.981,0.5} {32.57} & \cellcolor[rgb]{0.909,1,0.5} { 29.36} & \cellcolor[rgb]{0.844,1,0.5} {37.41} & \cellcolor[rgb]{0.877,1,0.5} {37.61} & \cellcolor[rgb]{1,0.955,0.5} {38.38} & \cellcolor[rgb]{1,0.902,0.5} { 42.86} & \cellcolor[rgb]{1,0.953,0.5} {42.21} & \cellcolor[rgb]{1,0.5,0.55} {  62.68} \\ \hline
						 \parbox[t]{2mm}{\multirow{8}{*}{\rotatebox[origin=c]{90}{expert}}} 
			& $\E$-NoImp & \cellcolor[rgb]{0.5,0.9,0.5} {\textbf{37.50}} & \cellcolor[rgb]{0.761,0.987,0.5} {37.29} & \cellcolor[rgb]{1,0.5,0.55} {35.47} & \cellcolor[rgb]{1,0.5,0.55} {-10.24} & \cellcolor[rgb]{1,0.5,0.55} {25.88} & \cellcolor[rgb]{1,0.5,0.55} {-107.88} & \cellcolor[rgb]{1,0.5,0.55} {31.30} & \cellcolor[rgb]{1,0.5,0.55} {  2.68} & \cellcolor[rgb]{0.552,0.917,0.5} {37.35} & \cellcolor[rgb]{0.708,0.969,0.5} {37.57} & \cellcolor[rgb]{1,0.5,0.55} {39.37} & \cellcolor[rgb]{1,0.5,0.55} { 76.70} & \cellcolor[rgb]{1,0.5,0.55} {51.83} & \cellcolor[rgb]{1,0.5,0.55} { 517.79} \\ 
			& $\E$-LinImp & \cellcolor[rgb]{0.5,0.9,0.5} {37.50} & \cellcolor[rgb]{0.604,0.935,0.5} {37.32} & \cellcolor[rgb]{0.989,1,0.5} {36.67} & \cellcolor[rgb]{0.879,1,0.5} { 32.60} & \cellcolor[rgb]{1,0.991,0.5} {29.64} & \cellcolor[rgb]{1,0.96,0.5} {  25.48} & \cellcolor[rgb]{0.739,0.98,0.5} {32.65} & \cellcolor[rgb]{0.821,1,0.5} { 29.38} & \cellcolor[rgb]{0.5,0.9,0.5} {\textbf{37.34}} & \cellcolor[rgb]{0.5,0.9,0.5} {\textbf{37.53}} & \cellcolor[rgb]{0.804,1,0.5} {38.27} & \cellcolor[rgb]{0.936,1,0.5} { 42.71} & \cellcolor[rgb]{0.878,1,0.5} {42.11} & \cellcolor[rgb]{1,0.5,0.55} {  63.76} \\ 
			& $\E$-LinImpMeff & \cellcolor[rgb]{0.808,1,0.5} {37.44} & \cellcolor[rgb]{0.5,0.9,0.5} {37.34} & \cellcolor[rgb]{0.765,0.988,0.5} {36.73} & \cellcolor[rgb]{1,0.782,0.5} { 32.35} & \cellcolor[rgb]{0.862,1,0.5} {29.68} & \cellcolor[rgb]{1,0.554,0.5} {  25.16} & \cellcolor[rgb]{1,0.991,0.5} {32.58} & \cellcolor[rgb]{1,0.769,0.5} { 29.13} & \cellcolor[rgb]{1,0.9,0.5} {37.57} & \cellcolor[rgb]{1,0.979,0.5} {37.67} & \cellcolor[rgb]{0.653,0.951,0.5} {38.24} & \cellcolor[rgb]{1,0.818,0.5} { 42.97} & \cellcolor[rgb]{1,0.945,0.5} {42.22} & \cellcolor[rgb]{1,0.5,0.55} {  69.79} \\ 
			& $\E$-Meff & \cellcolor[rgb]{0.843,1,0.5} {37.43} & \cellcolor[rgb]{0.5,0.9,0.5} {37.34} & \cellcolor[rgb]{0.553,0.918,0.5} {36.77} & \cellcolor[rgb]{0.5,0.9,0.5} {\textbf{32.67}} & \cellcolor[rgb]{0.5,0.9,0.5} {29.74} & \cellcolor[rgb]{0.652,0.951,0.5} {  25.57} & \cellcolor[rgb]{0.739,0.98,0.5} {32.65} & \cellcolor[rgb]{0.566,0.922,0.5} { 29.42} & \cellcolor[rgb]{1,0.9,0.5} {37.57} & \cellcolor[rgb]{1,0.979,0.5} {37.67} & \cellcolor[rgb]{0.5,0.9,0.5} {38.21} & \cellcolor[rgb]{0.5,0.9,0.5} {\textbf{42.60}} & \cellcolor[rgb]{0.5,0.9,0.5} {\textbf{42.02}} & \cellcolor[rgb]{0.659,0.953,0.5} {  61.41} \\ 
			& $\E$ & \cellcolor[rgb]{0.5,0.9,0.5} {37.50} & \cellcolor[rgb]{0.604,0.935,0.5} {37.32} & \cellcolor[rgb]{0.918,1,0.5} {36.69} & \cellcolor[rgb]{0.56,0.92,0.5} { 32.66} & \cellcolor[rgb]{0.95,1,0.5} {29.66} & \cellcolor[rgb]{0.803,1,0.5} {  25.55} & \cellcolor[rgb]{0.679,0.96,0.5} {32.66} & \cellcolor[rgb]{0.633,0.944,0.5} { 29.41} & \cellcolor[rgb]{0.5,0.9,0.5} {37.34} & \cellcolor[rgb]{0.5,0.9,0.5} {37.53} & \cellcolor[rgb]{0.755,0.985,0.5} {38.26} & \cellcolor[rgb]{0.637,0.946,0.5} { 42.63} & \cellcolor[rgb]{0.732,0.977,0.5} {42.07} & \cellcolor[rgb]{0.5,0.9,0.5} {\textbf{61.36}} \\ 
			& $\E$-$\var$-U & \cellcolor[rgb]{0.76,0.987,0.5} {37.45} & \cellcolor[rgb]{0.657,0.952,0.5} {37.31} & \cellcolor[rgb]{1,0.994,0.5} {36.66} & \cellcolor[rgb]{1,0.941,0.5} { 32.51} & \cellcolor[rgb]{1,0.947,0.5} {29.60} & \cellcolor[rgb]{1,0.719,0.5} {  25.29} & \cellcolor[rgb]{1,0.991,0.5} {32.58} & \cellcolor[rgb]{1,0.956,0.5} { 29.30} & \cellcolor[rgb]{0.809,1,0.5} {37.40} & \cellcolor[rgb]{0.708,0.969,0.5} {37.57} & \cellcolor[rgb]{1,0.998,0.5} {38.33} & \cellcolor[rgb]{1,0.5,0.55} { 57.16} & \cellcolor[rgb]{1,0.5,0.55} {44.96} & \cellcolor[rgb]{1,0.5,0.55} { 218.01} \\ 
			& VaR & \cellcolor[rgb]{0.656,0.952,0.5} {37.47} & \cellcolor[rgb]{0.552,0.917,0.5} {37.33} & \cellcolor[rgb]{0.918,1,0.5} {36.69} & \cellcolor[rgb]{1,0.951,0.5} { 32.52} & \cellcolor[rgb]{1,0.991,0.5} {29.64} & \cellcolor[rgb]{1,0.821,0.5} {  25.37} & \cellcolor[rgb]{0.918,1,0.5} {32.61} & \cellcolor[rgb]{1,0.945,0.5} { 29.29} & \cellcolor[rgb]{0.879,1,0.5} {37.42} & \cellcolor[rgb]{0.76,0.987,0.5} {37.58} & \cellcolor[rgb]{0.704,0.968,0.5} {38.25} & \cellcolor[rgb]{1,0.993,0.5} { 42.74} & \cellcolor[rgb]{0.971,1,0.5} {42.14} & \cellcolor[rgb]{1,0.666,0.5} {  62.18} \\ 
			& CVaR & \cellcolor[rgb]{0.708,0.969,0.5} {37.46} & \cellcolor[rgb]{0.657,0.952,0.5} {37.31} & \cellcolor[rgb]{0.918,1,0.5} {36.69} & \cellcolor[rgb]{0.958,1,0.5} { 32.58} & \cellcolor[rgb]{1,0.991,0.5} {29.64} & \cellcolor[rgb]{1,0.884,0.5} {  25.42} & \cellcolor[rgb]{0.918,1,0.5} {32.61} & \cellcolor[rgb]{0.865,1,0.5} { 29.37} & \cellcolor[rgb]{0.844,1,0.5} {37.41} & \cellcolor[rgb]{0.656,0.952,0.5} {37.56} & \cellcolor[rgb]{0.704,0.968,0.5} {38.25} & \cellcolor[rgb]{1,0.993,0.5} { 42.74} & \cellcolor[rgb]{0.94,1,0.5} {42.13} & \cellcolor[rgb]{1,0.766,0.5} {  61.99} \\ 
			\hline
		\end{tabular}
	\end{adjustbox}
	\endgroup
	\caption{Average actual gain $\overline{\wtilde{G}}$ (EUR/MWh) of the considered strategies as a new market player. Colour indicates the performance column-wise (the greener, the better). With bold, we depicted the best values in each column} 
	\label{tab:avg_gain_new_bids}
\end{table}

\begin{table}[t!]
	\centering
	\begingroup\small
	\begin{adjustbox}{max width=1\textwidth}
		\setlength{\tabcolsep}{1pt}
		\begin{tabular}{|cr|rrrrrrrr||rrrrrr|}
			\hline
			&  & \multicolumn{8}{c||}{Supply/Sell (the higher the price the better)} & \multicolumn{6}{c|}{Demand/Buy (the lower the price the better)} \\
			\rotatebox[origin=c]{90}{Model} & Strategy & \thead{1\\MW} & \thead{10\\MW} & \thead{100\\MW} & \thead{1000\\MW} & \thead{1\% of\\ wind} & \thead{5\% of\\ wind} & \thead{1\% of\\ solar} & \thead{5\% of\\ solar} & \thead{1\\MW} & \thead{10\\MW} & \thead{100\\MW} & \thead{1000\\MW} & \thead{1\% of\\ load} & \thead{5\% of\\ load} \\ 
			\hline
			&IA-only & \cellcolor[rgb]{1,0.84,0.5} {37.22} & \cellcolor[rgb]{1,0.71,0.5} {37.01} & \cellcolor[rgb]{1,0.5,0.55} {34.55} & \cellcolor[rgb]{1,0.5,0.55} {-37.59} & \cellcolor[rgb]{1,0.5,0.55} {23.69} & \cellcolor[rgb]{1,0.5,0.55} {-151.61} & \cellcolor[rgb]{1,0.5,0.55} {28.13} & \cellcolor[rgb]{1,0.5,0.55} {-52.56} & \cellcolor[rgb]{1,0.978,0.5} {37.47} & \cellcolor[rgb]{1,0.839,0.5} {37.70} & \cellcolor[rgb]{1,0.5,0.55} {40.08} & \cellcolor[rgb]{1,0.5,0.55} {100.42} & \cellcolor[rgb]{1,0.5,0.55} {59.55} & \cellcolor[rgb]{1,0.5,0.55} {989.44} \\ 
			&TC-min & \cellcolor[rgb]{0.843,1,0.5} {37.45} & \cellcolor[rgb]{0.552,0.917,0.5} {37.45} & \cellcolor[rgb]{0.843,1,0.5} {37.45} & \cellcolor[rgb]{1,0.706,0.5} { 37.45} & \cellcolor[rgb]{1,0.964,0.5} {30.98} & \cellcolor[rgb]{1,0.665,0.5} {  30.98} & \cellcolor[rgb]{0.831,1,0.5} {33.65} & \cellcolor[rgb]{1,0.841,0.5} { 33.66} & \cellcolor[rgb]{1,0.908,0.5} {37.55} & \cellcolor[rgb]{1,0.97,0.5} {37.55} & \cellcolor[rgb]{0.982,1,0.5} {37.55} & \cellcolor[rgb]{1,0.706,0.5} { 37.55} & \cellcolor[rgb]{1,0.831,0.5} {38.98} & \cellcolor[rgb]{1,0.5,0.55} { 38.97} \\ \hline
			\parbox[t]{2mm}{\multirow{8}{*}{\rotatebox[origin=c]{90}{naive}}} 
			& $\E$-NoImp & \cellcolor[rgb]{0.708,0.969,0.5} {37.48} & \cellcolor[rgb]{0.912,1,0.5} {37.37} & \cellcolor[rgb]{1,0.5,0.55} {36.13} & \cellcolor[rgb]{1,0.5,0.55} {  0.12} & \cellcolor[rgb]{1,0.5,0.55} {27.50} & \cellcolor[rgb]{1,0.5,0.55} { -67.09} & \cellcolor[rgb]{1,0.5,0.55} {31.19} & \cellcolor[rgb]{1,0.5,0.55} {-17.62} & \cellcolor[rgb]{0.709,0.97,0.5} {37.37} & \cellcolor[rgb]{0.913,1,0.5} {37.49} & \cellcolor[rgb]{1,0.5,0.55} {38.72} & \cellcolor[rgb]{1,0.5,0.55} { 68.95} & \cellcolor[rgb]{1,0.5,0.55} {49.55} & \cellcolor[rgb]{1,0.5,0.55} {647.03} \\
			& $\E$-LinImp & \cellcolor[rgb]{0.708,0.969,0.5} {37.48} & \cellcolor[rgb]{0.808,1,0.5} {37.40} & \cellcolor[rgb]{1,0.944,0.5} {37.34} & \cellcolor[rgb]{1,0.997,0.5} { 37.79} & \cellcolor[rgb]{1,0.839,0.5} {30.86} & \cellcolor[rgb]{1,0.893,0.5} {  31.20} & \cellcolor[rgb]{1,0.955,0.5} {33.56} & \cellcolor[rgb]{1,0.975,0.5} { 33.80} & \cellcolor[rgb]{0.709,0.97,0.5} {37.37} & \cellcolor[rgb]{0.843,1,0.5} {37.47} & \cellcolor[rgb]{1,0.952,0.5} {37.61} & \cellcolor[rgb]{0.88,1,0.5} { 37.18} & \cellcolor[rgb]{1,0.999,0.5} {38.78} & \cellcolor[rgb]{0.805,1,0.5} { 38.05} \\ 
			& $\E$-LinImpMeff & \cellcolor[rgb]{0.843,1,0.5} {37.45} & \cellcolor[rgb]{0.552,0.917,0.5} {37.45} & \cellcolor[rgb]{0.5,0.9,0.5} {\textbf{37.52}} & \cellcolor[rgb]{0.977,1,0.5} { 37.80} & \cellcolor[rgb]{0.5,0.9,0.5} {\textbf{31.11}} & \cellcolor[rgb]{1,0.976,0.5} {  31.28} & \cellcolor[rgb]{0.616,0.939,0.5} {33.69} & \cellcolor[rgb]{0.83,1,0.5} { 33.87} & \cellcolor[rgb]{1,0.908,0.5} {37.55} & \cellcolor[rgb]{1,0.97,0.5} {37.55} & \cellcolor[rgb]{0.5,0.9,0.5} {\textbf{37.44}} & \cellcolor[rgb]{0.71,0.97,0.5} { 37.14} & \cellcolor[rgb]{0.601,0.934,0.5} {38.68} & \cellcolor[rgb]{1,0.98,0.5} { 38.13} \\ 
			& $\E$-Meff & \cellcolor[rgb]{0.843,1,0.5} {37.45} & \cellcolor[rgb]{0.552,0.917,0.5} {37.45} & \cellcolor[rgb]{0.5,0.9,0.5} {37.52} & \cellcolor[rgb]{0.5,0.9,0.5} {\textbf{37.91}} & \cellcolor[rgb]{0.5,0.9,0.5} {31.11} & \cellcolor[rgb]{0.5,0.9,0.5} {\textbf{31.40}} & \cellcolor[rgb]{0.5,0.9,0.5} {\textbf{33.71}} & \cellcolor[rgb]{0.5,0.9,0.5} {\textbf{33.93}} & \cellcolor[rgb]{1,0.908,0.5} {37.55} & \cellcolor[rgb]{1,0.978,0.5} {37.54} & \cellcolor[rgb]{0.552,0.917,0.5} {37.45} & \cellcolor[rgb]{0.5,0.9,0.5} {\textbf{37.10}} & \cellcolor[rgb]{0.5,0.9,0.5} {\textbf{38.66}} & \cellcolor[rgb]{0.757,0.986,0.5} { 38.04} \\ 
			& $\E$ & \cellcolor[rgb]{0.708,0.969,0.5} {37.48} & \cellcolor[rgb]{0.76,0.987,0.5} {37.41} & \cellcolor[rgb]{1,0.987,0.5} {37.39} & \cellcolor[rgb]{0.757,0.986,0.5} { 37.86} & \cellcolor[rgb]{1,0.933,0.5} {30.95} & \cellcolor[rgb]{0.931,1,0.5} {  31.32} & \cellcolor[rgb]{0.909,1,0.5} {33.63} & \cellcolor[rgb]{0.73,0.977,0.5} { 33.89} & \cellcolor[rgb]{0.709,0.97,0.5} {37.37} & \cellcolor[rgb]{0.809,1,0.5} {37.46} & \cellcolor[rgb]{1,0.996,0.5} {37.56} & \cellcolor[rgb]{0.658,0.953,0.5} { 37.13} & \cellcolor[rgb]{0.752,0.984,0.5} {38.71} & \cellcolor[rgb]{0.84,1,0.5} { 38.06} \\ 
			& $\E$-$\var$-U & \cellcolor[rgb]{1,0.987,0.5} {37.39} & \cellcolor[rgb]{1,0.996,0.5} {37.34} & \cellcolor[rgb]{1,0.849,0.5} {37.23} & \cellcolor[rgb]{1,0.954,0.5} { 37.74} & \cellcolor[rgb]{1,0.87,0.5} {30.89} & \cellcolor[rgb]{1,0.934,0.5} {  31.24} & \cellcolor[rgb]{1,0.898,0.5} {33.50} & \cellcolor[rgb]{1,0.966,0.5} { 33.79} & \cellcolor[rgb]{1,0.943,0.5} {37.51} & \cellcolor[rgb]{1,0.978,0.5} {37.54} & \cellcolor[rgb]{1,0.874,0.5} {37.70} & \cellcolor[rgb]{1,0.5,0.55} { 87.15} & \cellcolor[rgb]{1,0.5,0.55} {51.92} & \cellcolor[rgb]{1,0.5,0.55} {381.80} \\ 
			& VaR & \cellcolor[rgb]{0.808,1,0.5} {37.46} & \cellcolor[rgb]{0.708,0.969,0.5} {37.42} & \cellcolor[rgb]{0.912,1,0.5} {37.43} & \cellcolor[rgb]{1,0.92,0.5} { 37.70} & \cellcolor[rgb]{1,0.954,0.5} {30.97} & \cellcolor[rgb]{1,0.862,0.5} {  31.17} & \cellcolor[rgb]{0.947,1,0.5} {33.62} & \cellcolor[rgb]{1,0.928,0.5} { 33.75} & \cellcolor[rgb]{0.948,1,0.5} {37.43} & \cellcolor[rgb]{0.878,1,0.5} {37.48} & \cellcolor[rgb]{0.913,1,0.5} {37.53} & \cellcolor[rgb]{1,0.942,0.5} { 37.28} & \cellcolor[rgb]{1,0.982,0.5} {38.80} & \cellcolor[rgb]{1,0.741,0.5} { 38.41} \\ 
			& CVaR & \cellcolor[rgb]{0.843,1,0.5} {37.45} & \cellcolor[rgb]{0.808,1,0.5} {37.40} & \cellcolor[rgb]{0.877,1,0.5} {37.44} & \cellcolor[rgb]{0.943,1,0.5} { 37.81} & \cellcolor[rgb]{0.976,1,0.5} {31.02} & \cellcolor[rgb]{1,0.996,0.5} {  31.30} & \cellcolor[rgb]{0.789,0.996,0.5} {33.66} & \cellcolor[rgb]{0.787,0.996,0.5} { 33.88} & \cellcolor[rgb]{0.809,1,0.5} {37.39} & \cellcolor[rgb]{0.843,1,0.5} {37.47} & \cellcolor[rgb]{0.982,1,0.5} {37.55} & \cellcolor[rgb]{1,0.986,0.5} { 37.23} & \cellcolor[rgb]{0.936,1,0.5} {38.76} & \cellcolor[rgb]{1,0.561,0.5} { 38.62} \\ \hline
			\parbox[t]{2mm}{\multirow{8}{*}{\rotatebox[origin=c]{90}{expert}}} 
			& $\E$-NoImp & \cellcolor[rgb]{0.5,0.9,0.5} {\textbf{37.52}} & \cellcolor[rgb]{0.76,0.987,0.5} {37.41} & \cellcolor[rgb]{1,0.5,0.55} {36.17} & \cellcolor[rgb]{1,0.5,0.55} {  1.84} & \cellcolor[rgb]{1,0.5,0.55} {27.17} & \cellcolor[rgb]{1,0.5,0.55} { -87.07} & \cellcolor[rgb]{1,0.5,0.55} {31.95} & \cellcolor[rgb]{1,0.5,0.55} { -8.77} & \cellcolor[rgb]{0.5,0.9,0.5} {\textbf{37.33}} & \cellcolor[rgb]{0.761,0.987,0.5} {37.45} & \cellcolor[rgb]{1,0.5,0.55} {38.65} & \cellcolor[rgb]{1,0.5,0.55} { 67.48} & \cellcolor[rgb]{1,0.5,0.55} {48.61} & \cellcolor[rgb]{1,0.5,0.55} {671.17} \\ 
			& $\E$-LinImp & \cellcolor[rgb]{0.5,0.9,0.5} {37.52} & \cellcolor[rgb]{0.604,0.935,0.5} {37.44} & \cellcolor[rgb]{0.946,1,0.5} {37.42} & \cellcolor[rgb]{0.84,1,0.5} { 37.84} & \cellcolor[rgb]{1,0.985,0.5} {31.00} & \cellcolor[rgb]{0.973,1,0.5} {  31.31} & \cellcolor[rgb]{0.789,0.996,0.5} {33.66} & \cellcolor[rgb]{0.83,1,0.5} { 33.87} & \cellcolor[rgb]{0.5,0.9,0.5} {37.33} & \cellcolor[rgb]{0.5,0.9,0.5} {\textbf{37.40}} & \cellcolor[rgb]{0.808,1,0.5} {37.50} & \cellcolor[rgb]{0.658,0.953,0.5} { 37.13} & \cellcolor[rgb]{0.752,0.984,0.5} {38.71} & \cellcolor[rgb]{0.976,1,0.5} { 38.10} \\ 
			& $\E$-LinImpMeff & \cellcolor[rgb]{0.843,1,0.5} {37.45} & \cellcolor[rgb]{0.5,0.9,0.5} {\textbf{37.46}} & \cellcolor[rgb]{0.5,0.9,0.5} {37.52} & \cellcolor[rgb]{1,0.937,0.5} { 37.72} & \cellcolor[rgb]{0.563,0.921,0.5} {31.10} & \cellcolor[rgb]{1,0.934,0.5} {  31.24} & \cellcolor[rgb]{0.616,0.939,0.5} {33.69} & \cellcolor[rgb]{1,0.966,0.5} { 33.79} & \cellcolor[rgb]{1,0.908,0.5} {37.55} & \cellcolor[rgb]{1,0.97,0.5} {37.55} & \cellcolor[rgb]{0.552,0.917,0.5} {37.45} & \cellcolor[rgb]{1,0.977,0.5} { 37.24} & \cellcolor[rgb]{0.835,1,0.5} {38.73} & \cellcolor[rgb]{1,0.5,0.507} { 38.70} \\ 
			& $\E$-Meff & \cellcolor[rgb]{0.843,1,0.5} {37.45} & \cellcolor[rgb]{0.5,0.9,0.5} {37.46} & \cellcolor[rgb]{0.5,0.9,0.5} {37.52} & \cellcolor[rgb]{0.5,0.9,0.5} { 37.91} & \cellcolor[rgb]{0.5,0.9,0.5} {31.11} & \cellcolor[rgb]{0.624,0.941,0.5} {  31.38} & \cellcolor[rgb]{0.5,0.9,0.5} {33.71} & \cellcolor[rgb]{0.557,0.919,0.5} { 33.92} & \cellcolor[rgb]{1,0.908,0.5} {37.55} & \cellcolor[rgb]{1,0.97,0.5} {37.55} & \cellcolor[rgb]{0.604,0.935,0.5} {37.46} & \cellcolor[rgb]{0.5,0.9,0.5} { 37.10} & \cellcolor[rgb]{0.5,0.9,0.5} {38.66} & \cellcolor[rgb]{0.5,0.9,0.5} {\textbf{37.99}} \\ 
			& $\E$ & \cellcolor[rgb]{0.5,0.9,0.5} {37.52} & \cellcolor[rgb]{0.604,0.935,0.5} {37.44} & \cellcolor[rgb]{0.877,1,0.5} {37.44} & \cellcolor[rgb]{0.654,0.951,0.5} { 37.88} & \cellcolor[rgb]{0.976,1,0.5} {31.02} & \cellcolor[rgb]{0.807,1,0.5} {  31.35} & \cellcolor[rgb]{0.731,0.977,0.5} {33.67} & \cellcolor[rgb]{0.73,0.977,0.5} { 33.89} & \cellcolor[rgb]{0.5,0.9,0.5} {37.33} & \cellcolor[rgb]{0.5,0.9,0.5} {37.40} & \cellcolor[rgb]{0.76,0.987,0.5} {37.49} & \cellcolor[rgb]{0.605,0.935,0.5} { 37.12} & \cellcolor[rgb]{0.651,0.95,0.5} {38.69} & \cellcolor[rgb]{0.5,0.9,0.5} { 37.99} \\  
			& $\E$-$\var$-U & \cellcolor[rgb]{0.76,0.987,0.5} {37.47} & \cellcolor[rgb]{0.552,0.917,0.5} {37.45} & \cellcolor[rgb]{0.808,1,0.5} {37.46} & \cellcolor[rgb]{0.977,1,0.5} { 37.80} & \cellcolor[rgb]{0.893,1,0.5} {31.04} & \cellcolor[rgb]{1,0.965,0.5} {  31.27} & \cellcolor[rgb]{0.789,0.996,0.5} {33.66} & \cellcolor[rgb]{0.907,1,0.5} { 33.85} & \cellcolor[rgb]{0.761,0.987,0.5} {37.38} & \cellcolor[rgb]{0.761,0.987,0.5} {37.45} & \cellcolor[rgb]{1,0.944,0.5} {37.62} & \cellcolor[rgb]{1,0.5,0.55} { 50.15} & \cellcolor[rgb]{1,0.5,0.55} {41.41} & \cellcolor[rgb]{1,0.5,0.55} {176.12} \\ 
			& VaR & \cellcolor[rgb]{0.708,0.969,0.5} {37.48} & \cellcolor[rgb]{0.552,0.917,0.5} {37.45} & \cellcolor[rgb]{0.76,0.987,0.5} {37.47} & \cellcolor[rgb]{1,0.971,0.5} { 37.76} & \cellcolor[rgb]{0.934,1,0.5} {31.03} & \cellcolor[rgb]{1,0.914,0.5} {  31.22} & \cellcolor[rgb]{0.789,0.996,0.5} {33.66} & \cellcolor[rgb]{1,0.975,0.5} { 33.80} & \cellcolor[rgb]{0.879,1,0.5} {37.41} & \cellcolor[rgb]{0.761,0.987,0.5} {37.45} & \cellcolor[rgb]{0.656,0.952,0.5} {37.47} & \cellcolor[rgb]{0.985,1,0.5} { 37.21} & \cellcolor[rgb]{0.903,1,0.5} {38.75} & \cellcolor[rgb]{1,0.878,0.5} { 38.25} \\  
			& CVaR & \cellcolor[rgb]{0.708,0.969,0.5} {37.48} & \cellcolor[rgb]{0.552,0.917,0.5} {37.45} & \cellcolor[rgb]{0.76,0.987,0.5} {37.47} & \cellcolor[rgb]{0.909,1,0.5} { 37.82} & \cellcolor[rgb]{0.851,1,0.5} {31.05} & \cellcolor[rgb]{0.931,1,0.5} {  31.32} & \cellcolor[rgb]{0.674,0.958,0.5} {33.68} & \cellcolor[rgb]{0.83,1,0.5} { 33.87} & \cellcolor[rgb]{0.809,1,0.5} {37.39} & \cellcolor[rgb]{0.656,0.952,0.5} {37.43} & \cellcolor[rgb]{0.552,0.917,0.5} {37.45} & \cellcolor[rgb]{0.845,1,0.5} { 37.17} & \cellcolor[rgb]{0.802,1,0.5} {38.72} & \cellcolor[rgb]{1,0.997,0.5} { 38.11} \\ 
			\hline
		\end{tabular}
	\end{adjustbox}
	\endgroup
	\caption{Average actual gain $\overline{\wtilde{G}}$ (EUR/MWh) of the considered strategies as an existing market player rebidding their portfolio $\bsv$. Colour indicates the performance row-wise (the greener, the better). With bold, we depicted the best values in each row} 
	\label{tab:avg_gain_rebidded}
\end{table}

\begin{figure}[t!]
	\centering
	\subfloat{\includegraphics[width = 0.98\linewidth]{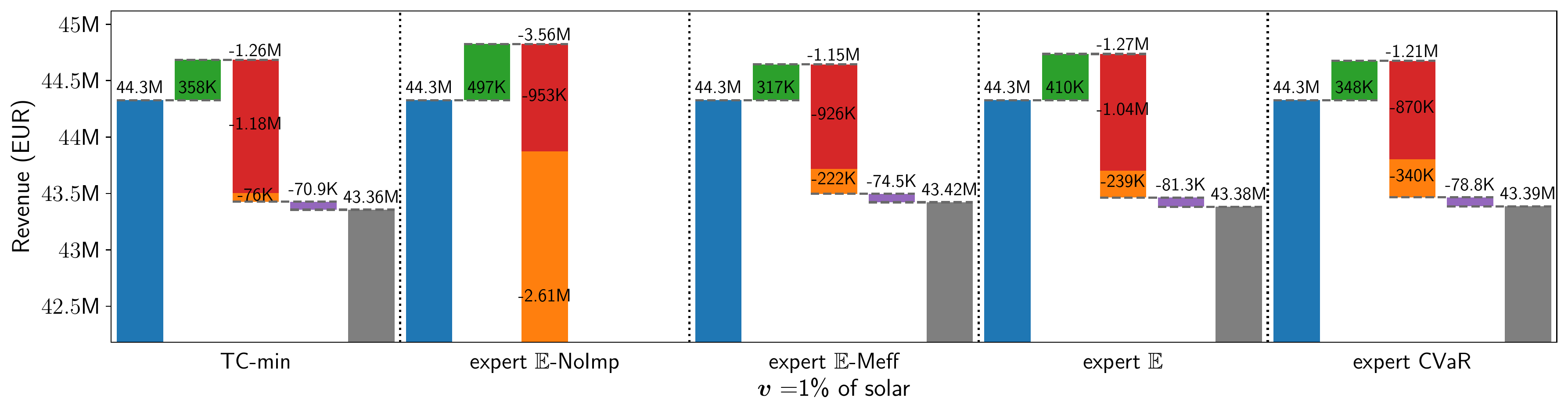}}\vspace{-0.5em}
	\subfloat{\includegraphics[width = 0.98\linewidth]{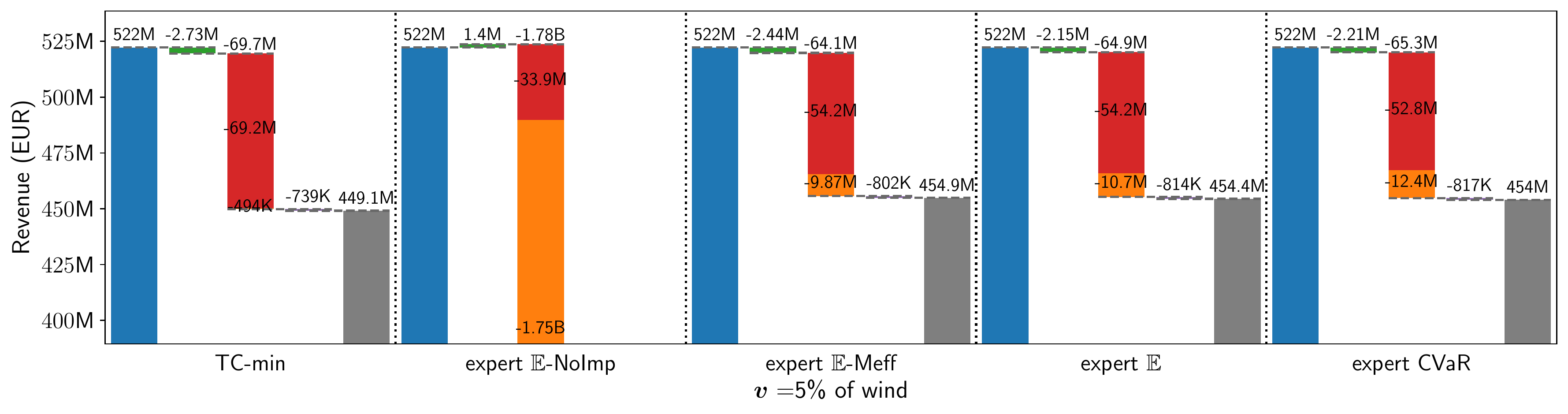}}\vspace{-0.5em}
	\subfloat{\includegraphics[width = 0.98\linewidth]{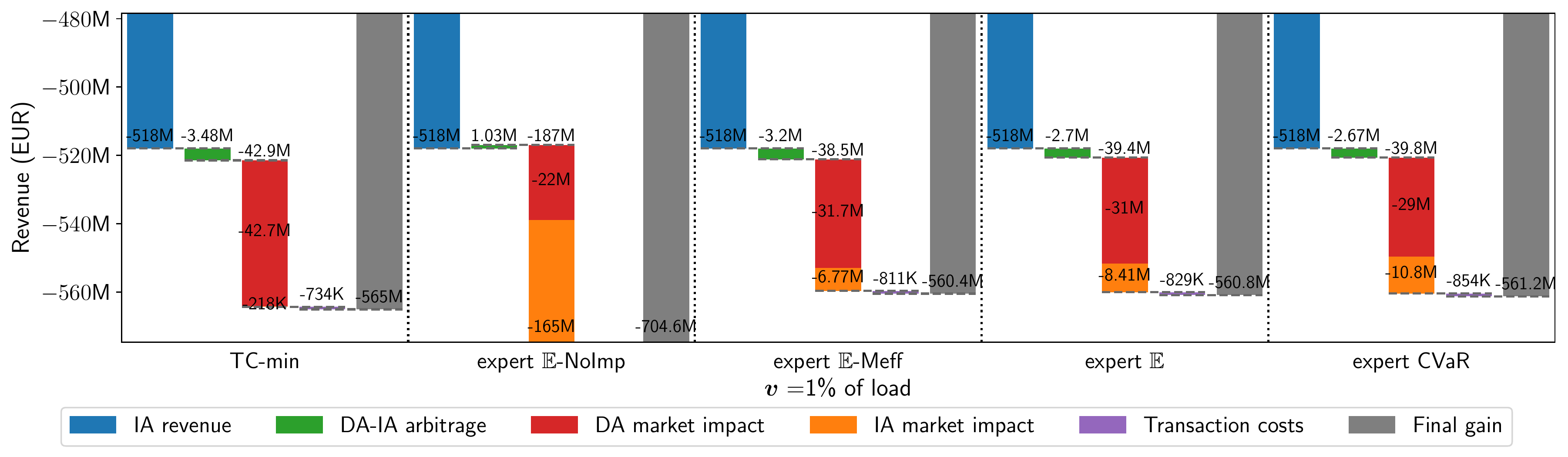}}
	\caption{Actual gain decomposition as in \eqref{eq_gain_gen_transaction_costs_decomp_interpret} for selected portfolios $\bsv$ and selected strategies in the setting of rebidding the portfolio. The impact bars of $\E$-\textbf{NoImp} strategy push the final gain to very low values. Therefore, they are not reported for the sake of legibility.}
	\label{fig:components_rebidded}
\end{figure}

Tables~\ref{tab:avg_gain_new_bids} and~\ref{tab:avg_gain_rebidded} present the average actual gain $\overline{\wtilde{G}}$ of all strategies $\bsb$ and portfolios $\bsv$ in both considered settings.
The tables are split to two parts -- the supply and the demand. In the first case, we want to maximize the price, in the second one we want to minimize it. We observe that overall the \textbf{TC-min} benchmark performs pretty well and the \textbf{IA-only} very bad, especially for the bigger portfolios. This is caused mainly by much lower liquidity in the IA market and thus a much higher price impact caused by the volume. Similarly, the $\E$-\textbf{NoImp} strategy fails to deliver satisfying results for bigger volumes. However, it performs best for small volumes where the no market impact assumption is met. Replacing the no market impact~\eqref{eq_no_market_impact_assumption} with the linear market impact~\eqref{eq_linear_market_impact_assumption} assumption keeps the very good performance for small volumes and improves it substantially for larger volumes. By not making any additional assumption on the price impact, i.e. considering the $\E$ strategy, we improve the performance for large volumes even more. However, the best strategy for large volumes is the one assuming the market efficiency~\eqref{eq_market_efficient} -- $\E$-\textbf{Meff}. Moreover, it is as good for the \textbf{naive} forecasts as for the \textbf{expert} ones. This is perfectly sensible as this strategy focuses on minimizing the overall price impact and transaction costs and ignores the possible gain from the market arbitrage. Figure~\ref{fig:components_rebidded} presents an evidence that it is far more profitable to minimize the price impact rather than maximize the arbitrage.
Let us remark that a much bigger improvement of the gain may be observed in the setting of rebidding the portfolios than in the setting of a new market player when compared to the \textbf{TC-min} strategy. This shows that our study may be particularly interesting for already existing market players. However, this also indicates that the assumption of $\delta = 1$ in equation~\eqref{eq:deltas} should be verified in future research.

The enormous difference in EPF performance between the \textbf{naive} and \textbf{expert} models does not always mean much better results in terms of trading gain. The difference in gain between the best \textbf{naive}-based and \textbf{expert}-based strategies is often a few cents, and under the market efficiency assumption it disappears. If we, however, consider the difference in total actual gain, it becomes clear that better price forecasts are advantageous. For example, if we consider the 5\% of wind portfolio in the rebidding setting and the \textbf{CVaR} strategy, the difference of 0.02 EUR/MWh in average may seem not very high. Taking into account the total gain this seeming small difference translates into over 290000~EUR of additional revenue in favour of the \textbf{expert} model over the analysed 1097 days. The additional revenue is more than 10 times greater if we compare it with the benchmark \textbf{TC-min} strategy. 

The potential advantage of utilizing perfect curve and price forecasts can be observed by comparing Tables~\ref{tab:avg_gain_new_bids} and~\ref{tab:avg_gain_rebidded} with Tables~\ref{tab:avg_gain_new_bids_perfect} and~\ref{tab:avg_gain_rebidded_perfect} from Appendix~C. Having an oracle forecast of the intersection curves $\bsC^{-1}_{d,h}$ brings further a few cents of additional gain, but the impact of oracle price forecast is far higher for all portfolios. In the example of 5\% of wind portfolio it is 0.02~EUR/MWh using oracle curve forecast and 0.50~EUR/MWh using additionally oracle price forecast. Thus, it is likely much more rewarding to improve the price models rather than the curve models, especially given the wide EPF literature.

\begin{figure}[b!]
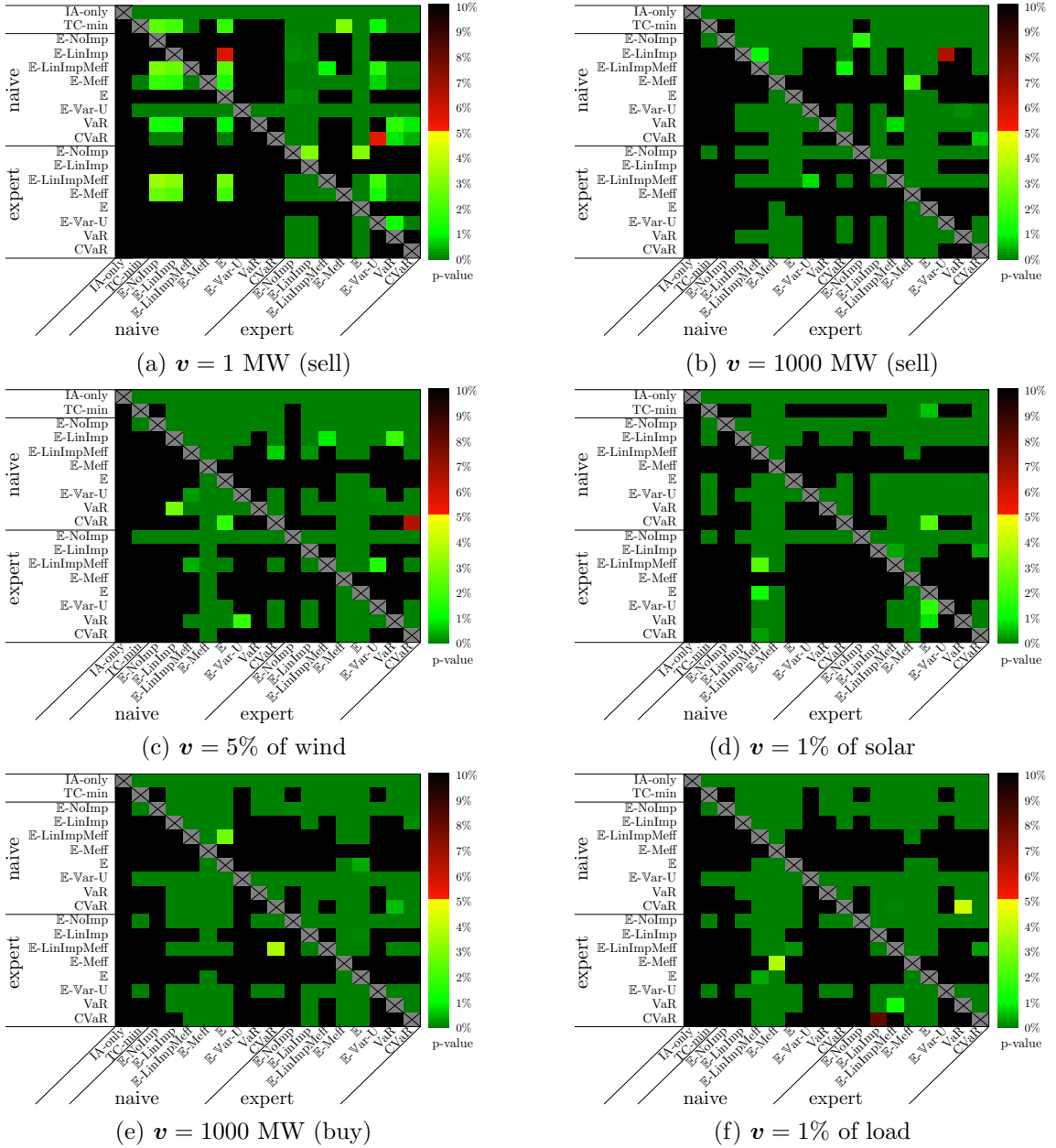

	\centering
	\subfloat[$\bsv = 1$ MW (sell)]{\resizebox{0.455\textwidth}{!}{\input{fig/ttest_adjusted_1.tex}}}\hfill
	\subfloat[$\bsv = 1000$ MW (sell)]{\resizebox{0.455\textwidth}{!}{\input{fig/ttest_adjusted_1000.tex}}}\\
	\subfloat[$\bsv = 5\%$ of wind]{\resizebox{0.455\textwidth}{!}{\input{fig/ttest_adjusted_5_of_wind.tex}}}\hfill
	\subfloat[$\bsv = 1\%$ of solar]{\resizebox{0.455\textwidth}{!}{\input{fig/ttest_adjusted_1_of_solar.tex}}}\\		\subfloat[$\bsv = 1000$ MW (buy)]{\resizebox{0.455\textwidth}{!}{\input{fig/ttest_adjusted_-1000.tex}}}\hfill
	\subfloat[$\bsv = 1\%$ of load]{\resizebox{0.455\textwidth}{!}{\input{fig/ttest_adjusted_1_of_load.tex}}}
	\caption{Results of the $\overline{\wtilde{G}}$ mean inequality test for selected portfolios $\bsv$ in the setting of rebidding the portfolio. The plots present p-values --- the closer they are to zero ($\to$ dark green), the more significant the difference is between gains of X-axis strategy (better) and gains of the Y-axis strategy (worse).}
	\label{fig:tests_rebidded}
\end{figure}

\begin{figure}[b!]
	\centering
	\subfloat{\includegraphics[width = 1\linewidth]{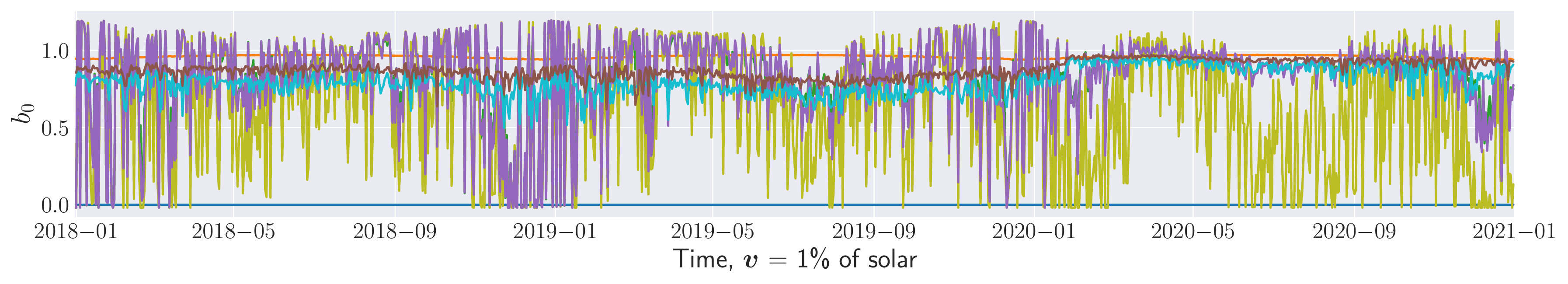}}\vspace{-0.5em}
	\subfloat{\includegraphics[width = 1\linewidth]{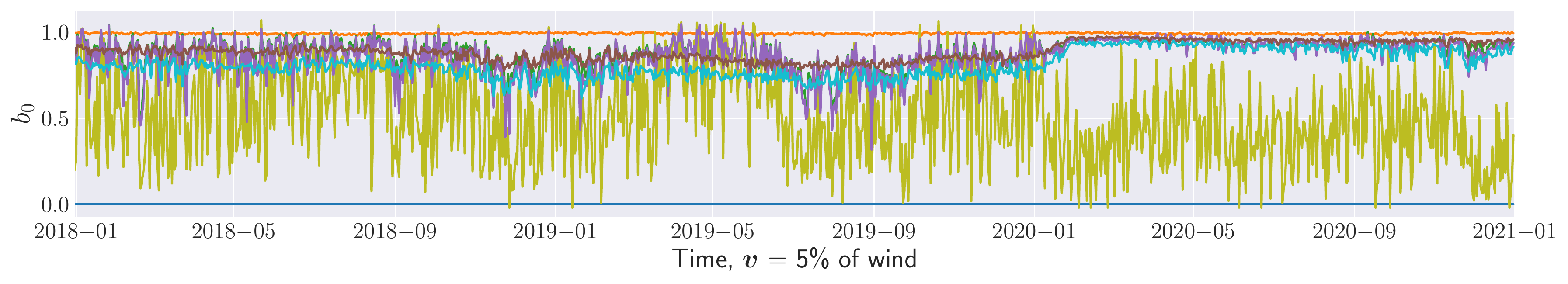}}\vspace{-0.5em}
	\subfloat{\includegraphics[width = 1\linewidth]{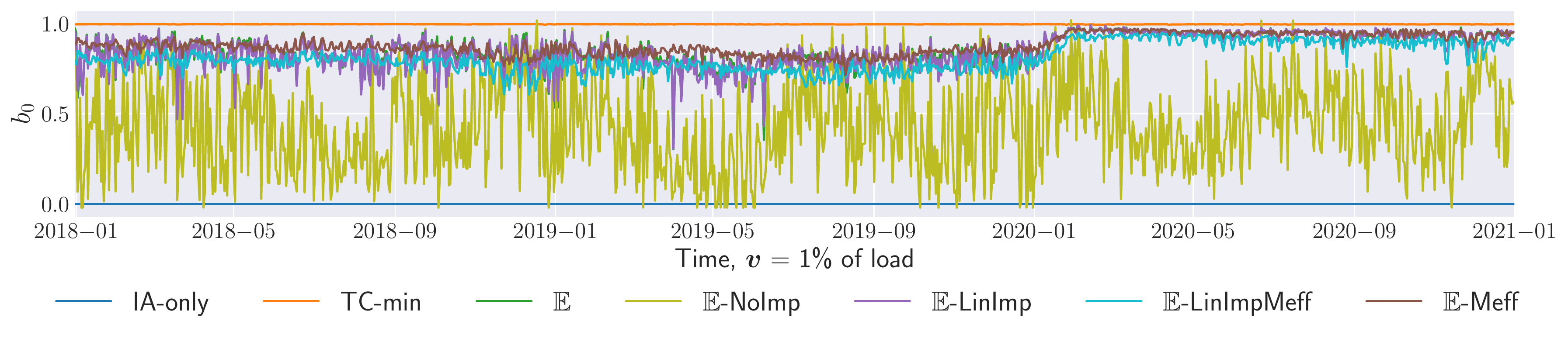}}\\
	\subfloat{\includegraphics[width = 1\linewidth]{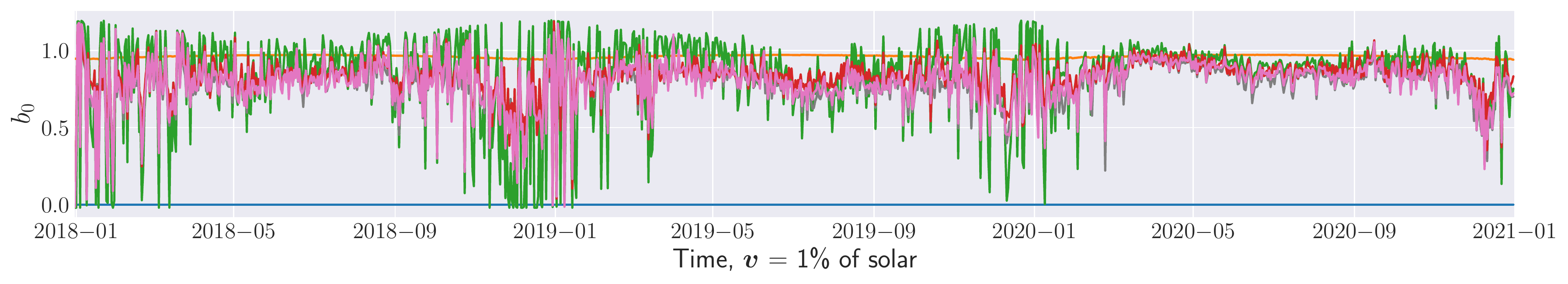}}\vspace{-0.5em}
	\subfloat{\includegraphics[width = 1\linewidth]{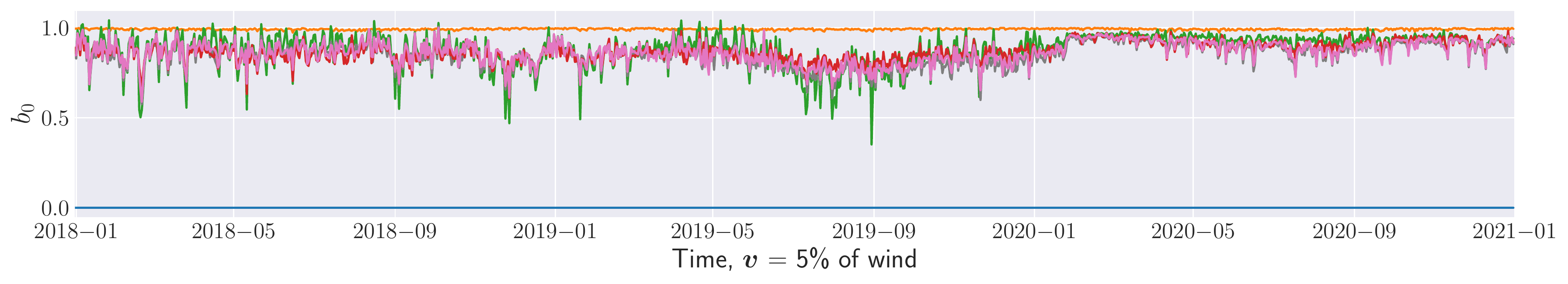}}\vspace{-0.5em}
	\subfloat{\includegraphics[width = 1\linewidth]{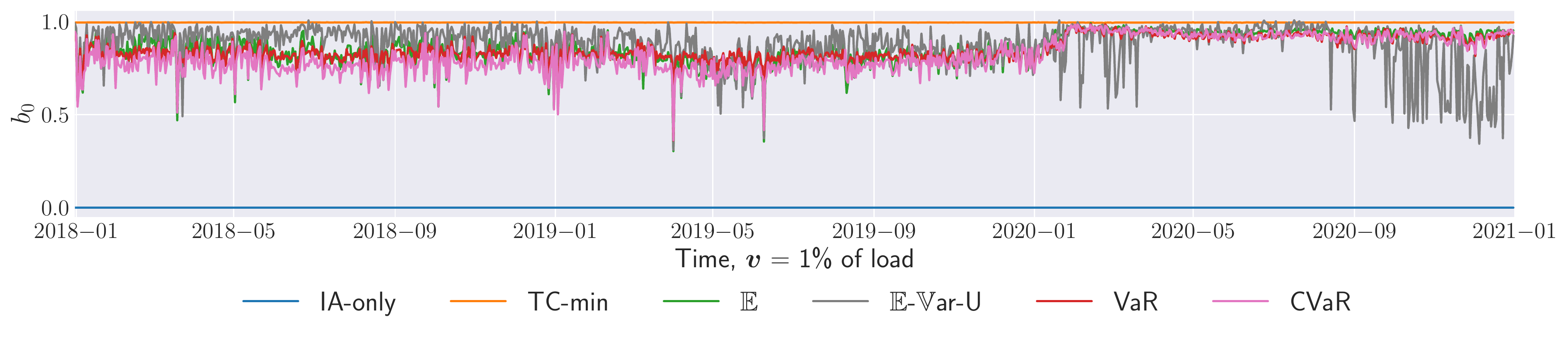}}
	\caption{The average daily weight of $b_0$ in relation to the whole $\bsb$ strategy for selected portfolios $\bsv$ in the setting of rebidding the portfolio. The \textbf{naive}-based strategies are excluded for better clarity}
	\label{fig:da_weights_rebidded}
\end{figure}

Figure~\ref{fig:tests_rebidded} shows the results of the significance tests for selected portfolios in the rebidding setting. The results for remaining portfolios can be found in Appendix~C. Strategy $\E$-\textbf{Meff} is in most cases significantly better or not significantly worse than the others. As seen before, its performance is rather undistinguishable between both price models. This means that in both settings the market participants could significantly improve their revenue. Figure~\ref{fig:da_weights_rebidded} presents the average daily weight of $b_0$ for selected portfolios in the rebidding setting. For better clarity, we plot the risk neutral and the risk averse strategies separately. Analogous plots for remaining portfolios can be found in Appendix~C. All strategies that do not neglect the market impact tend to vary higher between the considered markets when trading smaller portfolios. On the contrary, they put significantly higher weight to the more liquid DA market when trading bigger portfolios. Additionally, the strategies assuming market efficiency behave much smoother, i.e. they do not exhibit as high spikes as the other strategies. A structural change in the weights can be observed in the beginning of year 2020. The algorithms started then putting a much higher weight to the DA market what was caused by a significant decrease of the number of offers in the IA market. This lead to much higher impact on the prices of self bids in the IA market. Since the curves are forecasted using only $K=28$ last days, the algorithms could adjust their behaviour relatively fast. This shows a robustness of the proposed strategies for changing market conditions.

\section{Discussion on limitations and generalizations}\label{sec:discussion}

\subsection{Price-volume bids}
In the study we considered only unlimited bids. Thus, we assumed that the market player bids always $\bsp_{\min}$ on the supply side and $\bsp_{\max}$ on the demand side. However, of no less importance are the price-volume bids where the market participant sets a price limit on the bid. These could be particularly interesting for such electricity producers or consumers that can quite flexibly manage their production or consumption, e.g. hydropower and battery storages, or natural gas, wind and solar power plants. 
In order to use such price-volume bids, the setting requires a slight modification.

Let $(\bsb, \bsp_{\lim})$ denote the pair of vectors of bids with its corresponding price limits. Placing the limited bid in the market may induce a change in the potential price grids $\bPP^{\Sup}$ and $\bPP^{\Dem}$ as it might happen that $B^{\Sup}_{i}(p_{\lim,i}) = 0$ or $B^{\Dem}_{i}(p_{\lim,i}) = 0$ for $i \in \{0,\dots, 4\}$. Such bids result in changing the shape of the supply $\bsA_{\bsb}^{\Sup}(\bsp)$ and demand $\bsA_{\bsb}^{\Dem}(\bsp)$ curves, as they would be shifted in $\bsp = \bsp_{\lim}$. Thus, an adjustment of intersection curve estimation would be needed. Finally, to fulfil the imbalance constraint~\eqref{eq_rabap_constr} one may need to introduce also a price-limited volume vector $\bsv$. A model with multiple price-volume bids would increase the complexity of the trading problem even further.
 
\subsection{Imbalance constraint}
Our results rely heavily on the imbalance constraint.
If $\RR = \E$ and the imbalance constraint~\eqref{eq_rabap_constr} does not hold, then we receive 
\begin{equation}
\E[G(\bsb;\bsv)]= \E[\bsP_{\bsb}^*]'(\bss \odot \bsb)
- 
\bstau'\left(\bss \odot \bsb^{\abs}\right)
-  \left(\left(\bsv-\bsS'\bsb\right)^{\abs}\right)'\E[\bsR].
\end{equation}
If the imbalance $\bsv-\bsS'\bsb$ gets small, 
the imbalance penalty term gets small as well.
If 
the sign of expected imbalance price $\E[\bsR]$ is not in favour for the trader, they should have an intrinsic motivation to have $\bsv-\bsS'\bsb$ close to zero.
However, in general forcing
$\bsv-\bsS'\bsb=\bsnull$ does not lead to the global optimumm, even if $\bsR$ is independent of $\bsP_{\bsb}^*$.

\subsection{Stochastic trading volume -- imbalance uncertainty}	
The results for transaction cost minimal trading also hold for uncertain volumes to trade.
Note that the consideration of stochastic trading volumes should be the preferred choice if the trader faces uncertainty in production or consumption. For wind and solar power traders this is a natural situation, due to the meteorologically driven uncertainty in the production at time of the auctions.
If we trade stochastic volumes, denoted by the random vector $\bsV=(V_1,\ldots,V_4)$, then even the minimal transaction cost solution depends on risk~$\RR$. 

If we choose $\RR=\E$ the transaction cost minimal solution is
\begin{align}
b_0^{\text{opt}}(\bsV) = \argmin_{b_0\in \R} \E[\TT(b_0;\bsV)] = \argmin_{b_0\in \R}\left(\tau_0 |b_0| \sum^4_{i=1} \tau_{i} s_i \E \left|V_i -b_0\right| \right).
\end{align}
This can be regarded as
joint $\bstau \odot \bss$-weighted median of $(0,\bsV)'$.

We can also show the generalized version of Theorem~\ref{theorem}:
If we assume no market impact \eqref{eq_no_market_impact_assumption} 
then we receive for $\widetilde{\bsb} = (2b_0,\bsV )'-b_0\bsone$
\begin{align}
\E[G(\widetilde{\bsb})]
= \E\left[P^*_{\bsnull,0}b_0\right] + \frac{1}{4}
\sum^4_{i=1} \E\left[P^*_{\bsnull,i}(V_i-b_0)\right] -   \E[\TT(b_0)]
\end{align}
Similarly as in Section~\ref{sec:optimal_expected_profit_no_imp}, we disentangle the DA and IA component and 
get
\begin{align}
\E[G(\widetilde{\bsb})]
&= \E[P^*_{\bsnull,0}] b_0  + 
\frac{1}{4}\sum_{i=1}^4 \E[P^*_{\bsnull,i} (V_i-b_0)] -   \E[\TT(b_0)]
 \\
&=  \frac{1}{4}\sum_{i=1}^4 \left(\E[P^*_{\bsnull,i}] \E[V_i] + 
\cov[P^*_{\bsnull,i},V_i] \right)+
\MM(\bsP_{\bsnull}^*) b_0  
 -   \E[\TT(b_0)] .
 \label{eq_G_E_V}
\end{align}
by using
$\E[P^*_{\bsnull,i} (V_i-b_0)] = 
\E[P^*_{\bsnull,i}] \E[V_i]-b_0\E[P^*_{\bsnull,i}] + 
\cov[P^*_{\bsnull,i},V_i]$.
Again, the market efficiency assumption $\MM(\bsP_{\bsnull}^*)=0$ leads to Theorem~\ref{theorem}. Also 
without market efficiency assumption, the solution
$b_{0,\E\text{-NoImp}}$ has the same structure based on the transaction cost minimal solution as the $\E[P^*_{\bsnull,i}] \E[V_i]$ and $ 
\cov[P^*_{\bsnull,i},V_i]$ terms are not impacted by $b_0$.

The problematic part of this result is that it only holds under the imbalance constraint~\eqref{eq_rabap_constr}.
For non-deterministic random variables (no Dirac measures)
$\bsV-\bsS'\bsb = \bszero$ can never be satisfied.
To maintain the results from Theorem~\ref{theorem} for random volumes $\bsV$ for $\RR=\E$ we 
have to require that the expected imbalance constraint
 \begin{equation}
 \E[ (\bsV-\bsS'\bsb)^{\abs})' \bsR] = 
 \sum_{j=1}^4 \E\left[
\left|V_j- \sum_{i=0}^4 s_{i,j} b_i \right|R_j \right]   
\label{eq_expected_rebap}
 \end{equation}
does not depend on $b_0$. Unfortunately, 
assuming that \eqref{eq_expected_rebap} does not depend on $b_0$
 is a non-trivial assumption.
However, it holds
$$\E\left[
\left|V_j- \sum_{i=0}^4 s_{i,j} b_i \right|R_j \right]
=
\E \left|V_j- \sum_{i=0}^4 s_{i,j} b_i \right| \E[R_j] 
+ \cov\left[\left|V_j- \sum_{i=0}^4 s_{i,j} b_i \right|, R_j\right].
$$
Hence, if we bid $\bsb$ such that $\sum_{i=0}^4 s_{i,j} b_i = \text{med}(V_j) $ holds for the median volumes $\text{med}(V_j)$ for $j=1,\ldots,4$ it follows that 
\begin{align}
 \E \left|V_j- \sum_{i=0}^4 s_{i,j} b_i \right| = 0
\end{align}
holds.
Additionally, we require that the absolute imbalance volume
$|V_j- \sum_{i=0}^4 s_{i,j} b_i |$
has a correlation with $R_j$ that does not depend on our bids.
Note that it is not required that the imbalance volume is uncorrelated with the imbalance price.
Summarizing, Theorem~\ref{theorem} holds under the 
no impact and market efficiency assumptions
if
$\cor[ |V_j- \sum_{i=0}^4 s_{i,j} b_i |, R_j]$ does not depend on $b_0$ and 
$v_j = \text{med}(V_j)$ is chosen for trading.

However, again this solution is not the global optimum to the trading problem
$\E[ G(\bsb;\bsV)]$ even under no market impact assumption.
The general solution requires a deeper investigation of the imbalance price $\bsR$.

\subsection{Sequential impact of the markets}
We have accounted for the price impact caused by the sequential order of the DA and IA markets with the introduction of $\delta \ge 0$ parameter in equation~\eqref{eq:deltas}. The idea is that market participants can react in the IA market to the conditions that appeared in the previous one. In this paper we assumed $\delta = 1$, but the discrepancy in the results between the two considered settings suggests that a deeper investigation of this problem may be needed. We would rather suspect that this parameter is not equal for every quarter-hour (or delivery period in general). Moreover, it may also depend on the price level and possibly other factors. It is clear that its structure is not trivial and requires a thorough analysis. Therefore, in this study we limited ourselves to the assumption of $\delta = 1$, and we leave the deeper analysis for the future research.

\subsection{Beyond DA und IA}
The results of Theorem~\ref{theorem} without market impact can be generalized easily to other settings with two consecutive trading options
where the latter market allows trading on an all equally sized delivery periods.
In such a setting, we only have to adjust 
the summation matrix $\bsS$, such that $\bsS = (\bsS_{1},\bsS_{2})$ and $\bss = \bsS'\bsone / \bsone'\bsS \bsone$ for the two markets. Our setting results from choosing $\bsS_{1} = \bsone_4$ and $\bsS_{2} = \bsI_4$.
Similarly, we can model the setting in France and Great Britain where we currently have half-hourly Intraday opening auction. Here, 
we simply choose $\bsS_{1} = \bsone_2$ and $\bsS_{2} = \bsI_2$.
If we consider e.g. the future market with the day-ahead base product and the day-ahead auction, then we can model this by choosing $\bsS_{1} = \bsone_{24}$ and $\bsS_{2} = \bsI_{24}$. 
If we traded in the futures market the day-ahead base and peak product then $\bsS_{1} = (\bsone_{24}, \bsone_{\text{peak}})$ would be required with
$\bsone_{\text{peak}}=(\bsnull_{7},\bsone_{12},\bsnull_{5})$.
The theory remains basically the same, e.g. 
for $\bsb = (\bsb_1, \bsb_2)'$ the imbalance constraint~\eqref{eq_rabap_constr} leads to 
$\bsv = \bsS'\bsb = \bsS_1\bsb_1 + \bsS_2\bsb_2$ and for invertible $\bsS_2$ 
it holds
$\bsb_2= \bsS_2^{-1}(\bsv - \bsS_1 \bsb_1)$.
This 
implies $\bsb = (\bsb_1' , (\bsS_2^{-1}(\bsv-\bsS_1\bsb_1))')'$.
The transaction cost minimal solution can be derived in the same way by minimizing $\TT(\bsb_1).$
Again, this is also the optimal solution under no market impact and the market efficiency assumption which leads to
a generalized version of Theorem~\ref{theorem}.
Note that for more than two trading options the results are not easy to obtain as the problem has to be solved recursively. 

\section{Conclusion}\label{sec:conclusion}

The paper raised the issue of optimal bidding of various electricity portfolios between two auction-based markets. The analysis included the market impact estimation what is necessary for large market players. We considered also the transaction costs and provided theoretical insights regarding the minimal transaction costs strategy. The latter is optimal for risk neutral traders under the relatively plausible assumptions of market efficiency and no-market impacts if the volumes to trade are small. Additionally, we considered various strategies with no/linear/non-linear market impact assumption as well as with the (no) market efficiency assumption.
The conducted study contained a number of portfolios that mimic the majority of electricity market participants like wind and solar power producers -- from small to large ones. The results proved that even though we used very basic models to forecast the prices and curves, we could significantly improve the overall revenue for the majority of considered portfolios. Also, the analysis of gain components showed that the crucial part of gain maximization is the price impact minimization, especially for large volumes. The possible market arbitrage and the transaction costs are of marginal size compared to the impact.

We conducted an extensive analysis of all aspects of the raised problem. The price formation and trading in the European electricity auction markets together with possible extensions and challenges were discussed. The paper leaves many open questions for future research, and we believe it can be a solid foundation for that. Especially that the attention of researchers and practitioners may be brought to this topic additionally by the recent launch of intraday opening auctions in further European countries~\cite{epexIAintroduction}.

	\section*{Acknowledgments}
	This research article was partially supported by the German Research Foundation (DFG, Germany) and the National Science Center (NCN, Poland) through BEETHOVEN grant no. 2016/23/G/HS4/01005 (to FZ and MN), and 
	 the National Science Center (NCN, Poland) through MAESTRO grant No. 2018/30/A/HS4/00444 (to FZ)

\vspace{-5mm} 
\bibliographystyle{unsrtnat1}

\setlength{\bibsep}{0pt plus 0.3ex}
\bibliography{bibliography}	

%

%
        
	\section*{Appendix A}
	\renewcommand{\thesubsection}{\Alph{section}.\arabic{subsection}}
	\setcounter{section}{1}
	\subsection{Abbreviations}
	{ \small
		\begin{tabbing}
			\hspace{6em}\=\kill
			CVaR		\> Conditional value-at-risk\\
			DA 			\> Day-Ahead Auction\\
			EEX			\> European Energy Exchange \\
			EPEX		\> European Power Exchange \\
			EPF			\> Electricity price forecasting\\
			$\E\text{-}\var\text{-}\textbf{U}$		\> Mean-variance utility\\
			IA			\> Intraday Auction\\
			IA-only		\> strategy to bid only in the IA market\\
			LinImp		\> strategy assuming the linear price impact\\
			Meff		\> strategy assuming the market efficiency\\
			NoImp		\> strategy assuming no price impact\\
			REBAP		\> Cross-control area uniform balancing energy price (abbreviation from German)\\
			TC-min		\> strategy to bid at minimum transaction cost\\
			VaR			\> Value-at-risk\\
		\end{tabbing}
	}
		\subsection{Notation used in Sections~\ref{sec:price_formation}-\ref{sec:forecasting_model}}
			{ \small
			\begin{tabbing}
				\hspace{6em}\=\kill
			$\odot$		\> element-wise multiplication (Hadamard product)\\	
			$\bsz^{\abs}$		\> element-wise absolute value, i.e. $\bsz^{\abs} = \bsz^+ + \bsz^-$\\			
			$A^{\Dem/\Sup}_{\bsb,i}$	\> aggregated demand/supply curve impacted by own bid $\bsb$ in the $i$th market\\			
			$\widehat{\bsA}_{\bsnull,d,h}^{\Dem/\Sup}$ \> estimator for the not impacted aggregated demand/supply curve for day $d$ and hour $h$\\
			$a_i$		\> expected slope of the linear impact in the $i$th market\\
			$\bsa$		\> vector of expected slopes of the linear impact\\
			$\what{\bsa}_{d,h}$ \> estimator of the slope of the linear impact on day $d$ and hour $h$\\
			$\alpha$		\> risk aversion parameter of VaR and CVaR\\
			$B^{\Dem/\Sup}_{\bsb,i}$		\> non-negative demand/supply volume bids impacted by own bid $\bsb$\\
			$\bsB^{\Dem/\Sup}_{\bsb}$		\> vector of non-negative demand/supply volume bids impacted by own bid $\bsb$\\
			$b_i$		\> bid in the $i$th market\\
			$\bsb$		\> vector of bids\\
			$\bsb^{+/-}$	\> element-wise positive/negative part of $\bsb$\\
			$\widetilde{\bsb}$ \> $\widetilde{\bsb} = (2b_0,\bsv )'-b_0\bsone$ -- vector of bids linear in $b_0$\\
			$\what{\bsC}_{d,h}$ \> estimated intersection curve for day $d$ and hour $h$\\
			$\bsDelta_{\bsb}$		\> price impact due to the trading of volume $\bsb$\\
			$\what{\bsDelta}_{\bsb,d,h}^{*,m}$ \> estimated price impact of $m$th scenario due to the trading of volume $\bsb$\\
			$\delta$ 		\> market efficiency factor\\
			$\widehat{\bseps}_{d,h}^m$	\> $m$th drawn with replacement in-sample residual for day $d$ and hour $h$\\			
			$G$		\> gain function\\
			$\wtilde{G}$ \> actual gain function\\
			$\gamma$		\> risk aversion parameter of $\E\text{-}\var\text{-}\textbf{U}$ strategy\\
			$i$ 		\> $i \in \{0, \dots, 4\}$ is a market index with $0=\text{DA}$, $1, \dots, 4 = \text{IA}$\\
			$\MM(\bsP^*_{\bsnull})$ 		\> expected price difference between the DA and the IA\\
			$m$ 		\> $m = 1, \dots, M$ is a bootstrapping index\\
			$P^*_{\bsb,i}$		\> market clearing price impacted by bid $\bsb$\\
			$\bsP^*_{\bsb}$		\> vector of market clearing prices impacted by bid $\bsb$\\
			$\widehat{\bsP}_{\bsnull, d,h}^{*,m}$ \> vector $m$th forecasted prices for day $d$ and hour $h$\\		
			$\PP^{\Dem/\Sup}_i$		\> price grid in the considered auction\\
			$\bPP^{\Dem/\Sup}$		\> vector of price grids\\
			$p_{\max}$ 		\> maximum price bid, $p_{\max,\text{DA}}=p_{\max,\text{IA}}=3000$\\
			$\bsp_{\max}$ 		\> vector of maximum price bids\\			
			$p_{\min}$ 		\> minimum price bid, $p_{\min,\text{DA}}=-500$, $p_{\min,\text{IA}}=-3000$\\
			$\bsp_{\min}$ 		\> vector of minimum price bids\\			
			$\RR$		\> risk functional\\
			$R_j$		\> imbalance price in $j$th quarter-hour\\
			$\bsR$		\> imbalance (REBAP) price\\
			$\bsS$		\> $\bsS =(S_{i,j})= (\bsone_4,\bsI_4)'$ is a 5x4 dimensional summation matrix\\
			$\bss$ 		\>  $\bss=\bsS'\bsone_4/4 = (1,.25,.25,.25,.25)'$ is a summation vector converting MW to MWh\\
			$\mathcal{T}$		\> simplified transaction cost function\\
			$\tau_i$ 		\> transaction cost in the $i$th market\\
			$\bstau$ 		\> vector of transaction costs\\
			$V^*_{\bsb,i}$		\> market clearing volume impacted by bid $\bsb$ in the $i$th market\\
			$\bsV^*_{\bsb}$		\> vector of market clearing volumes impacted by bid $\bsb$\\
			$v_j$		\> volume to be traded in $j$th quarter-hour with $j \in \{1,\dots,4\}$\\
			$\bsv$		\> vector of volumes to be traded in the given hour\\			
			$\what{\bsxi}_{\bsnull,d,h}^{*,m}$ \> $m$th shift for the intersection adjustment for day $d$ and hour $h$\\

		\end{tabbing}
}
\section*{Appendix B}

We present the values of RMSE, MAE and CRPS which are strictly proper scoring rules for the mean, median and marginal distribution forecasts \cite{gneiting2007strictly}. These measures are commonly used by the practitioners and researchers \cite{narajewski2020ensemble}. We report also the bias of the forecasted price trajectories.
	The formulas are given by
	\begin{align}
		\text{RMSE}  &=  \sqrt{\frac{1}{24\cdot 5N } \sum_{h = 1}^{24} \sum_{d = 1}^{N} \sum_{i = 0}^{4}\left({P}_{\bsnull,i,d,h}^{*} - \frac{1}{M} \sum_{m=1}^{M} \widehat{P}_{\bsnull,i,d,h}^{*,m}\right)^2},
		\label{eq:rmse} \\
		\text{MAE} &=  \frac{1}{24\cdot 5N }   \sum_{h = 1}^{24} \sum_{d = 1}^{N} \sum_{i = 0}^{4} \left|{P}_{\bsnull,i,d,h}^{*} - \text{med}_{m = 1, \dots, M} \left(\widehat{P}_{\bsnull,i,d,h}^{*,m} \right) \right|
	\end{align}
	where $\widehat{P}_{\bsnull,i,d,h}^{*,m}$ is the $m$-th simulation of ${P}_{\bsnull,i,d,h}^{*}$ and $\text{med}_{m = 1, \dots, M} \left(\widehat{P}_{\bsnull,i,d,h}^{*,m} \right)$ is the median of $M$ simulated $\widehat{P}_{\bsnull,i,d,h}^{*,m}$ prices.
	
	We approximate the CRPS using the pinball loss
	\begin{equation}
		\text{CRPS}_{i,d,h} = \frac{1}{R} \sum_{\tau \in r} \text{PB}_{i,d,h}^{\tau}
	\end{equation}
	for a dense equidistant grid of probabilities $r$ between 0 and 1 of size $R$, see e.g.~\cite{nowotarski2018recent}. In this study, we consider $r = \{0.01, 0.02,\dots, 0.99\}$ of size $R = 99$. $\text{PB}_{i,d,h}^{\tau}$ is the pinball loss with respect to probability $\tau$. Its formula is given by
	\begin{equation}
		\text{PB}_{i,d,h}^{\tau} = \left(\tau - \mathds{1}_{\left\{ {P}_{\bsnull,i,d,h}^{*} < Q_{m = 1, \dots, M}^{\tau}\left(\widehat{P}_{\bsnull,i,d,h}^{*,m}\right)\right\}} \right) \left({P}_{\bsnull,i,d,h}^{*} - Q_{m = 1, \dots, M}^{\tau}\left(\widehat{P}_{\bsnull,i,d,h}^{*,m}\right) \right)  
	\end{equation}
	where $Q_{m = 1, \dots, M}^{\tau}\left(\widehat{P}_{\bsnull,i,d,h}^{*,m}\right)$ is the $\tau$-th quantile of $M$ simulated $\widehat{P}_{\bsnull,i,d,h}^{*,m}$ prices. To calculate the overall CRPS value we use a simple average
	\begin{equation}
		\text{CRPS} = \frac{1}{24\cdot 5N }   \sum_{h = 1}^{24} \sum_{d = 1}^{N} \sum_{i = 0}^{4} \text{CRPS}_{i,d,h}.
	\end{equation}

	\begin{table}[b!]
	\centering
	\begingroup\small
	\begin{tabular}{rrrrr}
		\hline
		& MAE & RMSE & CRPS & bias \\ 
		\hline
		\textbf{naive} & 10.73 & 16.64 & 4.17 & 0.08 \\ 
		\textbf{expert} & 6.00 & 8.68 & 2.22 & 0.22 \\ 
		\hline
	\end{tabular}
	\endgroup
	\caption{Error measures of the considered price models} 
	\label{tab:price_errors}
\end{table}

Table~\ref{tab:price_errors} shows the error measures of the two considered price models based on the whole out-of-sample data. Let us recall that the out-of-sample consists of 3 years of data (years 2018 to 2020). We observe a huge difference in the performance of the two models. The \textbf{expert} model reports lower errors, but is slightly more biased. Its performance is naturally not a surprise and is inline with the EPF literature \cite{weron2014electricity, ziel2018day, uniejewski2017variance, uniejewski2018efficient, narajewski2020econometric, uniejewski2019understanding}. Let us note that the forecasts could be easily improved using a higher number of regressors or more sophisticated estimation methods both for the point and probabilistic models.

\section*{Appendix C}	

\begin{table}[h!]
	\centering
\begingroup\small
\begin{adjustbox}{max width=1\textwidth}
	\setlength{\tabcolsep}{1pt}
	\begin{tabular}{|cr|rrrrrrrr||rrrrrr|}
		\hline
		&  & \multicolumn{8}{c||}{Supply/Sell (the higher the price the better)} & \multicolumn{6}{c|}{Demand/Buy (the lower the price the better)} \\
		\rotatebox[origin=c]{90}{Model} & Strategy & \thead{1\\MW} & \thead{10\\MW} & \thead{100\\MW} & \thead{1000\\MW} & \thead{1\% of\\ wind} & \thead{5\% of\\ wind} & \thead{1\% of\\ solar} & \thead{5\% of\\ solar} & \thead{1\\MW} & \thead{10\\MW} & \thead{100\\MW} & \thead{1000\\MW} & \thead{1\% of\\ load} & \thead{5\% of\\ load} \\ 
		\hline
		&IA-only & \cellcolor[rgb]{1,0.84,0.5} {37.20} & \cellcolor[rgb]{1,0.691,0.5} {36.87} & \cellcolor[rgb]{1,0.5,0.55} {33.68} & \cellcolor[rgb]{1,0.5,0.55} {-61.20} & \cellcolor[rgb]{1,0.5,0.55} {21.78} & \cellcolor[rgb]{1,0.5,0.55} {-214.07} & \cellcolor[rgb]{1,0.5,0.55} {26.69} & \cellcolor[rgb]{1,0.5,0.55} {-84.74} & \cellcolor[rgb]{1,0.978,0.5} {37.49} & \cellcolor[rgb]{1,0.814,0.5} {37.84} & \cellcolor[rgb]{1,0.5,0.55} {40.94} & \cellcolor[rgb]{1,0.5,0.55} {117.18} & \cellcolor[rgb]{1,0.5,0.55} {66.31} & \cellcolor[rgb]{1,0.5,0.55} {1055.64} \\ 
		&TC-min & \cellcolor[rgb]{0.843,1,0.5} {37.43} & \cellcolor[rgb]{0.5,0.9,0.5} \textbf{37.34} & \cellcolor[rgb]{0.553,0.918,0.5} {36.78} & \cellcolor[rgb]{1,0.931,0.5} { 32.55} & \cellcolor[rgb]{0.762,0.987,0.5} {29.73} & \cellcolor[rgb]{1,0.859,0.5} {  25.48} & \cellcolor[rgb]{0.5,0.9,0.5} {\textbf{32.69}} & \cellcolor[rgb]{0.821,1,0.5} { 29.38} & \cellcolor[rgb]{1,0.909,0.5} {37.57} & \cellcolor[rgb]{1,0.961,0.5} {37.67} & \cellcolor[rgb]{0.653,0.951,0.5} {38.23} & \cellcolor[rgb]{1,0.894,0.5} { 42.82} & \cellcolor[rgb]{1,0.992,0.5} {42.13} & \cellcolor[rgb]{1,0.5,0.55} {  63.10} \\   \hline
		\parbox[t]{2mm}{\multirow{8}{*}{\rotatebox[origin=c]{90}{naive}}} 
		& $\E$-NoImp & \cellcolor[rgb]{0.708,0.969,0.5} {37.46} & \cellcolor[rgb]{0.913,1,0.5} {37.25} & \cellcolor[rgb]{1,0.5,0.55} {35.41} & \cellcolor[rgb]{1,0.5,0.55} {-13.95} & \cellcolor[rgb]{1,0.5,0.55} {26.21} & \cellcolor[rgb]{1,0.5,0.55} { -85.85} & \cellcolor[rgb]{1,0.5,0.55} {30.30} & \cellcolor[rgb]{1,0.5,0.55} {-24.90} & \cellcolor[rgb]{0.709,0.97,0.5} {37.39} & \cellcolor[rgb]{0.981,1,0.5} {37.62} & \cellcolor[rgb]{1,0.5,0.55} {39.45} & \cellcolor[rgb]{1,0.5,0.55} { 79.09} & \cellcolor[rgb]{1,0.5,0.55} {53.30} & \cellcolor[rgb]{1,0.5,0.55} { 558.32} \\ 
		& $\E$-LinImp & \cellcolor[rgb]{0.708,0.969,0.5} {37.46} & \cellcolor[rgb]{0.809,1,0.5} {37.28} & \cellcolor[rgb]{1,0.923,0.5} {36.59} & \cellcolor[rgb]{1,0.673,0.5} { 32.29} & \cellcolor[rgb]{1,0.816,0.5} {29.51} & \cellcolor[rgb]{1,0.5,0.55} {  24.46} & \cellcolor[rgb]{1,0.931,0.5} {32.52} & \cellcolor[rgb]{1,0.758,0.5} { 29.12} & \cellcolor[rgb]{0.709,0.97,0.5} {37.39} & \cellcolor[rgb]{0.912,1,0.5} {37.60} & \cellcolor[rgb]{1,0.955,0.5} {38.37} & \cellcolor[rgb]{1,0.772,0.5} { 42.98} & \cellcolor[rgb]{1,0.868,0.5} {42.29} & \cellcolor[rgb]{1,0.5,0.55} {  66.25} \\ 
		& $\E$-LinImpMeff & \cellcolor[rgb]{0.843,1,0.5} {37.43} & \cellcolor[rgb]{0.552,0.917,0.5} {37.33} & \cellcolor[rgb]{0.812,1,0.5} {36.73} & \cellcolor[rgb]{1,0.623,0.5} { 32.24} & \cellcolor[rgb]{0.993,1,0.5} {29.68} & \cellcolor[rgb]{1,0.5,0.55} {  24.94} & \cellcolor[rgb]{1,0.991,0.5} {32.58} & \cellcolor[rgb]{1,0.647,0.5} { 29.02} & \cellcolor[rgb]{1,0.909,0.5} {37.57} & \cellcolor[rgb]{1,0.961,0.5} {37.67} & \cellcolor[rgb]{0.704,0.968,0.5} {38.24} & \cellcolor[rgb]{1,0.756,0.5} { 43.00} & \cellcolor[rgb]{1,0.922,0.5} {42.22} & \cellcolor[rgb]{1,0.5,0.55} {  68.91} \\ 
		& $\E$-Meff & \cellcolor[rgb]{0.877,1,0.5} {37.42} & \cellcolor[rgb]{0.657,0.952,0.5} {37.31} & \cellcolor[rgb]{0.553,0.918,0.5} {36.78} & \cellcolor[rgb]{0.738,0.979,0.5} { 32.68} & \cellcolor[rgb]{0.566,0.922,0.5} {29.76} & \cellcolor[rgb]{0.803,1,0.5} {  25.63} & \cellcolor[rgb]{0.619,0.94,0.5} {32.67} & \cellcolor[rgb]{0.5,0.9,0.5} { 29.43} & \cellcolor[rgb]{1,0.917,0.5} {37.56} & \cellcolor[rgb]{1,0.987,0.5} {37.64} & \cellcolor[rgb]{0.5,0.9,0.5} {38.20} & \cellcolor[rgb]{0.683,0.961,0.5} { 42.59} & \cellcolor[rgb]{0.593,0.931,0.5} {42.01} & \cellcolor[rgb]{1,0.945,0.5} {  61.20} \\ 
		& $\E$ & \cellcolor[rgb]{0.708,0.969,0.5} {37.46} & \cellcolor[rgb]{0.657,0.952,0.5} {37.31} & \cellcolor[rgb]{0.953,1,0.5} {36.69} & \cellcolor[rgb]{0.878,1,0.5} { 32.65} & \cellcolor[rgb]{1,0.98,0.5} {29.66} & \cellcolor[rgb]{1,0.986,0.5} {  25.58} & \cellcolor[rgb]{0.839,1,0.5} {32.63} & \cellcolor[rgb]{0.633,0.944,0.5} { 29.41} & \cellcolor[rgb]{0.709,0.97,0.5} {37.39} & \cellcolor[rgb]{0.843,1,0.5} {37.58} & \cellcolor[rgb]{0.804,1,0.5} {38.26} & \cellcolor[rgb]{0.814,1,0.5} { 42.62} & \cellcolor[rgb]{0.732,0.977,0.5} {42.04} & \cellcolor[rgb]{1,0.919,0.5} {  61.25} \\ 
		& $\E$-$\var$-U & \cellcolor[rgb]{1,0.987,0.5} {37.37} & \cellcolor[rgb]{1,0.978,0.5} {37.20} & \cellcolor[rgb]{1,0.835,0.5} {36.49} & \cellcolor[rgb]{1,0.842,0.5} { 32.46} & \cellcolor[rgb]{1,0.783,0.5} {29.48} & \cellcolor[rgb]{1,0.821,0.5} {  25.45} & \cellcolor[rgb]{1,0.822,0.5} {32.41} & \cellcolor[rgb]{1,0.89,0.5} { 29.24} & \cellcolor[rgb]{1,0.943,0.5} {37.53} & \cellcolor[rgb]{1,0.97,0.5} {37.66} & \cellcolor[rgb]{1,0.802,0.5} {38.55} & \cellcolor[rgb]{1,0.5,0.55} {101.01} & \cellcolor[rgb]{1,0.5,0.55} {58.61} & \cellcolor[rgb]{1,0.5,0.55} { 362.07} \\ 
		& VaR & \cellcolor[rgb]{0.808,1,0.5} {37.44} & \cellcolor[rgb]{0.761,0.987,0.5} {37.29} & \cellcolor[rgb]{1,0.976,0.5} {36.65} & \cellcolor[rgb]{1,0.832,0.5} { 32.45} & \cellcolor[rgb]{1,0.936,0.5} {29.62} & \cellcolor[rgb]{1,0.783,0.5} {  25.42} & \cellcolor[rgb]{1,0.951,0.5} {32.54} & \cellcolor[rgb]{1,0.857,0.5} { 29.21} & \cellcolor[rgb]{0.913,1,0.5} {37.44} & \cellcolor[rgb]{0.947,1,0.5} {37.61} & \cellcolor[rgb]{1,0.998,0.5} {38.32} & \cellcolor[rgb]{1,0.84,0.5} { 42.89} & \cellcolor[rgb]{1,0.937,0.5} {42.20} & \cellcolor[rgb]{1,0.5,0.55} {  62.85} \\ 
		& CVaR & \cellcolor[rgb]{0.843,1,0.5} {37.43} & \cellcolor[rgb]{0.844,1,0.5} {37.27} & \cellcolor[rgb]{1,0.976,0.5} {36.65} & \cellcolor[rgb]{1,0.931,0.5} { 32.55} & \cellcolor[rgb]{1,0.958,0.5} {29.64} & \cellcolor[rgb]{1,0.91,0.5} {  25.52} & \cellcolor[rgb]{1,0.981,0.5} {32.57} & \cellcolor[rgb]{1,0.979,0.5} { 29.32} & \cellcolor[rgb]{0.809,1,0.5} {37.41} & \cellcolor[rgb]{0.947,1,0.5} {37.61} & \cellcolor[rgb]{1,0.981,0.5} {38.34} & \cellcolor[rgb]{1,0.84,0.5} { 42.89} & \cellcolor[rgb]{1,0.945,0.5} {42.19} & \cellcolor[rgb]{1,0.5,0.55} {  63.86} \\  \hline
		\parbox[t]{2mm}{\multirow{8}{*}{\rotatebox[origin=c]{90}{expert}}} 
		& $\E$-NoImp & \cellcolor[rgb]{0.5,0.9,0.5} \textbf{37.50} & \cellcolor[rgb]{0.761,0.987,0.5} {37.29} & \cellcolor[rgb]{1,0.5,0.55} {35.47} & \cellcolor[rgb]{1,0.5,0.55} {-10.24} & \cellcolor[rgb]{1,0.5,0.55} {25.88} & \cellcolor[rgb]{1,0.5,0.55} {-107.88} & \cellcolor[rgb]{1,0.5,0.55} {31.30} & \cellcolor[rgb]{1,0.5,0.55} {  2.68} & \cellcolor[rgb]{0.5,0.9,0.5} {37.35} & \cellcolor[rgb]{0.808,1,0.5} {37.57} & \cellcolor[rgb]{1,0.5,0.55} {39.37} & \cellcolor[rgb]{1,0.5,0.55} { 76.70} & \cellcolor[rgb]{1,0.5,0.55} {51.83} & \cellcolor[rgb]{1,0.5,0.55} { 517.79} \\ 
		& $\E$-LinImp & \cellcolor[rgb]{0.5,0.9,0.5} {37.50} & \cellcolor[rgb]{0.604,0.935,0.5} {37.32} & \cellcolor[rgb]{1,0.994,0.5} {36.67} & \cellcolor[rgb]{1,0.752,0.5} { 32.37} & \cellcolor[rgb]{1,0.936,0.5} {29.62} & \cellcolor[rgb]{1,0.5,0.55} {  25.12} & \cellcolor[rgb]{0.798,0.999,0.5} {32.64} & \cellcolor[rgb]{1,0.901,0.5} { 29.25} & \cellcolor[rgb]{0.5,0.9,0.5} \textbf{37.35} & \cellcolor[rgb]{0.604,0.935,0.5} {37.53} & \cellcolor[rgb]{0.838,1,0.5} {38.27} & \cellcolor[rgb]{1,0.825,0.5} { 42.91} & \cellcolor[rgb]{1,0.922,0.5} {42.22} & \cellcolor[rgb]{1,0.5,0.55} {  68.17} \\
		& $\E$-LinImpMeff & \cellcolor[rgb]{0.843,1,0.5} {37.43} & \cellcolor[rgb]{0.5,0.9,0.5} {37.34} & \cellcolor[rgb]{0.812,1,0.5} {36.73} & \cellcolor[rgb]{1,0.653,0.5} { 32.27} & \cellcolor[rgb]{0.862,1,0.5} {29.71} & \cellcolor[rgb]{1,0.5,0.55} {  25.09} & \cellcolor[rgb]{1,0.981,0.5} {32.57} & \cellcolor[rgb]{1,0.658,0.5} { 29.03} & \cellcolor[rgb]{1,0.909,0.5} {37.57} & \cellcolor[rgb]{1,0.961,0.5} {37.67} & \cellcolor[rgb]{0.704,0.968,0.5} {38.24} & \cellcolor[rgb]{1,0.733,0.5} { 43.03} & \cellcolor[rgb]{1,0.914,0.5} {42.23} & \cellcolor[rgb]{1,0.5,0.55} {  71.24} \\ 
		& $\E$-Meff & \cellcolor[rgb]{0.843,1,0.5} {37.43} & \cellcolor[rgb]{0.552,0.917,0.5} {37.33} & \cellcolor[rgb]{0.5,0.9,0.5} {\textbf{36.79}} & \cellcolor[rgb]{0.5,0.9,0.5} {\textbf{32.72}} & \cellcolor[rgb]{0.5,0.9,0.5} {\textbf{29.77}} & \cellcolor[rgb]{0.5,0.9,0.5} {\textbf{25.67}} & \cellcolor[rgb]{0.619,0.94,0.5} {32.67} & \cellcolor[rgb]{0.5,0.9,0.5} \textbf{ 29.43} & \cellcolor[rgb]{1,0.909,0.5} {37.57} & \cellcolor[rgb]{1,0.97,0.5} {37.66} & \cellcolor[rgb]{0.5,0.9,0.5} \textbf{38.20} & \cellcolor[rgb]{0.5,0.9,0.5} {\textbf{42.55}} & \cellcolor[rgb]{0.5,0.9,0.5} {\textbf{41.99}} & \cellcolor[rgb]{0.5,0.9,0.5} {\textbf{60.91}} \\ 
		& $\E$ & \cellcolor[rgb]{0.5,0.9,0.5} {37.50} & \cellcolor[rgb]{0.5,0.9,0.5} {37.34} & \cellcolor[rgb]{0.812,1,0.5} {36.73} & \cellcolor[rgb]{0.619,0.94,0.5} { 32.70} & \cellcolor[rgb]{0.906,1,0.5} {29.70} & \cellcolor[rgb]{0.803,1,0.5} {  25.63} & \cellcolor[rgb]{0.619,0.94,0.5} {32.67} & \cellcolor[rgb]{0.566,0.922,0.5} { 29.42} & \cellcolor[rgb]{0.5,0.9,0.5} {37.35} & \cellcolor[rgb]{0.5,0.9,0.5} {\textbf{37.51}} & \cellcolor[rgb]{0.602,0.934,0.5} {38.22} & \cellcolor[rgb]{0.637,0.946,0.5} { 42.58} & \cellcolor[rgb]{0.686,0.962,0.5} {42.03} & \cellcolor[rgb]{0.596,0.932,0.5} {  60.94} \\ 
		& $\E$-$\var$-U & \cellcolor[rgb]{0.76,0.987,0.5} {37.45} & \cellcolor[rgb]{0.604,0.935,0.5} {37.32} & \cellcolor[rgb]{0.953,1,0.5} {36.69} & \cellcolor[rgb]{1,0.901,0.5} { 32.52} & \cellcolor[rgb]{0.993,1,0.5} {29.68} & \cellcolor[rgb]{1,0.834,0.5} {  25.46} & \cellcolor[rgb]{1,0.971,0.5} {32.56} & \cellcolor[rgb]{1,0.868,0.5} { 29.22} & \cellcolor[rgb]{0.761,0.987,0.5} {37.40} & \cellcolor[rgb]{0.808,1,0.5} {37.57} & \cellcolor[rgb]{1,0.998,0.5} {38.32} & \cellcolor[rgb]{1,0.5,0.55} { 60.66} & \cellcolor[rgb]{1,0.5,0.55} {45.93} & \cellcolor[rgb]{1,0.5,0.55} { 230.18} \\ 
		& VaR & \cellcolor[rgb]{0.708,0.969,0.5} {37.46} & \cellcolor[rgb]{0.552,0.917,0.5} {37.33} & \cellcolor[rgb]{0.883,1,0.5} {36.71} & \cellcolor[rgb]{1,0.911,0.5} { 32.53} & \cellcolor[rgb]{1,0.991,0.5} {29.67} & \cellcolor[rgb]{1,0.847,0.5} {  25.47} & \cellcolor[rgb]{0.918,1,0.5} {32.61} & \cellcolor[rgb]{1,0.89,0.5} { 29.24} & \cellcolor[rgb]{0.844,1,0.5} {37.42} & \cellcolor[rgb]{0.843,1,0.5} {37.58} & \cellcolor[rgb]{0.704,0.968,0.5} {38.24} & \cellcolor[rgb]{1,0.947,0.5} { 42.75} & \cellcolor[rgb]{1,0.999,0.5} {42.12} & \cellcolor[rgb]{1,0.5,0.55} {  62.17} \\ 
		& CVaR & \cellcolor[rgb]{0.708,0.969,0.5} {37.46} & \cellcolor[rgb]{0.604,0.935,0.5} {37.32} & \cellcolor[rgb]{0.883,1,0.5} {36.71} & \cellcolor[rgb]{1,0.971,0.5} { 32.59} & \cellcolor[rgb]{0.949,1,0.5} {29.69} & \cellcolor[rgb]{1,0.948,0.5} {  25.55} & \cellcolor[rgb]{0.878,1,0.5} {32.62} & \cellcolor[rgb]{1,0.99,0.5} { 29.33} & \cellcolor[rgb]{0.809,1,0.5} {37.41} & \cellcolor[rgb]{0.76,0.987,0.5} {37.56} & \cellcolor[rgb]{0.602,0.934,0.5} {38.22} & \cellcolor[rgb]{1,0.963,0.5} { 42.73} & \cellcolor[rgb]{0.91,1,0.5} {42.09} & \cellcolor[rgb]{1,0.524,0.5} {  61.99} \\ \hline
		& perfect forecast & \cellcolor[rgb]{0.5,0.9,0.5} {38.42} & \cellcolor[rgb]{0.5,0.9,0.5} {38.23} & \cellcolor[rgb]{0.5,0.9,0.5} {37.30} & \cellcolor[rgb]{0.5,0.9,0.5} { 33.01} & \cellcolor[rgb]{0.5,0.9,0.5} {30.31} & \cellcolor[rgb]{0.5,0.9,0.5} {  26.04} & \cellcolor[rgb]{0.5,0.9,0.5} {33.16} & \cellcolor[rgb]{0.5,0.9,0.5} { 29.81} & \cellcolor[rgb]{0.5,0.9,0.5} {36.43} & \cellcolor[rgb]{0.5,0.9,0.5} {36.64} & \cellcolor[rgb]{0.5,0.9,0.5} {37.61} & \cellcolor[rgb]{0.5,0.9,0.5} { 42.24} & \cellcolor[rgb]{0.5,0.9,0.5} {41.68} & \cellcolor[rgb]{0.5,0.9,0.5} {  59.39} \\ 
		\hline
	\end{tabular}
\end{adjustbox}
\endgroup
\caption{Average actual gain $\overline{\wtilde{G}}$ (EUR/MWh) of the considered strategies as a new market player with an oracle forecast of intersection curves. Colour indicates the performance row-wise (the greener, the better). With bold, we depicted the best values in each row.
} 
\label{tab:avg_gain_new_bids_perfect}
\end{table}
\begin{table}[h!]
	\centering
\begingroup\small
\begin{adjustbox}{max width=1\textwidth}
	\setlength{\tabcolsep}{1pt}
	\begin{tabular}{|cr|rrrrrrrr||rrrrrr|}
		\hline
		&  & \multicolumn{8}{c||}{Supply/Sell (the higher the price the better)} & \multicolumn{6}{c|}{Demand/Buy (the lower the price the better)} \\
		\rotatebox[origin=c]{90}{Model} & Strategy & \thead{1\\MW} & \thead{10\\MW} & \thead{100\\MW} & \thead{1000\\MW} & \thead{1\% of\\ wind} & \thead{5\% of\\ wind} & \thead{1\% of\\ solar} & \thead{5\% of\\ solar} & \thead{1\\MW} & \thead{10\\MW} & \thead{100\\MW} & \thead{1000\\MW} & \thead{1\% of\\ load} & \thead{5\% of\\ load} \\ 
		\hline
		&IA-only & \cellcolor[rgb]{1,0.84,0.5} {37.22} & \cellcolor[rgb]{1,0.71,0.5} {37.01} & \cellcolor[rgb]{1,0.5,0.55} {34.55} & \cellcolor[rgb]{1,0.5,0.55} {-37.59} & \cellcolor[rgb]{1,0.5,0.55} {23.69} & \cellcolor[rgb]{1,0.5,0.55} {-151.61} & \cellcolor[rgb]{1,0.5,0.55} {28.13} & \cellcolor[rgb]{1,0.5,0.55} {-52.56} & \cellcolor[rgb]{1,0.978,0.5} {37.47} & \cellcolor[rgb]{1,0.831,0.5} {37.70} & \cellcolor[rgb]{1,0.5,0.55} {40.08} & \cellcolor[rgb]{1,0.5,0.55} {100.42} & \cellcolor[rgb]{1,0.5,0.55} {59.55} & \cellcolor[rgb]{1,0.5,0.55} {989.44} \\ 
		&TC-min & \cellcolor[rgb]{0.843,1,0.5} {37.45} & \cellcolor[rgb]{0.552,0.917,0.5} {37.45} & \cellcolor[rgb]{0.843,1,0.5} {37.45} & \cellcolor[rgb]{1,0.689,0.5} { 37.45} & \cellcolor[rgb]{1,0.954,0.5} {30.98} & \cellcolor[rgb]{1,0.665,0.5} {  30.98} & \cellcolor[rgb]{0.789,0.996,0.5} {33.65} & \cellcolor[rgb]{1,0.86,0.5} { 33.66} & \cellcolor[rgb]{1,0.908,0.5} {37.55} & \cellcolor[rgb]{1,0.961,0.5} {37.55} & \cellcolor[rgb]{0.982,1,0.5} {37.55} & \cellcolor[rgb]{1,0.688,0.5} { 37.55} & \cellcolor[rgb]{1,0.839,0.5} {38.98} & \cellcolor[rgb]{1,0.5,0.55} { 38.97} \\   \hline
		\parbox[t]{2mm}{\multirow{8}{*}{\rotatebox[origin=c]{90}{naive}}} 
		& $\E$-NoImp & \cellcolor[rgb]{0.708,0.969,0.5} {37.48} & \cellcolor[rgb]{0.912,1,0.5} {37.37} & \cellcolor[rgb]{1,0.5,0.55} {36.13} & \cellcolor[rgb]{1,0.5,0.55} {  0.12} & \cellcolor[rgb]{1,0.5,0.55} {27.50} & \cellcolor[rgb]{1,0.5,0.55} { -67.09} & \cellcolor[rgb]{1,0.5,0.55} {31.19} & \cellcolor[rgb]{1,0.5,0.55} {-17.62} & \cellcolor[rgb]{0.709,0.97,0.5} {37.37} & \cellcolor[rgb]{0.948,1,0.5} {37.49} & \cellcolor[rgb]{1,0.5,0.55} {38.72} & \cellcolor[rgb]{1,0.5,0.55} { 68.95} & \cellcolor[rgb]{1,0.5,0.55} {49.55} & \cellcolor[rgb]{1,0.5,0.55} {647.03} \\ 
		& $\E$-LinImp & \cellcolor[rgb]{0.708,0.969,0.5} {37.48} & \cellcolor[rgb]{0.808,1,0.5} {37.40} & \cellcolor[rgb]{1,0.935,0.5} {37.33} & \cellcolor[rgb]{1,0.757,0.5} { 37.53} & \cellcolor[rgb]{1,0.808,0.5} {30.84} & \cellcolor[rgb]{1,0.5,0.55} {  30.28} & \cellcolor[rgb]{1,0.926,0.5} {33.52} & \cellcolor[rgb]{1,0.832,0.5} { 33.63} & \cellcolor[rgb]{0.709,0.97,0.5} {37.37} & \cellcolor[rgb]{0.878,1,0.5} {37.47} & \cellcolor[rgb]{1,0.952,0.5} {37.61} & \cellcolor[rgb]{1,0.872,0.5} { 37.34} & \cellcolor[rgb]{1,0.915,0.5} {38.89} & \cellcolor[rgb]{1,0.612,0.5} { 38.53} \\ 
		& $\E$-LinImpMeff & \cellcolor[rgb]{0.843,1,0.5} {37.45} & \cellcolor[rgb]{0.552,0.917,0.5} {37.45} & \cellcolor[rgb]{0.552,0.917,0.5} {37.51} & \cellcolor[rgb]{1,0.86,0.5} { 37.65} & \cellcolor[rgb]{0.688,0.963,0.5} {31.09} & \cellcolor[rgb]{1,0.572,0.5} {  30.89} & \cellcolor[rgb]{0.616,0.939,0.5} {33.68} & \cellcolor[rgb]{1,0.927,0.5} { 33.73} & \cellcolor[rgb]{1,0.908,0.5} {37.55} & \cellcolor[rgb]{1,0.97,0.5} {37.54} & \cellcolor[rgb]{0.552,0.917,0.5} {37.45} & \cellcolor[rgb]{1,0.942,0.5} { 37.26} & \cellcolor[rgb]{0.869,1,0.5} {38.75} & \cellcolor[rgb]{1,0.543,0.5} { 38.61} \\ 
		& $\E$-Meff & \cellcolor[rgb]{0.877,1,0.5} {37.44} & \cellcolor[rgb]{0.656,0.952,0.5} {37.43} & \cellcolor[rgb]{0.604,0.935,0.5} {37.50} & \cellcolor[rgb]{0.706,0.969,0.5} { 37.89} & \cellcolor[rgb]{0.751,0.984,0.5} {31.08} & \cellcolor[rgb]{0.562,0.921,0.5} {  31.39} & \cellcolor[rgb]{0.5,0.9,0.5} {33.70} & \cellcolor[rgb]{0.558,0.919,0.5} { 33.90} & \cellcolor[rgb]{1,0.917,0.5} {37.54} & \cellcolor[rgb]{1,0.987,0.5} {37.52} & \cellcolor[rgb]{0.656,0.952,0.5} {37.47} & \cellcolor[rgb]{0.605,0.935,0.5} { 37.10} & \cellcolor[rgb]{0.5,0.9,0.5} {38.67} & \cellcolor[rgb]{0.942,1,0.5} { 38.06} \\ 
		& $\E$ & \cellcolor[rgb]{0.708,0.969,0.5} {37.48} & \cellcolor[rgb]{0.656,0.952,0.5} {37.43} & \cellcolor[rgb]{0.946,1,0.5} {37.42} & \cellcolor[rgb]{0.874,1,0.5} { 37.85} & \cellcolor[rgb]{1,0.943,0.5} {30.97} & \cellcolor[rgb]{0.931,1,0.5} {  31.32} & \cellcolor[rgb]{0.831,1,0.5} {33.64} & \cellcolor[rgb]{0.73,0.977,0.5} { 33.87} & \cellcolor[rgb]{0.709,0.97,0.5} {37.37} & \cellcolor[rgb]{0.843,1,0.5} {37.46} & \cellcolor[rgb]{0.843,1,0.5} {37.51} & \cellcolor[rgb]{0.71,0.97,0.5} { 37.12} & \cellcolor[rgb]{0.651,0.95,0.5} {38.70} & \cellcolor[rgb]{0.977,1,0.5} { 38.07} \\ 
		& $\E$-$\var$-U & \cellcolor[rgb]{1,0.979,0.5} {37.38} & \cellcolor[rgb]{1,0.996,0.5} {37.34} & \cellcolor[rgb]{1,0.918,0.5} {37.31} & \cellcolor[rgb]{1,0.903,0.5} { 37.70} & \cellcolor[rgb]{1,0.891,0.5} {30.92} & \cellcolor[rgb]{1,0.976,0.5} {  31.28} & \cellcolor[rgb]{1,0.917,0.5} {33.51} & \cellcolor[rgb]{1,0.956,0.5} { 33.76} & \cellcolor[rgb]{1,0.943,0.5} {37.51} & \cellcolor[rgb]{1,0.97,0.5} {37.54} & \cellcolor[rgb]{1,0.727,0.5} {37.87} & \cellcolor[rgb]{1,0.5,0.55} { 88.43} & \cellcolor[rgb]{1,0.5,0.55} {53.17} & \cellcolor[rgb]{1,0.5,0.55} {331.24} \\ 
		& VaR & \cellcolor[rgb]{0.808,1,0.5} {37.46} & \cellcolor[rgb]{0.708,0.969,0.5} {37.42} & \cellcolor[rgb]{0.981,1,0.5} {37.41} & \cellcolor[rgb]{1,0.852,0.5} { 37.64} & \cellcolor[rgb]{1,0.943,0.5} {30.97} & \cellcolor[rgb]{1,0.893,0.5} {  31.20} & \cellcolor[rgb]{0.986,1,0.5} {33.60} & \cellcolor[rgb]{1,0.889,0.5} { 33.69} & \cellcolor[rgb]{0.913,1,0.5} {37.42} & \cellcolor[rgb]{0.913,1,0.5} {37.48} & \cellcolor[rgb]{0.947,1,0.5} {37.54} & \cellcolor[rgb]{1,0.872,0.5} { 37.34} & \cellcolor[rgb]{1,0.966,0.5} {38.83} & \cellcolor[rgb]{1,0.509,0.5} { 38.65} \\ 
		& CVaR & \cellcolor[rgb]{0.843,1,0.5} {37.45} & \cellcolor[rgb]{0.76,0.987,0.5} {37.41} & \cellcolor[rgb]{0.912,1,0.5} {37.43} & \cellcolor[rgb]{1,0.971,0.5} { 37.78} & \cellcolor[rgb]{1,0.985,0.5} {31.01} & \cellcolor[rgb]{0.931,1,0.5} {  31.32} & \cellcolor[rgb]{0.87,1,0.5} {33.63} & \cellcolor[rgb]{0.907,1,0.5} { 33.83} & \cellcolor[rgb]{0.809,1,0.5} {37.39} & \cellcolor[rgb]{0.878,1,0.5} {37.47} & \cellcolor[rgb]{0.947,1,0.5} {37.54} & \cellcolor[rgb]{1,0.925,0.5} { 37.28} & \cellcolor[rgb]{0.869,1,0.5} {38.75} & \cellcolor[rgb]{1,0.5,0.55} { 38.98} \\   \hline
		\parbox[t]{2mm}{\multirow{8}{*}{\rotatebox[origin=c]{90}{expert}}} 
		& $\E$-NoImp & \cellcolor[rgb]{0.5,0.9,0.5} \textbf{37.52} & \cellcolor[rgb]{0.76,0.987,0.5} {37.41} & \cellcolor[rgb]{1,0.5,0.55} {36.17} & \cellcolor[rgb]{1,0.5,0.55} {  1.84} & \cellcolor[rgb]{1,0.5,0.55} {27.17} & \cellcolor[rgb]{1,0.5,0.55} { -87.07} & \cellcolor[rgb]{1,0.5,0.55} {31.95} & \cellcolor[rgb]{1,0.5,0.55} { -8.77} & \cellcolor[rgb]{0.5,0.9,0.5} \textbf{37.33} & \cellcolor[rgb]{0.809,1,0.5} {37.45} & \cellcolor[rgb]{1,0.5,0.55} {38.65} & \cellcolor[rgb]{1,0.5,0.55} { 67.48} & \cellcolor[rgb]{1,0.5,0.55} {48.61} & \cellcolor[rgb]{1,0.5,0.55} {671.17} \\ 
		& $\E$-LinImp & \cellcolor[rgb]{0.5,0.9,0.5} {37.52} & \cellcolor[rgb]{0.552,0.917,0.5} {37.45} & \cellcolor[rgb]{0.981,1,0.5} {37.41} & \cellcolor[rgb]{1,0.826,0.5} { 37.61} & \cellcolor[rgb]{1,0.933,0.5} {30.96} & \cellcolor[rgb]{1,0.841,0.5} {  31.15} & \cellcolor[rgb]{0.87,1,0.5} {33.63} & \cellcolor[rgb]{1,0.947,0.5} { 33.75} & \cellcolor[rgb]{0.5,0.9,0.5} {37.33} & \cellcolor[rgb]{0.552,0.917,0.5} {37.40} & \cellcolor[rgb]{0.843,1,0.5} {37.51} & \cellcolor[rgb]{1,0.907,0.5} { 37.30} & \cellcolor[rgb]{1,0.991,0.5} {38.80} & \cellcolor[rgb]{1,0.5,0.55} { 39.30} \\ 
		& $\E$-LinImpMeff & \cellcolor[rgb]{0.843,1,0.5} {37.45} & \cellcolor[rgb]{0.5,0.9,0.5} \textbf{37.46} & \cellcolor[rgb]{0.5,0.9,0.5} {\textbf{37.52}} & \cellcolor[rgb]{1,0.834,0.5} { 37.62} & \cellcolor[rgb]{0.5,0.9,0.5} {\textbf{31.12}} & \cellcolor[rgb]{1,0.934,0.5} {  31.24} & \cellcolor[rgb]{0.674,0.958,0.5} {33.67} & \cellcolor[rgb]{1,0.975,0.5} { 33.78} & \cellcolor[rgb]{1,0.908,0.5} {37.55} & \cellcolor[rgb]{1,0.961,0.5} {37.55} & \cellcolor[rgb]{0.552,0.917,0.5} {37.45} & \cellcolor[rgb]{1,0.907,0.5} { 37.30} & \cellcolor[rgb]{0.835,1,0.5} {38.74} & \cellcolor[rgb]{1,0.5,0.55} { 39.92} \\ 
		& $\E$-Meff & \cellcolor[rgb]{0.843,1,0.5} {37.45} & \cellcolor[rgb]{0.552,0.917,0.5} {37.45} & \cellcolor[rgb]{0.552,0.917,0.5} {37.51} & \cellcolor[rgb]{0.5,0.9,0.5} {\textbf{37.93}} & \cellcolor[rgb]{0.688,0.963,0.5} {31.09} & \cellcolor[rgb]{0.5,0.9,0.5} {\textbf{31.4}} & \cellcolor[rgb]{0.5,0.9,0.5} \textbf{33.70} & \cellcolor[rgb]{0.5,0.9,0.5} {\textbf{33.91}} & \cellcolor[rgb]{1,0.908,0.5} {37.55} & \cellcolor[rgb]{1,0.97,0.5} {37.54} & \cellcolor[rgb]{0.708,0.969,0.5} {37.48} & \cellcolor[rgb]{0.5,0.9,0.5} {\textbf{37.08}} & \cellcolor[rgb]{0.5,0.9,0.5} \textbf{38.67} & \cellcolor[rgb]{0.551,0.917,0.5} { 37.97} \\ 
		& $\E$ & \cellcolor[rgb]{0.5,0.9,0.5} {37.52} & \cellcolor[rgb]{0.5,0.9,0.5} {37.46} & \cellcolor[rgb]{0.843,1,0.5} {37.45} & \cellcolor[rgb]{0.706,0.969,0.5} { 37.89} & \cellcolor[rgb]{0.976,1,0.5} {31.03} & \cellcolor[rgb]{0.686,0.962,0.5} {  31.37} & \cellcolor[rgb]{0.674,0.958,0.5} {33.67} & \cellcolor[rgb]{0.673,0.958,0.5} { 33.88} & \cellcolor[rgb]{0.5,0.9,0.5} {37.33} & \cellcolor[rgb]{0.5,0.9,0.5} {\textbf{37.39}} & \cellcolor[rgb]{0.708,0.969,0.5} {37.48} & \cellcolor[rgb]{0.605,0.935,0.5} { 37.10} & \cellcolor[rgb]{0.55,0.917,0.5} {38.68} & \cellcolor[rgb]{0.5,0.9,0.5} {\textbf{37.96}} \\ 
		& $\E$-$\var$-U & \cellcolor[rgb]{0.76,0.987,0.5} {37.47} & \cellcolor[rgb]{0.552,0.917,0.5} {37.45} & \cellcolor[rgb]{0.708,0.969,0.5} {37.48} & \cellcolor[rgb]{1,0.98,0.5} { 37.79} & \cellcolor[rgb]{0.751,0.984,0.5} {31.08} & \cellcolor[rgb]{1,0.976,0.5} {  31.28} & \cellcolor[rgb]{0.789,0.996,0.5} {33.65} & \cellcolor[rgb]{1,0.995,0.5} { 33.80} & \cellcolor[rgb]{0.761,0.987,0.5} {37.38} & \cellcolor[rgb]{0.809,1,0.5} {37.45} & \cellcolor[rgb]{1,0.935,0.5} {37.63} & \cellcolor[rgb]{1,0.5,0.55} { 51.50} & \cellcolor[rgb]{1,0.5,0.55} {41.93} & \cellcolor[rgb]{1,0.5,0.55} {176.09} \\ 
		& VaR & \cellcolor[rgb]{0.76,0.987,0.5} {37.47} & \cellcolor[rgb]{0.552,0.917,0.5} {37.45} & \cellcolor[rgb]{0.76,0.987,0.5} {37.47} & \cellcolor[rgb]{1,0.894,0.5} { 37.69} & \cellcolor[rgb]{0.976,1,0.5} {31.03} & \cellcolor[rgb]{1,0.903,0.5} {  31.21} & \cellcolor[rgb]{0.831,1,0.5} {33.64} & \cellcolor[rgb]{1,0.918,0.5} { 33.72} & \cellcolor[rgb]{0.879,1,0.5} {37.41} & \cellcolor[rgb]{0.843,1,0.5} {37.46} & \cellcolor[rgb]{0.656,0.952,0.5} {37.47} & \cellcolor[rgb]{1,0.942,0.5} { 37.26} & \cellcolor[rgb]{0.97,1,0.5} {38.78} & \cellcolor[rgb]{1,0.706,0.5} { 38.42} \\ 
		& CVaR & \cellcolor[rgb]{0.708,0.969,0.5} {37.48} & \cellcolor[rgb]{0.552,0.917,0.5} {37.45} & \cellcolor[rgb]{0.708,0.969,0.5} {37.48} & \cellcolor[rgb]{1,0.971,0.5} { 37.78} & \cellcolor[rgb]{0.809,1,0.5} {31.07} & \cellcolor[rgb]{0.973,1,0.5} {  31.31} & \cellcolor[rgb]{0.674,0.958,0.5} {33.67} & \cellcolor[rgb]{0.945,1,0.5} { 33.82} & \cellcolor[rgb]{0.809,1,0.5} {37.39} & \cellcolor[rgb]{0.709,0.97,0.5} {37.43} & \cellcolor[rgb]{0.5,0.9,0.5} {\textbf{37.44}} & \cellcolor[rgb]{1,0.995,0.5} { 37.20} & \cellcolor[rgb]{0.752,0.984,0.5} {38.72} & \cellcolor[rgb]{1,0.835,0.5} { 38.27} \\  \hline
		& perfect forecast & \cellcolor[rgb]{0.5,0.9,0.5} {38.44} & \cellcolor[rgb]{0.5,0.9,0.5} {38.35} & \cellcolor[rgb]{0.5,0.9,0.5} {38.09} & \cellcolor[rgb]{0.5,0.9,0.5} { 38.29} & \cellcolor[rgb]{0.5,0.9,0.5} {31.73} & \cellcolor[rgb]{0.5,0.9,0.5} {  31.87} & \cellcolor[rgb]{0.5,0.9,0.5} {34.23} & \cellcolor[rgb]{0.5,0.9,0.5} { 34.33} & \cellcolor[rgb]{0.5,0.9,0.5} {36.41} & \cellcolor[rgb]{0.5,0.9,0.5} {36.50} & \cellcolor[rgb]{0.5,0.9,0.5} {36.81} & \cellcolor[rgb]{0.5,0.9,0.5} { 36.69} & \cellcolor[rgb]{0.5,0.9,0.5} {38.26} & \cellcolor[rgb]{0.5,0.9,0.5} { 37.40} \\  
		\hline
	\end{tabular}
\end{adjustbox}
\endgroup
\caption{Average actual gain $\overline{\wtilde{G}}$ (EUR/MWh) of the considered strategies as an existing market player rebidding their portfolio $\bsv$ with an oracle forecast of intersection curves. Colour indicates the performance row-wise (the greener, the better). With bold, we depicted the best values in each row.
} 
\label{tab:avg_gain_rebidded_perfect}
\end{table}

\begin{figure}[h!]
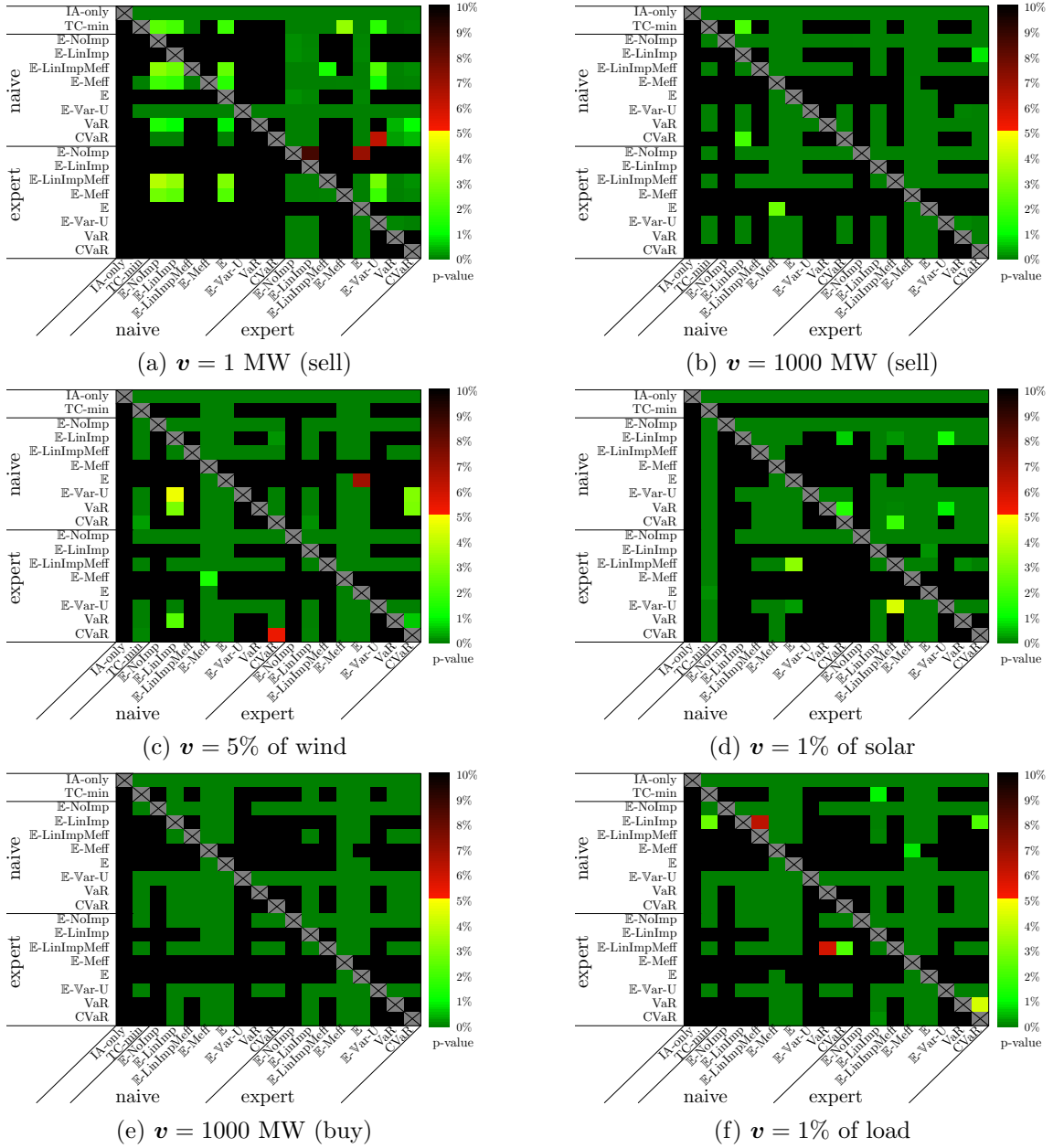

	\centering
	\subfloat[$\bsv = 1$ MW (sell)]{\resizebox{0.455\textwidth}{!}{\input{fig/ttest_standard_1.tex}}}\hfill
	\subfloat[$\bsv = 1000$ MW (sell)]{\resizebox{0.455\textwidth}{!}{\input{fig/ttest_standard_1000.tex}}}\\
	\subfloat[$\bsv = 5\%$ of wind]{\resizebox{0.455\textwidth}{!}{\input{fig/ttest_standard_5_of_wind.tex}}}\hfill
	\subfloat[$\bsv = 1\%$ of solar]{\resizebox{0.455\textwidth}{!}{\input{fig/ttest_standard_1_of_solar.tex}}}\\		\subfloat[$\bsv = 1000$ MW (buy)]{\resizebox{0.455\textwidth}{!}{\input{fig/ttest_standard_-1000.tex}}}\hfill
	\subfloat[$\bsv = 1\%$ of load]{\resizebox{0.455\textwidth}{!}{\input{fig/ttest_standard_1_of_load.tex}}}
	\caption{Results of the $\overline{\wtilde{G}}$ mean inequality test for remaining portfolios $\bsv$ in the setting of a new market player. The plots present p-values --- the closer they are to zero ($\to$ dark green), the more significant the difference is between gains of X-axis strategy (better) and gains of the Y-axis strategy (worse).}
	\label{fig:tests_new_bids}
\end{figure}

\begin{figure}[h!]
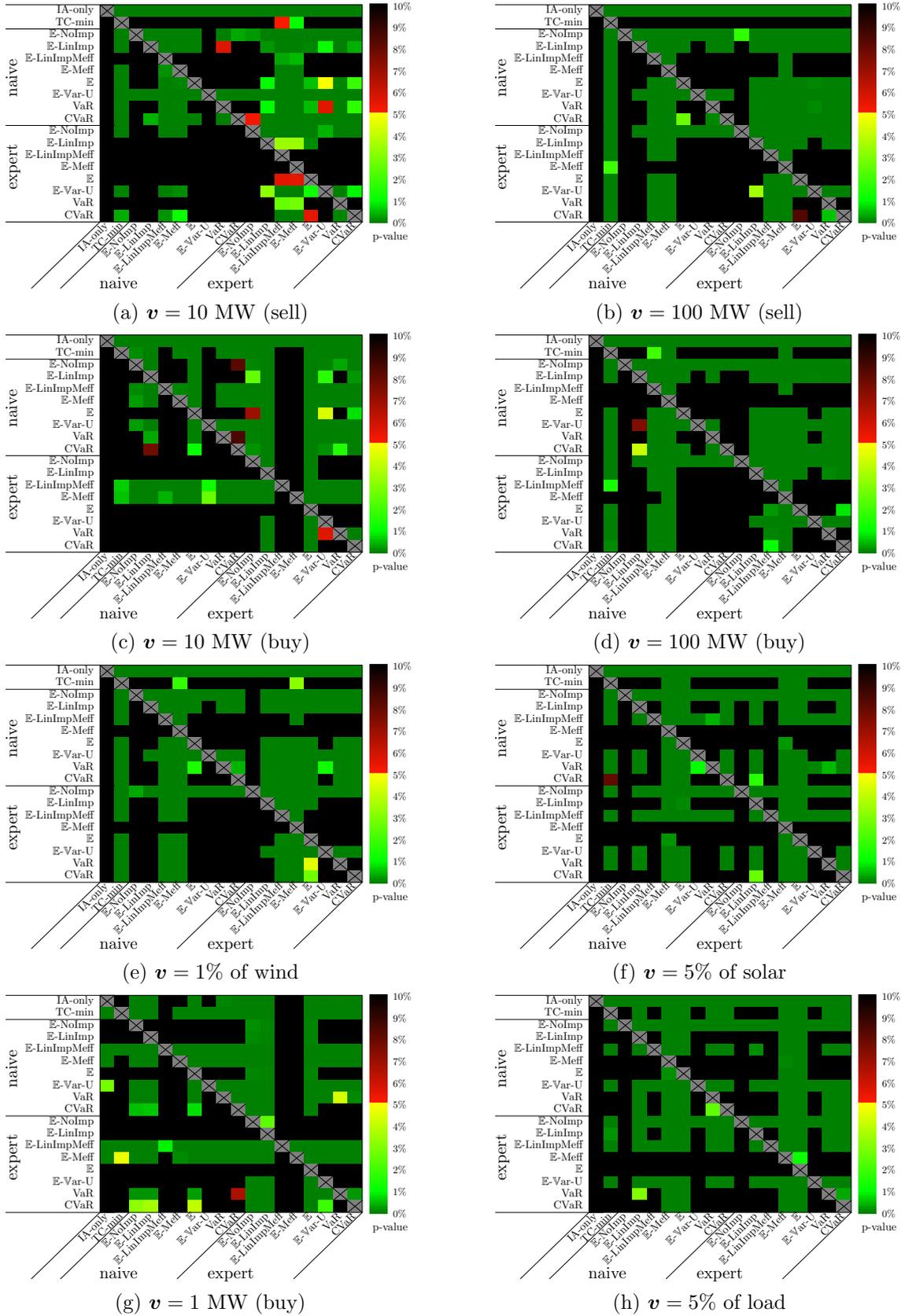

	\centering
	\subfloat[$\bsv = 10$ MW (sell)]{\resizebox{0.455\textwidth}{!}{\input{fig/ttest_standard_10.tex}}}\hfill
	\subfloat[$\bsv = 100$ MW (sell)]{\resizebox{0.455\textwidth}{!}{\input{fig/ttest_standard_100.tex}}}\\
	\subfloat[$\bsv = 10$ MW (buy)]{\resizebox{0.455\textwidth}{!}{\input{fig/ttest_standard_-10.tex}}}\hfill
	\subfloat[$\bsv = 100$ MW (buy)]{\resizebox{0.455\textwidth}{!}{\input{fig/ttest_standard_-100.tex}}}\\
	\subfloat[$\bsv = 1\%$ of wind]{\resizebox{0.455\textwidth}{!}{\input{fig/ttest_standard_1_of_wind.tex}}}\hfill
	\subfloat[$\bsv = 5\%$ of solar]{\resizebox{0.455\textwidth}{!}{\input{fig/ttest_standard_5_of_solar.tex}}}\\		\subfloat[$\bsv = 1$ MW (buy)]{\resizebox{0.455\textwidth}{!}{\input{fig/ttest_standard_-1.tex}}}\hfill
	\subfloat[$\bsv = 5\%$ of load]{\resizebox{0.455\textwidth}{!}{\input{fig/ttest_standard_5_of_load.tex}}}
	\caption{Results of the $\overline{\wtilde{G}}$ mean inequality test for remaining portfolios $\bsv$ in the setting of a new market player. For details on interpretation see Figure~\ref{fig:tests_new_bids}.}
	\label{fig:tests_new_bids2}
\end{figure}

\begin{figure}[h!]
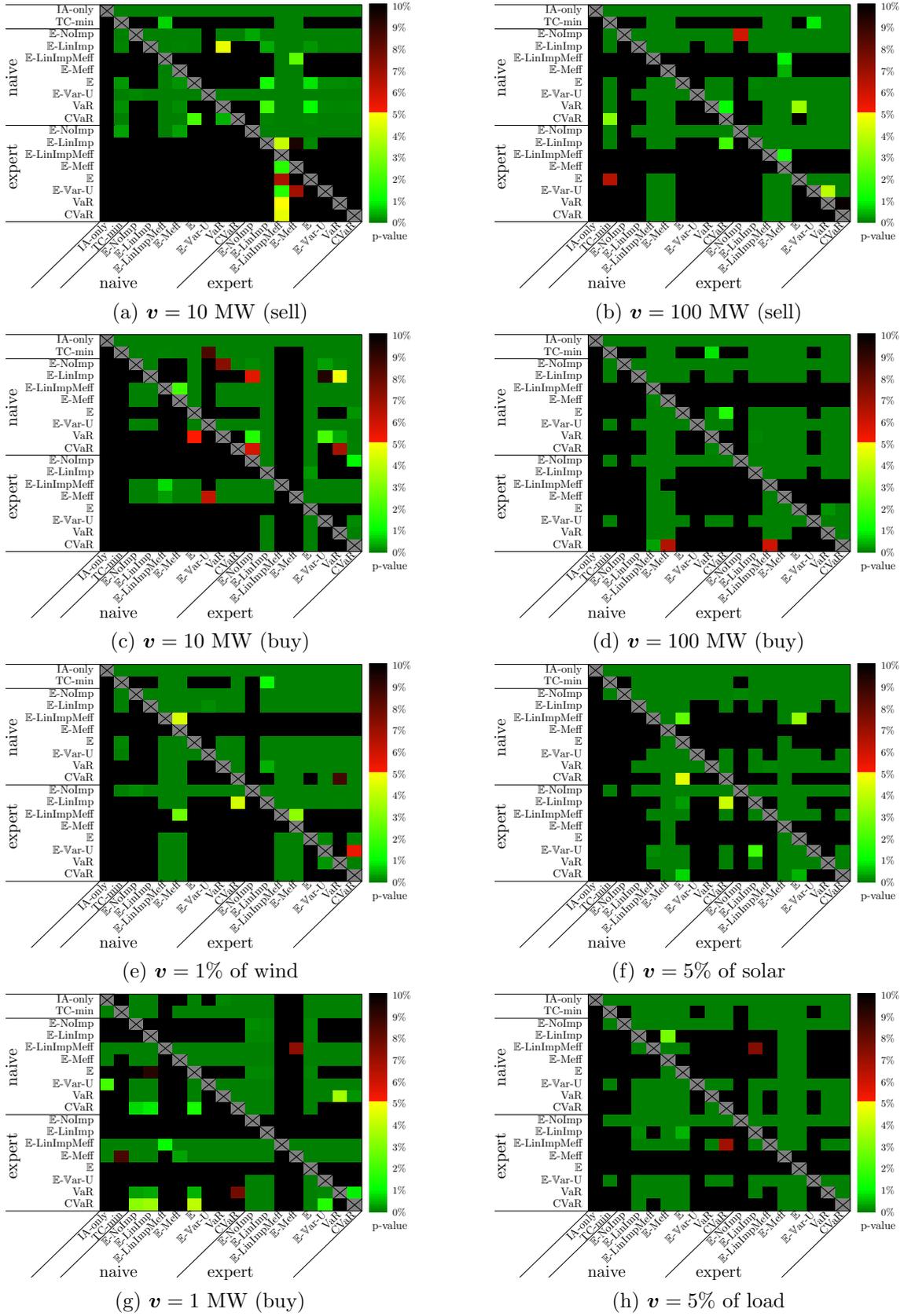

	\centering
	\subfloat[$\bsv = 10$ MW (sell)]{\resizebox{0.455\textwidth}{!}{\input{fig/ttest_adjusted_10.tex}}}\hfill
	\subfloat[$\bsv = 100$ MW (sell)]{\resizebox{0.455\textwidth}{!}{\input{fig/ttest_adjusted_100.tex}}}\\
	\subfloat[$\bsv = 10$ MW (buy)]{\resizebox{0.455\textwidth}{!}{\input{fig/ttest_adjusted_-10.tex}}}\hfill
	\subfloat[$\bsv = 100$ MW (buy)]{\resizebox{0.455\textwidth}{!}{\input{fig/ttest_adjusted_-100.tex}}}\\
	\subfloat[$\bsv = 1\%$ of wind]{\resizebox{0.455\textwidth}{!}{\input{fig/ttest_adjusted_1_of_wind.tex}}}\hfill
	\subfloat[$\bsv = 5\%$ of solar]{\resizebox{0.455\textwidth}{!}{\input{fig/ttest_adjusted_5_of_solar.tex}}}\\		\subfloat[$\bsv = 1$ MW (buy)]{\resizebox{0.455\textwidth}{!}{\input{fig/ttest_adjusted_-1.tex}}}\hfill
	\subfloat[$\bsv = 5\%$ of load]{\resizebox{0.455\textwidth}{!}{\input{fig/ttest_adjusted_5_of_load.tex}}}
	
	\caption{Results of the $\overline{\wtilde{G}}$ mean inequality test for remaining portfolios $\bsv$ in the setting of rebidding the portfolio. For details on interpretation see Figure~\ref{fig:tests_new_bids}.
	}
	\label{fig:tests_rebidded2}
\end{figure}

\begin{figure}[h!]
	\centering
	\subfloat{\includegraphics[width = 1\linewidth]{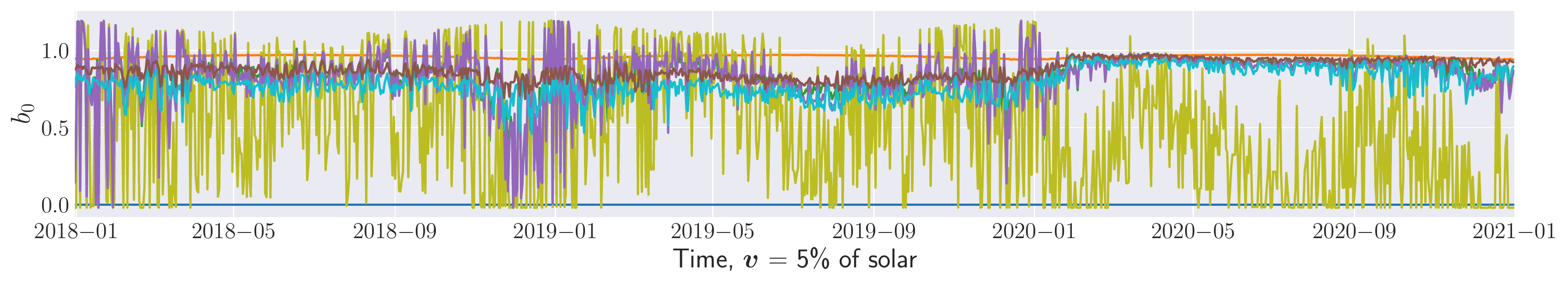}}\vspace{-0.5em}
	\subfloat{\includegraphics[width = 1\linewidth]{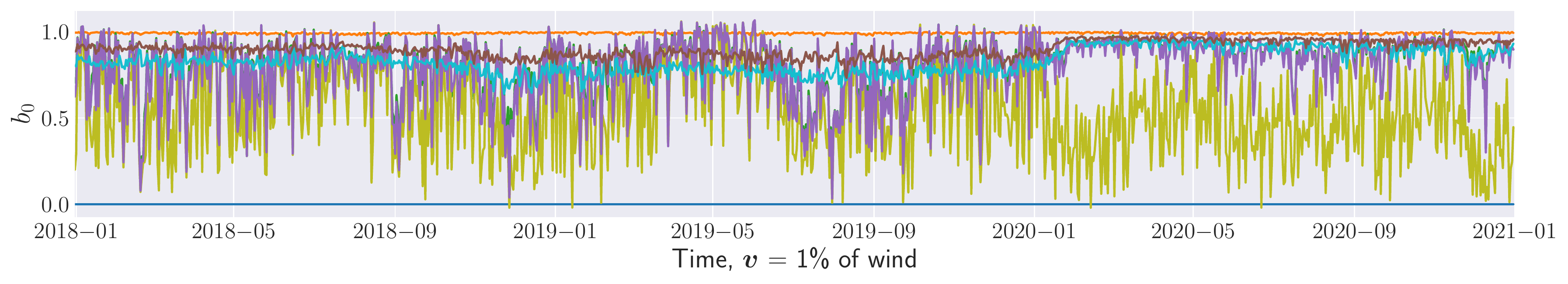}}\vspace{-0.5em}
	\subfloat{\includegraphics[width = 1\linewidth]{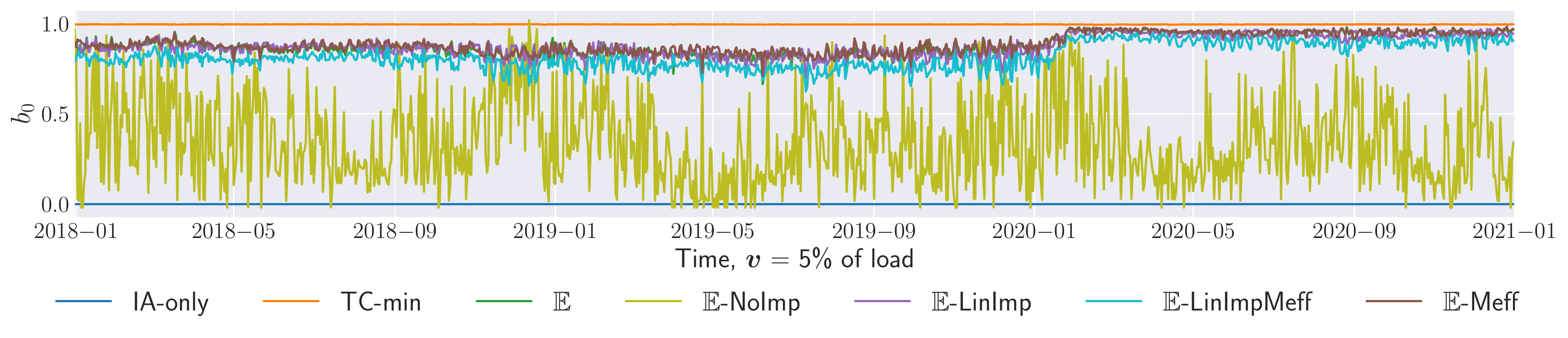}}\\
	\subfloat{\includegraphics[width = 1\linewidth]{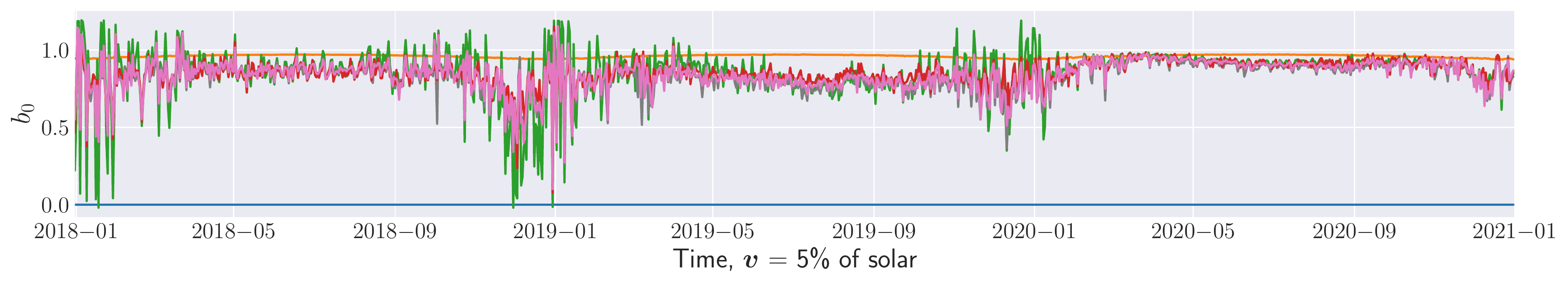}}\vspace{-0.5em}
	\subfloat{\includegraphics[width = 1\linewidth]{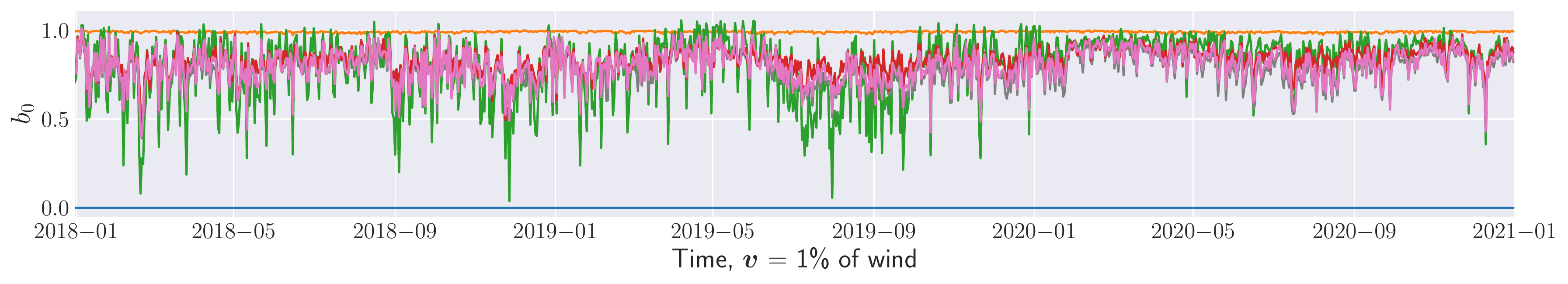}}\vspace{-0.5em}
	\subfloat{\includegraphics[width = 1\linewidth]{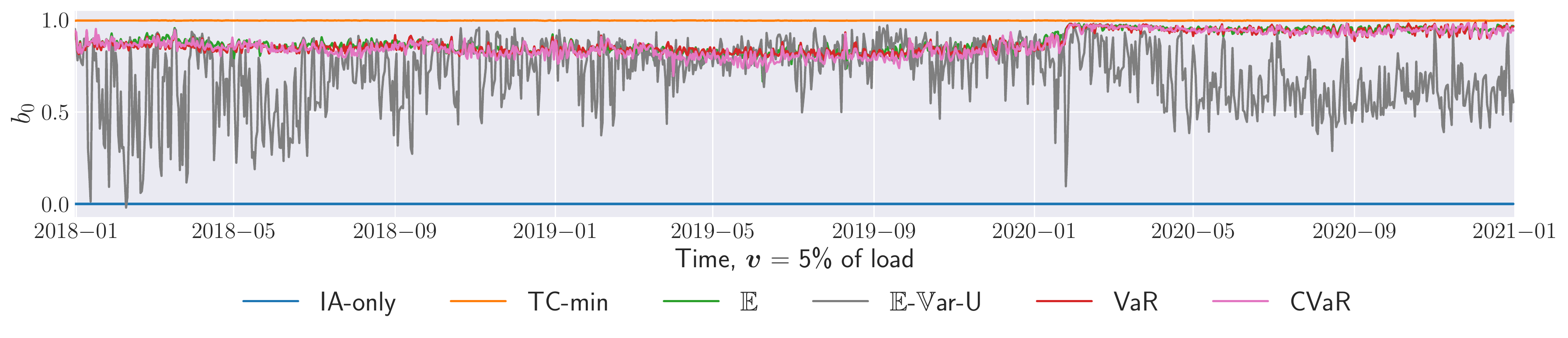}}
	\caption{The average daily weight of $b_0$ in relation to the whole $\bsb$ strategy for remaining portfolios $\bsv$ in the setting of rebidding the portfolio. The \textbf{naive}-based strategies are excluded for better clarity}
	\label{fig:da_weights_rebidded2}
\end{figure}

\begin{figure}[h!]
	\centering
	\subfloat{\includegraphics[width = 1\linewidth]{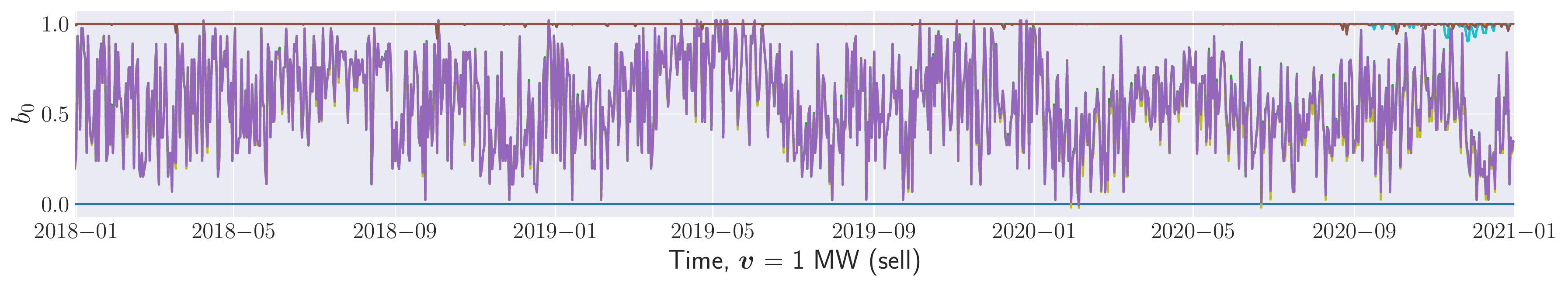}}\vspace{-0.5em}
	\subfloat{\includegraphics[width = 1\linewidth]{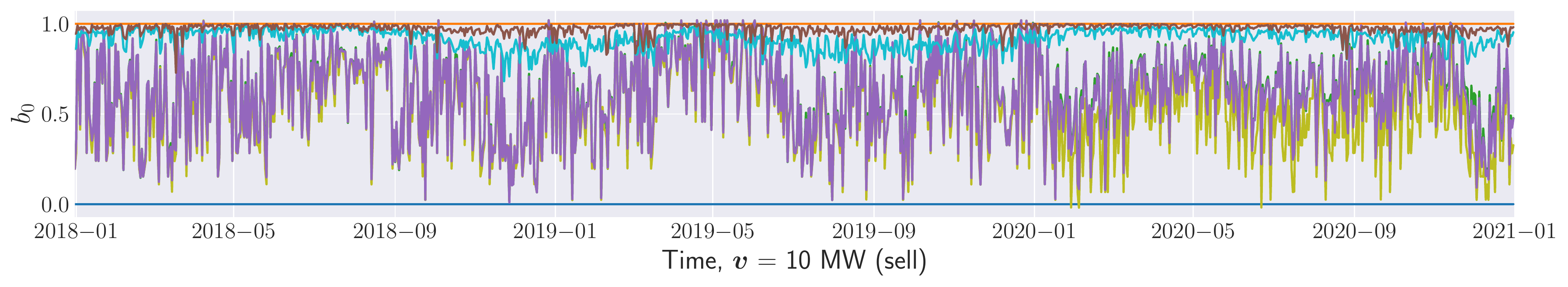}}\vspace{-0.5em}
	\subfloat{\includegraphics[width = 1\linewidth]{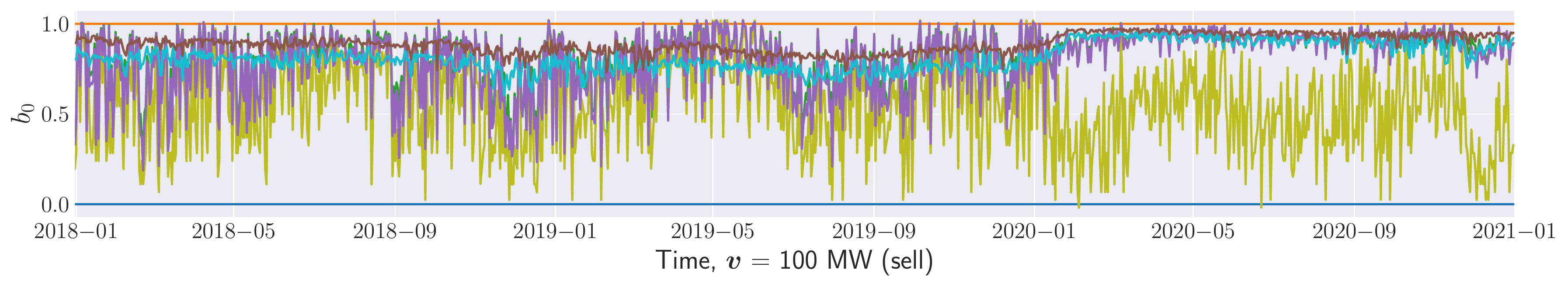}}\vspace{-0.5em}
	\subfloat{\includegraphics[width = 1\linewidth]{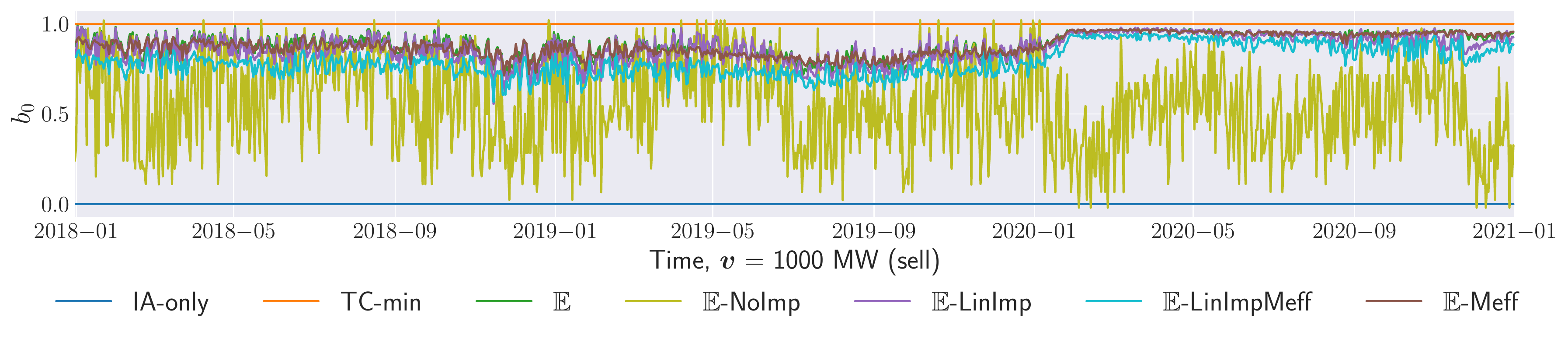}}\\
	\subfloat{\includegraphics[width = 1\linewidth]{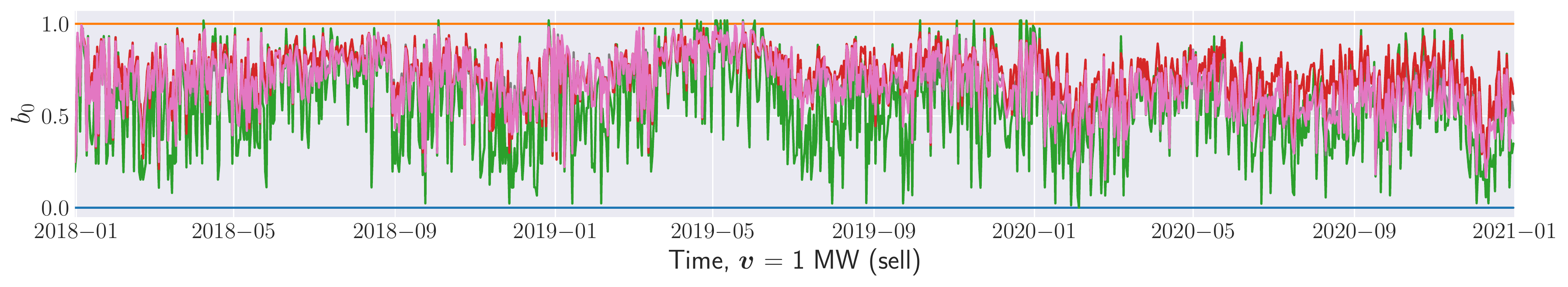}}\vspace{-0.5em}
	\subfloat{\includegraphics[width = 1\linewidth]{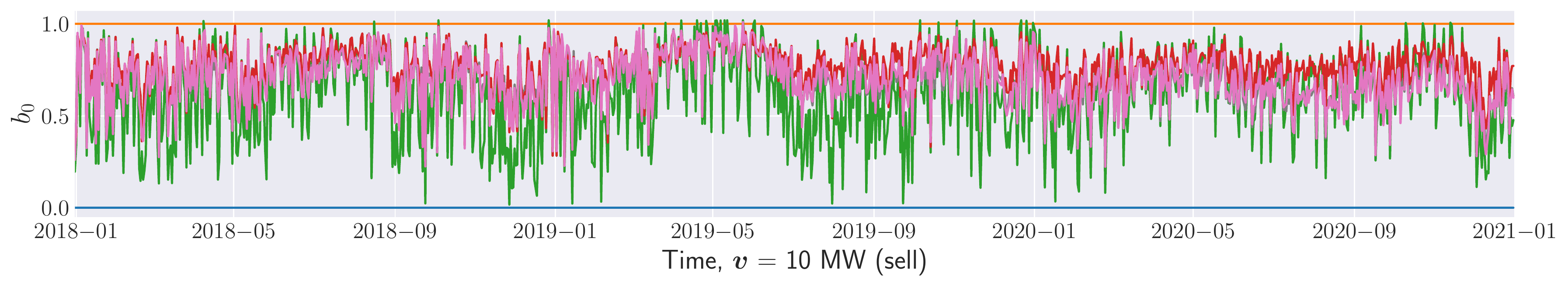}}\vspace{-0.5em}
	\subfloat{\includegraphics[width = 1\linewidth]{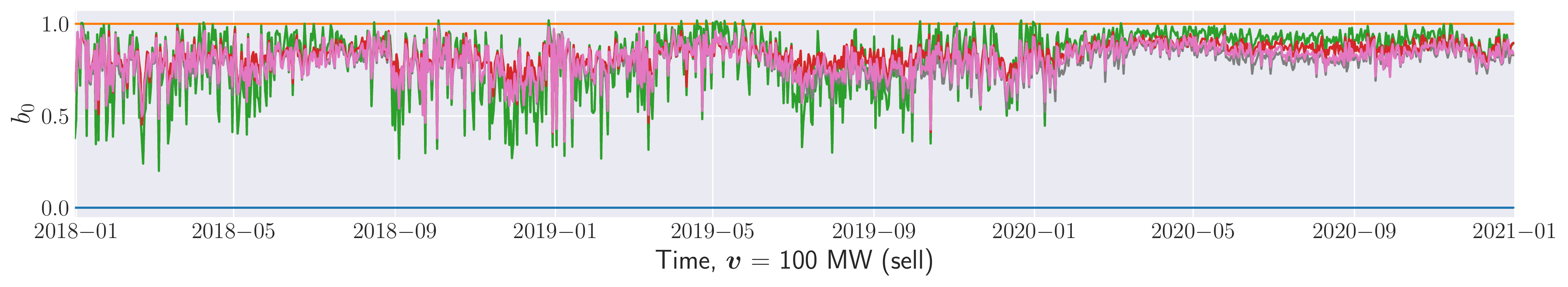}}\vspace{-0.5em}
	\subfloat{\includegraphics[width = 1\linewidth]{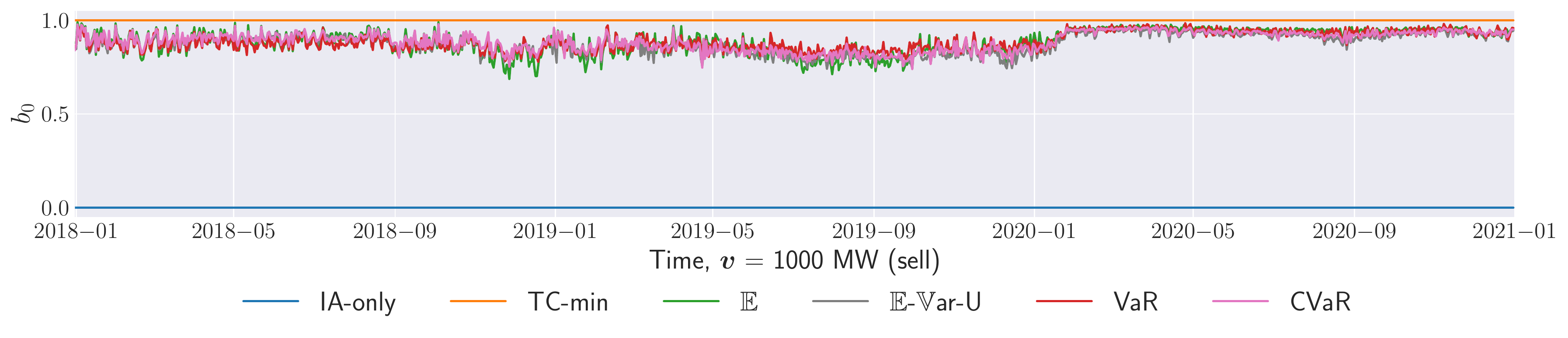}}	
	\caption{The average daily weight of $b_0$ in relation to the whole $\bsb$ strategy for remaining portfolios $\bsv$ in the setting of rebidding the portfolio. The \textbf{naive}-based strategies are excluded for better clarity}
	\label{fig:da_weights_rebidded3}
\end{figure}

\begin{figure}[h!]
	\centering
	\subfloat{\includegraphics[width = 1\linewidth]{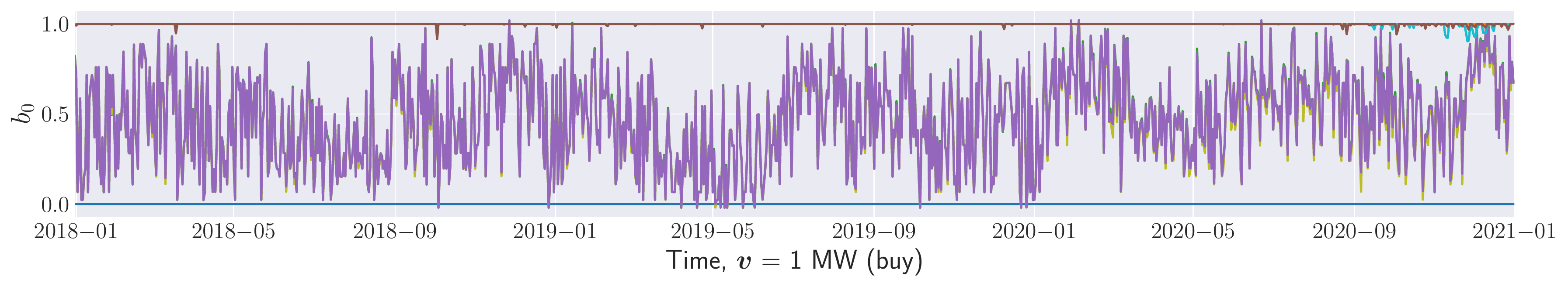}}\vspace{-0.5em}
	\subfloat{\includegraphics[width = 1\linewidth]{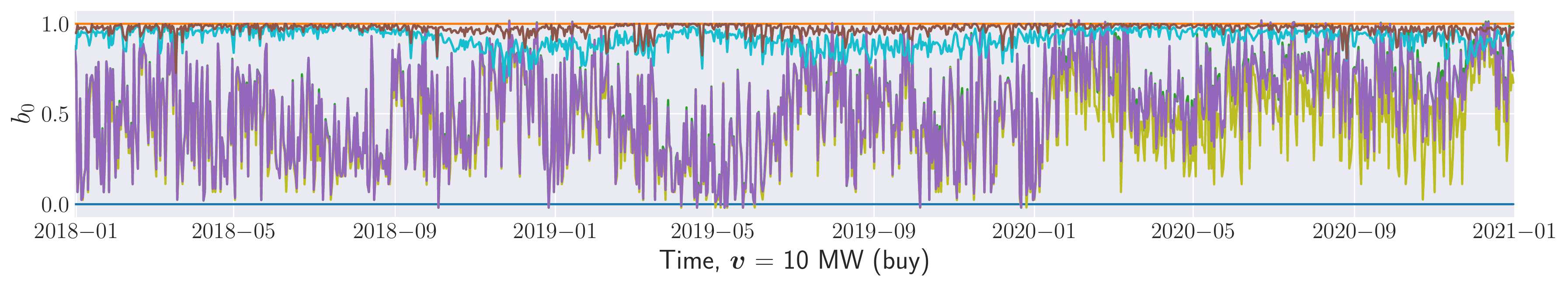}}\vspace{-0.5em}
	\subfloat{\includegraphics[width = 1\linewidth]{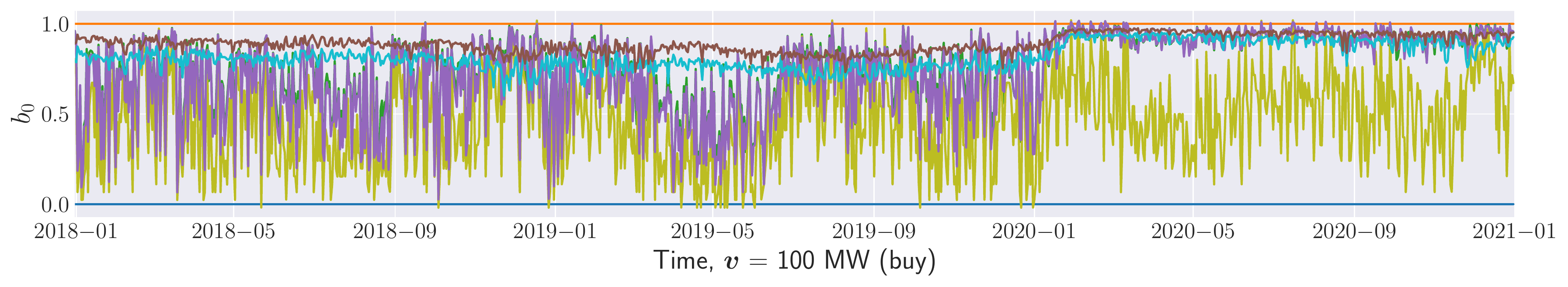}}\vspace{-0.5em}
	\subfloat{\includegraphics[width = 1\linewidth]{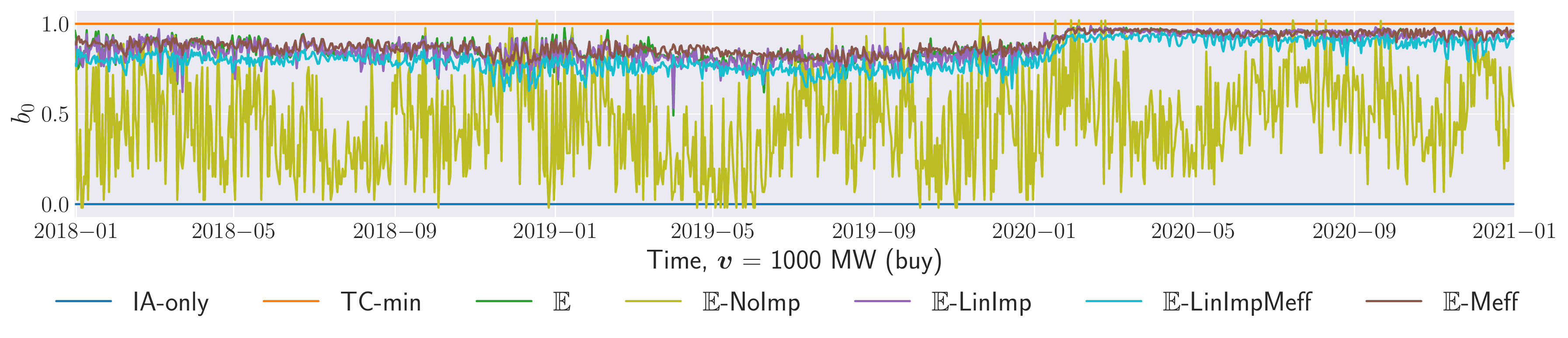}}\\
	\subfloat{\includegraphics[width = 1\linewidth]{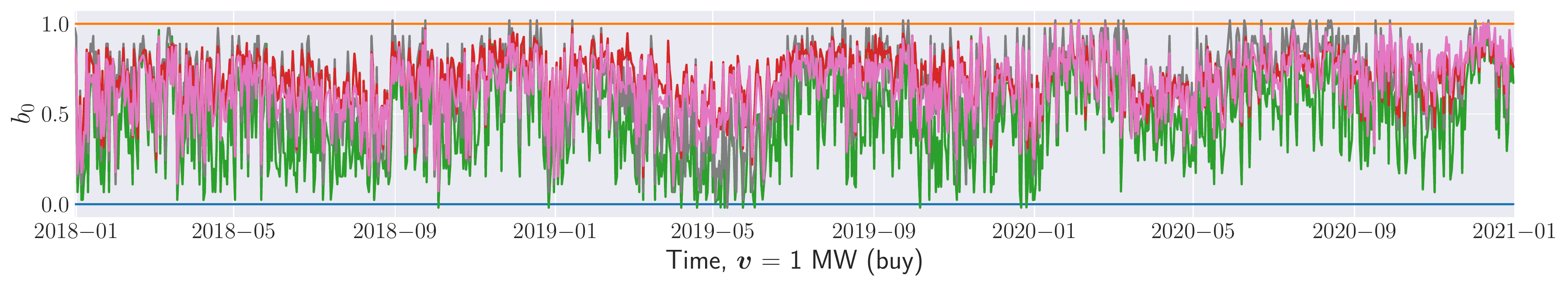}}\vspace{-0.5em}
	\subfloat{\includegraphics[width = 1\linewidth]{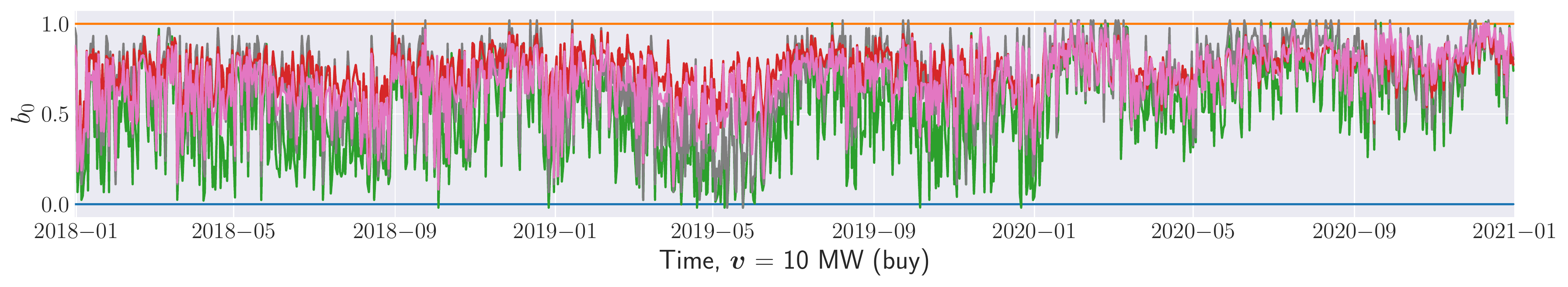}}\vspace{-0.5em}
	\subfloat{\includegraphics[width = 1\linewidth]{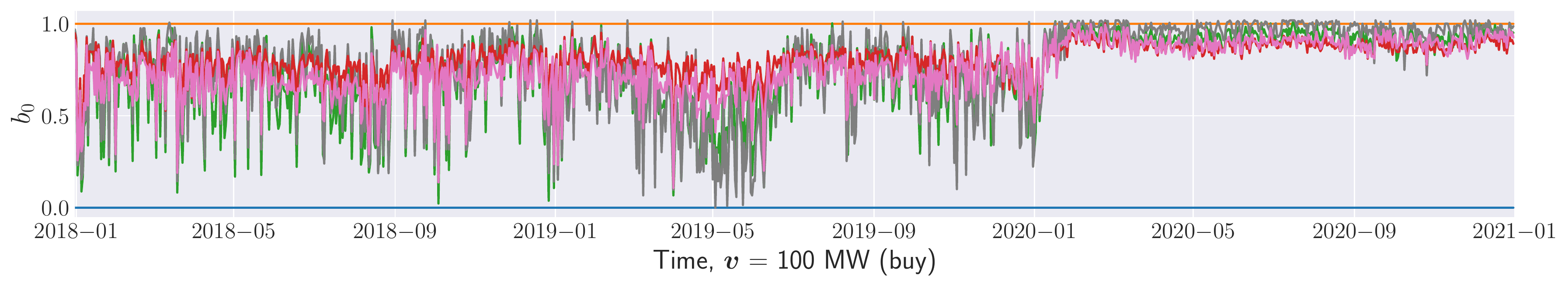}}\vspace{-0.5em}
	\subfloat{\includegraphics[width = 1\linewidth]{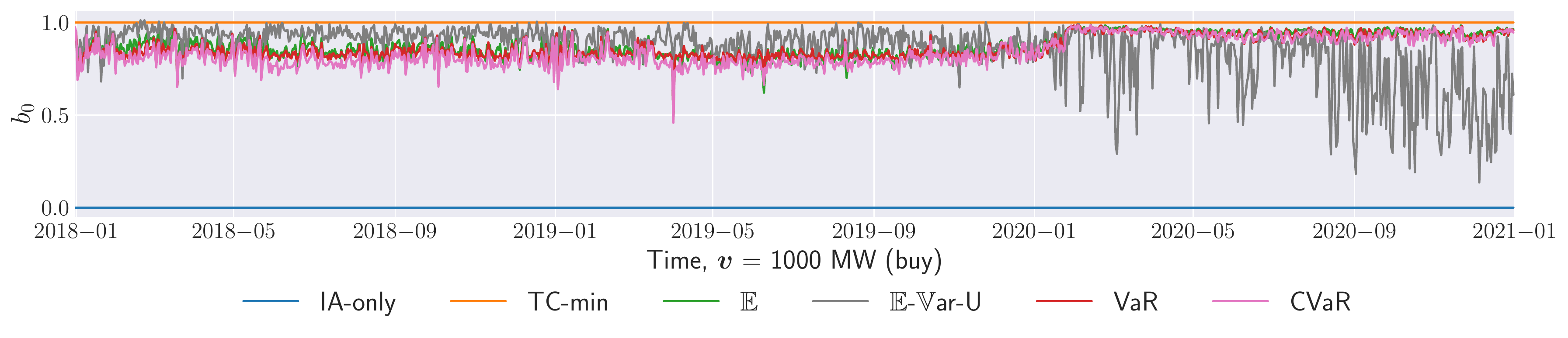}}	
	\caption{The average daily weight of $b_0$ in relation to the whole $\bsb$ strategy for remaining portfolios $\bsv$ in the setting of rebidding the portfolio. The \textbf{naive}-based strategies are excluded for better clarity}
	\label{fig:da_weights_rebidded4}
\end{figure}

\begin{figure}[h!]
	\centering
	\subfloat{\includegraphics[width = 1\linewidth]{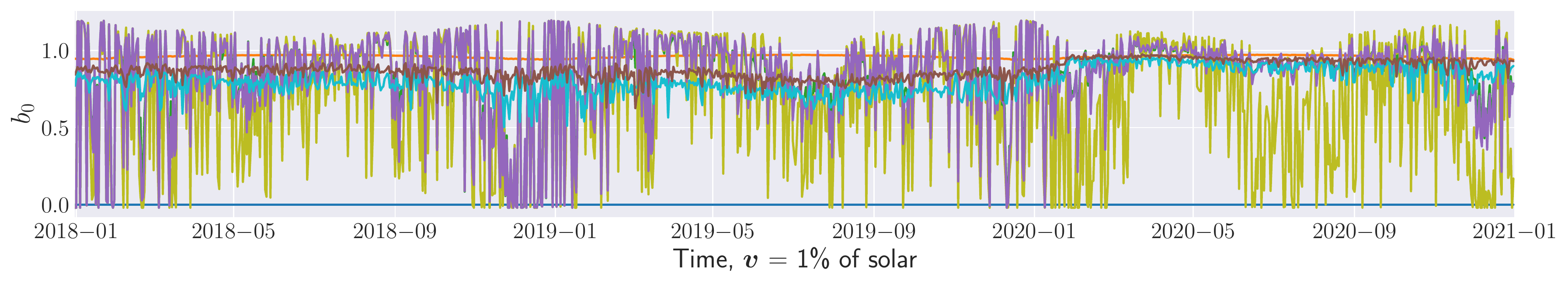}}\vspace{-0.5em}
	\subfloat{\includegraphics[width = 1\linewidth]{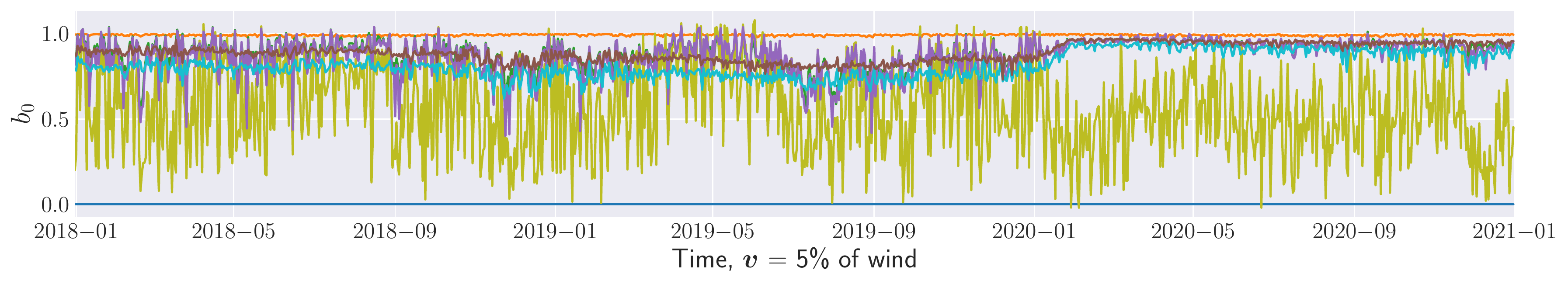}}\vspace{-0.5em}
	\subfloat{\includegraphics[width = 1\linewidth]{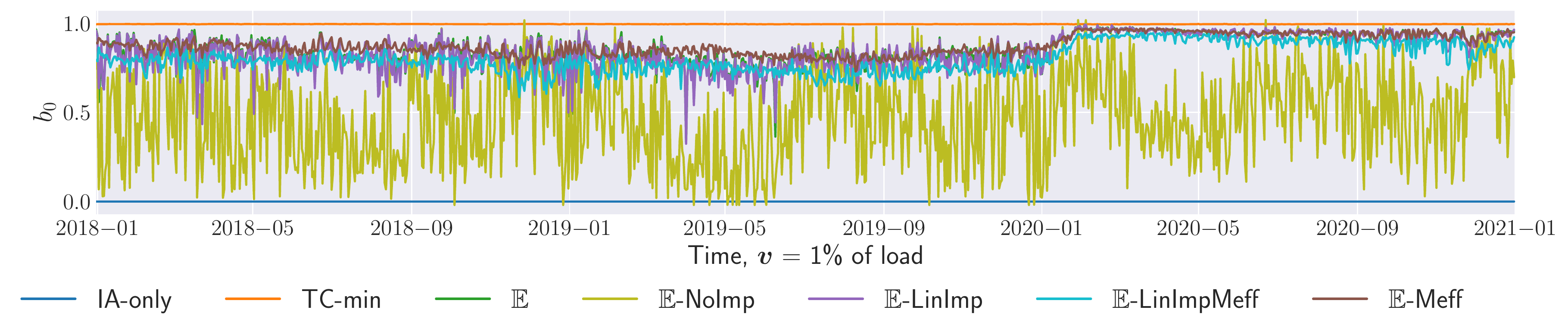}}\\
	\subfloat{\includegraphics[width = 1\linewidth]{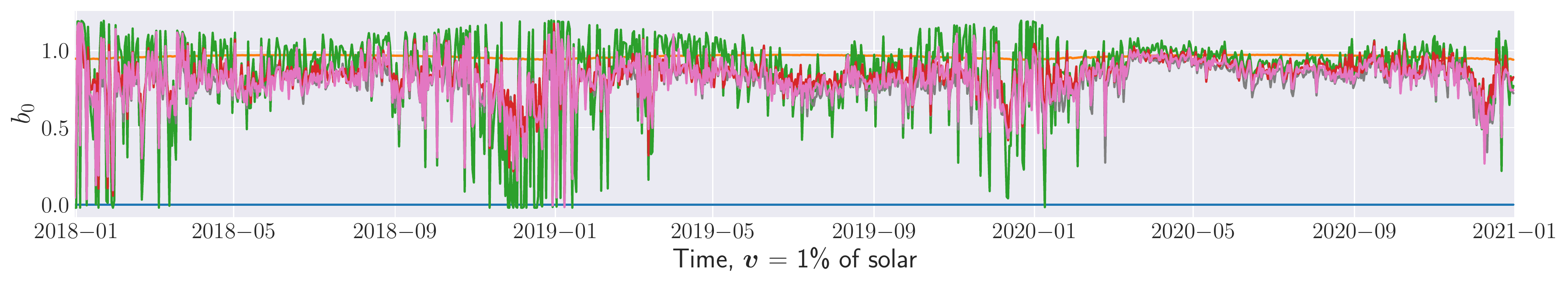}}\vspace{-0.5em}
	\subfloat{\includegraphics[width = 1\linewidth]{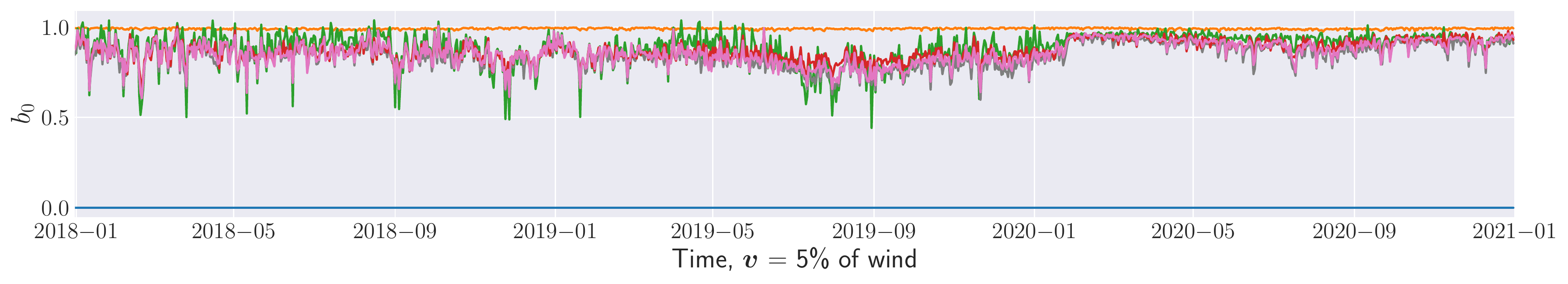}}\vspace{-0.5em}
	\subfloat{\includegraphics[width = 1\linewidth]{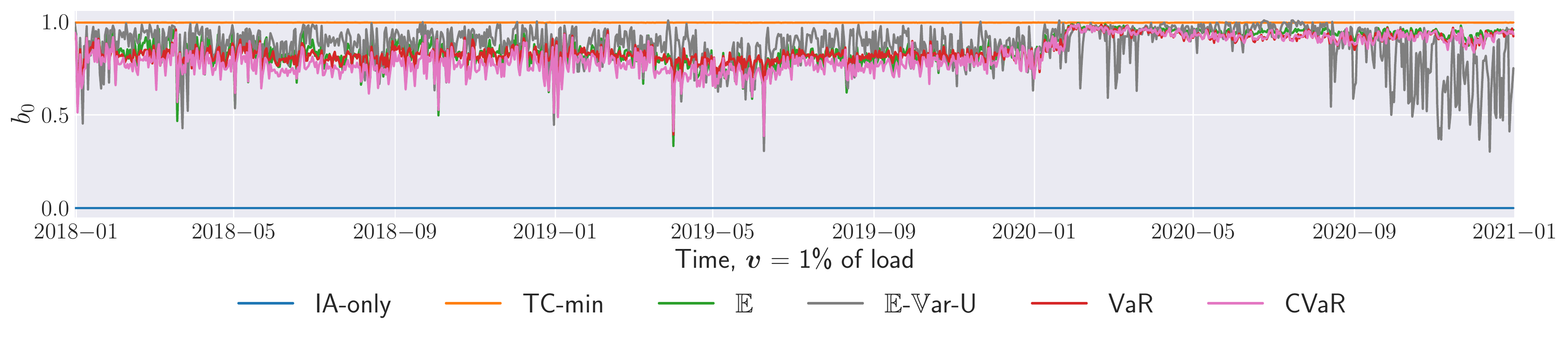}}
	\caption{The average daily weight of $b_0$ in relation to the whole $\bsb$ strategy for remaining portfolios $\bsv$ in the setting of a new market player. The \textbf{naive}-based strategies are excluded for better clarity}
	\label{fig:da_weights_newbids}
\end{figure}

\begin{figure}[h!]
	\centering
	\subfloat{\includegraphics[width = 1\linewidth]{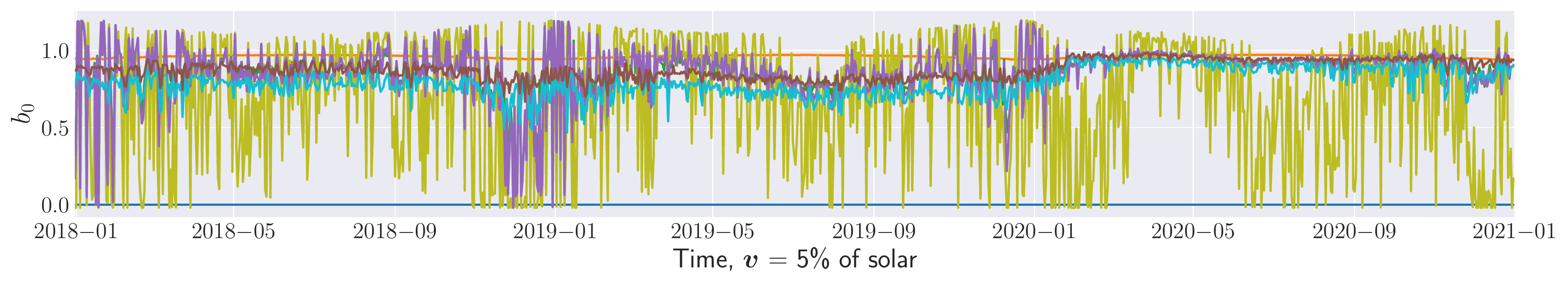}}\vspace{-0.5em}
	\subfloat{\includegraphics[width = 1\linewidth]{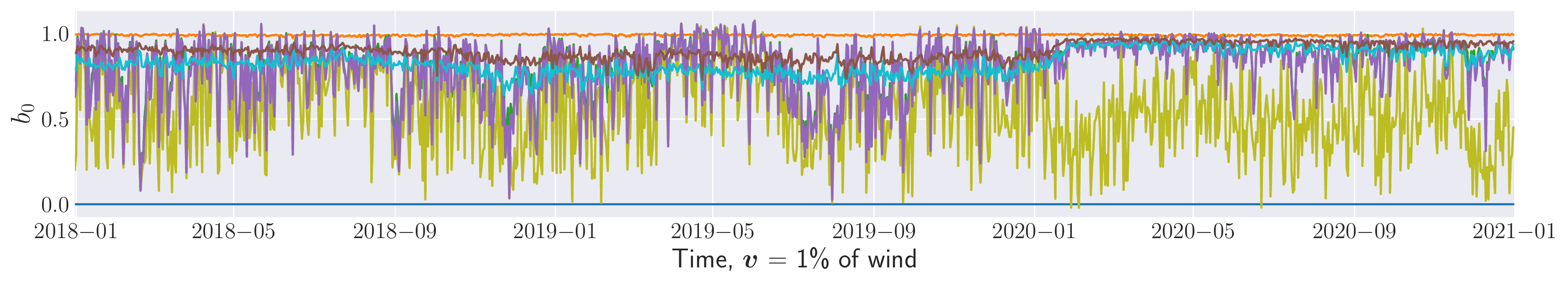}}\vspace{-0.5em}
	\subfloat{\includegraphics[width = 1\linewidth]{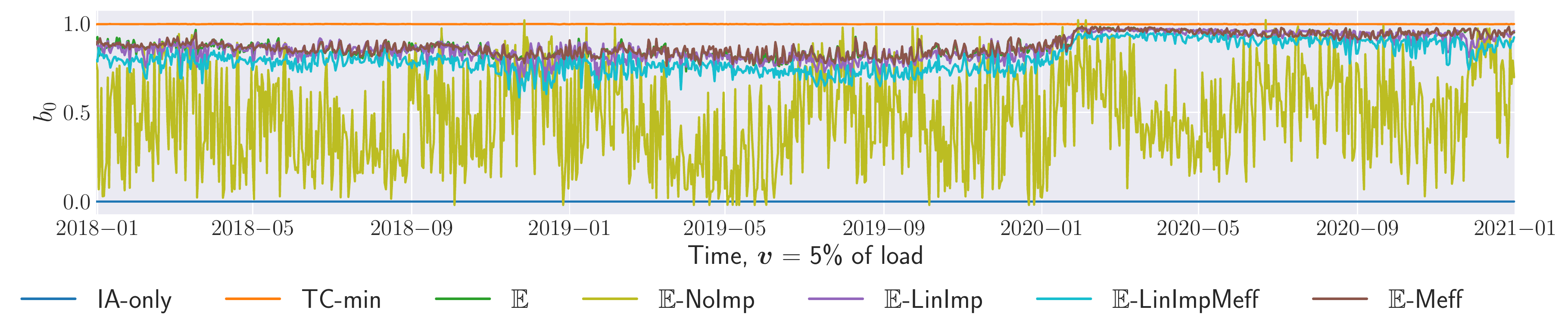}}\\
	\subfloat{\includegraphics[width = 1\linewidth]{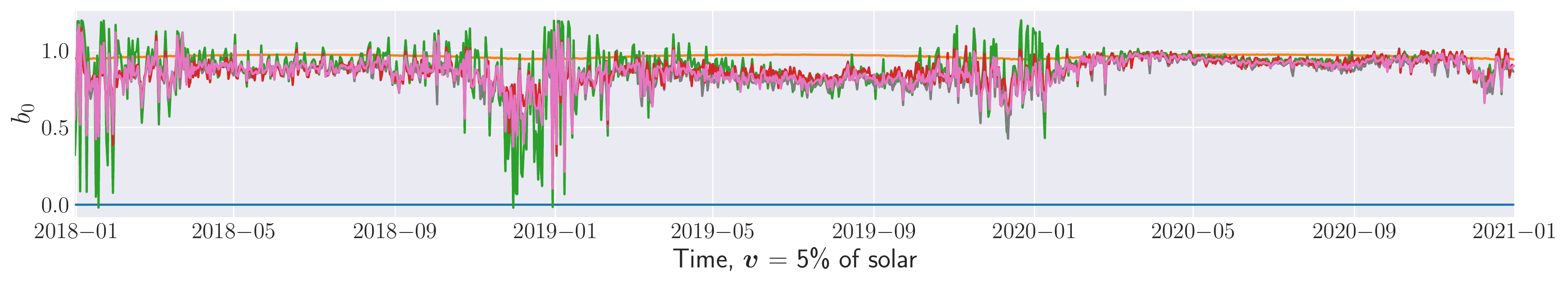}}\vspace{-0.5em}
	\subfloat{\includegraphics[width = 1\linewidth]{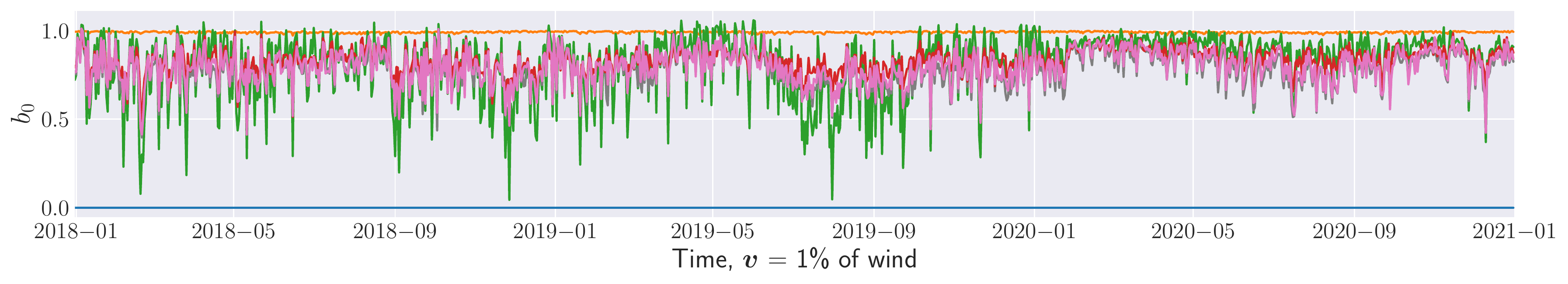}}\vspace{-0.5em}
	\subfloat{\includegraphics[width = 1\linewidth]{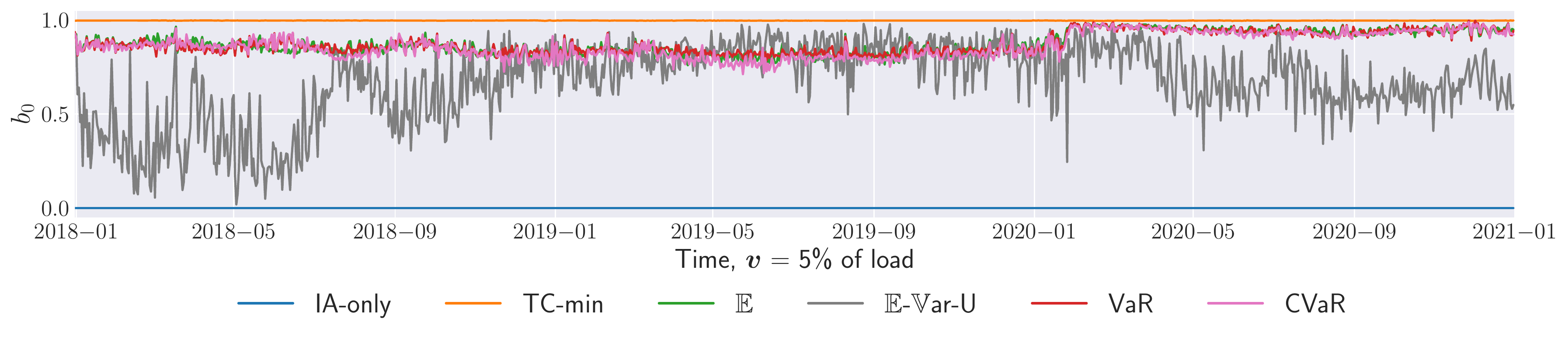}}
	\caption{The average daily weight of $b_0$ in relation to the whole $\bsb$ strategy for remaining portfolios $\bsv$ in the setting of a new market player. The \textbf{naive}-based strategies are excluded for better clarity}
	\label{fig:da_weights_newbids2}
\end{figure}

\begin{figure}[h!]
	\centering
	\subfloat{\includegraphics[width = 1\linewidth]{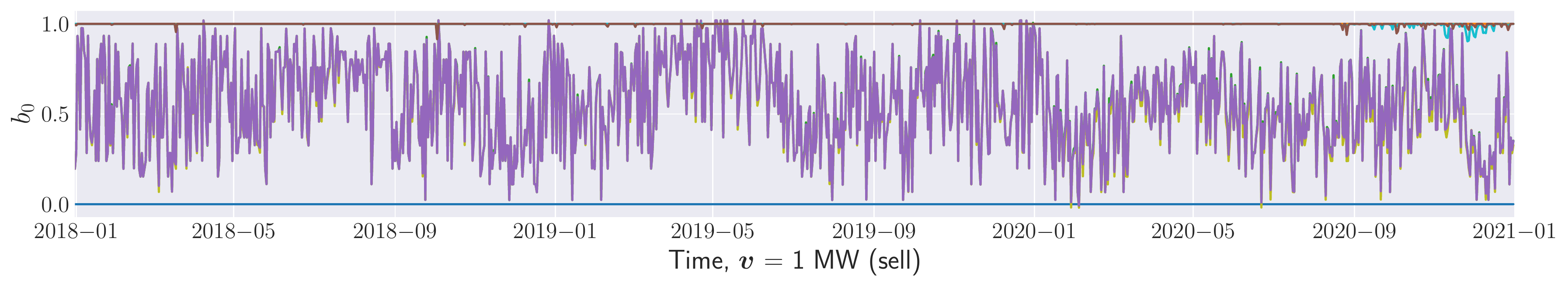}}\vspace{-0.5em}
	\subfloat{\includegraphics[width = 1\linewidth]{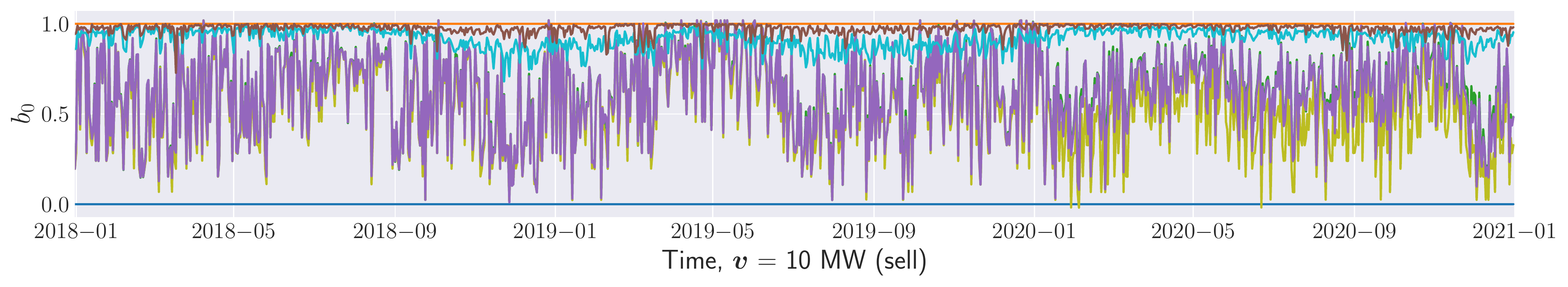}}\vspace{-0.5em}
	\subfloat{\includegraphics[width = 1\linewidth]{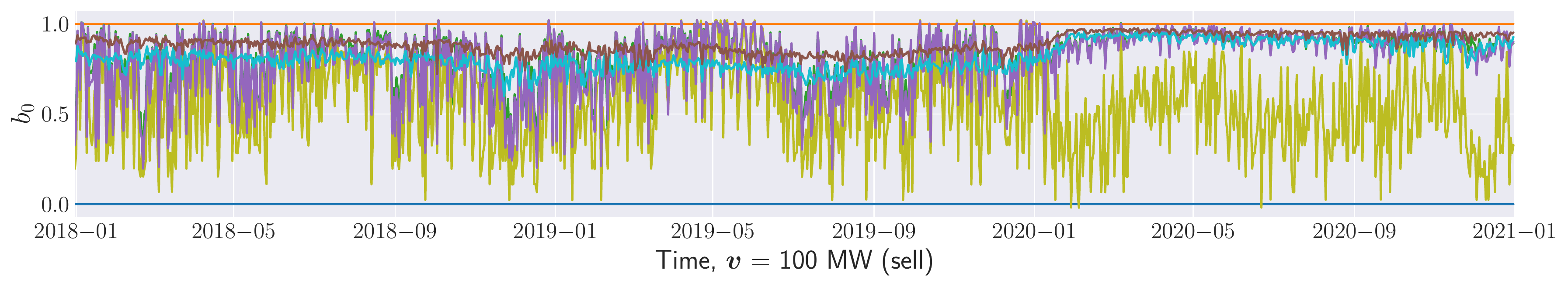}}\vspace{-0.5em}
	\subfloat{\includegraphics[width = 1\linewidth]{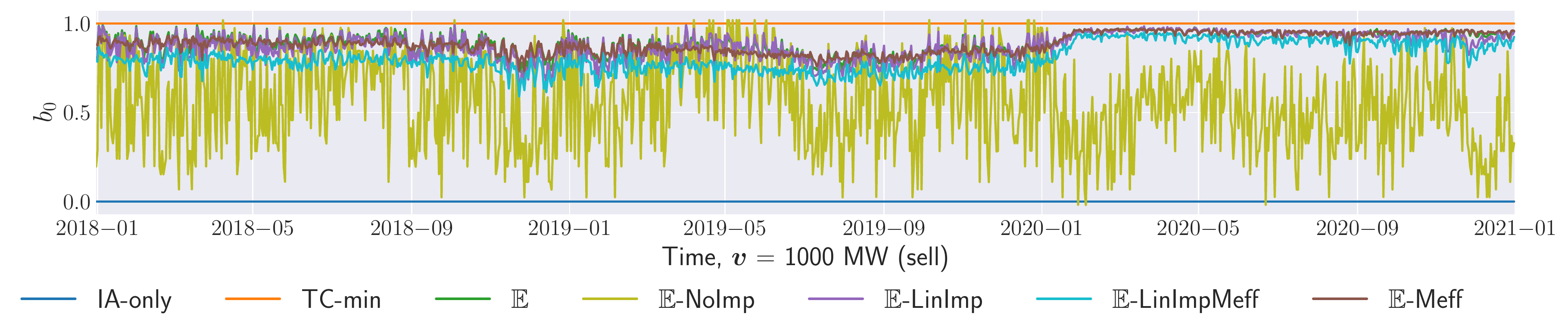}}\\
	\subfloat{\includegraphics[width = 1\linewidth]{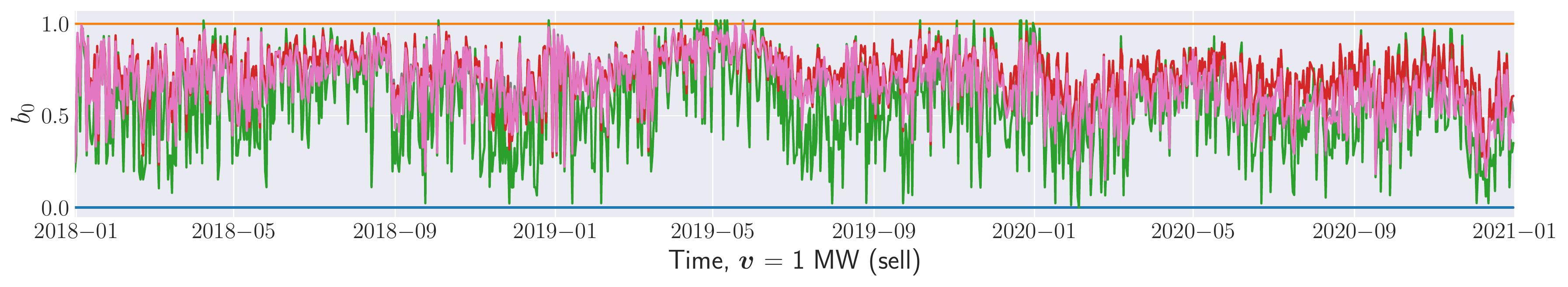}}\vspace{-0.5em}
	\subfloat{\includegraphics[width = 1\linewidth]{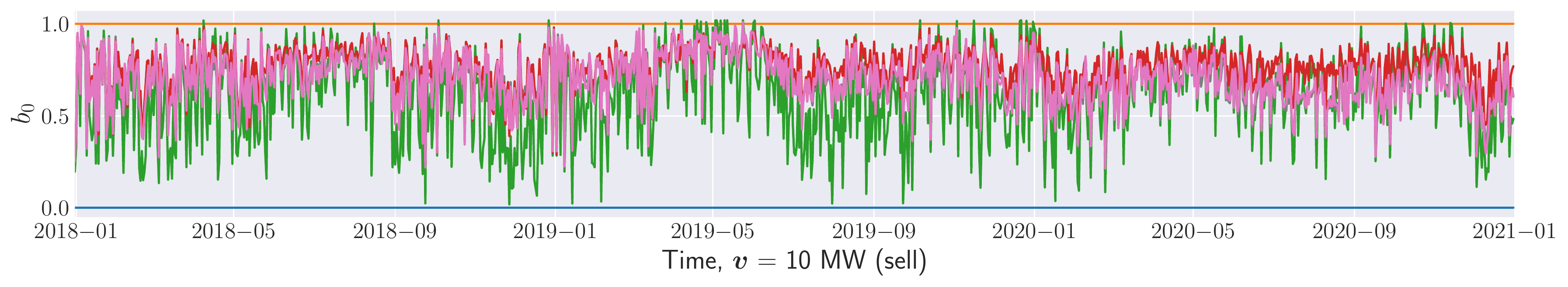}}\vspace{-0.5em}
	\subfloat{\includegraphics[width = 1\linewidth]{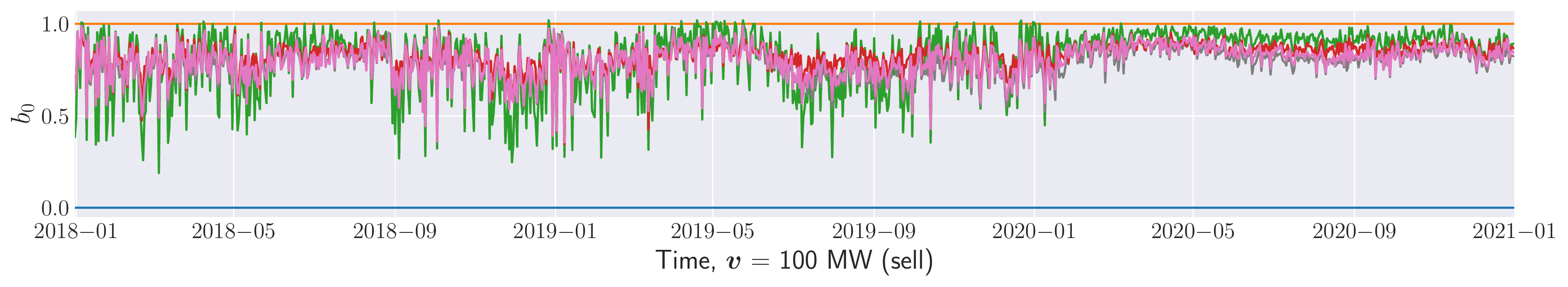}}\vspace{-0.5em}
	\subfloat{\includegraphics[width = 1\linewidth]{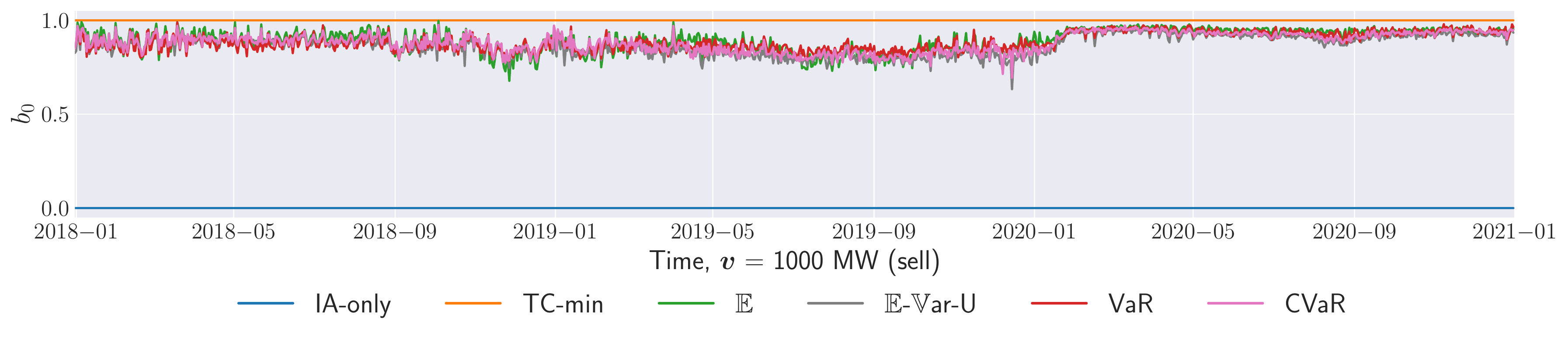}}	
	\caption{The average daily weight of $b_0$ in relation to the whole $\bsb$ strategy for remaining portfolios $\bsv$ in the setting of a new market player. The \textbf{naive}-based strategies are excluded for better clarity}
	\label{fig:da_weights_newbids3}
\end{figure}

\begin{figure}[h!]
	\centering
	\subfloat{\includegraphics[width = 1\linewidth]{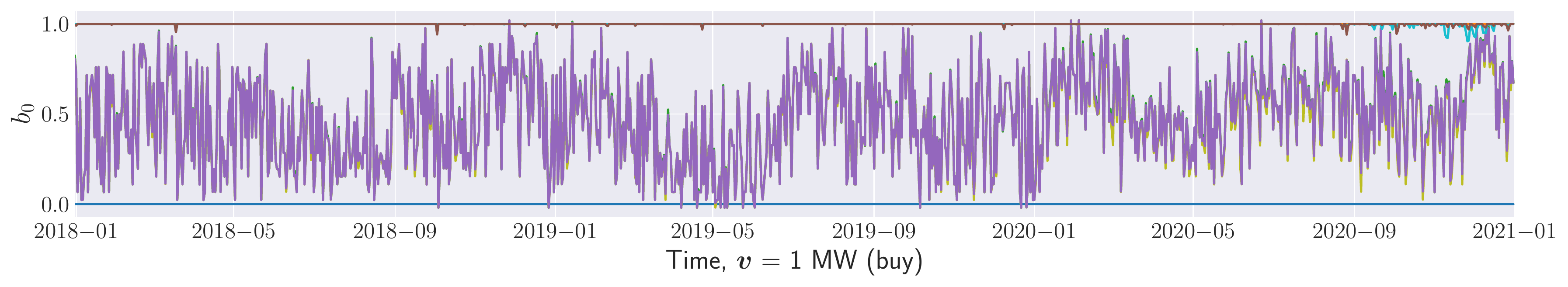}}\vspace{-0.5em}
	\subfloat{\includegraphics[width = 1\linewidth]{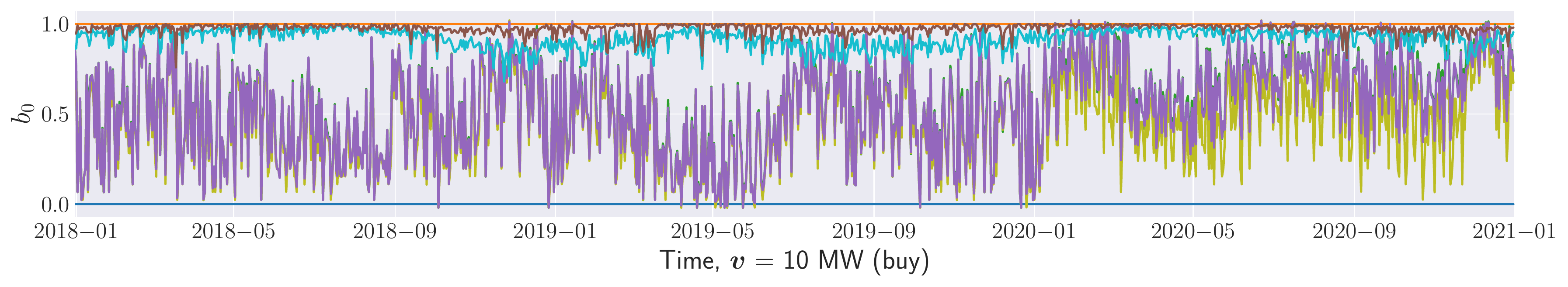}}\vspace{-0.5em}
	\subfloat{\includegraphics[width = 1\linewidth]{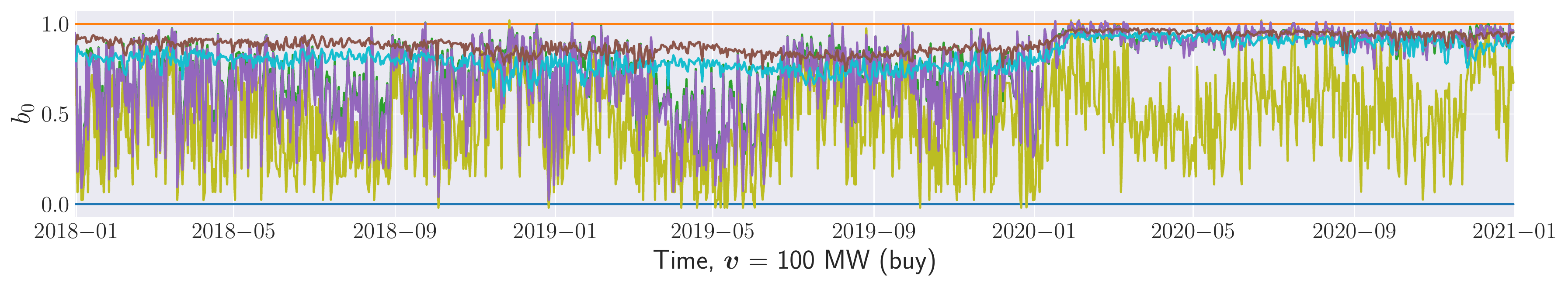}}\vspace{-0.5em}
	\subfloat{\includegraphics[width = 1\linewidth]{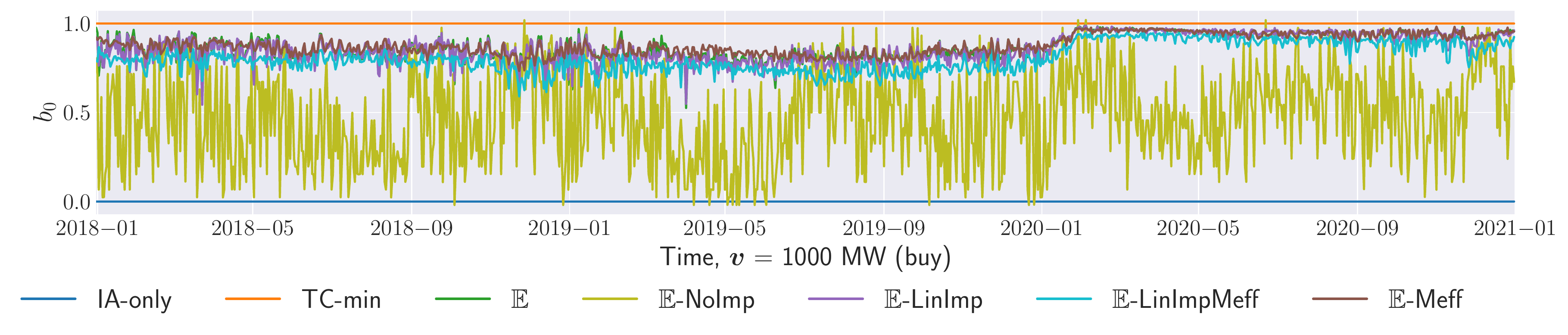}}\\
	\subfloat{\includegraphics[width = 1\linewidth]{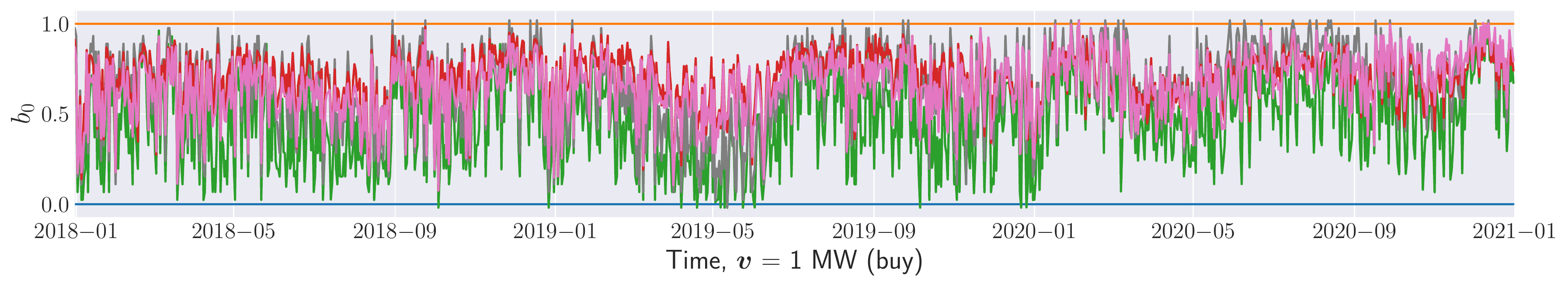}}\vspace{-0.5em}
	\subfloat{\includegraphics[width = 1\linewidth]{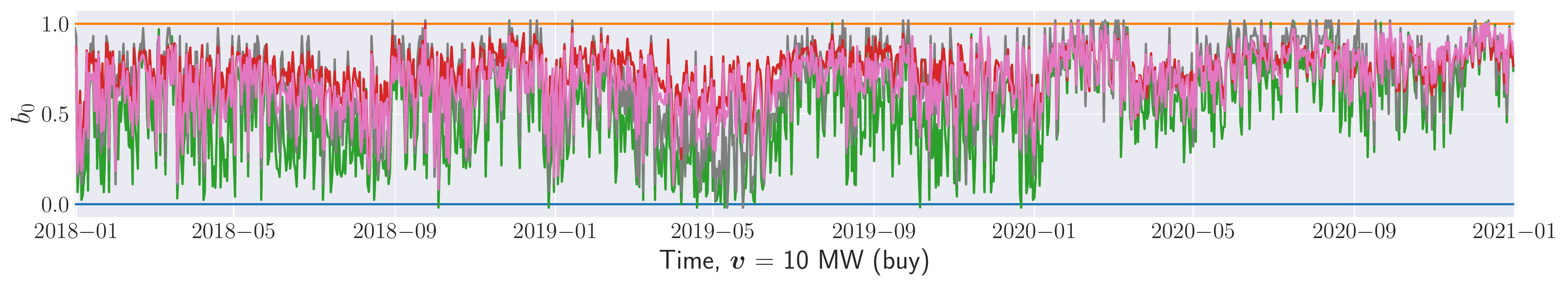}}\vspace{-0.5em}
	\subfloat{\includegraphics[width = 1\linewidth]{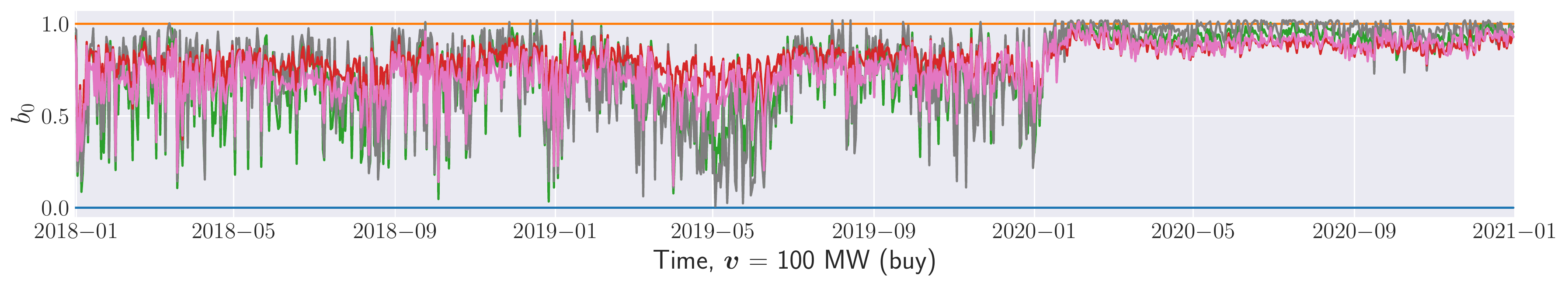}}\vspace{-0.5em}
	\subfloat{\includegraphics[width = 1\linewidth]{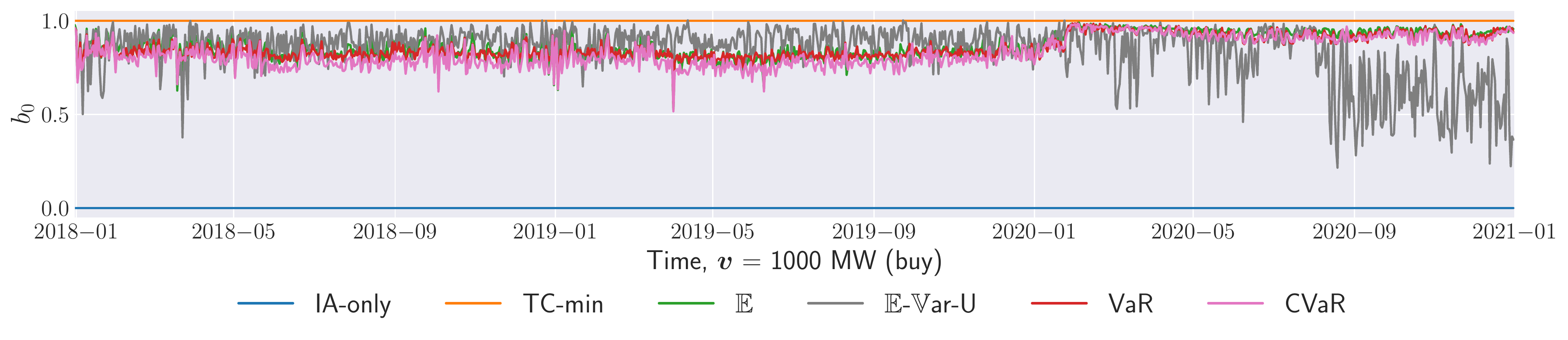}}	
	\caption{The average daily weight of $b_0$ in relation to the whole $\bsb$ strategy for remaining portfolios $\bsv$ in the setting of a new market player. The \textbf{naive}-based strategies are excluded for better clarity}
	\label{fig:da_weights_newbids4}
\end{figure}

\end{document}